# CONVERSATIONAL CONCURRENCY

TONY GARNOCK-JONES

Submitted in partial fulfillment of the requirements
for the degree of Doctor of Philosophy

College of Computer and Information Science
Northeastern University

2017

Northeastern University
College of Computer and Information Science

**PhD Thesis Approval**

*THESIS TITLE:* Conversational Concurrency

*AUTHOR:* Tony Garnock-Jones

*Ph.D. Thesis Approved to complete all degree requirements for the Ph.D. Degree in Computer Science.*

| | |
|---|---|
| M. Felleisen | 8 Dec 2017 |
| *Thesis Advisor* | *Date* |
| [signature] | 8 Dec 2017 |
| *Thesis Reader* | *Date* |
| Mitchell Wand | 12/8/2017 |
| *Thesis Reader* | *Date* |
| [signature] | 12/8/17 |
| *Thesis Reader* | *Date* |
| | |
| *Thesis Reader* | *Date* |

*GRADUATE SCHOOL APPROVAL:*

| | |
|---|---|
| [signature] | 12/12/17 |
| *Director, Graduate School* | *Date* |

*COPY RECEIVED IN GRADUATE SCHOOL OFFICE:*

| | |
|---|---|
| Megan E Barry | 12/12/17 |
| *Recipient's Signature* | *Date* |

*Distribution: Once completed, this form should be scanned and attached to the front of the electronic dissertation document (page 1). An electronic version of the document can then be uploaded to the Northeastern University-UMI website.*

# *Abstract*

Concurrent computations resemble conversations. In a conversation, participants direct utterances at others and, as the conversation evolves, exploit the known common context to advance the conversation. Similarly, collaborating software components share knowledge with each other in order to make progress as a group towards a common goal.

This dissertation studies concurrency from the perspective of cooperative knowledge-sharing, taking the conversational exchange of knowledge as a central concern in the design of concurrent programming languages. In doing so, it makes five contributions:

1. It develops the idea of a common *dataspace* as a medium for knowledge exchange among concurrent components, enabling a new approach to concurrent programming.

   While dataspaces loosely resemble both "fact spaces" from the world of Linda-style languages and Erlang's collaborative model, they significantly differ in many details.

2. It offers the first crisp formulation of cooperative, conversational knowledge-exchange as a mathematical model.

3. It describes two faithful implementations of the model for two quite different languages.

4. It proposes a completely novel suite of linguistic constructs for organizing the internal structure of individual actors in a conversational setting.

   The combination of dataspaces with these constructs is dubbed Syndicate.

5. It presents and analyzes evidence suggesting that the proposed techniques and constructs combine to simplify concurrent programming.

The dataspace concept stands alone in its focus on representation and manipulation of conversational frames and conversational state and in its integral use of explicit epistemic knowledge. The design is particularly suited to integration of general-purpose I/O with otherwise-functional languages, but also applies to actor-like settings more generally.

# *Acknowledgments*

> Networking is interprocess communication.
>
> —Robert Metcalfe, 1972, quoted in Day (2008)

I am deeply grateful to the many, many people who have supported, taught, and encouraged me over the past seven years.

My heartfelt thanks to my advisor, Matthias Felleisen. Matthias, it has been an absolute privilege to be your student. Without your patience, insight and willingness to let me get the crazy ideas out of my system, this work would not have been possible. My gratitude also to the members of my thesis committee, Mitch Wand, Sam Tobin-Hochstadt, and Jan Vitek. Sam in particular helped me convince Matthias that there might be something worth looking into in this concurrency business. I would also like to thank Olin Shivers for providing early guidance during my studies.

Thanks also to my friends and colleagues from the Programming Research Lab, including Claire Alvis, Leif Andersen, William Bowman, Dan Brown, Sam Caldwell, Stephen Chang, Ben Chung, Andrew Cobb, Ryan Culpepper, Christos Dimoulas, Carl Eastlund, Spencer Florence, Oli Flückiger, Dee Glaze, Ben Greenman, Brian LaChance, Ben Lerner, Paley Li, Max New, Jamie Perconti, Gabriel Scherer, Jonathan Schuster, Justin Slepak, Vincent St-Amour, Paul Stansifer, Stevie Strickland, Asumu Takikawa, Jesse Tov, and Aaron Turon. Sam Caldwell deserves particular thanks for being the second ever Syndicate programmer and for being willing to pick up the ideas of Syndicate and run with them.

Many thanks to Alex Warth and Yoshiki Ohshima, who invited me to intern at CDG Labs with a wonderful research group during summer and fall 2014, and to John Day, whose book helped motivate me to return to academia. Thanks also to the DARPA CRASH program and to several NSF grants that helped to fund my PhD research.

I wouldn't have made it here without crucial interventions over the past few decades from a wide range of people. Nigel Bree hooked me on Scheme in the early '90s, igniting a lifelong interest in functional programming. A decade later, while working at a company called LShift, my education as a computer scientist truly began when Matthias Radestock and Greg Meredith introduced me to the $\pi$-calculus and many related ideas. Andy Wilson broadened my mind with music, philosophy and political ideas both new and old. A few years later, Alexis Richardson showed me the depth and importance of distributed systems as we developed new ideas about messaging middleware and programming languages while working together on RabbitMQ. My colleagues at LShift were instrumental to the development of the ideas that ultimately led to this work. My thanks to all of you. In particular, I owe an enormous debt of gratitude to my good friend Michael Bridgen. Michael, the discussions we have had over the years contributed to this work in so many ways that I'm still figuring some of them out.

Life in Boston wouldn't have been the same without the friendship and hospitality of Scott and Megs Stevens. Thank you both.

Finally, I'm grateful to my family. The depth of my feeling prevents me from adequately conveying quite how grateful I am. Thank you Mum, Dad, Karly, Casey, Sabrina, and Blyss. Each of you has made an essential contribution to the person I've become, and I love you all. Thank you to the Yates family and to Warren, Holden and Felix for much-needed distraction and moments of zen in the midst of the write-up. But most of all, thank you to Donna. You're my person.

Tony Garnock-Jones
Boston, Massachusetts
December 2017

# *Contents*

# *List of Figures*

# Part I

# BACKGROUND

# 1

# *Introduction*

Concurrency and its constant companions, communication and coordination, are ubiquitous in computing. From warehouse-sized datacenters through multi-processor operating systems to interactive or multi-threaded programs, coroutines, and even the humble function, every computation exists in some context and must exchange information with that context in a prescribed manner at a prescribed time. Functions receive inputs from and transmit outputs to their callers; impure functions may access or update a mutable store; threads update shared memory and transfer control via locks; and network services send and receive messages to and from their peers.

Each of these acts of communication contributes to a shared understanding of the relevant knowledge required to undertake some task common to the involved parties. That is, the purpose of communication is to share state: to *replicate* information from peer to peer. After all, a communication that does not affect a receiver's view of the world literally has no effect. Put differently, each task shared by a group of components entails various acts of communication in the frame of an overall conversation, each of which conveys knowledge to components that need it. Each act of communication contributes to the overall *conversational state* involved in the shared task. Some of this conversational state relates to *what* must be or has been done; some relates to *when* it must be done. Traditionally, the "what" corresponds closely to "communication," and the "when" to "coordination."

The central challenge in programming for a concurrent world is the *unpredictability* of a component's interactions with its context. Pure, total functions are the only computations whose interactions are completely predictable: a single value in leads to a terminating computation which yields a single value out. Introduction of effects such as non-termination, exceptions, or mutability makes function output unpredictable. Broadening our perspective to coroutines makes even the *inputs* to a component unpredictable: an input may arrive at an unexpected time or may not arrive at all. Threads may observe shared memory in an unexpected state, or may manipulate locks in an unexpected order. Networks may corrupt, discard, duplicate, or reorder messages; network services may delegate tasks to third parties, transmit out-of-date information, or simply never reply to a request.

This seeming chaos is intrinsic: unpredictability is a defining characteristic of concurrency. To remove the one would eliminate the other. However, we shall not declare defeat. If we cannot eliminate harmful unpredictability, we may try to minimize it on one hand, and to cope with it on the other. We may seek a model of computation that helps programmers eliminate some forms of unpredictability and understand those that remain.

To this end, I have developed new programming language design, Syndicate, which rests on a new model of concurrent computation, the *dataspace model*. In this dissertation I will defend the thesis that

> Syndicate provides a new, effective, realizable linguistic mechanism for sharing state in a concurrent setting.

This claim must be broken down before it can be understood.

**mechanism for sharing state.** The dataspace model is, at heart, a mechanism for sharing state among neighboring concurrent components. The design focuses on mechanisms for sharing state because effective mechanisms for communication and coordination follow as special cases. Chapter 2 motivates the Syndicate design, and chapter 3 surveys a number of existing linguistic approaches to coordination and communication, outlining the multi-dimensional design space which results. Chapter 4 then presents a vocabulary for and formal model of dataspaces along with basic correctness theorems.

**linguistic mechanism.** The dataspace model, taken alone, explains communication and coordination among components but does not offer the programmer any assistance in structuring the internals of components. The full Syndicate design presents the primitives of the dataspace model to the programmer by way of new language constructs. These constructs extend the underlying programming language used to write a component, bridging between the language's own computational model and the style of interaction offered by the dataspace model. Chapter 5 presents these new constructs along with an example of their application to a simple programming language.

**realizability.** A design that cannot be implemented is useless; likewise an implementation that cannot be made performant enough to be fit-for-purpose. Chapter 6 examines an example of the integration of the Syndicate design with an existing host language. Chapter 7 discusses the key data structures, algorithms, and implementation techniques that allowed construction of the two Syndicate prototypes, Syndicate/rkt and Syndicate/js.

**effectiveness.** Chapter 8 argues informally for the effectiveness of the programming model by explaining idiomatic Syndicate style through dissection of example protocols and programs. Chapter 9 goes further, arguing that Syndicate eliminates various *patterns* prevalent in concurrent programming, thereby *simplifying* programming tasks. Chapter 10 discusses the performance of the Syndicate design, first in terms of the needs of the programmer and second in terms of the actual measured characteristics of the prototype implementations.

**novelty.** Chapter 11 places Syndicate within the map sketched in chapter 3, showing that it occupies a point in design space not covered by other models of concurrency.

Concurrency is ubiquitous in computing, from the very smallest scales to the very largest. This dissertation presents Syndicate as an approach to concurrency within a *non-distributed* program.[1] However, the design has consequences that may be of use in broader settings such as distributed systems, network architecture, or even operating system design. Chapter 12 concludes the dissertation, sketching possible connections between Syndicate and these areas that may be examined more closely in future work.

1 That is, Syndicate does not yet address the issues of unreliable or congested media, uncontrollable latency or scheduling, or secure separation of powers familiar from Deutsch's "fallacies of distributed computing" (Rotem-Gal-Oz 2006).

# 2

# *Philosophy and Overview of the* Syndicate *Design*

Computer Scientists don't do philosophy.
—Mitch Wand

Taking seriously the idea that concurrency is fundamentally about knowledge-sharing has consequences for programming language design. In this chapter I will explore the ramifications of the idea and outline a mechanism for communication among and coordination of concurrent components that stems directly from it.

Concurrency demands special support from our programming languages. Often specific communication mechanisms like message-passing or shared memory are baked in to a language. Sometimes additional coordination mechanisms such as locks, condition variables, or transactions are provided; in other cases, such as in the actor model, the chosen communication mechanisms double as coordination mechanisms. In some situations, the provided coordination mechanisms are even disguised: the event handlers of browser-based JavaScript programs are carefully sequenced by the system, showing that even sequential programming languages exhibit internal concurrency and must face issues arising from the unpredictability of the outside world.[1]

Let us step back from consideration of specific conversational mechanisms, and take a broader viewpoint. Seen from a distance, all these approaches to communication and coordination appear to be means to an end: namely, they are means by which *relevant knowledge* is shared among *cooperating components*. Knowledge-sharing is then simply the means by which they cooperate in performing their common task.

Focusing on knowledge-sharing allows us to ask high-level questions that are unavailable to us when we consider specific communication and coordination mechanisms alone:

K1 What does it mean to *cooperate* by sharing knowledge?

K2 What general sorts of facts do components know?

K3 What do they need to know to do their jobs?

It also allows us to frame the inherent unpredictability of concurrent systems in terms of knowledge. Unpredictability arises in many different ways. Components may crash, or suffer errors or exceptions during their operation. They may freeze, deadlock, enter unintentional infinite

1 This example reinforces the useful distinction of *concurrency* from *parallelism*: the former results when multiple independent ongoing activities exist; the latter, when more than one can be pursued *simultaneously*.

loops, or merely take an unreasonable length of time to reply. Their actions may interleave arbitrarily. New components may join and existing components may leave the group without warning. Connections to the outside world may fail. Demand for shared resources may wax and wane. Considering all these issues in terms of knowledge-sharing allows us to ask:

K4 Which forms of knowledge-sharing are robust in the face of such unpredictability?

K5 What knowledge helps the programmer mitigate such unpredictability?

Beyond the unpredictability of the operation of a concurrent system, the task the system is intended to perform can itself change in unpredictable ways. Unforeseen program change requests may arrive. New features may be invented, demanding new components, new knowledge, and new connections and relationships between existing components. Existing relationships between components may be altered. Again, our knowledge-sharing perspective allows us to raise the question:

K6 Which forms of knowledge-sharing are robust to and help mitigate the impact of changes in the goals of a program?

In the remainder of this chapter, I will examine these questions generally and will outline SYNDICATE's position on them in particular, concluding with an overview of the SYNDICATE approach to concurrency. We will revisit these questions in chapter 3 when we make a detailed examination of and comparison with other forms of knowledge-sharing embodied in various programming languages and systems.

## 2.1 COOPERATING BY SHARING KNOWLEDGE

We have identified conversation among concurrent components abstractly as a mechanism for knowledge-sharing, which itself is the means by which components work together on a common task. However, taken alone, the mere exchange of knowledge is insufficient to judge whether an interaction is cooperative, neutral, or perhaps even malicious. As programmers, we will frequently wish to orchestrate multiple components, all of which are under our control, to cooperate with each other. From time to time, we must equip our programs with the means for responding to non-cooperative, possibly-malicious interactions with components that are *not* under our control. To achieve these goals, an understanding of what it is to be cooperative is required.

H. Paul Grice, a philosopher of language, proposed the *cooperative principle* of conversation in order to make sense of the meanings people derive from utterances they hear:

COOPERATIVE PRINCIPLE (CP). Make your conversational contribution such as is required, at the stage at which it occurs, by the accepted purpose or direction of the talk exchange in which you are engaged. (Grice 1975)

He further proposed four *conversational maxims*[2] as corollaries to the CP, presented in figure 1. It is important to note the character of these maxims:

2 As opposed to other kinds of maxims, "aesthetic, social, or moral in nature" (Grice 1975, p. 47)

QUANTITY.

1. Make your contribution as informative as required (for the current purposes of the exchange).
2. Do not make your contribution more informative than is required.

QUALITY. Try to make your contribution one that is true.

1. Do not say what you believe to be false.
2. Do not say that for which you lack adequate evidence.

RELATION. Be relevant.

MANNER. Be perspicuous.

1. Avoid obscurity of expression.
2. Avoid ambiguity.
3. Be brief (avoid unnecessary prolixity).
4. Be orderly.

Figure 1: Grice's Conversational Maxims (Grice 1975)

> They are not sociological generalizations about speech, nor they are moral prescriptions or proscriptions on what to say or communicate. Although Grice presented them in the form of guidelines for how to communicate successfully, I think they are better construed as presumptions about utterances, presumptions that we as listeners rely on and as speakers exploit. (Bach 2005)

Grice's principle and maxims can help us tackle question K1 in two ways. First, they can be read directly as constructive advice for designing conversational protocols for cooperative interchange of information. Second, they can attune us to particular families of design mistakes in such protocols that result from cases in which these "presumptions" are invalid. This can in turn help us come up with guidelines for protocol design that help us avoid such mistakes. Thus, we may use these maxims to judge a given protocol among concurrent components, asking ourselves whether each communication that a component makes lives up to the demands of each maxim.

Grice introduces various ways of failing to fulfill a maxim, and their consequences:

1. Unostentatious violation of a maxim, which can mislead peers.
2. Explicit opting-out of participation in a maxim or even the Cooperative Principle in general, making plain a deliberate lack of cooperation.
3. Conflict between maxims: for example, there may be tension between speaking some necessary (Quantity(1)) truth (Quality(1)), and a lack of evidence in support of it (Quality(2)), which may lead to shaky conclusions down the line.

4. *Flouting* of a maxim: blatant, obviously deliberate violation of a conversational maxim, which "*exploits*" the maxim, with the intent to force a hearer out of the usual frame of the conversation and into an analysis of some higher-order conversational context.

Many, but not all, of these can be connected to analogous features of computer communication protocols. In this dissertation, I am primarily assuming a setting involving components that deliberately *aim* to cooperate. We will not dwell on deliberate violation of conversational maxims. However, we will from time to time see that consideration of *accidental* violation of conversational maxims is relevant to the design and analysis of computer protocols. For example, Grice writes that

> [the] second maxim [of Quantity] is disputable; it might be said that to be overinformative is not a transgression of the [Cooperative Principle] but merely a waste of time. However, it might be answered that such overinformativeness may be confusing in that it is liable to raise side issues; and there may also be an indirect effect, in that the hearers may be misled as a result of thinking that there is some particular *point* in the provision of the excess of information. (Grice 1975)

This directly connects to (perhaps accidental) excessive bandwidth use ("waste of time") as well as programmer errors arising from exactly the misunderstanding that Grice describes.

It may seem surprising to bring ideas from philosophy of language to bear in the setting of cooperating concurrent computerized components. However, Grice himself makes the connection between his specific conversational maxims and "their analogues in the sphere of transactions that are not talk exchanges," drawing on examples of shared tasks such as cooking and car repair, so it does not seem out of place to apply them to the design and analysis of our *conversational* computer protocols. This is particularly the case in light of Grice's ambition to explain the Cooperative Principle as "something that it is *reasonable* for us to follow, that we *should not* abandon." (Grice 1975, p. 48; emphasis in original)

The CP makes mention of the "purpose or direction" of a given conversation. We may view the fulfillment of the task shared by the group of collaborating components as the purpose of the conversation. Each individual component in the group has its own role to play and, therefore, its own "personal" goals in working toward successful completion of the shared task. Kitcher (1990), writing in the context of the social structure of scientific collaboration, introduces the notions of *personal* and *impersonal epistemic intention*.[3] We may adapt these ideas to our setting, explicitly drawing out the notion of a role within a conversational protocol. A cooperative component "wishes" for the group as a whole to succeed: this is its "impersonal" epistemic intention. It also has goals for itself, "personal" epistemic intentions, namely to successfully perform its roles within the group.

Finally, the CP is a specific example of the general idea of *epistemic reasoning*, logical reasoning incorporating knowledge and beliefs about one's own knowledge and beliefs, and about the knowledge and beliefs of other parties (Fagin et al. 2004; Hendricks and Symons 2015; van Ditmarsch, van der Hoek and Kooi 2017). However, epistemic reasoning has further applications in the design of conversational protocols among concurrent components, which brings us to our next topic.

3 See also Dunn (2017) who places Kitcher's work in a wider context.

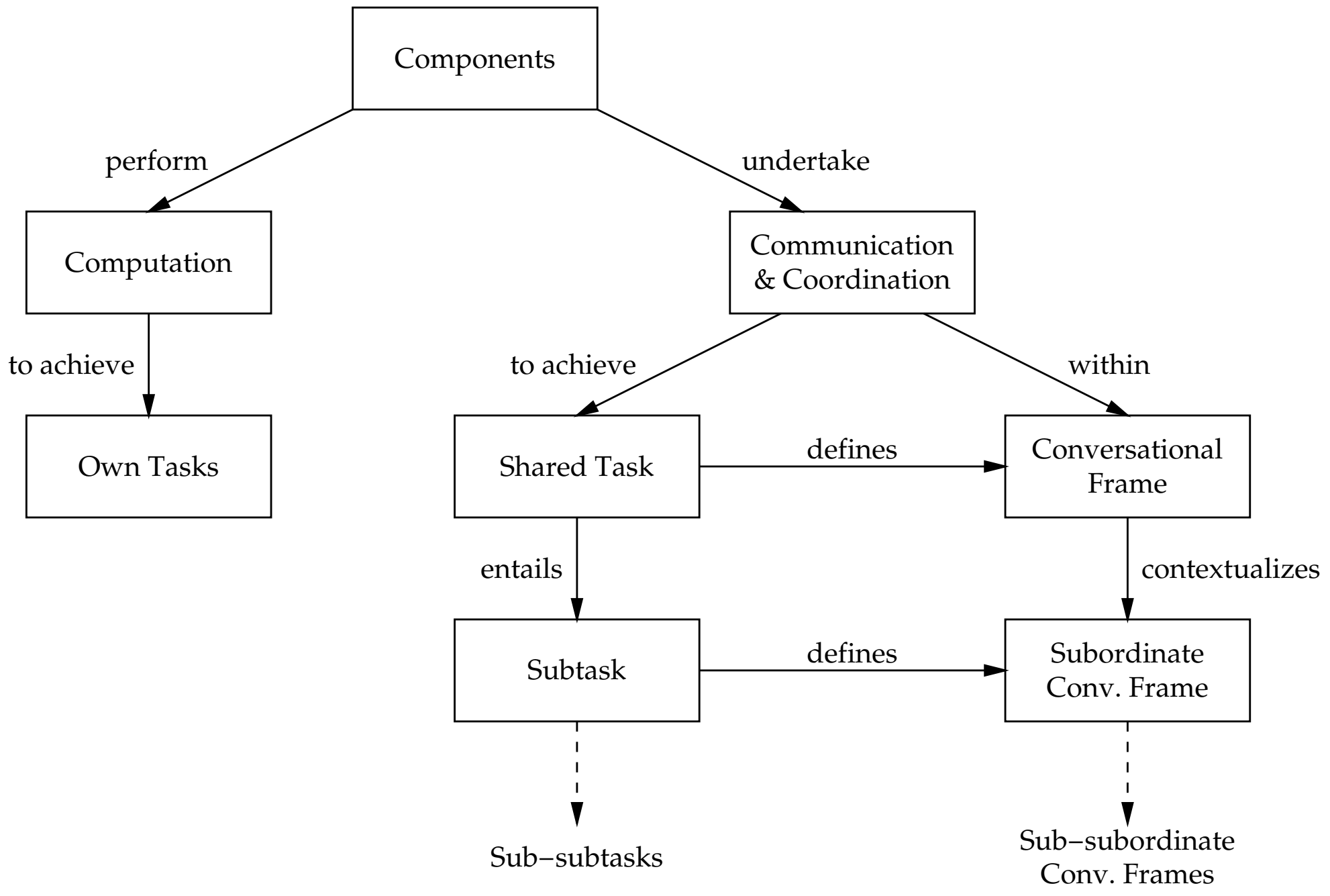

Figure 2: Components, tasks, and conversational structure

## 2.2 KNOWLEDGE TYPES AND KNOWLEDGE FLOW

The conversational state that accumulates as part of a collaboration among components can be thought of as a collection of facts. First, there are those facts that define the *frame* of a conversation. These are exactly the facts that identify the task at hand; we label them "framing knowledge", and taken together, they are the "conversational frame" for the conversation whose purpose is completion of a particular shared task. Just as tasks can be broken down into more finely-focused subtasks, so can conversations be broken down into sub-conversations. In these cases, part of the conversational state of an overarching interaction will describe a frame for each sub-conversation, within which corresponding sub-conversational state exists. The knowledge framing a conversation acts as a bridge between it and its wider context, defining its "purpose" in the sense of the CP. Figure 2 schematically depicts these relationships.

Some facts define conversational frames, but *every* shared fact is contextualized *within* some conversational frame. Within a frame, then, some facts will pertain directly to the task at hand. These, we label "domain knowledge". Generally, such facts describe global aspects of the common problem that remain valid as we shift our perspective from participant to participant. Other facts describe the knowledge or beliefs of particular components. These, we label "epistemic knowledge".

For example, as a file transfer progresses, the actual content of the file does not change: it remains a global fact that byte number 300 (say) has value 255, no matter whether the transfer has reached that position or not. The content of the file is thus "domain knowledge". However, as the transfer proceeds and acknowledgements of receipt stream from the recipient to the transmitter, the transmitter's beliefs about the receiver's knowledge change. Each successive acknowledgement leads the transmitter to believe that the receiver has learned a little more of the file's content. Information on the progress of the transfer is thus "epistemic knowledge".[4]

If domain knowledge is "what is true in the world", and epistemic knowledge is "who knows what", the third piece of the puzzle is "who *needs* to know what" in order to effectively make a contribution to the shared task at hand. We will use the term "interests" as a name for those facts that describe knowledge that a component needs to learn. Knowledge of the various interests in a group allows collaborators to plan their communication acts according to the needs of individual components and the group as a whole. In conversations among people, interests are expressed as *questions*; in a computational setting, they are conveyed by *requests*, *queries*, or *subscriptions*.[5]

The interests of components in a concurrent system thus direct the flow of knowledge within the system. The interests of a group may be constant, or may vary with time.

When interest is fixed, remaining the same for a certain class of shared task, the programmer can plan paths for communication up front. For example, in the context of a single TCP connection, the interests of the two parties involved are always the same: each peer wishes to learn what the other has to say. As a consequence, libraries implementing TCP can bake in the assumption that clients will wish to access received data. As another example, a programmer charged with implementing a request counter in a web server may choose to use a simple global integer variable, safe in the knowledge that the only possible item of interest is the current value of the counter.

A changing, dynamic set of interests, however, demands development of a vocabulary for communicating changes in interest during a conversation. For example, the query language of a SQL database is just such a vocabulary. The server's initial interest is in *what the client is interested in*, and is static, but the client's own interests vary with each request, and must be conveyed anew in the context of each separate interaction. Knowledge *about* dynamically-varying interests allows a group of collaborating components to change its interaction patterns on the fly.[6]

With this ontology in hand, we may answer questions K2 and K3. Each task is delimited by a conversational frame. Within that frame, components share knowledge related to the domain of the task at hand, and knowledge related to the knowledge, beliefs, needs, and interests of the various participants in the collaborative group. Conversations are recursively structured by shared knowledge of (sub-)conversational frames, defined in terms of any or all of the types of knowledge we have discussed. Some conversations take place at different levels within a larger frame, bridging between tasks and their subtasks. Components are frequently engaged in multiple tasks, and thus often participate in multiple conversations at once. The knowledge

4 Is the receiver telling the truth, or has it been discarding the received data, falsely acknowledging safe receipt of it? This is where the Cooperative Principle comes in. Acting as if the transmitter's beliefs are in fact knowledge trusts that the receiver is properly cooperating.

5 The fact of a "need to know" is also perhaps a form of epistemic knowledge, as it expresses a claim about the knowledge of a particular component: namely, that it does *not* know some specific thing or things.

6 This perspective lines up very well with the Cooperative Principle, in that an expressed interest—a question or query—strongly suggests an immediately relevant, appropriate, required conversational contribution.

a component needs to do its job is provided to it when it is created, or later supplied to it in response to its interests.

## 2.3 UNPREDICTABILITY AT RUN-TIME

A full answer to question K4 must wait until the survey of communication and coordination mechanisms of chapter 3. However, this dissertation will show that at least one form of knowledge-sharing, the SYNDICATE design, encourages robust handling of many kinds of concurrency-related unpredictability.

The epistemological approach we have taken to questions K1–K3 suggests some initial steps toward an answer to question K5. In order for a program to be robust in the face of unpredictable events, it must first be able to detect these events, and second be able to muster an appropriate response to them. Certain kinds of events can be reliably detected and signaled, such as component crashes and exceptions, and arrivals and departures of components in the group. Others cannot easily be detected reliably, such as nontermination, excessive slowness, or certain kinds of deadlock and datalock. Half-measures such as use of timeouts must suffice for the latter sort. Still other kinds of unpredictability such as memory races or message races may be explicitly worked around via careful protocol design, perhaps including information tracking causality or provenance of a piece of knowledge or arranging for extra coordination to serialize certain sensitive operations.

No matter the source of the unpredictability, once detected it must be signaled to interested parties. Our epistemic, knowledge-sharing focus allows us to treat the *facts* of an unpredictable event as knowledge within the system. Often, such a fact will have an epistemic consequence. For example, learning that a component has crashed will allow us to discount any partial results we may have learned from it, and to discard any records we may have been keeping of the state of the failed component itself. Generally speaking, an epistemological perspective can help each component untangle intact from damaged or potentially untrustworthy pieces of knowledge. Having classified its records into "salvageable" and "unrecoverable", it may discard items as necessary and engage with the remaining portion of the group in actions to repair the damage and continue toward the ultimate goal.

One particular strategy is to retry a failed action. Consideration of the *roles* involved in a shared task can help determine the scope of the action to retry. For example, the idea of *supervision* that features so prominently in Erlang programming (Armstrong 2003) is to restart entire failing components from a specification of their roles. Here, consideration of the epistemic intentions of components can be seen to help the programmer design a system robust to certain forms of unpredictable failure.

## 2.4 UNPREDICTABILITY IN THE DESIGN PROCESS

Programs are seldom "finished". Change must be accommodated at every stage of a program's life cycle, from the earliest phases of development to, in many cases, long after a program is deployed. When concurrency is involved, such change often involves emendations to protocol definitions and shifts in the roles and relationships within a group of components. Just as with question K4, a full examination of question K6 must wait for chapter 3. However, approaching the question in the abstract, we may identify a few desirable characteristics of linguistic support for concurrent programming.

First, debugging of concurrent programs can be extremely difficult. A language should have tools for helping programmers gain insight into the intricacies of the interactions among each program's components. Such tools depend on information gleaned from the knowledge-sharing mechanism of the language. As such, a mechanism that generates trace information that matches the mental model of the programmer is desirable.

Second, changes to programs often introduce new interactions among existing components. A knowledge-sharing mechanism should allow for straightforward composition of pieces of program code describing (sub)conversations that a component is to engage in. It should be possible to introduce an existing component to a new conversation without heavy revision of the code implementing the conversations the component already supports.

Finally, service programs must often run for long periods of time without interruption. In cases where new features or important bug-fixes must be introduced, it is desirable to be able to replace or upgrade program components without interrupting service availability. Similar concerns arise even for user-facing graphical applications, where upgrades to program code must preserve various aspects of program state and configuration across the change.

## 2.5 SYNDICATE'S APPROACH TO CONCURRENCY

SYNDICATE places knowledge front and center in its design in the form of *assertions*. An assertion is a representation of an item of knowledge that one component wishes to communicate to another. Assertions may represent framing knowledge, domain knowledge, and epistemic knowledge, as a component sees fit. Each component in a group exists within a *dataspace* which both keeps track of the group's current set of assertions and schedules execution of its constituent components. Components add and remove assertions from the dataspace freely, and the dataspace ensures that components are kept informed of *relevant* assertions according to their *declared interests*.

In order to perform this task, SYNDICATE dataspaces place just one constraint on the interpretation of assertions: there must exist, in a dataspace implementation, a distinct piece of syntax for constructing assertions that will mean *interest in some other assertion*. For example, if "the color of the boat is blue" is an assertion, then so is "there exists some interest in the color of the boat being blue". A component that asserts interest in a set of other assertions will be kept informed as members of that set appear and disappear in the dataspace through the actions of the component or its peers.

SYNDICATE makes extensive use of *wildcards* for generating large—in fact, often infinite—sets of assertions. For example, "interest in the color of the boat being anything at all" is a valid and useful set of assertions, generated from a piece of syntax with a wildcard marker in the position where a specific color would usually reside. Concretely, we might write

*interestExists*(*color*(*boat*, $\star$)), which generates the set of assertions *interestExists*(*color*(*boat*, $x$)), with $x$ ranging over the entire universe of assertions.[7]

The design of the dataspace model thus far seems similar to the tuplespace model (Gelernter 1985; Gelernter and Carriero 1992; Carriero et al. 1994). There are two vital distinctions. The first is that tuples in the tuplespace model are "generative", taking on independent existence once placed in the shared space, whereas assertions in the dataspace model are not. Assertions in a dataspace never outlive the component that is currently asserting them;[8] when a component terminates, *all its assertions are retracted* from the shared space. This occurs whether termination was normal or the result of a crash or an exception. The second key difference is that multiple *copies* of a particular tuple may exist in a tuplespace, while redundant assertions in a dataspace cannot be distinguished by observers. If two components separately place an assertion $x$ into their common dataspace, a peer that has previously asserted interest in $x$ is informed merely that $x$ has been asserted, not *how many times* it has been asserted. If one redundant assertion of $x$ is subsequently withdrawn, the observer will not be notified; only when every assertion of $x$ is retracted is the observer notified that $x$ is no longer present in the dataspace. Observers are shown only a *set* view on an underlying *bag* of assertions. In other words, producing a tuple is non-idempotent, while making an assertion is idempotent.

Even more closely related is the *fact space model* (Mostinckx et al. 2007; Mostinckx, Lombide Carreton and De Meuter 2008), an approach to middleware for connecting programs in mobile networks. The model is based on an underlying tuplespace, interpreting tuples as *logical facts* by working around the generativity and poor fault-tolerance properties of the tuplespace mechanism in two ways. First, tuples are recorded alongside the identity of the program that produced them. This provenance information allows tuples to be removed when their producer crashes or is otherwise disconnected from the network. Second, tuples can be interpreted in an idempotent way by programs. This allows programs to ignore redundant tuples, recovering a set view from the bag of tuples they observe. While the motivations and foundations of the two works differ, in many ways the dataspace and fact space models address similar concerns. Conceptually, the dataspace model can be viewed as an adaptation and integration of the fact space model into a programming language setting. The fact space model focuses on scaling *up* to distributed systems, while our focus is instead on a mechanism that scales *down* to concurrency in the small. In addition, the dataspace model separates itself from the fact space model in its explicit, central epistemic constructions and its emphasis on conversational frames.

The dataspace model maintains a strict isolation between components in a dataspace, forcing all interactions between peers through the shared dataspace. Components access and update the dataspace solely via message passing. Shared memory in the sense of multi-threaded models is ruled out. In this way, the dataspace model seems similar to the actor model (Hewitt, Bishop and Steiger 1973; Agha 1986; Agha et al. 1997; De Koster et al. 2016). The core distinction between the models is that components in the dataspace model communicate *indirectly* by making and retracting assertions in the shared store which are observed by other components, while actors in the actor model communicate *directly* by exchange of messages which are

7 We defer selection of a specific universe of assertions to chapter 4.
8 Assertions thus have an additional, intrinsic epistemic character: the existence of an assertion implies the existence of an asserter.

addressed to other actors. Assertions in a dataspace are routed according to the intersection between sets of assertions and sets of asserted *interests* in assertions, while messages in the actor model are each routed to an explicitly-named target actor.

The similarities between the dataspace model and the actor, tuplespace, and fact space models are strong enough that we borrow terminology from them to describe concepts in SYNDICATE. Specifically, we borrow the term "actor" to denote a SYNDICATE component. What the actor model calls a "configuration" we fold into our idea of a "dataspace", a term which also denotes the shared knowledge store common to a group of actors. The term "dataspace" itself was chosen to highlight this latter denotation, making a connection to fact spaces and tuplespaces.

We will touch again on the similarities and differences among these models in chapter 3, examining details in chapter 11. In the remainder of this subsection, let us consider SYNDICATE's relationship to questions K1–K6.

COOPERATION, KNOWLEDGE & CONVERSATION. The SYNDICATE design takes questions K1–K3 to heart, placing them at the core of its choice of sharing mechanism and the concomitant approach to protocol design. Actors exchange knowledge encoded as assertions via a shared dataspace. *All* shared state in a SYNDICATE program is represented as assertions: this includes domain knowledge, epistemic knowledge, and frame knowledge. Key to SYNDICATE's functioning is the use of a special form of epistemic knowledge, namely assertions of interest. It is these assertions that drive knowledge *flow* in a program from parties asserting some fact to parties asserting interest in that fact.

Viewing an interaction among actors as a *conversation* and shared assertions as *conversational state* allows programmers to employ the linguistic tools discussed in section 2.1, taking steps toward a pragmatics of computer protocols.[9] SYNDICATE encourages programmers to design conversational protocols directly in terms of roles and to map conversational contributions onto the assertion and retraction of assertions in the shared space. Grice's maxims offer high-level guidance for defining the meaning of each assertion: the maxims of quantity guide the design of the individual records included in each assertion; those of quality and relevance help determine the criteria for when an assertion should be made and when it should be retracted; and those of manner shape a vocabulary of primitive assertions with precisely-defined meanings that *compose* when simultaneously expressed to yield complex derived meanings.

SYNDICATE's assertions of interest determine the movement of knowledge in a system. They define, in effect, the set of facts an actor is "listening" for. All communication mechanisms must have some equivalent feature, used to route information from place to place. Unusually, however, SYNDICATE allows actors to *react to* these assertions of interest, in that assertions of interest are ordinary assertions like any other. Actors may act based on their *knowledge* of the

9 In linguistics, 'pragmatics' means something slightly different to its meaning in the field of programming languages:

> Pragmatics is sometimes characterized as dealing with the effects of context [...] if one collectively refers to all the facts that can vary from utterance to utterance as 'context.' (Korta and Perry 2015)

Mey (2001) defines pragmatics as the subfield of linguistics which "studies the use of language in human communication as determined by the conditions of society". Broadening its scope to include *computer* languages in *software* communication as determined by the conditions of *the system as a whole* takes us into a somewhat speculative area.

way knowledge moves in a system by expressing interest in interest and deducing *implicatures* from the discovered facts. Mey (2001) defines a *conversational implicature* as "something which is implied in conversation, that is, something which is left implicit in actual language use." Grice (1975) makes three statements helpful in pinning down the idea of conversational implicature: 1. "To assume the presence of a conversational implicature, we have to assume that at least the Cooperative Principle is being observed." 2. "Conversational implicata are not part of the meaning of the expressions to the employment of which they attach." This is what distinguishes *implicature* from *implication*. 3. "To calculate a conversational implicature is to calculate what has to be supposed in order to preserve the supposition that the Cooperative Principle is being observed."

For example, imagine an actor $F$ responsible for answering questions about factorials. The assertion $\mathit{fact}(8, 40320)$ means that the factorial of 8 is 40320. If $F$ learns that some peer has asserted $\mathit{interestExists}(\mathit{fact}(8, \star))$, which is to be interpreted as interest in the set of facts describing all *potential* answers to the question "what is the factorial of 8?," it can act on this knowledge to compute a suitable answer and can then assert $\mathit{fact}(8, 40320)$ in response. Once it learns that interest in the factorial of 8 is no longer present in the group, it can retract its own assertion and release the corresponding storage resources.[10] Knowledge of interest in a topic acts as a signal of *demand* for some resource: here, computation (directly) and storage (indirectly). The raw fact of the interest itself has the direct semantic meaning "please convey to me any assertions matching this pattern", but has an indirect, unspoken, pragmatic meaning—an implicature—in our imagined protocol of *"please compute the answer to this question."*[11]

The idea of implicature finds use beyond assertions of interest. For example, the process of deducing an implicature may be used to reconstruct temporarily- or permanently-unavailable information "from context," based on the underlying assumption that the parties involved are following the Cooperative Principle. For example, a message describing successful fulfillment of an order carries an implicature of the existence of the order. A hearer of the message may infer the order's existence on this basis. Similarly, a reply implicates the existence of a request.

Finally, the mechanism that Syndicate provides for conveying assertions from actor to actor via the dataspace allows reasoning about *common knowledge* (Fagin et al. 2004). An actor placing some assertion into the dataspace knows both that all interested peers will automatically learn of the assertion and that each such peer *knows* that all others will learn of the assertion. Providing this guarantee at the language level encourages the use of epistemic reasoning in protocol design while avoiding the risks of implementing the necessary state-management substrate by hand.

Run-time unpredictability. Recall from section 2.3 that robust treatment of unpredictability requires that we must be able to either detect and respond to or forestall the occurrence of the various unpredictable situations inherent to concurrent programming. The dataspace model is the foundation of Syndicate's approach to questions K4 and K5, offering

10 See section 8.7 for more on "procedure calls" and associated resource management.

11 The semantic meaning of the assertion is general across Syndicate programs: interest in an assertion has a fixed meaning *to Syndicate* no matter the domain of the protocol concerned. Implicatures deduced from assertions, however, have meaning only within a specific protocol.

a means for signaling and detection of such events. However, by itself the dataspace model is not enough. The picture is completed with linguistic features for structuring state and control flow within each individual actor. These features allow programmers to concisely express appropriate responses to unexpected events. Finally, SYNDICATE's knowledge-based approach suggests techniques for protocol design which can help avoid certain forms of unpredictability by construction.

The dataspace model constrains the means by which SYNDICATE programs may communicate events within a group, including communication of unpredictable events. All communication must be expressed as changes in the set of assertions in the dataspace. Therefore, an obvious approach is to use assertions to express such ideas as demand for some service, membership of some group, presence in some context, availability of some resource, and so on. Actors expressing interest in such assertions will receive notifications as matching assertions come and go, including when they vanish unexpectedly. Combining this approach with the guarantee that the dataspace removes all assertions of a failing actor from the dataspace yields a form of exception propagation.

For example, consider a protocol where actors assert *userMessage*$(S)$, where $S$ is a message for the user, in order to cause a user interface element to appear on the user's display. The actor responsible for reacting to such assertions, creating and destroying graphical user interface elements, will react to *retraction* of a *userMessage* assertion by removing the associated graphical element. The actor that *asserts* some *userMessage* may deliberately retract it when it is no longer relevant for the user. However, it may also crash. If it does, the dataspace model ensures that its assertions are all retracted. Since this includes the *userMessage* assertion, the actor managing the display learns automatically that its services are no longer required.

Another example may be seen in the *fact* example discussed above. The client asserting *interestExists*(*fact*$(8, \star)$) may "lose interest" before it receives an answer, or of course may crash unexpectedly. From the perspective of actor $F$, the two situations are identical: $F$ is informed of the retraction, concludes that no interest in the factorial of 8 remains, and may then choose to abandon the computation. The request implicated by assertion of *interestExists*(*fact*$(8, \star)$) is effectively *canceled* by retraction, whether this is caused by some active decision on the part of the requestor or is an automatic consequence of its unexpected failure.

The dataspace model thus offers a mechanism for using changes in assertions to express changes in demand for some resource, including both expected and unpredictable changes. Building on this mechanism, SYNDICATE offers linguistic tools for *responding* appropriately to such changes. Assertions describing a demand or a request act as *framing knowledge* and thus delimit a conversation about the specific demand or request concerned. For example, the presence of *userMessage*$(S)$ for each particular $S$ corresponds to one particular "topic of conversation". Likewise, the assertion *interestExists*(*fact*$(8, \star)$) corresponds to a particular "call frame" invoking the services of actor $F$. Actors need tools for describing such conversational frames, associating local conversational state, relevant event handlers, and any conversation-specific assertions that need to be made with each conversational frame created.

Syndicate introduces a language construct called a *facet* for this purpose.[12] Each actor is composed of multiple facets; each facet represents a particular conversation that the actor is engaged in. A facet both scopes and specifies conversational responses to incoming events. Each facet includes private state variables related to the conversation concerned, as well as a bundle of assertions and event handlers. Each event handler has a pattern over assertions associated with it. Each of these patterns is translated into an assertion of interest and combined with the other assertions of the facet to form the overall contribution that the facet makes to the shared dataspace. An analogy to objects in object-oriented languages can be drawn. Like an object, a facet has private state. Its event handlers are akin to an object's methods. Unique to facets, though, is their contribution to the shared state in the dataspace: objects lack a means to automatically convey changes in their local state to interested peers.[13]

Facets may be nested. This can be used to reflect nested sub-conversations via nested facets. When a containing facet is terminated, its contained facets are also terminated, and when an actor has no facets left, the actor itself terminates. Of course, if the actor crashes or is explicitly shut down, all its facets are removed along with it. These termination-related aspects correspond to the idea that a thread of conversation that logically depends on some overarching discussion context clearly becomes irrelevant when the broader discussion is abandoned.

The combination of Syndicate's facets and its assertion-centric approach to state replication yields a mechanism for robustly detecting and responding to certain kinds of unpredictable event. However, not all forms of unpredictability lend themselves to explicit modeling as shared assertions. For these, we require an alternative approach.

Consider unpredictable interleavings of events: for example, UDP datagrams may be reordered arbitrarily by the network. If some datagram $B$ can only be interpreted after datagram $A$ has been interpreted, a datagram receiver $R$ must arrange to buffer packets when they are received out of order, reconstructing an appropriate order to perform its task. The same applies to messages passed between actors in the actor model. The observation that datagram $A$ establishes necessary context for the subsequent message $B$ suggests an approach we may take in Syndicate. If instead of messages we model $A$ and $B$ as assertions, then we may write our program $R$ as follows:

1. Express interest in $A$. Wait until notified that $A$ has been asserted.
2. Express interest in $B$. Wait until notified that $B$ has been asserted.
3. Process $A$ and $B$ as usual.
4. Withdraw the previously-asserted interests in $A$ and $B$.

This program will function correctly no matter whether $A$ is asserted before $B$ or vice versa. The structure of program $R$ reflects the observation that $A$ supplies a *frame* within which $B$ is to be understood by paying attention to $B$ only after having learned $A$. Use of assertions instead of messages allows an interpreter of knowledge to decouple itself from the precise order of events

12 The term "facet" is borrowed from a related use in the language E (Miller 2006, section 6.2), which seems to have taken the name in turn from the language Joule (Agorics, Inc. 1995, chapter 3).

13 Almost all object-oriented languages turn to the *observer pattern* (Gamma et al. 1994) to simulate this ability.

in which knowledge is acquired and shared, concentrating instead on the logical dependency ordering among items of knowledge.[14]

Finally, certain forms of unpredictability cannot be effectively detected or forestalled. For example, no system can distinguish nontermination from mere slowness in practice. In cases such as these, timeouts can be used in SYNDICATE just as in other languages. Modeling time as a protocol involving assertions *laterThan*$(t)$ in the dataspace allows us to smoothly incorporate time with other protocols, treating it as just like any other kind of knowledge about the world.

UNPREDICTABILITY IN THE DESIGN PROCESS. Section 2.4, expanding on question K6, introduced the challenges of debuggability, flexibility, and upgradeability. The dataspace model contributes to debuggability, while facets and hierarchical layering of dataspaces contribute to flexibility. While this dissertation does not offer more than a cursory investigation of upgradeability, the limited exploration of the topic so far completed does suggest that it could be smoothly integrated with the SYNDICATE design.

The dataspace model leads the programmer to reason about the group of collaborating actors as a whole in terms of two kinds of *change*: actions that alter the set of assertions in the dataspace, and events delivered to individual actors as a consequence of such actions. This suggests a natural tracing mechanism. There is nothing to the model other than events and actions, so capturing and displaying the sequence of actions and events not only accurately reflects the operation of a dataspace program, but directly connects to the programmer's mental model as well.

Facets can be seen as atomic units of interaction. They allow decomposition of an actor's relationships and conversations into small, self-contained pieces with well-defined boundaries. As the overall goals of the system change, its actors can be evolved to match by making alterations to groups of related facets in related actors. Altering, adding, or removing one facet while leaving others in an actor alone makes perfect sense.

The dataspace model is *hierarchical*. Each dataspace is modeled as a component in some wider context: as an actor in another, *outer* dataspace. This applies recursively. Certain assertions in the dataspace may be marked with a special constructor that causes them to be *relayed* to the next containing dataspace in the hierarchy, yielding cross-dataspace interaction. Peers in a particular dataspace are given no means of detecting whether their collaborators are simple actors or entire nested dataspaces with rich internal structure. This frees the program designer to decompose an actor into a nested dataspace with multiple contained actors, without affecting other actors in the system at large. This recursive, hierarchical (dis)aggregation of actors also contributes to the flexibility of a SYNDICATE program as time goes by and requirements change.

Code upgrade is a challenging problem for any system. Replacing a unit of code involves the old code marshaling its state into a bundle of information to be delivered to the new code. In other words, the actor involved sends a message to its "future self". Systems like Erlang (Armstrong 2003) incorporate sophisticated language- and library-level mechanisms for supporting such code replacement. SYNDICATE shares with Erlang some common ideas from the actor model. The strong isolation between actors allows each to be treated separately

14 Program R recovers a form of logical monotonicity for the small protocol fragment it is engaging in. An interesting connection can be made here to the CALM principle of Alvaro et al. (2011).

when it comes to code replacement. Logically, each is running an independent codebase. By casting all interactions among actors in terms of a protocol, both Erlang and Syndicate offer the possibility of protocol-mediated upgrades and reboots affecting anything from a small part to the entirety of a running system.[15]

## 2.6 Syndicate design principles

In upcoming chapters, we will see concrete details of the Syndicate design and its implementation and use. Before we leave the high-level perspective on concurrency, however, a few words on general principles of the design of concurrent and distributed systems are in order. I have taken these guidelines as principles to be encouraged in Syndicate and in Syndicate programs. To be clear, they are my own conjectures about what makes good software. I developed them both through my experiences with early Syndicate prototypes and my experiences of development of large-scale commercial software in my career before beginning this project. In some cases, the guidelines influenced the Syndicate design, having an indirect but universal effect on Syndicate programs. In others, they form a set of background assumptions intended to directly shape the protocols designed by Syndicate programmers.

**Exclude implementation concepts from domain ontologies.** When working with a Syndicate implementation, programmers must design conversational protocols that capture relevant aspects of the domain each program is intended to address. The most important overarching principle is that Syndicate programs and protocols should *make their domain manifest*, and *hide implementation constructs*. Generally, each domain will include an ontology of its own, relating to concepts largely internal to the domain. Such an ontology will seldom or never include concepts from the host language or even Syndicate-specific ideas.

Following this principle, Syndicate takes care to avoid polluting a programmer's domain models with implementation- and programming-language-level concepts. As far as possible, the structure and meaning of each assertion is left to the programmer. Syndicate implementations reserve the contents of a dataspace for domain-level concepts. Access to information in the domain of *programs*, relevant to debugging, tracing and otherwise reflecting on the operation of a running program, is offered by other (non-dataspace, non-assertion) means. This separation of domain from implementation mechanism manifests in several specific corollaries:

1. Do not propagate host-language exception values across a dataspace.

   An actor that raises an uncaught exception is terminated and removed from the dataspace, but the details of the exception (stack traces, error messages, error codes etc.) are *not* made available to peers via the dataspace. After all, exceptions describe some aspect of a running computer program, and do not in general relate to the program's domain.

15 The content of a given dataspace is just the union of the assertions currently maintained by its contained actors. Each connected actor usually maintains a complete picture of its own assertions. When *all* the actors in a group do this, the *dataspace* underpinning the group could in principle be rebooted or upgraded seamlessly without disrupting the work of the group as a whole, reconstructing dataspace state from the records of the actors themselves.

Instead, a special reflective mechanism is made available for host-language programs to access such information for debugging and other similar purposes. Actors in a dataspace do not use this mechanism when operating normally. As a rule, they instead depend on *domain-level* signaling of failures in terms of the (automatic) removal of domain-level assertions on failure, and do not depend on host-language exceptions to signal domain-level exceptional situations.[16]

2. Make internal actor identifiers completely invisible.

   The notion of a (programming-language) actor is almost never part of the application domain; this goes double for the notion of an actor's internal identifier (a.k.a. pointer, "pid", or similar). Where identity of specific parties is relevant to a domain, SYNDICATE requires the protocol to explicitly specify and manage such identities, and they remain distinct from the internal identities of actors in a running SYNDICATE program. Again, during debugging, the identities of specific actors are relevant to the programmer, but this is because the programmer is operating in a different domain from that of the program under study.

   Explicit treatment of identity unlocks two desirable abilities:

   a) One (implementation-level) actor can transparently perform multiple (domain-level) roles. Having decoupled implementation-level identity from domain-level information, we are free to choose arbitrary relations connecting them.

   b) One actor can transparently delegate portions of its responsibilities to others. Explicit management of identity allows actors to *share* a domain-level identity without needing to share an implementation-level identity. Peers interacting with such actors remain unaware of the particulars of any delegation being employed.

3. Multicast communication should be the norm; point-to-point, a special case.

   Conversational interactions can involve any number of participants. In languages where the implementation-provided medium of conversation always involves exactly two participants, programmers have to encode $n$-party domain-level conversations using the two-party mechanism. Because of this, messages between components have to mention implementation-level conversation endpoints such as channel or actor IDs, polluting otherwise domain-specific ontologies with implementation-level constructs. In order to keep implementation ideas out of domain ontologies, SYNDICATE does not define any kind of value-level representation of a conversation. Instead, it leaves the choice of scheme for naming conversations up to the programmer.

4. Equivalences on messages, assertions and other forms of shared state should be in terms of the domain, not in terms of implementation constructs.

16 SYNDICATE distinguishes itself from Erlang here. Erlang's failure-signaling primitives, links and monitors, necessarily operate in terms of actor IDs, so it is no great step to include stack traces and error messages alongside an actor ID in a failure description record.

For example, consider deduplication of received messages. In some protocols, in order to make message receipt idempotent, a table of previously-seen messages must be maintained. To decide membership of this table, a particular equivalence must be chosen. Forcing this equivalence to involve implementation-level constructs entails a need for the programmer to explicitly normalize messages to ensure that the implementation-level equivalence reflects the desired domain-level equivalence. To be even more specific:

a) If a transport includes message sequence numbers, message identifiers, timestamps etc., then these items of information from the transport should not form part of the equivalence used.

b) Sender identity should not form part of the equivalence used. If a particular protocol needs to know the identity of the sender of a message, it should explicitly include a definition of the relevant notion of identity (not necessarily the implementation-level identity of the sender) and explicitly include it in message type definitions.

SUPPORT RESOURCE MANAGEMENT DECISIONS. Concurrent programs in all their forms rely on being able to scope the size and lifetime of allocations of internal resources made in response to external demand. "Demand" and "resource" are extremely general ideas. As a result, resource management decisions appear in many different guises, and give rise to a number of related principles:

1. *Demand-matching* should be well-supported.

   Demand-matching is the process of automatic allocation *and release* of some resource in response to detected need elsewhere in a program. The concept applies in many different places.

   For example, in response to the *demand* of an incoming TCP connection, a server may allocate resources including a pair of memory buffers and a new thread. The buffers, combined with TCP back-pressure, give control over memory usage, and the thread gives control over compute resources as well as offering a convenient language construct to attach other kinds of resource-allocation and -release decisions to. When the connection closes, the server may terminate the thread, release other associated resources, and finalize its state.

   Another example can be found in graphical user interfaces, where various widgets manifest in response to the needs of the program. An entry in a "buddy list" in a chat program may be added in response to presence of a contact, making the "demand" the presence of the contact and the "resource" the resulting list entry widget. When the contact disconnects, the "demand" for the "resource" vanishes, and the list entry widget should be removed.

2. Service presence (Konieczny et al. 2009) and presence information generally should be well-supported.

   Consider linking multiple independent services together to form a concurrent application. A web-server may depend on a database: it "demands" the services of the database, which acts as a "resource". The web-server and database may in turn depend upon a logging service. Each service cannot start its work before its dependencies are ready: it observes the *presence* of its dependencies as part of its initialization.

Similarly, in a publish-subscribe system, it may be expensive to collect and broadcast a certain statistic. A publisher may use the availability of subscriber information to decide whether or not the statistic needs to be maintained. Consumers of the statistic act as "demand", and the resource is the entirety of the activity of producing the statistic, along with the statistic itself. Presence of consumers is used to manage resource commitment.

Finally, the AMQP messaging middleware protocol (The AMQP Working Group 2008) includes special flags named "immediate" and "mandatory" on each published message. They cause a special "return to sender" feature to be activated, triggering a notification to the sender only when *no receiver is present* for the message at the time of its publication. This form of presence allows a sender to take alternative action in case no peer is available to attend to its urgent message.

SUPPORT DIRECT COMMUNICATION OF PUBLIC ASPECTS OF COMPONENT STATE. This is a generalization of the notion of presence, which is just one portion of overall state.

AVOID DEPENDENCE ON TIMEOUTS. In a distributed system, a failed component is indistinguishable from a slow one and from a network failure. Timeouts are a pragmatic solution to the problem in a distributed setting. Here, however, we have the luxury of a *non*-distributed design, and we may make use of specific forms of "demand" information or presence in order to communicate failure. Timeouts are still required for inter-operation with external systems, but are seldom needed as a normal part of greenfield SYNDICATE protocol design.

REDUCE DEPENDENCE ON ORDER-OF-OPERATIONS. The language should be designed to make programs robust by default to reordering of signals. As part of this, idempotent signals should be the default where possible.

1. Event-handlers should be written as if they were to be run in a (pseudo-) random order, even if a particular implementation does not rearrange them randomly. This is similar to the thinking behind the random event selection in CML's choice mechanism (Reppy 1992, page 131).

2. Questions of deduplication, equivalence, and identity must be placed at the heart of each SYNDICATE protocol design, even if only at an abstract level.

ESCHEW TRANSFER OF HIGHER-ORDER DATA. Mathematical and computational structures enjoy an enormous amount of freedom not available to structures that must be realized in the physical world. Similarly, patterns of interaction that can be realized in a non-distributed setting are often inappropriate, unworkable, or impossible to translate to a distributed setting.

One example of this concerns *higher-order* data, by which I mean certain kinds of closure,[17] mutable data structures, and any other stateful kind of entity.

Syndicate is not a distributed programming language, but was heavily inspired by my experience of distributed programming and by limitations of existing programming languages employed in a distributed setting. Furthermore, certain features of the design suggest that it may lead to a useful distributed programming model in future. With this in mind, certain principles relate to a form of physical realizability; chief among them, the idea of limiting information exchange to first-order data wherever possible. The language should encourage programmers to act as if transfer of higher-order data between peers in a dataspace were impossible. While non-distributed implementations of Syndicate can offer support for transfer of functions, objects containing mutable references, and so on, stepping to a distributed setting limits programs to exchange of first-order data only, since real physical communication networks are necessarily first-order. Transfer of higher-order data involves a hidden use/mention distinction. Higher-order data may be *encoded*, but cannot directly be *transmitted*.

With that said, however, notions of stateful location or place are important to certain domains, and the ontologies of such domains may well naturally include references to such domain-relevant location information. It is host-language higher-order data that Syndicate discourages, not domain-level references to location and located state.

ARRANGE ACTORS HIERARCHICALLY. Many experiments in structuring groups of (actor model) actors have been performed over the past few decades. Some employ *hierarchies* of actors, that is, the overall system is structured as a tree, with each actor or group existing in exactly one group (e.g. Varela and Agha 1999). Others allow actors to be placed in more than one group at once, yielding a *graph* of actors (e.g. Callsen and Agha 1994).

Syndicate limits actor composition to tree-shaped hierarchies of actors, again inspired by physical realizability. Graph-like connectivity is encoded in terms of protocols layered atop the hierarchical medium provided. Recursive groupings of computational entities in real systems tend to be hierarchical: threads within processes within containers managed by a kernel running under a hypervisor on a core within a CPU within a machine in a datacenter.

## 2.7 ON THE NAME "SYNDICATE"

Now that we have seen an outline of the Syndicate design, the following definitions may shed light on the choice of the name "Syndicate":

> A **syndicate** is a self-organizing group of individuals, companies, corporations or entities formed to transact some specific business, to pursue or promote a shared interest.
>
> — Wikipedia[18]

17 Specifically, closures closing over *mutable* state; "pure" closures are in some sense not higher-order. See also Miller's work on "spores" (Miller, Haller and Odersky 2014; Miller et al. 2016).

18 Definition retrieved from Wikipedia, https://en.wikipedia.org/wiki/Syndicate, on 23 August 2017.

> **Syndicate**, n.
>
> 1. A group of individuals or organizations combined to promote a common interest.
>
>    1.1 An association or agency supplying material simultaneously to a number of newspapers or periodicals.
>
> **Syndicate**, v.tr.
>
> ...
>
>    1.1 Publish or broadcast (material) simultaneously in a number of newspapers, television stations, etc.
>
> — Oxford Dictionary[19]

An additional relevant observation is that a syndicate can be a group of companies, and a company can be a group of actors.

19 Definition retrieved from the online Oxford Living Dictionaries, https://en.oxforddictionaries.com/definition/syndicate on 23 August 2017. The full Oxford English Dictionary entries for "syndicate" are much longer and do not make such a pleasing connection to the language design idea.

# 3

# *Approaches to Coordination*

Our analysis of communication and coordination so far has yielded a high-level, abstract view on concurrency, taking *knowledge-sharing* as the linchpin of cooperation among components. The previous chapter raised several questions, answering some in general terms, and leaving others for investigation in the context of specific *mechanisms* for sharing knowledge. In this chapter, we explore these remaining questions. To do so, we survey the paradigmatic approaches to communication and coordination. Our focus is on the needs of programmers and the operational issues that arise in concurrent programming. That is, we look at ways in which an approach helps or hinders achievement of a program's goals in a way that is robust to unpredictability and change.

## 3.1 A concurrency design landscape

The outstanding questions from chapter 2 define a multi-dimensional landscape within which we place different approaches to concurrency. A given concurrency model can be assigned to a point in this landscape based on its properties as seen through the lens of these questions. Each point represents a particular set of trade-offs with respect to the needs of programmers.

To recap, the questions left for later discussion were:

K4 Which forms of knowledge-sharing are robust in the face of the unpredictability intrinsic to concurrency?

K6 Which forms of knowledge-sharing are robust to and help mitigate the impact of changes in the goals of a program?

In addition, the investigation of question K3 ("what do concurrent components need to know to do their jobs?") concluded with a picture of domain knowledge, epistemic knowledge, framing knowledge, and knowledge flow within a group of components. However, it left unaddressed the question of *mechanism*, giving rise to a follow-up question:

K3*bis* How do components learn what they need to know as time goes by?

In short, the three questions relate to *robustness*, *operability* and *mechanism*, respectively. The rest of the chapter is structured around an informal investigation of characteristics refining these categories.

MECHANISM (K3*bis*). A central characteristic of a given concurrency model is its mechanism for exchange of knowledge among program components. Each mechanism yields a different set of possibilities for how concurrent conversations evolve. First, a conversation may have arbitrarily many participants, and a participant may engage in multiple conversations at once. Hence, models and language designs must be examined as to

C1 how they support various *conversation group sizes* and

C2 how they support *correlation and demultiplexing* of incoming events.

Second, conversations come with associated *state*. Each participating component must find out about changes to this state and must *integrate those changes* with its local view. The component may also wish to *change* conversational state; such changes must be *signaled* to relevant peers. A mechanism can thus be analyzed in terms of

C3 how it supports *integration of state changes* with a component's local view and

C4 how it arranges for state changes to be *signaled* to conversational peers.

ROBUSTNESS (K4). Each concurrency model offers a different level of support to the programmer for addressing the unpredictability intrinsic to concurrent programming. Programs rely on the *integrity* of each participant's view of overall conversational state; this may entail consideration of *consistency* among different views of the shared state in the presence of unpredictable latency in change propagation. These lead to investigation of

C5 how a model helps *maintain integrity* of conversational state and

C6 how it helps *ensure consistency* of state as a program executes.

In addition, viewing a conversation as a series of events describing changes in conversational state has direct implications for the connection between data flow and control flow. Clearly, the arrival of a notification (data) at a participant ought to reliably trigger control flow; but conversely, the creation and termination of components must also be able to reliably trigger notifications to peers. This includes exceptions and other forms of partial failure. Hence, we may ask

C7 how data flow leads to control flow in programs and

C8 how control flow, such as start-up or termination of a component, leads to data flow.

Finally, robust programs demand effective strategies for management of computational, storage and other types of resources, leading us to inquire

C9 how a concurrency model supports *resource management* during execution.

OPERABILITY (K6). The notion of operability is broad, including attributes pertaining to the ease of working with the model at design, development, debugging and deployment time. We will focus on the ability of a model to support

C10 *debuggability* and *visualizability* of interactions and relationships among components;

C11 *evolvability* of the pattern of interactions within a program; and

C12 *durability* of long-lived state as code evolves and features come and go.

Figure 3: Characteristics of approaches to concurrency

Characteristics C1–C12 in figure 3 will act as a lens through which we will examine three broad families of concurrency: shared memory models, message-passing models, and tuplespaces and external databases. In addition, we will analyze the fact space model briefly mentioned in the previous chapter.

We illustrate our points throughout with a *chat server* that connects an arbitrary number of participants. It relays text typed by a user to all others and generates announcements about the arrival and departure of peers. A client may thus display a list of active users. The chat server involves chat-room state—the membership of the room—and demands many-to-many communication among the concurrent agents representing connected users. Each such agent receives events from two sources: its peers in the chat-room and the TCP connection to its user. If a user disconnects or a programming error causes a failure in the agent code, resources such as TCP sockets must be cleaned up correctly, and appropriate notifications must be sent to the remaining agents and users.

## 3.2 SHARED MEMORY

Shared memory languages are those where *threads* communicate via modifications to *shared memory*, usually *synchronized* via constructs such as monitors (Gosling et al. 2014; IEEE 2009; ISO 2014). Figure 4 sketches the heart of a chat room implementation using a monitor (Brinch Hansen 1993) to protect the shared members variable.

(C1; C3; C4) Mutable memory tracks shared state and also acts as a communications mechanism. Buffers and routing information for messages between threads are explicitly encoded as part of the conversational state, which naturally accommodates the multi-party conversations of our chat server. However, announcing changes in conversational state to peers—a connection or disconnection, for example—requires construction of a broadcast mechanism out of low-level primitives.

(C2) To engage in multiple conversations at once, a thread must monitor multiple regions of memory for changes. Languages with powerful memory transactions make this easy; the combination of "retry" and "orelse" gives the requisite power (Harris et al. 2005). Absent such transactions, and ruling out polling, threads must explicitly signal each other when making changes. If a thread must wait for any one of several possible events, it is necessary to reinvent multiplexing based on condition variables and write code to perform associated book-keeping.

(C5) Maintaining the integrity of shared state is famously difficult. The burden of correctly placing transaction boundaries or locks and correctly ordering updates falls squarely on the programmer. It is reflected in figure 4 not only in the use of the monitor concept itself, but also in the careful ordering of events in the `connect` and `disconnect` methods. In particular, the call to `announce` (line 13) must follow the removal of `user` (line 12), because otherwise, the system may invoke `callback` for the disconnected `user`. Similarly, cloning the `members` map (line 15) is necessary so that a disconnecting user (line 17) does not change the collection mid-iteration. Moreover, even with transactions and correct locking discipline, care must be taken to maintain logical invariants of an application. For example, if a chat user's thread terminates unexpectedly without calling `disconnect`, the system continues to send *output* to

```
1  class Chatroom
2    private Map<String, (String->())> members

3    public synchronized connect(user, callback)
4      for (existingUser, _) in members
5        callback(existingUser + " arrived")
6      members.put(user, callback)
7      announce(user + " arrived")

8    public synchronized speak(user, text)
9      announce(user + ": " + text)

10   public synchronized disconnect(user)
11     if (!members.containsKey(user)) { return }
12     members.remove(user)
13     announce(user + " left")

14   private announce(what)
15     for (user, callback) in members.clone()
16       try { callback(what) }
17       catch (exn) { disconnect(user) }
```

Figure 4: Monitor-style chat room

the associated TCP socket indefinitely, even though *input* from the socket is no longer being handled, meaning `members` has become logically corrupted. Conversely, a seemingly-correct program may call `disconnect` twice in corner cases, which explains the check (line 11) for preventing double departure announcements.

(C7; C8) Memory transactions with "retry" allow control flow to follow directly from changes to shared data; otherwise, however, data flow is completely decoupled from inter-thread control flow. The latter is provided via synchronization primitives, which are only coincidentally associated with changes to the shared store. Coming from the opposite direction, control flow is also decoupled from data flow. For example, exceptions do not automatically trigger a clean-up of shared state or signal the termination of the thread to the relevant group of peers.[1] Determining responsibility for a failure and deciding on appropriate recovery actions is challenging. Consider an action by user A that leads to a call to announce. If the callback associated with user B (line 16) throws an exception, the handler on line 17 catches it. To deal with this situation, the developer must reason in terms of three separate, stateful entities with non-trivial responsibilities: the agents for A and B plus the chat room itself. If the exception propagates, it may not only damage the monitor's state but terminate the thread representing A, even though it is the fault of B's callback. Contrast the problems seen in this situation with the call to the callback in connect (line 5); it does not need an exception handler, because the data flow resulting from the natural control flow of exception propagation is appropriate.

---

1 This line of reasoning recalls the explanation offered by Sun (now Oracle) for why the Java method `Thread.stop` is deprecated. http://docs.oracle.com/javase/1.5.0/docs/guide/misc/threadPrimitiveDeprecation.html

(C9) The thread model also demands the manual management of resources for a given conversation. For example, disposal of unwanted or broken TCP sockets must be coded explicitly in every program.

(C6) On the bright side, because it is common to have a single copy of any given piece of information, with all threads sharing access to that copy, explicit consideration of consistency among replicas is seldom necessary.

The many interlocking problems described above are difficult to discover in realistic programs, either through testing or formal verification. To reach line 17, a callback must fail midway through an announcement caused by a different user. The need for the `.clone()` on line 15 is not directly obvious. To truly gain confidence in the implementation, one must consider cases where multiple failures occur during one announcement, including the scenario where a failure during `speak` causes `disconnect` and another failure occurs during the resulting announcement. The interactions between the various locks, loops, callbacks, exception handlers, and pieces of mutable state are manifold and non-obvious.

(C10; C11; C12) Because shared memory languages allow unconstrained access to shared memory, not connected to any kind of scoping construct or protocol description, recovering a clear picture of the relationships and interactions among threads is extremely challenging. Similarly, as discussed for character C2, modifying a component to engage in multiple conversations at once or expanding the scope of a conversation to include multiple components is in general invasive. Finally, the lack of a clear linguistic specification of the structure of the shared memory and its relationship to a program's threads largely precludes automated support for orthogonal persistence and code upgrade.

An important variation on shared memory is the single-threaded, event-based style of JavaScript (ECMA 2015). While use of explicit locking is reduced in such cases, most of the analysis of the threaded approach continues to hold.

## 3.3 MESSAGE-PASSING

Message-passing models of concurrency include languages using Hoare's CSP channels (Hoare 1985) or channels from the $\pi$-calculus (Milner 1999), and those based on the actor model (Hewitt, Bishop and Steiger 1973; Agha 1986; Agha et al. 1997; De Koster et al. 2016). Channel languages include CML (Donnelly and Fluet 2008; Reppy 1991), Go, and Rust, which all use channels in a shared-memory setting, and the Join Calculus (Fournet and Gonthier 2000), which assumes an isolated-process setting. This section concentrates on isolated processes because channel-based systems using shared memory are like those discussed in section 3.2. Actor languages include Erlang (Armstrong 2003), Scala (Haller and Odersky 2009), AmbientTalk (Van Cutsem et al. 2014), and E (Miller, Tribble and Shapiro 2005).

Channel- and actor-based models are closely related (Fowler, Lindley and Wadler 2016). An actor receives input exclusively via a mailbox (Agha 1986), and messages are explicitly addressed by the sending actor to a specific recipient. In channel-based languages, messages are explicitly addressed to particular channels; each message goes to a single recipient, even when a channel's receive capability is shared among a group of threads.

```
1  def chatroom()
2    members = new Hashtable()
3    while True
4      match receiveMessage()
5        case Connect(user, PID)
6          monitor(PID)  // Erlang-style "link"
7          for peer in members.keys
8            send(PID, ChatOutput(peer + " arrived"))
9          members.put(user, PID)
10         announce(members, user + " arrived")
11       case EXIT_SIGNAL(PID)
12         user = members.findKeyForValue(PID)
13         members.remove(user)
14         announce(members, user + " left")
15       case Speak(user, text)
16         announce(members, user + ": " + text)

17 def announce(members, what)
18   for PID in members.values
19     send(PID, ChatOutput(what))
```

Figure 5: Actor-style chat room

(C1) Both actor- and channel-based languages force an encoding of the chat room's one-to-many medium in terms of built-in point-to-point communication constructs.[2] Compare figure 5, which expresses the chat room as a process-style actor, with figure 6, which presents pseudocode for a channel-based implementation. In figure 5, the actor embodying the chat room's broadcast medium responds to `Speak` messages (line 15) by sending `ChatOutput` messages to actors representing users in the room. In figure 6, the thread running the `chatroom()` procedure responds similarly to `Speak` instructions received on its control channel (line 13).

(C2) Languages with channels often provide a "select" construct, so that programs can wait for events on any of a group of channels. Such constructs implement automatic demultiplexing by channel identity. For example, a thread acting as a user agent might await input from the chat room *or* the thread's TCP connection (figure 7a). The language runtime takes care to atomically resolve the transaction. In these languages, a channel reference can stand directly for a specific conversational context. By contrast, actor languages lack such a direct representation of a conversation. Actors retrieve messages from their own *private* mailbox and then demultiplex manually by inspecting received messages for correlation identifiers (figure 7b). While the channel-based approach forces use of an implementation-level correlator—the channel reference—explicit pattern-based demultiplexing allows domain-level information in each received message to determine the relevant conversational context. The E language (Miller 2006; De Koster, Van Cutsem and De Meuter 2016) is a hybrid of the two approaches, offering

2 AmbientTalk is unusual among actor languages for the depth of its consideration for multicast communication and coordination, offering n-way primitives alongside point-to-point communication. We discuss AmbientTalk further in section 3.5.

```
def chatroom(ch)
  members = new Hashtable()
  while True
    match ch.get()
      case Connect(user, callbackCh)

        for peer in members.keys
          callbackCh <- peer + " arrived"
        members.put(user, callbackCh)
        announce(members, user + " arrived")
      case Disconnect(user)

        members.remove(user)
        announce(members, user + " left")
      case Speak(user, text)
        announce(members, user + ": " + text)

def announce(members, what)
  for callbackCh in members.values
    callbackCh <- what
```

Figure 6: Channel-style chat room

```
select {
  case line <- callbackCh:
    tcpOutputCh <- line
  case line <- tcpInputCh:
    chatroomCh <- Speak(myName, line)
}
```

(a) channel-style

```
match receiveMessage() {
  case ChatOutput(line):
    socket.write(line)
  case TcpInput(_, line):
    send(ChatroomPID, Speak(myName, line))
}
```

(b) Actor-style

Figure 7: Demultiplexing multiple conversations.

object references to denote specific conversations within the heap of a given actor, and employs method dispatch as a limited pattern matcher over received messages.

(C3; C4; C5) With either actors or channels, only a small amount of conversational state is managed by the language runtime. In actor systems, it is the routing table mapping actor IDs to mailbox addresses; in channel-based systems, the implementation of channel references and buffers performs the analogous role. Developers implement other kinds of shared state using message passing. This approach to conversational state demands explicit programming of updates to a local replica of the state based on received messages. Conversely, when an agent decides that a change to conversational state is needed, it must broadcast the change to the relevant parties. Correct notification of changes is crucial to maintaining integrity of conversational state. Most other aspects of integrity maintenance become local problems due to the isolation of individual replicas. In particular, a crashing agent cannot corrupt peers.

(C5) Still, the programmer is not freed from having to consider execution order when it comes to maintaining local state. Consider the initial announcement of already-present peers to an arriving user in figure 5 (lines 7–8). Many subtle variations on this code arise from moving the addition of the new user (line 9) elsewhere in the `Connect` handler clause; some omit self-announcement or announce the user's appearance twice.

(C7; C8) Both models make it impossible to have data flow *between* agents without associated control flow. As Hewitt, Bishop and Steiger (1973) write, "control flow and data flow are inseparable" in the actor model. However, control flow *within* an agent may not coincide with an appropriate flow of data to peers, especially when an exception is raised and crashes an agent. Channel references are not exclusively owned by threads, meaning we cannot generally close channels in case of a crashing thread. Furthermore, most channel-based languages are *synchronous*, meaning a send blocks if no recipient is ready to receive. If a thread servicing a channel crashes, then the next send to that channel may never complete. In our chat server, a crashed user agent thread can deadlock the whole system: the chatroom thread may get stuck during `callbacks` (lines 7 and 17 in figure 6). In general, synchronous channel languages preclude local reasoning about potential deadlocks; interaction with some party can lead to deadlock via a long chain of dependencies. Global, synchronous thinking has to be brought to bear in protocol design for such languages: the programmer must consider *scheduling* in addition to data flow. Actors can do better. Sends are asynchronous, introducing latency and buffering but avoiding deadlock, and mailboxes are owned by exactly one actor. If that actor crashes, further communication to or from that actor is hopeless. Indeed, Erlang offers *monitors* and *exit signals*, i.e., an actor may subscribe to a peer's lifecycle events (line 6 in figure 5). Such subscriptions allow the chat room to combine error handling with normal disconnection. No matter whether a user agent actor terminates normally or abnormally, the `EXIT_SIGNAL` handler (lines 12–14) runs, announcing the departure to the remaining peers. The E language allows references to remote objects to *break* when the associated remote vat exits, crashes, or disconnects, providing a hybrid of channel-style demultiplexing with Erlang-style exit signaling.

(C6) Where many replicas of a piece of state exist alongside communications delays, the problem of maintaining consistency among replicas arises. Neither channels nor actors have any support to offer here. Channels, and synchronous communication in general, seem to prioritize (without guaranteeing) consistency at the expense of deadlock-proneness; asynchronous communication avoids deadlock, but risks inconsistency through the introduction of latency.

(C9) Exit signals are a step toward automatically managing resource deallocation. While actors must manually allocate resources, the exit signal mechanism may be used to tie the lifetime of a resource, such as a TCP socket, to the lifetime of an actor. If fine-grained control is needed, it must be programmed manually. Additionally, in asynchronous (buffered) communication, problems with resource control arise in a different way: it is easy to overload a component, causing its input buffer or mailbox to grow potentially without bound.

(C10) Enforced isolation between components, and forcing all communication to occur via message-passing, makes the provision of tooling for visualizing execution traces possible. Languages such as Erlang include debug trace facilities in the core runtime, and make good use of them for lightweight capturing of traces even in production. However, the possibility of mes-sage races complicates reasoning and debugging; programmers are often left to analyze the live behavior of their programs, if tooling is unavailable or inadequate. Modification of programs to capture ad-hoc trace information frequently causes problematic races to disappear, further complicating such analysis.

(C11) As figure 7 makes clear, modifying a component to engage in multiple simultaneous conversations can be straightforward, if *all* I/O goes through a single syntactic location. However, if communication is hidden away in calls to library routines, such modifications demand non-local program transformations. Similarly, adding a new participant to an existing conversation can require non-local changes. In instances where a two-party conversation must now include three or more participants, this often results in reification of the communications medium into a program component in its own right.

(C12) Erlang encourages adherence to a "tail-call to next I/O action" convention allowing easy upgrade of running code. Strictly-immutable local data and functional programming combine with this convention to allow a module backing a process to be upgraded across such tail-calls, seamlessly transitioning to a new version of the code. In effect, all actor state is held in accumulator data structures explicitly threaded through actor implementations. Other actor languages without such strong conventions cannot offer such a smooth path to live code upgrade. Channel-based languages could include similar conventions; in practice, I am not aware of any that do so.

## 3.4 TUPLESPACES AND DATABASES

Finally, *hybrid* models exist, where a shared, mutable store is the medium of communication, but the store itself is accessed and components are synchronized via message passing. These models are database-like in nature. Languages employing such models include *tuplespace*-based languages such as Linda (Gelernter 1985; Carriero et al. 1994), Lime (Murphy, Picco and Roman 2006), and TOTAM (Scholliers, González Boix and De Meuter 2009; Scholliers et al. 2010; González Boix 2012; González Boix et al. 2014), as well as languages that depend solely on an external DBMS for inter-agent communication, such as PHP (Tatroe, MacIntyre and Lerdorf 2013).

Tuplespace languages have in common the notion of a "blackboard" data structure, a *tuplespace*, shared among a group of agents. Data items, called *tuples*, are written to the shared area and retrieved by pattern matching.[3] Once published to the space, tuples take on independent existence. Similarly, reading a tuple from the space may *move* it from the shared area to an agent's private store.

The original tuplespace model provided three essential primitives: `out`, `in`, and `rd`. The first *writes* tuples to the store; the other two *move* and *copy* tuples from the store to an agent, respectively. Both `in` and `rd` are blocking operations; if multiple tuples match an operation's pattern, an arbitrary single matching tuple is moved or copied. Later work extended this austere model with, for example, `copy-collect` (Rowstron and Wood 1996), which allows copying

3 Compare to shared-memory or message-passing communications media, where items are retrieved either in queue order or by memory location.

```
class UserAgent
  public run(name, socket)
    new Reaction(Present(_), fn(who) { socket.println(who + " arrived") })
    new Reaction(Absent(_), fn(who) { socket.println(who + " left") })
    new Reaction(Message(_,_), fn(who, what) { socket.println(who + ": " + what) })

    previousLine = null

    try
      inp(Absent(name))
      out(Present(name))

      while (line = socket.readLine()) != null
        if previousLine != null: in(Message(name, previousLine))
        out(Message(name, line))
        previousLine = line

    finally
      if previousLine != null: in(Message(name, previousLine))
      in(Present(name))
      out(Absent(name))
```

Figure 8: Tuplespace-style chat room user agent, modeled on `LChat.java` from the Lime 1.06 distribution.

of *all* matching tuples rather than the arbitrary single match yielded by `rd`. Such extensions add essential expressiveness to the system (Busi and Zavattaro 2001; Felleisen 1991). Lime goes further yet, offering not only non-blocking operations `inp` and `rdp`, but also *reactions*, which are effectively callbacks, executed once per matching tuple. Upon creation of a reaction, existing tuples trigger execution of the callback. When subsequent tuples are inserted, any matches to the reaction's pattern cause additional callback invocations. This moves tuplespace programming toward programming with publish/subscribe middleware (Eugster et al. 2003). TOTAM takes Lime's reactions even further, allowing reaction to *removal* of a previously-seen tuple.

External DBMS systems share many characteristics with tuplespaces: they allow storage of relations; stored items are persistent; retrieval by pattern-matching is common; and many modern systems can be extended with *triggers*, code to be executed upon insertion, update, or removal of matching data. One difference is the notion of *transactionality*, standard in DBMS settings but far from settled in tuplespaces (Bakken and Schlichting 1995; Papadopoulos and Arbab 1998). Another is the decoupling of notions of process from the DBMS itself, where tuplespace systems integrate process control with other aspects of the coordination mechanism.

Figure 8 presents a pseudo-code tuplespace implementation of a user agent, combining Java-like constructs with Lime-like reactions. Previous sketches have concentrated on appropriate implementation of the shared medium connecting user agents; here, we concentrate on the agents themselves, because tuplespaces are already sufficiently expressive to support broadcasting.[4]

4 Our chat server problem is challenging to solve using the original Linda primitives alone. The introduction of `copy-collect` and reactions removes these obstacles.

(C1; C2; C3) Tuplespaces naturally yield multi-party communication. All communication happens indirectly through manipulation of shared state. Inserted tuples are visible to all participants.[5] With reactions, programmers may directly express the relationship between appearance of tuples matching a pattern and execution of a code fragment, allowing a richer kind of demultiplexing of conversations than channel-based models. For example, the reactions in figure 8 (lines 3–5) manifestly associate conversations about presence, absence and utterances with specific responses, respectively; the tuplespace automatically selects the correct code to execute as events are received. By contrast, in tuplespace languages without reactions, the blocking natures of `in` and `rd` lead to multiplexing problems similar to those seen with shared memory and monitors.

(C1) Tuples are persistent, hence the need to retract each message before inserting the next (line 11). An unfortunate side effect is that if a new participant joins mid-conversation, it receives the most recent utterance from each existing peer, even though that utterance may have been made a long time ago.

(C7; C8; C4; C5) Data flow usually occurs concomitantly with control flow in a tuplespace; `in` and `rd` are blocking operations, and reactions trigger code execution in response to a received event. Control flow, however, does not always trigger associated data flow. Because manipulation of the tuplespace is imperative, no mechanism exists within the core tuplespace model to connect the lifetime of tuples in the space with the lifetime of the agent responsible for them. This can lead to difficulty maintaining application-level invariants, even though the system ensures data-structure-level integrity of the tuplespace itself. For an example, see the explicit clean-up action as the process prepares to exit (lines 15–17). In addition, the effect of exceptions inside reactions remains unclear in all tuplespace languages. Turning to external DBMS, we see that the situation is worse. There, setting aside the possibility of abusing triggers for the purpose, changes in state do not directly have an effect on the flow of control in the system. Connections between programs and the DBMS are viewed as entirely transient and records inserted are viewed as sacrosanct once committed.

(C8) Tuplespaces take a wide variety of approaches to failure-handling (Bakken and Schlichting 1995; Rowstron 2000). In Lime, in particular, tuples are localized to tuplespace *fragments* associated with individual agents. These fragments automatically combine when agents find themselves in a common context. Agent failure or disconnection removes its tuplespace fragment from the aggregate whole. While Lime does not offer the ability to react to removal of individual tuples, it can be configured to insert `_host_gone` tuples into the space when it detects a disconnection. By reacting to *appearance* of `_host_gone` tuples, applications can perform coarse-grained cleaning of the knowledgebase after disconnection or failure. Separately, TOTAM's per-tuple *leases* (González Boix et al. 2014) give an upper bound on tuple lifetime. Our example chat room is written in an imaginary tuplespace dialect lacking fine-grained reactions to tuple withdrawal, and thus inserts `Absent` records upon termination (lines 4, 8, and 17 in figure 8) to maintain its invariants.

(C6) Reactions and `copy-collect` allow maintenance of eventually-consistent views and production of consistent snapshots of the contents of a tuplespace, respectively. However, opera-

5 Systems like TOTAM introduce rule-based visibility constraints for tuples.

tions like `rd` are not only non-deterministic but non-atomic in the sense that by the time the existence of a particular tuple is signaled, that tuple may have been removed by a third party. Tuplespaces, then, offer some mechanisms by which the consistency of the various local replicas of tuplespace contents may be maintained and reasoned about. In contrast, most DBMS systems do not offer such mechanisms for reasoning about and maintaining a client's local copy of data authoritatively stored at a server. Instead, a common approach is to use transactions to atomically query and then alter information. The effect of this is to bound the lifetime of local views on global state, ensuring that while they exist, they fit in to the transactional framework on offer, and that after their containing transaction is over, they cannot escape to directly influence further computation.

(C9) Detection of demand for some resource can be done using tuples indicating demand and corresponding reactions. The associated callback can allocate and offer access to the demanded resource. In systems like TOTAM, retraction of a demand tuple can be interpreted as the end of the need for the resource it describes; in less advanced tuplespaces, release of resources must be arranged by other means.

(C10) Both tuplespaces and external databases give excellent visibility into application state, on the condition that the tuplespace or database is the *sole* locus of such state. In cases where this assumption holds, the entirety of the state of the group is visible as the current contents of the shared store. This unlocks the possibility of rich tooling for querying and modifying this state. Such tooling is a well-integrated part of existing DBMS ecosystems. In principle, recording and display of traces of interactions with the shared store could also be produced and used in visualization or debugging.

(C11) The original tuplespace model of Linda lacked non-blocking operations, leading it to suffer from usability flaws well-known from the context of synchronous IPC. As Elphinstone and Heiser write,

> While certainly minimal, and simple conceptually and in implementation, experience taught us significant drawbacks of [the model of synchronous IPC as the only mechanism]: it forces a multi-threaded design onto otherwise simple systems, with the resulting synchronisation complexities. (Elphinstone and Heiser 2013)

These problems are significantly mitigated by the addition of LIME's reactions and the later developments of TOTAM's context-aware tuplespace programming. Generally speaking, tuplespace-based designs have moved from synchronous early approaches toward asynchronous operations, and this has had benefits for extending the interactions of a given component as well as extending the scope of a given conversation. External DBMS systems are generally neutral when it comes to programming APIs, but many popular client libraries offer synchronous query facilities only, lack support for asynchronous operations, and offer only limited support for triggers.

(C12) External DBMS systems offer outstanding support for long-lived application state, making partial restarts and partial code upgrades a normal part of life with a DBMS application. Transactionality helps ensure that application restarts do not corrupt shared state. Tuplespaces in principle offer similarly good support.

Finally, the two models, viewed abstractly, suffer from a lack of proper integration with host languages. The original presentation of tuplespaces positions the idea as a complete, independent language design; in reality, tuplespaces tend to show up as libraries for existing languages. Databases are also almost always accessed via a library. As a result, developers must often follow design patterns to close the gap between the linguistic capabilities of the language and their programming needs. Worse, they also have to deploy several different coordination mechanisms, without support from their chosen language and without a clear way of resolving any incompatibilities.

## 3.5 THE FACT SPACE MODEL

The *fact space model* (Mostinckx et al. 2007) synthesizes rule-based systems and a rich tuplespace model with actor-based programming in a mobile, ad-hoc networking setting to yield a powerful form of context-aware programming. The initial implementation of the model, dubbed CRIME, integrates a RETE-based rule engine (Forgy 1982) with the TOTAM tuplespace and a functional reactive programming (FRP) library (Elliott and Hudak 1997; Bainomugisha et al. 2013) atop AmbientTalk, an object-oriented actor language in the style of E (Van Cutsem et al. 2014). AmbientTalk is unusual among actor languages for its consideration of multicast communication and coordination. In its role as "language laboratory", it has incorporated ideas from many other programming paradigms. AmbientTalk adds distributed service discovery, error handling, anycast and multicast to an actor-style core language intended for a mobile, ad-hoc network context; TOTAM supplements this with a distributed database, and the rule engine brings logic programming to the table.

In the words of Mostinckx et al.,

> The Fact Space model is a coordination model which provides applications with a federated fact space: a distributed knowledge base containing logic facts which are implicitly made available for all devices within reach. [...] [T]he Fact Space model combines the notion of a federated fact space with a logic coordination language. (Mostinckx, Lombide Carreton and De Meuter 2008)

Tuples placed within the TOTAM tuplespace are interpreted as ground facts in the Prolog logic-programming sense. Insertions correspond to Prolog's `assert`; removals to `retract`. TOTAM's reactions, which unlike LIME may be triggered on either insertion or removal of tuples, allow connection of changes in the tuplespace to the inputs to the RETE-based rule engine, yielding forward-chaining logic programming driven by activity in the common space.[6]

Figure 9 sketches a pseudo-code user agent program. An actor running `userAgent` is created for each connecting user. As it starts up, it registers two reactions. The first (lines 2–3) reacts to appearance *and disappearance* of `Present` tuples. The second (line 4) reacts to each `Message` tuple appearing in the space. Line 5 places a `Present` tuple representing the current user in the tuplespace, where it will be detected by peers. Lines 6–10 enter a loop, waiting for user input and replacing the user's previous `Message`, if any, with a new one.

6 When facts in the space are chosen to correspond to observable aspects of a program's context, this yields *context-aware programming*: programs react to relevant changes in their environment.

```
def userAgent(name, socket)
  whenever: [Present, ?who] read: { socket.println(who + " arrived") }
                  outOfContext: { socket.println(who + " left") }
  whenever: [Message, ?who, ?what] read: { socket.println(who + ": " + what) }

  publish: [Present, name]

  previousLine = nil
  while (line = socket.readLine()) != nil
    if previousLine != nil: inp([Message, name, previousLine])
    publish: [Message, name, line]
    previousLine = line
```

Figure 9: Fact space style chat room user agent

(C1; C2; C7) The TOTAM tuplespace offers multi-party communication, and the rule engine allows installation of pattern-based reactions to events, resulting in automatic demultiplexing and making for a natural connection from data flow to associated control flow. Where an interaction serves to open a conversational frame for a sub-conversation, additional reactions may be installed; however, there is no linguistic representation of such conversational frames, meaning that any logical association between conversations must be manually expressed and maintained.

(C3; C4) AmbientTalk's *reactive context-aware collections* (Mostinckx, Lombide Carreton and De Meuter 2008) allow automatic integration of conclusions drawn by the rule engine with collection objects such as sets and hash tables. Each collection object is manifested as a *behavior* in FRP terminology, meaning that changes to the collection can in turn trigger downstream reactions depending on the collection's value. However, achieving the effect of propagating changes in local variables as changes to tuples in the shared space is left to programmers.

(C5; C6; C8) The Fact Space model removes tuples upon component failure. Conclusions drawn from rules depending on removed facts are withdrawn in turn. Programs thereby enjoy logical consistency after partial failure. However, automatic retraction of tuples is performed *only* in cases of disconnection. When a running component is engaged in multiple conversations, and one of them comes to a close, there is no mechanism provided by which facts relating to the terminated conversation may be automatically cleaned up. Programmers manually delete obsolescent facts or turn to a strategy borrowed from the E language, namely creation of a separate actor for each sub-conversation. If they choose this option, however, the interactions among the resulting plethora of actors may increase overall system complexity.

(C9) The ability to react to removal as well as insertion of tuples allows programs to match supply of some service to demand, by interpreting particular assertions as demand for some resource. This can, in principle, allow automatic resource management; however, this is only true if all allocation of and interaction with such resources is done via the tuple space. For example, if the actor sketched in figure 9 were to crash, then absent explicit exception-handling code, the connected socket would leak, remaining open.[7] Additionally, in situations where

7 https://soft.vub.ac.be/amop/crime/download

```
1  def userAgent(name, socket)
2    presentUsers = set()
3    whenever: [Present, ?who, ?status]
4      read: {
5        if who not in presentUsers
6          presentUsers.add(who)
7          socket.println(who + " arrived")
8        socket.println(who + " status: " + status)
9      }
10     outOfContext: {
11       if rd([Present, who, ?anyStatus]) == nil
12         presentUsers.remove(who)
13         socket.println(who + " left")
14     }
15   ...
```

Figure 10: Aggregating distinct facts

tuples may be interpreted simultaneously at a coarse-grained and fine-grained level, some care must be taken in interpreting tuple arrival and departure events. For example, imagine a slight enhancement of our example program, where we include a user-chosen *status* message in our `Present` tuples. In order to react both to appearance and disappearance of a user as well as a change in a user's status, we must interpret `Present` tuples as sketched in figure 10. There, `Present` tuples are aggregated by their `who` fields, ignoring their `status` fields, in addition to being interpreted entire. The `presentUsers` collection serves as intermediate state for a kind of `SELECT DISTINCT` operation, indicating whether any `Present` tuples for a particular user exist at all in the tuplespace. In the retraction handler (lines 10–14) we explicitly check whether any `Present` tuples for the user concerned remain in the space, only updating `presentUsers` if none are left. This avoids incorrectly claiming that a user has left the chat room when they have merely altered their status message.

An alternative approach to the problem is to make use of a feature of CRIME not yet described. The CRIME implementation of the fact space model exposes the surface syntax of the included rule engine to the programmer, allowing logic program fragments to be written using a Prolog-like syntax and integrated with a main program written in AmbientTalk. This could allow a small program

```
UserPresent(?who) :- Present(?who,?status).
```

to augment the tuplespace with `UserPresent` tuples whenever any `Present` tuple for a given user exists at all. On the AmbientTalk side, programs would then react separately to appearance and disappearance of `UserPresent` and `Present` tuples.

(C10) Like tuplespaces generally, the fact space model has great potential for tool support and system state visualization. However, only those aspects of a program communicating via the underlying tuplespace benefit from its invariants. In the case of the CRIME implementation based on AmbientTalk, only selected inter-component interactions travel via the tuplespace and rule engine, leaving other interactions out of reach of potential fact-space-based tools. Program-mers must carefully combine reasoning based on the invariants of the fact space model with the properties of the other mechanisms available for programmer use, such as AmbientTalk's own inter-actor message delivery, service discovery and broadcast facilities.

(C11) Extending a conversation to new components and introducing an existing component to an additional conversation are both readily supported by the fact space model as implemented in CRIME. However, because no automatic support for release of conversation-associated state exists (other than outright termination of an entire actor), programmers must carefully consider the interactions among individual components. When one of an actor's conversations comes to a close but other conversations remain active, the programmer must make sure to release local conversational state and remove associated shared tuples, but only when they are provably inaccessible to the remaining conversations.

(C12) CRIME's AmbientTalk foundation is inspired by E, and can benefit directly from research done on persistence and object upgrade in E-like settings (Yoo et al. 2012; Miller, Van Cutsem and Tulloh 2013).

## 3.6 SURVEYING THE LANDSCAPE

Figure 11 summarizes this chapter's analysis. Each of the first four columns in the table shows, from the programmer's point of view, the support they can expect from a programming language taking the corresponding approach to concurrency. Each row corresponds to one of the properties of concurrency models introduced in figure 3. A few terms used in the table require explanation. An entry of "manual" indicates that the programmer is offered no special support for the property. An entry of "semi-automatic" indicates that some form of support for the property is available, at least for specialized cases, but that general support is again left to the programmer. For example, channel-based languages can automatically demultiplex conversations, but only so long as channels correspond one-to-one to conversations, and the fact space model automatically preserves integrity of conversational state, but only where the end of an actor's participation in a conversation is marked by disconnection from the shared space. Finally, an entry of "automatic" indicates that an approach to concurrency offers strong, general support for the property. An example is the fact space model's ability to integrate changes in the shared space with local variables via its reactive context-aware collections.

While the first four columns address the properties of existing models of concurrency, the final column of the table identifies an "ideal" point in design space for us to aim towards in the design of new models.

(C1; C2; C3; C4) We would like a flexible communications mechanism accommodating many-to-many as well as one-to-one conversations. A component should be able to engage in multiple conversations, without having to jump through hoops to do so. Events should map to event handlers directly in terms of their domain-level meaning. Since conversations come with conversational frames, and conversational frames scope state and behavior, such frames and their interrelationships should be explicit in program code. As conversations proceed, the associated conversational state evolves. Changes to that state should automatically be integrated with lo-cal views on it, and changes in local state should be able to be straightforwardly shared with peers. Agents should be offered the opportunity to react to all kinds of state changes.

| K3*bis* Mechanism | Shared memory | Message-passing | Tuplespaces | Fact spaces | Ideal |
|---|---|---|---|---|---|
| C1 Conversation group size | arbitrary | point-to-point | arbitrary | arbitrary | arbitrary |
| C2 Correlation/demultiplexing | manual | semi-automatic | semi-automatic | semi-automatic | automatic |
| C3 Integration of state change | automatic | manual | semi-automatic | automatic | automatic |
| C4 Signaling of state change | manual | manual | manual | manual | automatic |

| K4 Robustness | Shared memory | Message-passing | Tuplespaces | Fact spaces | Ideal |
|---|---|---|---|---|---|
| C5 Maintain state integrity | manual | manual | manual | semi-automatic | automatic |
| C6 Ensure replica consistency | trivial | manual | semi-automatic | semi-automatic | automatic |
| C7 Data $\Longrightarrow$ control flow | no | yes | yes | yes | yes |
| C8 Control $\Longrightarrow$ data flow | no | partial | no | coarse-grained | fine-grained |
| C9 Resource management | manual | manual | manual | coarse-grained | fine-grained |

| K6 Operability | Shared memory | Message-passing | Tuplespaces | Fact spaces | Ideal |
|---|---|---|---|---|---|
| C10 Debuggability/visualizability | poor | wide range | potentially good | potentially good | good |
| C11 Evolvability | poor | moderate | moderate | good | good |
| C12 Durability | poor | good | moderate/good | good | good |

Figure 11: Surveying the landscape

(C5; C6) We would like to automatically enforce application-level invariants regarding shared, conversational state. In case of partial failure, we should be able to identify and discard damaged portions of conversational state. Where replicas of a piece of conversational state exist, we would like to be able to reason about their mutual consistency. (C7; C8) Hewitt's criterion that "control and data flow are inseparable" should hold as far as possible, both in terms of control flow being manifestly influenced by data flow and in terms of translation of control effects such as exceptions into visible changes in the common conversational context. (C9) Since conversations often involve associated resources, we would like to be able to connect allocation and release of resources with the lifecycles of conversational frames.

(C10; C11; C12) Given the complexity of concurrent programming, we would like the ability to build tools to gain insight into system state and to visualize both correct and abnormal behavior for debugging and development purposes. Modification of our programs should easily accommodate changes in the scope of a given conversation among components, as well as changes to the set of interactions a given component is engaged in. Finally, robustness involves tolerance of partial failure and partial restarts; where long-lived application state exists, support for code upgrades should also be offered.

Part II

# THEORY

# Overview

SYNDICATE is a design in two parts. The first part is called the *dataspace model*. This model offers a mechanism for communication and coordination within groups of concurrent components, plus a mechanism for organizing such groups and relating them to each other in hierarchical assemblies. The second part is called the *facet model*. This model introduces new language features to address the challenges of describing an actor's participation in multiple simultaneous conversations.

Chapter 4 fleshes out the informal description of the dataspace model of section 2.5 with a formal semantics. The semantics describes a hierarchical structure of components in the shape of a tree. Intermediate nodes in the tree are called *dataspaces*. From the perspective of the dataspace model, leaf nodes in the tree are modeled as (pure) event-transducer functions; their internal structure is abstracted away.

Chapter 5 describes the facet model part of the SYNDICATE design, addressing the internal structure of the leaf actors of the dataspace model. Several possible means of interfacing a programming language to a dataspace exist. The simplest approach is to directly encode the primitives of the model in the language of choice, but this forces the programmer to attend to much detail that can be handled automatically by a suitable set of linguistic constructs. The chapter proposes such constructs, augments a generic imperative language model with them, and gives a formal semantics for the result. Together, the dataspace and facet models form a complete design for extending a non-concurrent host language with concurrency.

# 4

# *Computational Model I: The Dataspace Model*

This chapter describes the dataspace model using mathematical syntax and semantics, including theorems about the model's key properties. The goal of the model presented here is to articulate a language design idea. We wish to show how to construct a concurrent language from a generic base language via the addition of a fixed communication layer. The details of the base language are not important, and are thus largely abstracted away. We demand that the language be able to interpret dataspace events, encode dataspace actions, map between its internal data representations and the assertion values of the dataspace model, and confine its computational behavior to that expressible with a total mathematical function. We make the state of each actor programmed in the (extended) base language explicit, require that its behavior be specified as a state-transition function, and demand that it interacts with its peers exclusively via the exchange of immutable messages—not by way of effects. This strict enforcement of message-passing discipline does not prevent us from using an imperative base language, as long as its effects do not leak. In other words, the base could be a purely functional language such as Haskell, a higher-order imperative language such as Racket, or an object-oriented language such as JavaScript.

The dataspace model began life under the moniker "Network Calculus" (NC) (Garnock-Jones, Tobin-Hochstadt and Felleisen 2014), a formal model of publish-subscribe networking incorporating elements of *presence* as such, rather than the more general state-replication system described in the follow-up paper (Garnock-Jones and Felleisen 2016) and refined in this dissertation. The presentation in this chapter draws heavily on that of the latter paper, amending it in certain areas to address issues that were not evident at the time.

## 4.1 ABSTRACT DATASPACE MODEL SYNTAX AND INFORMAL SEMANTICS

Figure 12 displays the syntax of dataspace model programs. Each program $\mathsf{P}$ is an instruction to create a single actor: either a *leaf actor* or a *dataspace actor*. A leaf actor has the shape $\mathsf{actor}\ f_{boot}\ \pi$. Its initial assertions are described by the set $\pi$, while its boot function $f_{boot}$ embodies the first few computations the actor will perform. The boot function usually yields an $\mathsf{init}(\cdot)$ record specifying a sequence of initial actions $\overrightarrow{a} \in \overrightarrow{\mathbf{Act}}$ along with an existentially-quantified package $\mathsf{pack}\ \langle \tau, (f_{beh}, u) \rangle$. This latter specifies the type $\tau$ of the actor's private state, the initial private state value $u \in \tau$, and the actor's permanent event-transducing behavior function $f_{beh} \in \mathcal{F}_\tau$. Alternatively, the boot function may decide that the actor should immediately terminate, in which case it yields an $\mathsf{exit}(\cdot)$ record bearing a sequence of *final* actions $\overrightarrow{a} \in \overrightarrow{\mathbf{Act}}$ for the short-

$$\begin{aligned}
\text{Programs } P \in \textbf{Prog} &::= \mathsf{actor}\ f_{boot}\ \pi \mid \mathsf{dataspace}\ \overrightarrow{P} \\
\text{Events } e \in \textbf{Evt} &::= \langle c \rangle \mid \pi \\
\text{Actions } a \in \textbf{Act} &::= \langle c \rangle \mid \pi \mid P
\end{aligned}$$

$$\begin{aligned}
\text{Boot functions } f_{boot} \in \textbf{Boot} &= \mathbf{1} \to \mathsf{init}(\overrightarrow{\textbf{Act}} \times \exists\tau.(\mathcal{F}_\tau \times \tau)) + \mathsf{exit}(\overrightarrow{\textbf{Act}}) \\
\text{Behavior functions } f_{beh} \in \mathcal{F}_\tau &= \textbf{Evt} \times \tau \to \mathsf{continue}(\overrightarrow{\textbf{Act}} \times \tau) + \mathsf{exit}(\overrightarrow{\textbf{Act}})
\end{aligned}$$

$$\begin{aligned}
\text{Assertion/Message values } v, c \in \textbf{Val} &::= b \mid (c, \dots) \\
\text{Assertion sets } \pi \in \textbf{ASet} &= \mathcal{P}(\textbf{Val}) \\
\text{Base values } b \in \textbf{BVal} &= \text{Atoms, incl. strings, symbols, numbers, etc.}
\end{aligned}$$

$$\begin{aligned}
?c &\triangleq (\mathtt{observe}, c) \\
\downharpoonleft c &\triangleq (\mathtt{outbound}, c) \\
\upharpoonleft c &\triangleq (\mathtt{inbound}, c)
\end{aligned}$$

Figure 12: Syntax of dataspace model programs

lived actor to perform before it becomes permanently inert.[1] A dataspace actor has the shape $\mathsf{dataspace}\ \overrightarrow{P}$ and creates a group of communicating actors sharing a new assertion store. Each $P$ in the sequence of programs contained in the definition of a dataspace actor will form one of the initial actors placed in the group as it starts its existence.

Each leaf actor behavior function consumes an event plus its actor's current state. The function computes either a $\mathsf{continue}(\cdot)$ record, namely a sequence of desired actions plus an updated state value, or an $\mathsf{exit}(\cdot)$ record carrying a sequence of desired *final* actions alone in case the actor decides to request its own termination. We require that such behavior functions be total. If the base language supports exceptions, any uncaught exceptions or similar must be translated into an explicit termination request. If this happens, we say that the actor has crashed, even though it returned a valid termination request in an orderly way.

In the λ-calculus, a program is usually a combination of an inert part—a function value—and an input value. In the dataspace model, delivering an event to an actor is analogous to such an application. However, the pure λ-calculus has no analogue of the actions produced by dataspace model actors.

---

1 Design note: An alternative, roughly equivalent design omits **Boot** in favor of $\mathsf{actor}$ carrying some $\tau$, $f_{beh} \in \mathcal{F}_\tau$, and $u \in \tau$ directly, with that $f_{beh}$ receiving a distinct, one-time *startup* event. Yet another option is to define $\mathcal{F}_\tau$ to yield $\mathsf{continue}(\overrightarrow{\textbf{Act}} \times \exists\tau.(\mathcal{F}_\tau \times \tau))$, giving a "become"-like semantics (Agha et al. 1997). Neither variation simplifies the presentation. I have chosen the variation described because it seems to me to capture the idea of a one-time, staged startup computation without sacrificing a fixed behavior function or introducing a startup pseudo-event.

A dataspace model actor may produce actions like those in the traditional actor model, namely sending messages $\langle c \rangle$ and spawning new actors $P$, but it may also produce *state change notifications (SCNs)* $\pi$. These convey sets of assertions an actor wishes to publish to its containing dataspace.

As a dataspace interprets an SCN action, it updates its assertion store. It tracks every assertion made by each contained actor. It not only maps each actor to its current assertions, but each active assertion to the set of actors asserting it. The assertions of each actor, when combined with the assertions of its peers, form the overall set of assertions present in the dataspace.

When an actor issues an SCN action, the new assertion set completely replaces all previous assertions made by that actor. To retract an assertion, the actor issues a state change notification action lacking the assertion concerned. For example, imagine an actor whose most-recently-issued SCN action conveyed the assertion set $\{a, b, c\}$. By issuing an SCN action $\{a, b\}$, the actor would achieve the effect of retracting the assertion $c$. Alternatively, issuing an SCN $\{a, b, c, d\}$ would augment the actor's assertion set in the assertion store with a new assertion $d$. Finally, the SCN $\{a, b, d\}$ describes assertion of $d$ simultaneous with retraction of $c$.

We take the liberty of using wildcard $\star$ as a form of assertion set comprehension. For now, when we write expressions such as $\{(\mathtt{a}, \star)\}$, we mean the set of all pairs having the atom `a` on the left. In addition, we use three syntactic shorthands as constructors for commonly-used structures: $?c$, $\downharpoonleft c$ and $\upharpoonleft c$ are abbreviations for tuples of the atoms `observe`, `outbound` and `inbound`, respectively, with the value $c$. Thus, $\{?\star\}$ means $\{?c \mid c \in \mathbf{Val}\}$.[2]

When an actor issues an assertion of shape $?c$, it expresses an interest in being informed of all assertions $c$. In other words, an assertion $?c$ acts as a subscription to $c$. Similarly, $??c$ specifies interest in being informed about assertions of shape $?c$, and so on. The dataspace sends a state change notification *event* to an actor each time the set of assertions matching the actor's interests changes.

An actor's subscriptions are assertions like any other.[3] State change notifications thus give an actor control over its subscriptions as well as over any other information it wishes to make available to its peers or acquire from them.

DATASPACE "ISWIM". The examples in this chapter use a mathematical notation to highlight the essential aspects of the coordination abilities of the dataspace model without dwelling on base language details. While the notation used is not a real language (if you see what I mean (Landin 1966)), it does have implemented counterparts in the prototypes of the dataspace model that incorporate Racket and JavaScript as base languages. These implementations were used to write programs which in turn helped build intuition and serve as a foundation for the full SYNDICATE design.

We use *italic* text to denote Dataspace ISWIM variables and `monospace` to denote literal atoms and strings. In places where the model demands a sequence of values, for example the actions returned from a behavior function, our language supplies a single list value $[a_1, ..., a_n]$. We include list comprehensions $[a \mid a \in \mathbf{Act}, P(a), ...]$ because actors frequently need to construct,

2 Clearly, implementers must take pains to keep representations of sets specified in this manner tractable. We discuss this issue in more detail in section 7.1.

3 However, unlike most other assertions, they directly represent epistemic knowledge.

filter, and transform sequences of values. Similarly, we add syntax for sets $\{c_1, ..., c_n\}$, including set comprehensions $\{c \,|\, c \in \mathbf{Val}, P(c), ...\}$, and for tuples $(v_1, ..., v_n)$, to represent the sets and tuples needed by the model.

We define functions using patterns over the language's values. For example, the leaf behavior function definition

$$\mathit{box}\ (\langle(\mathtt{set}, \mathit{id}, v_c)\rangle, v_o)\ = \mathsf{continue}([\{?(\mathtt{set}, \mathit{id}, \star), (\mathtt{value}, \mathit{id}, v_c)\}], v_c)$$

introduces a function *box* that expects two arguments: a message and an arbitrary value. The $\langle(\mathtt{set}, \mathit{id}, v_c)\rangle$ pattern for the former says it must consist of a triple with the atom `set` on the left and arbitrary values in the center and right field. The function yields a $\mathsf{continue}(\cdot)$ record—it wishes to continue running—containing a pair whose left field is a sequence of actions and whose right field is the actor's new state value $v_c$. The sequence of actions consists of only one element: a state change notification action bearing an assertion set. The assertion set is written in part using a wildcard denoting an infinite set, and in part using a simple value. The resulting assertion set thus contains not only the triple $(\mathtt{value}, \mathit{id}, v_c)$ but also the infinite set of all ?-labeled triples with `set` on the left and with *id* in the middle.

**Example 4.1.** Suppose we wish to create an actor X with an interest in the price of milk. Here is how it might be written:

$$\mathsf{actor}\ f_{\mathit{bootX}}\ \{?(\mathtt{price}, \mathtt{milk}, \star)\}$$

The comprehension defining its initial assertion set is interpreted to denote the set

$$\{?(\mathtt{price}, \mathtt{milk}, c) \mid c \in \mathbf{Val}\}$$

If some peer Y previously asserted $(\mathtt{price}, \mathtt{milk}, 1.17)$, this assertion is immediately delivered to X in a state change notification event. Infinite sets of interests thus act as *query patterns* over the shared dataspace.

Redundant assertions do not cause change notifications. If actor Z subsequently also asserts $(\mathtt{price}, \mathtt{milk}, 1.17)$, no notification is sent to X, since X has already been informed that $(\mathtt{price}, \mathtt{milk}, 1.17)$ has been asserted. However, if Z instead asserts $(\mathtt{price}, \mathtt{milk}, 9.25)$, then a change notification is sent to X containing *both* asserted prices.

Symmetrically, it is not until the last assertion of shape $(\mathtt{price}, \mathtt{milk}, p)$ for some particular $p$ is retracted from the dataspace that X is sent a notification about the lack of assertions of shape $(\mathtt{price}, \mathtt{milk}, p)$.

When an actor crashes, all its assertions are automatically retracted. By implication, if no other actor is making the same assertions at the time, then peers interested in the crashing actor's assertions are sent a state change notification event informing them of the retraction(s). ♢

**Example 4.2.** For a different example, consider an actor representing a shared mutable reference cell. A new box (initially containing 0) is created by choosing a name *id* and launching the actor

$$\mathsf{actor}\ (\lambda().\mathsf{init}([], \mathsf{pack}\ \langle\mathbf{Val}, (\mathit{box}, 0)\rangle))\ \{?(\mathtt{set}, \mathit{id}, \star), (\mathtt{value}, \mathit{id}, 0)\}$$

The new actor's initial assertion set includes assertions of interest in `set` messages labeled with *id* as well as of the fact that the `value` of box *id* is currently 0. Its behavior is given by the function *box* whose definition we saw earlier, its initial actions by the empty sequence, and its initial state is just 0. Upon receipt of a `set` message bearing a new value $\nu_c$, we may read off its response by consulting the definition of *box* above. The actor replaces its private state value with $\nu_c$ and constructs a single action specifying the new set of facts the actor wants to assert. This new set of facts includes the unchanged `set`-message subscription as well as a new `value` fact, thereby replacing $\nu_o$ with $\nu_c$ in the shared dataspace.

To read the value of the box, clients either include an appropriate assertion in their initially declared interests or issue it later:

$$\mathsf{actor}\ \ (\lambda().\mathsf{init}([], \mathsf{pack}\,\langle \mathbf{1}, (\mathit{boxClient}, ())\rangle))\ \ \{?(\mathtt{value}, \mathit{id}, \star)\}$$

As corresponding facts come and go in response to actions taken by the box actor they are forwarded to interested parties. For example, the *boxClient* behavior function responds to notification of a change in the contents of the box by issuing an instruction to update the box:

$$\mathit{boxClient}\ (\{(\mathtt{value}, \mathit{id}, \nu)\}, ())\ \ = \mathsf{continue}([\langle(\mathtt{set}, \mathit{id}, \nu + 1)\rangle], ())$$

The behavior of the *box* and *boxClient* actors, when run together in a dataspace, is to repeatedly increment the number held in the *box*. ◊

**Example 4.3.** Our next example demonstrates *demand matching*. The need to measure demand for some service and allocate resources in response appears in different guises in a wide variety of concurrent systems. Here, we imagine a client, `A`, beginning a conversation with some service by adding (`hello`, `A`) to the shared dataspace. In response, the service should create a worker actor to talk to `A`.

The "listening" part of the service is spawned as follows:

$$\mathsf{actor}\ \ (\lambda().\mathsf{init}([], \mathsf{pack}\,\langle \mathbf{ASet}, (\mathit{demandMatcher}, \emptyset)\rangle))\ \ \{?(\mathtt{hello}, \star)\}$$

Its behavior function is defined as follows:

$$\mathit{demandMatcher}\ (\pi_{new}, \pi_{old})\ \ = \mathsf{continue}([\mathit{mkWorker}\, x \mid (\mathtt{hello}, x) \in \pi_{new} - \pi_{old}], \pi_{new})$$

The actor-private state of *demandMatcher*, $\pi_{old}$, is the (initially empty) set of currently-asserted `hello` tuples.[4] The incoming event, $\pi_{new}$, is the newest version of that set from the environment. The demand matcher performs set subtraction to determine newly-appeared requests and calls a helper function *mkWorker* to produce a matching service actor for each:

$$\begin{aligned} \mathit{mkWorker}\, x &= \mathsf{actor}\ \ (\lambda().\mathsf{init}(\mathit{initialActionsFor}\, x, \mathsf{pack}\,\langle \tau, (\mathit{worker}, s)\rangle))\ \ \emptyset \\ &\quad \text{where } s = \mathit{initialStateFor}\, x \in \tau \text{ and } \mathit{worker} \in \mathcal{F}_\tau \end{aligned}$$

Thus, when (`hello`, `A`) first appears as a member of $\pi_{new}$, the demand matcher invokes *mkWorker* with `A` as an argument, which yields a request to create a new worker actor that talks to client

4 Implementations of the dataspace model to date internalize assertion sets as *tries* (section 7.1)

`A`. The conversation between `A` and the new worker proceeds from there. A more sophisticated implementation of demand matching might maintain a pool of workers, allocating incoming conversation requests as necessary. ◊

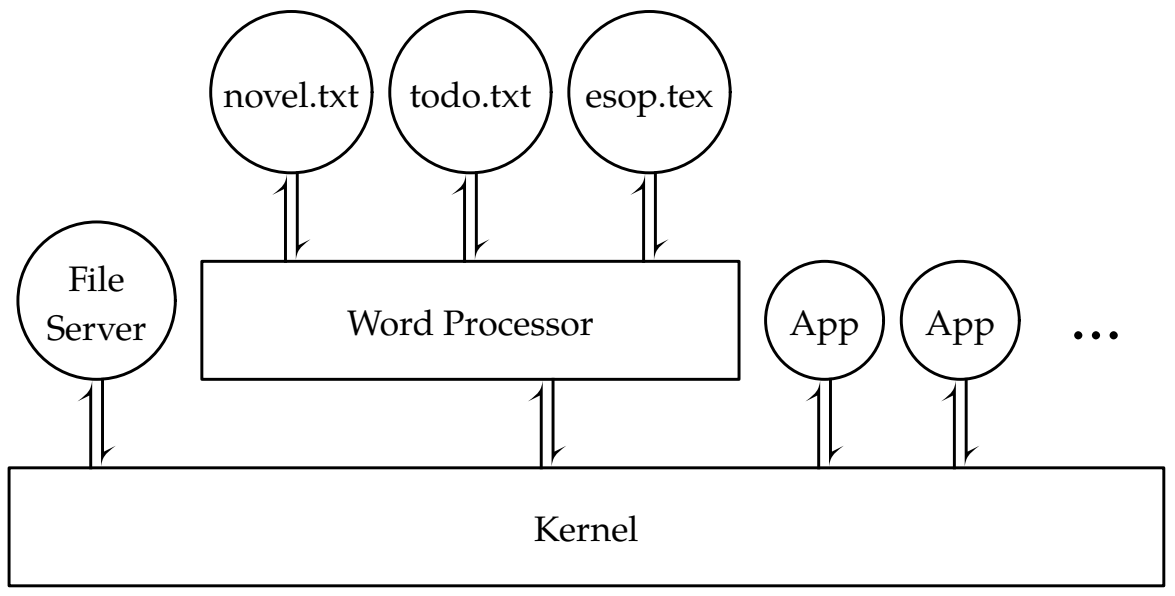

Figure 13: Layered File Server / Word Processor architecture

**Example 4.4.** Our final example demonstrates an architectural pattern seen in operating systems, web browsers, and cloud computing. Figure 13 sketches the architecture of a program implementing a word processing application with multiple open documents, alongside other applications and a *file server* actor. The "Kernel" dataspace is at the bottom of this tree-like representation of containment.

The hierarchical nature of the dataspace model means that each dataspace has a containing dataspace in turn. Actors may interrogate and augment assertions held in containing dataspaces by prefixing assertions relating to the $n$th relative dataspace layer with $n$ "`outbound`" markers $\downharpoonleft$. Dataspaces relay $\downharpoonleft$-labeled assertions outward. Some of these assertions may describe interest in assertions existing at an outer layer. Any assertions matching such interests are relayed back in by the dataspace, which prefixes them with an "`inbound`" marker $\upharpoonleft$ to distinguish them from local assertions.

In this example, actors representing open documents communicate directly with each other via a local dataspace scoped to the word processor, but only indirectly with other actors in the system. When the actor for a document decides that it is time to save its content to the file system, it issues a message such as

$$\langle \downharpoonleft (\texttt{save}, \texttt{"novel.txt"}, \texttt{"Call me Ishmael."}) \rangle$$

into its local dataspace. The harpoon ($\downharpoonleft$) signals that, like a system call in regular software applications, the message is intended to be relayed to the *next outermost* dataspace—the medium connecting the word processing application as a whole to its peers. Once the message is relayed, the message

$$\langle (\texttt{save}, \texttt{"novel.txt"}, \texttt{"Call me Ishmael."}) \rangle$$

is issued into the outer dataspace, where it may be processed by the file server. The harpoon is removed as part of the relaying operation, and no further harpoons remain, indicating that the message should be processed *here*, at this dataspace.

The file server responds to two protocols, one for writing files and one for reading file contents and broadcasting changes to files as they happen. These protocols are articulated as two subscriptions:

$$\{?(\texttt{save},\star,\star),??(\texttt{contents},\star,\star)\}$$

The first indicates interest in `save` messages. When a `save` message is received, the server stores the updated file content.

The second indicates interest in *subscriptions* in the shared dataspace, an interest in *interest* in file contents. This is how the server learns that peers wish to be kept informed of the contents of files under its control. The file server is told each time some peer asserts interest in the contents of a file. In response, it asserts facts of the form

$$(\texttt{contents},\texttt{"novel.txt"},\texttt{"Call me Ishmael."})$$

and keeps them up-to-date as `save` commands are received, finally retracting them when it learns that peers are no longer interested. In this way, the shared dataspace not only acts as a kind of cache for the files maintained on disk, but also doubles as an `inotify`-like mechanism (Love 2005) for signaling changes in files. ♢

Our examples illustrate the key properties of the dataspace model and their unique combination. Firstly, the box and demand-matcher examples show that conversations may naturally involve many parties, generalizing the actor model's point-to-point conversations. At the same time, the file server example shows that conversations are more precisely bounded than those of traditional actors. Each of its dataspaces crisply delimits its contained conversations, each of which may therefore use a task-appropriate language of discourse.

Secondly, all three examples demonstrate the shared-dataspace aspect of the model. Assertions made by one actor can influence other actors, but cannot directly alter or remove assertions made by others. The box's content is made visible through an assertion in the dataspace, and any actor that knows *id* can retrieve the assertion. The demand-matcher responds to changes in the dataspace that denote the existence of new conversations. The file server makes file contents available through assertions in the (outer) dataspace, in response to clients placing subscriptions in that dataspace.

Finally, the model places an upper bound on the lifetimes of entries in each shared space. Items may be asserted and retracted by actors at will in response to incoming events, but when an actor crashes, all of its assertions are automatically retracted.[5] If the box actor were to crash during a computation, the assertion describing its content would be visibly withdrawn, and peers could take some compensating action. The demand matcher can be enhanced to monitor *supply* as well as demand and to take corrective action if some worker instance exits unexpectedly. The combination of this temporal bound on assertions with the model's state change notifications gives good failure-signaling and fault-tolerance properties, improving on those seen in Erlang (Armstrong 2003).

$$
\begin{array}{rll}
\text{Dataspaces } C \in \mathbf{Cfg} ::= [\overrightarrow{q}; R; \overrightarrow{A}] & & C_I \in \mathbf{Cfg_I} ::= [\cdot; R; \overrightarrow{A_I}] \\
\text{Actors } A \in \mathbf{Actor} ::= \ell \mapsto \Sigma & A_Q \in \mathbf{Actor_Q} ::= \ell \mapsto \Sigma_Q & A_I \in \mathbf{Actor_I} ::= \ell \mapsto \Sigma_I \\
\text{States } \Sigma \in \mathbf{State} ::= \langle \overrightarrow{e} \triangleright B \triangleright \overrightarrow{a} \rangle & \Sigma_Q \in \mathbf{State_Q} ::= \langle \overrightarrow{e} \triangleright B \triangleright \cdot \rangle & \Sigma_I \in \mathbf{State_I} ::= \langle \cdot \triangleright B_I \triangleright \cdot \rangle \\
\text{Behaviors } B \in \mathbf{Beh} = \exists \tau.(\mathcal{F}_\tau \times \tau) \cup \mathbf{Cfg} & & B_I \in \mathbf{Beh_I} = \exists \tau.(\mathcal{F}_\tau \times \tau) \cup \mathbf{Cfg_I} \\
& \underbrace{\qquad\qquad\qquad\qquad}_{\text{Quiescent}} & \underbrace{\qquad\qquad\qquad\qquad}_{\text{Inert}}
\end{array}
$$

$$
\begin{array}{r}
\text{Queued Actions } q \in \mathbf{QAct} ::= (k, a) \\
\text{Dataspace Contents } R \in \mathbf{Space} = \mathcal{P}(\mathbf{ID} \times \mathbf{Val}) \\
\text{Peer Identifiers } j, k \in \mathbf{ID} ::= \ell \mid \downharpoonleft \\
\text{Locations } \ell \in \mathbf{Loc} = \mathbb{N}
\end{array}
$$

$$
\mathsf{boot} : \ \mathbf{Prog} \to \mathbf{State} \times \mathbf{ASet}
$$

$$
\mathsf{boot}\ (\mathsf{actor}\ f_{boot}\ \pi) = \begin{cases} (\langle \cdot \triangleright \mathsf{pack}\ \langle \tau, (f_{beh}, u) \rangle \triangleright \overrightarrow{a} \rangle, \pi) & \text{when } f_{boot}() = \mathsf{init}(\overrightarrow{a}, \mathsf{pack}\ \langle \tau, (f_{beh}, u) \rangle) \\ (\langle \cdot \triangleright \mathsf{pack}\ \langle \mathbf{1}, (\mathsf{noop}, ()) \rangle \triangleright \emptyset \overrightarrow{a} \rangle, \pi) & \text{when } f_{boot}() = \mathsf{exit}(\overrightarrow{a}) \end{cases}
$$

$$
\mathsf{boot}\ (\mathsf{dataspace}\ \overrightarrow{P}) = (\langle \cdot \triangleright [\overrightarrow{(\downharpoonleft, P)}; \emptyset; \cdot] \triangleright \cdot \rangle, \emptyset)
$$

$$
\begin{array}{r}
\mathsf{noop} : \ \mathcal{F}_{\mathbf{1}} \\
\mathsf{noop}\ \ (e, ()) = \mathsf{continue}(\cdot\ , ())
\end{array}
$$

Figure 14: Evaluation Syntax and Inert and Quiescent Terms

## 4.2 formal semantics of the dataspace model

The semantics of the dataspace model is most easily understood via an abstract machine. Figure 14 shows the syntax of machine configurations, plus a metafunction boot, which loads programs in **Prog** into starting machine states in **State**, and an inert behavior function noop.

The reduction relation operates on actor states $\Sigma = \langle \vec{e} \triangleright B \triangleright \vec{a} \rangle$, which are triples of a queue of events $\vec{e}$ destined for the actor, the actor's behavior and internal state B, and a queue of actions $\vec{a}$ issued by the actor and destined for processing by its containing dataspace. An actor's behavior and state B can take on one of two forms. For a leaf actor, behavior and state are kept together with the type of the actor's private state value in an existential package $B = \mathsf{pack}\,\langle \tau, (f_{beh}, u) \rangle \in \exists\tau.(\mathcal{F}_\tau \times \tau)$. For a dataspace actor, behavior is determined by the reduction rules of the model, and its state is a configuration $B \in$ **Cfg**.

Dataspace configurations C comprise three registers: a queue of actions to be performed $\vec{q}$, each labeled with some *identifier* denoting the origin of the action; the current contents of the assertion store R; and a sequence of actors $\overrightarrow{\ell \mapsto \Sigma}$ residing within the configuration. Each actor is assigned a local *label* $\ell$, also called a *location*, scoped strictly to the configuration and meaningless outside. Labels are required to be locally-unique within a given configuration. They are never made visible to leaf actors: labels are an internal matter, used solely as part of the behavior of dataspace actors. The identifiers marking each queued action in the configuration are either the labels of some contained actor or the special identifier $\lrcorner$ denoting an action resulting from some external force, such as an event arriving from the configuration's containing configuration.

reduction relation. The reduction relation drives actors toward *quiescent* and even *inert* states. Figure 14 defines these syntactic classes, which are roughly analogous to values in the call-by-value λ-calculus. A state Σ is quiescent when its sequence of actions is empty, and it is inert when, besides being quiescent, it has no more events to process and cannot take any further internal reductions.

The reductions of the dataspace model are defined by the following rules. For convenient reference, the rules are also shown together in figure 15. Rules notify-leaf and quit deliver an event to a leaf actor and update its state based on the results. Rule notify-ds delivers an event to a *dataspace* actor. Rule gather collects actions produced by contained actors in a dataspace to a central queue, and rules newtable, message, and spawn interpret previously-gathered actions. Finally, rule schedule allows contained actors to take a step if they are not already inert.

**Definition 4.5** (Rule notify-leaf)**.** A leaf actor's behavior function, given event $e_0$ and private state value u, may yield a continue() instruction, i.e. $f_{beh}(e_0, u) = \mathsf{continue}(\vec{a}\,', u')$. In this case, the actor's state is updated in place and newly-produced actions are enqueued for processing:

$$\langle \vec{e}\, e_0 \triangleright \mathsf{pack}\,\langle \tau, (f_{beh}, u) \rangle \triangleright \vec{a} \rangle \longrightarrow \langle \vec{e} \triangleright \mathsf{pack}\,\langle \tau, (f_{beh}, u') \rangle \triangleright \vec{a}\,' \vec{a} \rangle$$

5 This is a concept well-known in the networking community as *fate-sharing* (Clark 1988).

$$\langle \vec{e}\, e_0 \triangleright \mathsf{pack}\,\langle \tau, (f_{beh}, u)\rangle \triangleright \vec{a}\rangle \longrightarrow \langle \vec{e} \triangleright \mathsf{pack}\,\langle \tau, (f_{beh}, u')\rangle \triangleright \vec{a}\,'\vec{a}\rangle \quad \text{when } f_{beh}(e_0, u) = \mathsf{continue}(\vec{a}\,', u') \quad \text{(notify-leaf)}$$

$$\langle \vec{e}\, e_0 \triangleright \mathsf{pack}\,\langle \tau, (f_{beh}, u)\rangle \triangleright \vec{a}\rangle \longrightarrow \langle \vec{e} \triangleright \mathsf{pack}\,\langle \mathbf{1}, (\mathsf{noop}, ())\rangle \triangleright \emptyset\,\vec{a}\,'\vec{a}\rangle \quad \text{when } f_{beh}(e_0, u) = \mathsf{exit}(\vec{a}\,') \quad \text{(quit)}$$

$$\langle \vec{e}\, e_0 \triangleright [\cdot; R; \vec{A_I}] \triangleright \vec{a}\rangle \longrightarrow \langle \vec{e} \triangleright [(\downarrow, \mathsf{inp}\ e_0); R; \vec{A_I}] \triangleright \vec{a}\rangle \quad \text{(notify-ds)}$$

$$\begin{aligned} &\langle \vec{e} \triangleright [\ \vec{q}; R; \vec{A_Q}(\ell \mapsto \langle \vec{e}\,' \triangleright B \triangleright \vec{a}\,'a''\rangle)\vec{A}] \triangleright \vec{a}\rangle \\ \longrightarrow\ &\langle \vec{e} \triangleright [(\ell, a'')\vec{q}; R; \vec{A_Q}(\ell \mapsto \langle \vec{e}\,' \triangleright B \triangleright \vec{a}\,'\rangle)\vec{A}] \triangleright \vec{a}\rangle \end{aligned} \quad \text{(gather)}$$

$$\begin{aligned} &\langle \vec{e} \triangleright [\vec{q}(k, \pi); R; \vec{A_Q}] \triangleright \vec{a}\rangle \\ \longrightarrow\ &\langle \vec{e} \triangleright [\vec{q}; R \oplus (k, \pi); \overrightarrow{\mathsf{bc}\ k\ \pi\ R\ A_Q}] \triangleright (\mathsf{out}\ k\ \pi\ R)\vec{a}\rangle \end{aligned} \quad \text{(newtable)}$$

$$\begin{aligned} &\langle \vec{e} \triangleright [\vec{q}(k, \langle c\rangle); R; \vec{A_Q}] \triangleright \vec{a}\rangle \\ \longrightarrow\ &\langle \vec{e} \triangleright [\vec{q}; R; \overrightarrow{\mathsf{bc}\ k\ \langle c\rangle\ R\ A_Q}] \triangleright (\mathsf{out}\ k\ \langle c\rangle\ R)\vec{a}\rangle \end{aligned} \quad \text{(message)}$$

$$\begin{aligned} &\langle \vec{e} \triangleright [\vec{q}(k, P); R; \vec{A_Q}] \triangleright \vec{a}\rangle \\ \longrightarrow\ &\langle \vec{e} \triangleright [\vec{q}(\ell, \pi); R; \vec{A_Q}(\ell \mapsto \Sigma)] \triangleright \vec{a}\rangle \\ &\text{where } \ell = 1 + \max\{j \mid (j \mapsto \Sigma') \in \vec{A_Q}\} \text{ and } (\Sigma, \pi) = \mathsf{boot}\ P \end{aligned} \quad \text{(spawn)}$$

$$\frac{\Sigma_Q \longrightarrow \Sigma'}{\langle \vec{e} \triangleright [\cdot; R; \vec{A_I}(\ell \mapsto \Sigma_Q)\vec{A_Q}] \triangleright \vec{a}\rangle \longrightarrow \langle \vec{e} \triangleright [\cdot; R; \vec{A_Q}\vec{A_I}(\ell \mapsto \Sigma')] \triangleright \vec{a}\rangle} \quad \text{(schedule)}$$

Figure 15: Reduction semantics of the dataspace model

**Definition 4.6** (Rule quit)**.** Alternatively, a leaf actor's behavior function may yield an exit() instruction in response to event $e_0$, i.e. $f_{beh}(e_0, u) = \mathsf{exit}(\overrightarrow{a}')$. In this case, the terminating actor is replaced with a noop behavior and its final few actions are enqueued:

$$\langle \overrightarrow{e}\, e_0 \triangleright \mathsf{pack}\, \langle \tau, (f_{beh}, u) \rangle \triangleright \overrightarrow{a} \rangle \longrightarrow \langle \overrightarrow{e} \triangleright \mathsf{pack}\, \langle \mathbf{1}, (\mathsf{noop}, ()) \rangle \triangleright \emptyset\, \overrightarrow{a}'\, \overrightarrow{a} \rangle$$

Finally, a synthesized SCN action $\emptyset$ is enqueued. The result is the permanent retraction of the actor's remaining assertions. This rule covers both deliberate and exceptional termination.[6]

**Definition 4.7** (Rule notify-ds)**.** When an event $e_0$ arrives for a dataspace, it is labeled with the special location $\downharpoonleft$ and enqueued for subsequent interpretation.

$$\langle \overrightarrow{e}\, e_0 \triangleright [\cdot; R; \overrightarrow{A_I}] \triangleright \overrightarrow{a} \rangle \longrightarrow \langle \overrightarrow{e} \triangleright [(\downharpoonleft, \mathsf{inp}\ e_0); R; \overrightarrow{A_I}] \triangleright \overrightarrow{a} \rangle$$

**Definition 4.8** (Inbound event transformation)**.** The metafunction inp transforms each such incoming event by prepending an "inbound" marker $\upharpoonleft$ to each assertion contained in the event. This marks the assertions as pertaining to the next outermost dataspace, rather than to the local dataspace.

$$\begin{aligned} \mathsf{inp} &: \mathbf{Evt} \to \mathbf{Act} \\ \mathsf{inp}\ \pi &= \{\upharpoonleft c \mid c \in \pi\} \\ \mathsf{inp}\ \langle c \rangle &= \langle \upharpoonleft c \rangle \end{aligned}$$

**Definition 4.9** (Rule gather)**.** The gather rule reads from the queue of actions produced by a particular actor for interpretation by its dataspace. It marks each action with the label of the actor before enqueueing it in the dataspace's pending action queue for processing.

$$\langle \overrightarrow{e} \triangleright [\overrightarrow{q}; R; \overrightarrow{A_Q}(\ell \mapsto \langle \overrightarrow{e}' \triangleright B \triangleright \overrightarrow{a}' a'' \rangle) \overrightarrow{A}] \triangleright \overrightarrow{a} \rangle \longrightarrow \langle \overrightarrow{e} \triangleright [(\ell, a'') \overrightarrow{q}; R; \overrightarrow{A_Q}(\ell \mapsto \langle \overrightarrow{e}' \triangleright B \triangleright \overrightarrow{a}' \rangle) \overrightarrow{A}] \triangleright \overrightarrow{a} \rangle$$

Now that we have considered event delivery and action production and collection, we may turn to action *interpretation*. The newtable and message rules are central. They both depend on metafunctions bc (short for "broadcast") and out to transform queued actions into pending events for local actors and the containing dataspace, respectively. Before we examine the supporting metafunctions, we will examine the two rules themselves.

**Definition 4.10** (Dataspace update)**.** The assertions of a party labeled $k$ are replaced in a dataspace's contents $R$ by an assertion set $\pi$ using the $\oplus$ operator:

$$R \oplus (k, \pi) = \{(j, c) \mid (j, c) \in R, j \neq k\} \cup \{(k, c) \mid c \in \pi\}$$

6 Terminated actors remain in their configurations indefinitely with the reduction relation as written. In the same way that the CESK machine can be equipped with reduction rules for garbage collection (Felleisen, Findler and Flatt 2009, ch. 9), a rule for removing inert actors with no assertions can be added to our reduction relation if we wish.

**Definition 4.11** (Rule newtable)**.** A queued state change notification action $(k, \pi)$ not only completely replaces the assertions associated with $k$ in the shared dataspace but also inserts a state change notification *event* into the event queues of interested local actors via bc. Because $k$ may have made "outbound" assertions labeled with $\downharpoonright$, newtable also prepares a state change notification for the wider environment, using out.

$$\langle \overrightarrow{e} \triangleright [\overrightarrow{q}(k,\pi); R; \overrightarrow{A_Q}] \triangleright \overrightarrow{a} \rangle \longrightarrow \langle \overrightarrow{e} \triangleright [\overrightarrow{q}; R \oplus (k,\pi); \overrightarrow{\mathsf{bc}\ k\ \pi\ R\ A_Q}] \triangleright (\mathsf{out}\ k\ \pi\ R)\overrightarrow{a} \rangle$$

*Remark.* This is the only rule to update a dataspace's $R$. In addition, because $k$'s assertion set is completely replaced, it is here that *retraction* of previously-asserted items takes effect.

**Definition 4.12** (Rule message)**.** The message rule interprets send-message actions $\langle c \rangle$. The bc metafunction is again used to deliver the message to interested peers, and out relays the message on to the containing dataspace if it happens to be "outbound"-labeled with $\downharpoonright$.

$$\langle \overrightarrow{e} \triangleright [\overrightarrow{q}(k,\langle c \rangle); R; \overrightarrow{A_Q}] \triangleright \overrightarrow{a} \rangle \longrightarrow \langle \overrightarrow{e} \triangleright [\overrightarrow{q}; R; \overrightarrow{\mathsf{bc}\ k\ \langle c \rangle\ R\ A_Q}] \triangleright (\mathsf{out}\ k\ \langle c \rangle\ R)\overrightarrow{a} \rangle$$

**Definition 4.13** (Event broadcast)**.** The bc metafunction computes the consequences for an actor labeled $\ell$ of an action performed by another party labeled $k$. When it deals with a state change notification action $\pi$, the entire aggregate shared dataspace is projected according to the asserted interests of $\ell$. The results of the projection are assembled into a state change notification event, but are enqueued only if the event would convey new information to $\ell$. When bc deals with a message action $\langle c \rangle$, a corresponding message event is enqueued for $\ell$ only if $\ell$ has previously asserted interest in $c$.

$$\mathsf{bc} : \mathbf{ID} \times \mathbf{Evt} \times \mathbf{Space} \times \mathbf{Actor_Q} \to \mathbf{Actor}$$

$$\mathsf{bc}\ k\ \pi\ R_{old}\ \ (\ell \mapsto \langle \overrightarrow{e} \triangleright B \triangleright \cdot \rangle) = \begin{cases} \ell \mapsto \langle \pi_{new} \overrightarrow{e} \triangleright B \triangleright \cdot \rangle & \text{when } \pi_{new} \neq \pi_{old} \\ \ell \mapsto \langle \overrightarrow{e} \triangleright B \triangleright \cdot \rangle & \text{when } \pi_{new} = \pi_{old} \end{cases}$$

$$\begin{aligned} \text{where } R_{new} &= R_{old} \oplus (k,\pi) \\ \pi_{new} &= \{c \mid (j,c) \in R_{new}, (\ell, ?c) \in R_{new}\} \\ \pi_{old} &= \{c \mid (j,c) \in R_{old}, (\ell, ?c) \in R_{old}\} \end{aligned}$$

$$\mathsf{bc}\ k\ \langle c \rangle\ R_{old}\ \ (\ell \mapsto \langle \overrightarrow{e} \triangleright B \triangleright \cdot \rangle) = \begin{cases} \ell \mapsto \langle \langle c \rangle \overrightarrow{e} \triangleright B \triangleright \cdot \rangle & \text{when } (\ell, ?c) \in R_{old} \\ \ell \mapsto \langle \overrightarrow{e} \triangleright B \triangleright \cdot \rangle & \text{otherwise} \end{cases}$$

**Definition 4.14** (Outbound action transformation)**.** The metafunction out is analogous to bc, but for determining information to be relayed to a containing dataspace as a consequence of a local action.

$$\mathsf{out} : \mathbf{ID} \times \mathbf{Evt} \times \mathbf{Space} \to \overrightarrow{\mathbf{Act}}$$

$$\begin{aligned} \mathsf{out}\ \downharpoonright\ e\ R &= \cdot \qquad \text{(empty sequence of actions)} \\ \mathsf{out}\ \ell\ \pi\ R &= \{c \mid (j, \downharpoonright c) \in R \oplus (\ell,\pi)\} \cup \{?c \mid (j, ?\upharpoonright c) \in R \oplus (\ell,\pi)\} \end{aligned}$$

$$\mathsf{out}\ \ell\ \langle c \rangle\ R = \begin{cases} \langle d \rangle & \text{when } c = \downharpoonright d \\ \cdot & \text{otherwise} \end{cases}$$

The first clause ensures that the out metafunction never produces an action for transmission to the outer dataspace when the cause of the call to out is an action from the outer dataspace. Without this rule, configurations would never become inert.

**Definition 4.15** (Rule spawn)**.** The spawn rule allocates a fresh label $\ell$ and places a newly-spawned actor into the collection of local actors, alongside its siblings. The new label $\ell$ is chosen to be distinct from $k$, from every element of $\left\{k' \mid (k', a') \in \overrightarrow{q}\right\}$, and from the labels of every $\overrightarrow{A_Q}$. Any deterministic[7] allocation strategy will do; we will choose $\ell = 1 + \max\{j \mid (j \mapsto \Sigma') \in \overrightarrow{A_Q}\}$. The new actor's initial state $\Sigma$ and initial assertions $\pi$ are computed from the actor specification $P$ by $(\Sigma, \pi) = \mathsf{boot}\ P$.

$$\langle \overrightarrow{e} \triangleright [\overrightarrow{q}(k, P); R; \overrightarrow{A_Q}] \triangleright \overrightarrow{a} \rangle \longrightarrow \langle \overrightarrow{e} \triangleright [\overrightarrow{q}(\ell, \pi); R; \overrightarrow{A_Q}(\ell \mapsto \Sigma)] \triangleright \overrightarrow{a} \rangle$$

*Remark.* The rule takes care to ensure that a new actor's initial *assertions* $\pi$ are processed ahead of other queued actions $\overrightarrow{q}$, even though the new actor's initial *actions* will be placed at the end of the queue and processed in order as usual. This allows a spawning actor to atomically delegate responsibility to a new actor by issuing a state-change notification immediately following the actor action. Assertions indicating to the world that the spawning party has "taken responsibility" for some task may be placed in the new actor's initial assertion set and omitted from the subsequent state-change notification. This eliminates any possibility of an intervening moment in which a peer might see a retraction of the assertions concerned. Furthermore, even if the new actor crashes during boot, there will be a guaranteed moment in time before its termination when its initial assertion set was visible to peers. Because the computation of the initial assertion set happens in the execution context of the spawning actor, an uncaught exception raised during that computation correctly blames the spawning actor for the failure. However, the computation of the initial *actions* is performed in the context of the spawned actor, and an exception at that moment correctly blames the spawned actor.[8]

**Definition 4.16** (Rule schedule)**.** Finally, the schedule rule allows quiescent, non-inert contained actors to take a step. It rotates the sequence of actors as it does so.[9]

$$\frac{\Sigma_Q \longrightarrow \Sigma'}{\langle \overrightarrow{e} \triangleright [\cdot; R; \overrightarrow{A_I}(\ell \mapsto \Sigma_Q)\overrightarrow{A_Q}] \triangleright \overrightarrow{a} \rangle \longrightarrow \langle \overrightarrow{e} \triangleright [\cdot; R; \overrightarrow{A_Q}\overrightarrow{A_I}(\ell \mapsto \Sigma')] \triangleright \overrightarrow{a} \rangle}$$

Variations on this rule can express different scheduling policies. For example, sorting the sequence decreasing by event queue length prioritizes heavily-loaded actors.

7 Non-deterministic allocation strategies affect theorem 4.20 but are otherwise harmless, so long as they preserve local uniqueness of labels.

8 An alternative approach to spawning could involve "fork" and "exec" operations analogous to those of the same name offered by Unix kernels. An actor could "fork", leading to two (almost-) identical copies, both retaining the set of assertions current at the time of the fork. One copy would immediately perform an "exec" to replace its behavior function. Both actors would then tailor their assertion sets to their separate domains of responsibility.

9 This scheduling policy, in conjunction with the determinism of the system (theorem 4.20) and the totality of leaf actor behavior functions, yields fairness (Clinger 1981).

## 4.3 CROSS-LAYER COMMUNICATION

Actors label assertions and message bodies with $\downharpoonleft$ to address them to the dataspace's own containing dataspace, but there is no corresponding means of addressing an assertion or message to a *contained* dataspace or actor. Actors may reach out, but not in. Because there is always a unique containing dataspace, reserving specific names for referring to it—the harpoon marks $\downharpoonleft$ and $\upharpoonleft$—is reasonable. These two reserved constructors bootstrap arbitrary cross-layer communication arrangements. Actors draw communications inward by reaching out. They establish subscriptions at outer layers which cause relevant messages and assertions to be relayed towards the inner requesting layer. In effect, they "pull" rather than having peers "push" information.

Directing communications to specific siblings requires a name for each actor. Actor IDs are, as a matter of principle,[10] not made available to the programmer. In cases where "pushing" information inward is desired and useful, and where the resulting sensitive dependence on the topological structure of the overall configuration is acceptable, the dataspace model leaves the specific naming scheme chosen up to the programmer, offering a mechanism ($\downharpoonleft$ and $\upharpoonleft$) but remaining neutral on policy.

## 4.4 MESSAGES VERSUS ASSERTIONS

We have included message-sending actions $\langle c \rangle$ as primitive operations. However, message transmission can be usefully viewed as a derived construct, as a special case of assertion signaling. We may achieve substantially the same effect as $\langle c \rangle$ by asserting $c$, holding the assertion for "long enough" for it to register with interested peers, and then retracting $c$ again. A message, then, can be imagined as a *transient* assertion.

There are two interesting corner-cases to consider when thinking about messages in this way. The reduction rules as written have no trouble delivering messages of the form $\langle ?c \rangle$, despite the effect that an assertion of $?c$ would have; and a message $\langle c' \rangle$ will be delivered to interested recipients even if some neighboring actor is asserting the value $c'$ at the same time. In a variation on the dataspace model lacking primitive message-sending actions, neither situation works quite as expected.

First, consider the assertion-based analogue of the message $\langle ?c \rangle$. The sender would briefly assert $?c$ before retracting it again. However, $?c$ asserts interest in $c$. For the duration of the assertion, it would have the effect of drawing matching assertions $c$ toward the sending actor. Primitive support for messages, by contrast, imagines that the "assertion" of the message lasts for an infinitesimal duration. This applies equally to "assertions" of messages that appear to denote interest in other assertions. By the time the events triggered by the message are to be delivered, it is as if the assertion of interest has *already been retracted*, so no events describing assertions $c$ make their way toward the sender.

Second, consider performing the action $\langle c' \rangle$ when $c'$ is already being asserted by some *other* peer. The assertion-based analogue of $\langle c' \rangle$ is to briefly assert $c'$ and then to retract it. However, redundant assertions do not cause perceptible changes in state. The net effect of the fleeting

10 See discussion in section 2.6.

assertion of $c'$ is zero; no events are delivered.[11] Again, by incorporating messages primitively, we side-step the problem. Strictly speaking, the message rule should have a side-condition forbidding its application (or perhaps making it a no-op) when $(j, c') \in R$ for some $j$. This would imply that sending certain messages at certain times would lead reduction to become stuck. Certain data would be reserved for use in message-sending; others, for more long-lived assertions and retractions. Were this to be elaborated into a type system, each dataspace would have a type representing its protocol. This type would classify values as either message-like or assertion-like.[12] Judgments would connect with the type system of the base language to ensure that the classifications were respected by produced actions.[13]

## 4.5 PROPERTIES

A handful of theorems capture invariants that support the design of and reasoning about effective protocols for dataspace model programs. Theorem 4.17 ensures that the dataspace does not get stuck, even though individual actors within the dataspace may behave unpredictably. Theorem 4.20 ensures deterministic reduction of the system. Theorem 4.23 assures programmers that the dataspace does not reorder an actor's actions or any of the resulting events. Theorem 4.35 makes a causal connection between the actions of an actor and the events it subsequently receives. It expresses the *purpose* of the dataspace: to keep actors informed of exactly the assertions and messages relevant to their interests as those interests change. Tests constructed in Redex (Felleisen, Findler and Flatt 2009) and proofs written for Coq (Coq development team 2004) confirm theorems 4.17 and 4.20.

**Theorem 4.17** (Soundness)**.** *A state* $\Sigma \in \mathbf{State}$ *is either inert* ($\Sigma \in \mathbf{State_I}$) *or there exists some* $\Sigma'$ *such that* $\Sigma \longrightarrow \Sigma'$.

*Proof (Sketch).* We employ the Wright/Felleisen technique (Wright and Felleisen 1994) with the progress lemma below. The proof makes use of the fact that all leaf actor behavior functions are total. □

**Definition 4.18** (Height)**.** Let the *height* of a behavior be defined as follows:

$$\begin{aligned} \mathit{height} &: \mathbf{Beh} \rightarrow \mathbb{N} \\ \mathit{height}\ \mathsf{pack}\ \langle \tau, (f_{beh}, u) \rangle &= 0 \\ \mathit{height}\ [\overrightarrow{q}; R; \overrightarrow{\ell \mapsto \langle \overrightarrow{e} \triangleright B \triangleright \overrightarrow{a} \rangle}] &= 1 + \mathit{max}(\overrightarrow{\mathit{height}\ B}) \end{aligned}$$

**Lemma 4.19** (Progress)**.** *For all* $C \in \mathbf{Cfg}$ *and* $H \in \mathbb{N}$ *such that* $\mathit{height}(C) \leqslant H$, $C$ *is either inert* ($C \in \mathbf{Cfg_I}$) *or there exists some* $C'$, $\overrightarrow{a}$ *such that* $\langle \cdot \triangleright C \triangleright \cdot \rangle \longrightarrow \langle \cdot \triangleright C' \triangleright \overrightarrow{a} \rangle$.

11 This is in some ways similar to the idea of *medium access control.* Multiple stations transmitting at the same time "corrupt" each others' messages. Some means of ensuring the separation of overlapping transmissions in space or in time is required.

12 If message sending were a derived concept, such a type system would not suffice to ensure two peers did not simultaneously try to "send message c" by briefly asserting c.

13 Research into the design of such a type system is ongoing (Caldwell, Garnock-Jones and Felleisen 2017).

*Proof (Sketch).* By nested induction on the height bound and structure of C. □

**Theorem 4.20** (Deterministic Evaluation)**.** *For any $\Sigma$ there exists at most one $\Sigma'$ such that $\Sigma \longrightarrow \Sigma'$.*

*Proof.* The reduction relation is structured to ensure at most one applicable rule in any situation. Either

- $\Sigma = \langle \vec{e}\, e_0 \triangleright B_I \triangleright \vec{a} \rangle$, in which case event $e_0$ is consumed by $B_I$ (rules notify-leaf, notify-ds, and quit); or
- $\Sigma = \langle \vec{e} \triangleright [\vec{q}; R; \overrightarrow{A_Q}(\ell \mapsto \langle \vec{e}\,' \triangleright B \triangleright \vec{a}\,' a'' \rangle)\vec{A}] \triangleright \vec{a} \rangle$, in which case $a''$ is gathered onto $\vec{q}$ (rule gather); or
- $\Sigma = \langle \vec{e} \triangleright [\vec{q}(k, a''); R; \overrightarrow{A_Q}] \triangleright \vec{a} \rangle$, in which case $a''$ is interpreted (newtable, message, and spawn); or
- $\Sigma = \langle \vec{e} \triangleright [\cdot; R; \overrightarrow{A_I}(\ell \mapsto \Sigma_Q)\overrightarrow{A_Q}'] \triangleright \vec{a} \rangle$ and $\Sigma_Q \longrightarrow \Sigma'$, in which case actor $\ell$ takes a step (rule schedule).

Observe that the cases are disjoint: the first demands a $B_I$, but in the others the configuration is not inert; the second demands some non-quiescent actor; the third demands a queued action and *only* quiescent actors; the fourth demands *no* queued actions and *only* quiescent actors. Therefore, assume there exists distinct $\Sigma'$ and $\Sigma''$ such that $\Sigma \longrightarrow \Sigma'$ and $\Sigma \longrightarrow \Sigma''$. We may then show a contradiction by nested induction on the two instances of the reduction relation and systematic elimination of possible sources of difference between $\Sigma'$ and $\Sigma''$. □

*Remark* 4.21 (Concurrency and determinism)*.* Despite appearances, theorem 4.20 does not sacrifice concurrency; recall from chapter 2 the argument that sequential programs frequently include internal concurrency. Concurrency does not entail nondeterminism. Even with deterministic reduction rules as written, many sources of unpredictability remain. For example, programs might interact with the outside world, including external clocks of various kinds, leading to fine variation in timing of events; code written by one person might make use of "black box" library code written by another, without precisely-documented timing specifications; or fine details of the implementation of some component could change, leading to subtly different interleavings. Introduction of nondeterminism by, say, varying the schedule rule or relaxing some of the quiescence or inertness constraints in the other rules would merely introduce another source of unpredictability. The essential properties of the dataspace model survive such changes unharmed.

*Remark* 4.22 (Dataspace reliability)*.* While individual leaf actors may exit at any time, *dataspace* actors cannot terminate at all: no means for voluntary exit is provided, and theorem 4.17 assures us that a dataspace will not crash. In a correct implementation of the dataspace model, dataspace actors will likewise not crash. If the implementation is buggy enough that a dataspace does in fact crash, but not so buggy that it takes its *containing* dataspace down with it, the usual removal of an actor's assertions allows peers of the failing dataspace actor to observe the consequences of its termination. Abrupt failure of a dataspace is analogous to a crash of an entire computer: there is no opportunity for a clean shutdown of the programs the computer is running; instead, the entire computer simply vanishes offline from the perspective of its peers.

**Theorem 4.23** (Order Preservation)**.** *If an actor produces action A before action B, then A is interpreted by the dataspace before B. Events are enqueued atomically with interpretation of the action that causes them. If event C for actor $\ell$ is enqueued before event D, also for $\ell$, then C is delivered before D.*

*Proof (Sketch).* The reduction rules consistently move items one-at-a-time from the front of one queue to the back of another, and events are only enqueued during action interpretation. □

Our final theorem (4.35) guarantees the programmer that each actor receives "the truth, the whole truth, and nothing but the truth" from the dataspace, according to the declared interests of the actor, keeping in mind that there may be updates to the actor's interest set pending in the pipeline. It ensures that the dataspace conveys *every* relevant assertion and *only* relevant assertions,[14] and shows that the dataspace is being *cooperative* in the sense of Grice's Cooperative Principle and Conversational Maxims (section 2.1 and figure 1). The theorem directly addresses the maxims of Quantity, Quality, and Relation.

Before we are able to formally state the theorem, we must define several concepts.

**Definition 4.24** (Paths)**.** A *path* $p \in \mathbf{Path} = \overrightarrow{\mathbf{Loc}} \ni \vec{\ell}$ is a possibly-empty sequence of locations. A path resolves to a **State** by the partial recursive function *resolvePath*:

$$\begin{aligned} \mathit{resolvePath} &: \mathbf{State} \times \mathbf{Path} \rightharpoonup \mathbf{State} \\ \mathit{resolvePath}\ \Sigma\ \cdot &= \Sigma \\ \mathit{resolvePath}\ \langle \vec{e} \triangleright [\vec{q}\,; R; \vec{A}\,(\ell \mapsto \Sigma)\vec{A'}] \triangleright \vec{a} \rangle\ \ (\ell\ p) &= \mathit{resolvePath}\ \Sigma\ p \\ \mathit{resolvePath}\ \langle \vec{e} \triangleright [\vec{q}\,; R; \vec{A}] \triangleright \vec{a} \rangle\ \ (\ell\ p) &\quad \text{undefined when there is no actor labeled } \ell \text{ in } \vec{A} \end{aligned}$$

The definition of *resolvePath* makes it clear that locations in a path are ordered leftmost-outermost and rightmost-innermost with respect to a nested dataspace configuration. When *resolvePath* $\Sigma$ $p$ is defined, we say $p$ is in $\Sigma$, and write $p \in \Sigma$; otherwise, $p$ is not in $\Sigma$, $p \notin \Sigma$.

**Definition 4.25** (Dataspace contents for a path)**.** We write $R_\Sigma^p$ to denote the contents of the shared dataspace immediately surrounding the actor denoted by nonempty path $p = (p'\ \ell)$ in $\Sigma$. That is,

$$R_\Sigma^p = R \text{ where } \mathit{resolvePath}\ \Sigma\ p' = \langle \vec{e} \triangleright [\vec{q}\,; R; \vec{A}\ (\ell \mapsto \Sigma')\ \vec{A'}] \triangleright \vec{a} \rangle$$

**Definition 4.26** (Current interest set of $p$ in $\Sigma$)**.** The *current interest set* of the actor denoted by nonempty path $p = (p'\ \ell)$ in a given state $\Sigma$ is

$$\mathit{interestsOf}\,(p, \Sigma) \triangleq \left\{ c \mid (\ell, ?c) \in R_\Sigma^p \right\}$$

**Definition 4.27** (Syllabus)**.** The *syllabus* of an actor with nonempty path $p$ at state $\Sigma$ is

$$p \lessdot \Sigma \triangleq \left\{ c \mid (j, c) \in R_\Sigma^p, c \in \mathit{interestsOf}\,(p, \Sigma) \right\}$$

The syllabus describes the dataspace's understanding of *what* $p$ *needs to know* from the dataspace, as of the moment captured by the state $\Sigma$; the notation is chosen to connote the idea of $p$ "reading" from $\Sigma$. The syllabus of $p$ at $\Sigma$ will guide the dataspace as it constructs events conveying changed knowledge to $p$.

14 While theorem 4.35 captures many important properties of the dataspace model, it remains future work to extend it to soundness and completeness properties for assertions relayed across nested dataspace layers.

**Definition 4.28** (Reduction sequences)**.** We use the notation $\mathcal{S}(P) \in \overrightarrow{\textbf{State}}$ to denote a finite sequence of states corresponding to a prefix of the sequence of reductions of a program $P \in$ **Prog**. We write $\mathcal{S}(P)_i \in$ **State** to denote the ith element of the sequence. A sequence $\mathcal{S}(P)$ starts with $\mathcal{S}(P)_0 = \Sigma$ where $(\Sigma, \emptyset) = \mathsf{boot}\,(\mathsf{dataspace}\ P)$. Subsequent states in $\mathcal{S}(P)$ are pairwise related by the reduction relation; that is, $\mathcal{S}(P)_0 \longrightarrow \mathcal{S}(P)_1 \longrightarrow \cdots \longrightarrow \mathcal{S}(P)_{|\mathcal{S}(P)|}$. We say that a path $p$ is in $\mathcal{S}(P)$ if $p \in \mathcal{S}(P)_i$ for some $i$.

**Definition 4.29** (Enqueued events)**.** We write *enqueuedAt* $(\mathcal{S}(P), i, p, e)$ when event $e$ is enqueued for eventual delivery to actor $p$ in reduction sequence $\mathcal{S}(P)$ at the transition $\mathcal{S}(P)_i \longrightarrow \mathcal{S}(P)_{i+1}$:

$$\begin{aligned} \mathit{enqueuedAt}\,(\mathcal{S}(P), i, p, e) \iff \big(&\mathit{resolvePath}\ \mathcal{S}(P)_i\ p = \langle \overrightarrow{e}' \triangleright B \triangleright \overrightarrow{a} \rangle \\ &\wedge \mathit{resolvePath}\ \mathcal{S}(P)_{i+1}\ p = \langle e\,\overrightarrow{e}' \triangleright B \triangleright \overrightarrow{a} \rangle\big) \end{aligned}$$

**Definition 4.30** (Truthfulness)**.** An assertion set $\pi$ is called *truthful* with respect to a dataspace whose contents are $R$ if it contains only assertions actually present in $R$. That is, $\pi$ is truthful if $\pi \subseteq \{c \mid (j, c) \in R\}$.[15]

**Definition 4.31** (Relevance)**.** An assertion set $\pi$ is called *relevant* to an actor $\ell$ in a dataspace whose contents are $R$ if it contains only assertions of interest to $\ell$; i.e., if $\pi \subseteq \{c \mid (\ell, ?c) \in R\}$.

**Definition 4.32** (Soundness)**.** An assertion set $\pi$ is called *sound* for an actor $\ell$ in a dataspace whose contents are $R$ if it is both truthful w.r.t $R$ and relevant w.r.t. $\ell$ and $R$.

**Definition 4.33** (Completeness)**.** An assertion set $\pi$ is called *complete* for an actor named $\ell$ in a dataspace whose contents are $R$ if it contains every assertion both actually present and of interest to $\ell$; that is, if $\pi \supseteq (\{c \mid (j, c) \in R\} \cap \{c \mid (\ell, ?c) \in R\})$.

**Definition 4.34** (Most recent SCN event)**.** Let $p$ be a path of an actor, $\mathcal{S}(P)$ be a reduction sequence, and $i$ in index to a state in $\mathcal{S}(P)$. The most recent SCN event enqueued for $p$ as it exists within $\mathcal{S}(P)_i$, written $\pi_i^{\mathcal{S}(P),p}$, is computed by

$$\pi_i^{\mathcal{S}(P),p} = \begin{cases} \pi' & \text{if } \mathit{enqueuedAt}\,(\mathcal{S}(P), i-1, p, \pi')\text{; otherwise,} \\ \pi_{i-1}^{\mathcal{S}(P),p} & \text{if } \mathit{resolvePath}\ \mathcal{S}(P)_i\ p \text{ is defined; otherwise,} \\ \emptyset & \end{cases}$$

**Theorem 4.35** (Conversational Soundness and Completeness)**.** *Let $\mathcal{S}(P)$ be a reduction sequence. For every actor denoted by a nonempty path $p = (p'\ \ell)$ in $\mathcal{S}(P)$, at every step $i$,*

1. *interestsOf* $(p, \mathcal{S}(P)_i)$ *depends solely on successive SCN actions issued by actor* $p$.

2. $p \lessdot \mathcal{S}(P)_i \neq p \lessdot \mathcal{S}(P)_{i+1}$ *iff there exists $\pi$ such that enqueuedAt* $(\mathcal{S}(P), i, p, \pi)$.

3. *enqueuedAt* $(\mathcal{S}(P), i, p, \pi)$ *implies that $\pi$ is sound and complete for $\ell$ and* $R^{p}_{\mathcal{S}(P)_{i+1}}$.

15 Here, as elsewhere in this chapter, $\{c \mid (j, c) \in R\}$ is interpreted as $\{c \mid \exists j.\ (j, c) \in R\}$ when $j$ is free.

4. *resolvePath* $\mathcal{S}(P)_i\ p' = \langle \vec{e} \triangleright [\vec{q}(k, \langle c \rangle); R^{p}_{\mathcal{S}(P)_i}; \vec{A}\ (\ell \mapsto \Sigma)\ \vec{A}'] \triangleright \vec{a} \rangle \wedge c \in$ *interestsOf* $(p, \mathcal{S}(P)_i)$ $\iff$ *enqueuedAt* $(\mathcal{S}(P), i, p, \langle c \rangle)$.

That is: (1) the dataspace's understanding of the interests of p (which shape its syllabus) is solely determined by the actions of p; (2) every time the syllabus of p changes, an SCN event is enqueued for p, and every SCN event enqueued for p results from a change in its syllabus; (3) every SCN event for p is sound and complete with respect to the interests of p and the contents of the dataspace of p; and (4) every message action of interest to p results in a message event for p, and no other message events are produced for p.

*Proof.*
(1) By lemma 4.36, the gather rule, and theorem 4.23.
(2) Forward direction: by lemma 4.40, $\pi_i^{\mathcal{S}(P),p} \neq \pi_{i+1}^{\mathcal{S}(P),p}$. Let $\pi = \pi_{i+1}^{\mathcal{S}(P),p}$, and the conclusion follows from lemma 4.37. Reverse direction: we are given some $\pi$ s.t. *enqueuedAt* $(\mathcal{S}(P), i, p, \pi)$. By definition, then, $\pi_{i+1}^{\mathcal{S}(P),p} = \pi$; by lemma 4.42, $\pi_i^{\mathcal{S}(P),p} \neq \pi$. Combining these facts, $\pi_i^{\mathcal{S}(P),p} \neq \pi_{i+1}^{\mathcal{S}(P),p}$; now, apply lemma 4.40 and we are done.
(3) By definition, in conjunction with our premises, $\pi_{i+1}^{\mathcal{S}(P),p} = \pi$; lemma 4.41 yields our result.
(4) Forward direction: rule message is the only applicable rule; the conclusion follows by definition of bc for message routing. Reverse direction: likewise, because rule message is the only rule that enqueues message events. □

**Lemma 4.36.** *Let* $p = (p'\ \ell)$ *be a nonempty path of an actor in some* $\mathcal{S}(P)$*. Wherever interestsOf* $(p, \mathcal{S}(P)_i) \neq$ *interestsOf* $(p, \mathcal{S}(P)_{i+1})$*, we have that:*

1. $\vec{q} = \vec{q}'(\ell, \pi)$*, where resolvePath* $\mathcal{S}(P)_i\ p' = \langle \vec{e} \triangleright [\vec{q}; R^{p}_{\mathcal{S}(P)_i}; \vec{A}\ (\ell \mapsto \Sigma)\ \vec{A}'] \triangleright \vec{a} \rangle$*, and*
2. *interestsOf* $(p, \mathcal{S}(P)_{i+1}) = \{c \mid ?c \in \pi\}$.

*Proof.* Direct from the facts that rule newtable is the only possible rule that can apply as $\mathcal{S}(P)_i \longrightarrow \mathcal{S}(P)_{i+1}$ and that newtable replaces $R^{p}_{\mathcal{S}(P)_i}$ in the containing dataspace of p with $R^{p}_{\mathcal{S}(P)_{i+1}} = R \oplus (\ell, \pi)$. □

**Lemma 4.37.** $\pi_i^{\mathcal{S}(P),p} \neq \pi_{i+1}^{\mathcal{S}(P),p} \implies$ *enqueuedAt* $\left(\mathcal{S}(P), i, p, \pi_{i+1}^{\mathcal{S}(P),p}\right)$.

*Proof.* Straightforward consequence of definition 4.34. □

**Definition 4.38.** The notation $\Sigma_a\ [\Sigma_b] \xrightarrow{p:r} \Sigma_c\ [\Sigma_d]$ is interpreted as a relation defined by:

$$\Sigma_a\ [\Sigma_a] \xrightarrow{\cdot:r} \Sigma_c\ [\Sigma_c] \iff \Sigma_a \longrightarrow \Sigma_c \text{ by rule } r$$

$$\Sigma_a\ [\Sigma_b] \xrightarrow{(\ell\ p):r} \Sigma_c\ [\Sigma_d] \iff \big(\Sigma_a \longrightarrow \Sigma_c \text{ by rule schedule}$$
$$\wedge\ (\mathit{resolvePath}\ \Sigma_a\ \ell)\ [\Sigma_b] \xrightarrow{p:r} (\mathit{resolvePath}\ \Sigma_c\ \ell)\ [\Sigma_d]\big)$$

**Lemma 4.39.** *Let* $p = (p'\ \ell)$ *be a nonempty path and* $\mathcal{S}(P)$ *be a reduction sequence. If* $\pi_i^{\mathcal{S}(P),p} \neq \pi_{i+1}^{\mathcal{S}(P),p}$*,*

1. $\mathcal{S}(P)_i\,[\Sigma'] \overset{p':\mathit{newtable}}{\longrightarrow} \mathcal{S}(P)_{i+1}\,[\Sigma'']$ *for some* $\Sigma', \Sigma''$

2. $\pi_{i+1}^{\mathcal{S}(P),p} = \{c \mid (j,c) \in R^p_{\mathcal{S}(P)_{i+1}}, (\ell, ?c) \in R^p_{\mathcal{S}(P)_{i+1}}\} = p \lessdot \mathcal{S}(P)_{i+1}$

*Proof.* 1. By lemma 4.37, an SCN event must be enqueued for p at this step; no other rule than newtable enqueues SCN events. 2. Metafunction bc is the source of the new SCN event, which is equal to $\pi_{i+1}^{\mathcal{S}(P),p}$ by definition. The first case of bc must apply in order for some event to be enqueued; following the definitions and the use of bc in the newtable rule gives us our result. □

**Lemma 4.40.** *Let* $p = (p'\ \ell)$ *be a nonempty path and* $\mathcal{S}(P)$ *be a reduction sequence. For every* $i$, $\pi_i^{\mathcal{S}(P),p} = p \lessdot \mathcal{S}(P)_i$.

*Proof.* By induction on i.

- Case $i = 0$. Recall that $(\mathcal{S}(P)_0, \emptyset) = \mathsf{boot}\,(\mathsf{dataspace}\ P) = (\langle \cdot \triangleright [(\downarrow, P); \emptyset; \cdot] \triangleright \cdot \rangle, \emptyset)$. Vacuously true, because both $R^p_{\mathcal{S}(P)_0}$ and $p \lessdot \mathcal{S}(P)_0$ are undefined for all p.

- Case $i > 0$. If $\pi_{i-1}^{\mathcal{S}(P),p} \neq \pi_i^{\mathcal{S}(P),p}$, the result is immediate, by lemma 4.39. Otherwise, $\pi_{i-1}^{\mathcal{S}(P),p} = \pi_i^{\mathcal{S}(P),p}$; combining this with the induction hypothesis, we learn that $\pi_i^{\mathcal{S}(P),p} = p \lessdot \mathcal{S}(P)_{i-1}$.

  There are two cases to consider: either the dataspace containing actor p steps by rule newtable, or some other kind of reduction takes place.

  – If for some $\Sigma', \Sigma''$ we have that $\mathcal{S}(P)_{i-1}\,[\Sigma'] \overset{p':\mathsf{newtable}}{\longrightarrow} \mathcal{S}(P)_i\,[\Sigma'']$, then we know that an actor $\left(\ell \mapsto \langle \overrightarrow{e}\,' \triangleright B \triangleright \cdot \rangle\right)$ is an immediate child of the dataspace configuration in $\Sigma'$. Furthermore we know that $\left(\ell \mapsto \langle \overrightarrow{e}\,' \triangleright B \triangleright \cdot \rangle\right)$ must also be an immediate child of the dataspace configuration in $\Sigma''$, because otherwise $\pi_i^{\mathcal{S}(P),p}$ would differ from $\pi_{i-1}^{\mathcal{S}(P),p}$. It follows then that the second case of bc must apply for actor $\ell$ in this reduction step, and so bc's $\pi_{new} = \pi_{old}$, meaning that $p \lessdot \mathcal{S}(P)_i = p \lessdot \mathcal{S}(P)_{i-1}$. By $\pi_i^{\mathcal{S}(P),p} = p \lessdot \mathcal{S}(P)_{i-1}$, we are done.

  – Otherwise, it must be the case that $R^p_{\mathcal{S}(P)_{i-1}} = R^p_{\mathcal{S}(P)_i}$, because no other reduction step can possibly affect the R register of the dataspace containing actor p. Applying this to prove $p \lessdot \mathcal{S}(P)_i = p \lessdot \mathcal{S}(P)_{i-1}$ gives our result by $\pi_i^{\mathcal{S}(P),p} = p \lessdot \mathcal{S}(P)_{i-1}$.

□

**Lemma 4.41.** *Let* $p = (p'\ \ell)$ *be a nonempty path and* $\mathcal{S}(P)$ *be a reduction sequence. For every* $i$, $\pi_i^{\mathcal{S}(P),p}$ *is both sound and complete w.r.t.* $\ell$ *and* $R^p_{\mathcal{S}(P)_i}$.

*Proof.* By lemma 4.40, $\pi_i^{\mathcal{S}(P),p} = p \lessdot \mathcal{S}(P)_i$. We must show:

- Soundness demands truthfulness, $\pi_i^{\mathcal{S}(P),p} \subseteq \{c \mid (j,c) \in R^p_{\mathcal{S}(P)_i}\}$ and relevance, $\pi_i^{\mathcal{S}(P),p} \subseteq \{c \mid (\ell, ?c) \in R^p_{\mathcal{S}(P)_i}\}$. Both these properties are immediate from the definition of $p \lessdot \mathcal{S}(P)_i$.

- Completeness demands $\pi_i^{\mathcal{S}(P),p} \supseteq \left(\{c \mid (j,c) \in R^p_{\mathcal{S}(P)_i}\} \cap \{c \mid (\ell, ?c) \in R^p_{\mathcal{S}(P)_i}\}\right)$; this is also immediate from the definition of $p \lessdot \mathcal{S}(P)_i$.

□

**Lemma 4.42** (Necessity). *enqueuedAt* $(\mathcal{S}(P), i, p, \pi') \implies \pi_i^{\mathcal{S}(P),p} \neq \pi'$.

*Proof.* Only rule newtable can enqueue an event $\pi'$ for $p$. But it will only do so if, in bc, $R_{new} \neq R_{old}$ ; that is, if

$$\{c \mid (j,c) \in R^p_{\mathcal{S}(P)_{i+1}}, (\ell, ?c) \in R^p_{\mathcal{S}(P)_{i+1}}\} \neq \{c \mid (j,c) \in R^p_{\mathcal{S}(P)_i}, (\ell, ?c) \in R^p_{\mathcal{S}(P)_i}\}$$

Applying the definition of syllabus, this is $p \lessdot \mathcal{S}(P)_{i+1} \neq p \lessdot \mathcal{S}(P)_i$, and lemma 4.40 gives $\pi_{i+1}^{\mathcal{S}(P),p} \neq \pi_i^{\mathcal{S}(P),p}$. By definition, $\pi_{i+1}^{\mathcal{S}(P),p} = \pi'$ because *enqueuedAt* $(\mathcal{S}(P), i, p, \pi')$, and so we know that $\pi' \neq \pi_i^{\mathcal{S}(P),p}$. □

The "soundness" properties of theorem 4.35 forbid *overapproximation* of the interests of the actor; communicated assertions and messages must be genuinely relevant. However, when taken alone, they permit omission of information. The "completeness" properties ensure timely communication of *all* relevant assertions and messages, but taken alone permit inclusion of irrelevancies. It is only when both kinds of property are taken together that we obtain a practical result.

It is interesting to consider variations on the model that weaken these properties. A dataspace allowing inclusion of assertions not in R (violation of *truthfulness*) would be harmful: it would violate Grice's maxims of Quality, and hence risk being branded uncooperative. Likewise, a dataspace *omitting* assertions in R it knows to be of interest (violation of *completeness*) would also be harmful: this violates the first maxim of Quantity and the maxim of Relation. By contrast, allowing inclusion of assertions not in the interest set of a given actor (violation of *relevance*) would not be harmful, and may even be useful, even though strictly this overinformativeness would be a violation of the second maxim of Quantity. For example, it may be more convenient or more efficient for a dataspace to convey "all sizes are available" than the collection of separate facts "size 4 is available", "size 6 is available" and "size 7 is available" to some actor expressing interest only in the specific sizes 4, 6 and 7. As another example, use of a narrow probabilistic overapproximation of an actor's interest (e.g. a Bloom filter (Bloom 1970)) could save significant memory and CPU resources in a dataspace implementation while placing only the modest burden of discarding irrelevant assertions on each individual actor.

All this is true only in situations where *secrecy* is not a concern. If it is important that actors be forbidden from learning the contents of certain assertions, then the relevance aspect of soundness suddenly becomes crucial. For example, consider a system using unguessable IDs as capabilities. Clearly, it would be wrong to send an actor spurious assertions mentioning capabilities that it does not legitimately hold. Secrecy is further discussed in section 11.3.

## 4.6 Incremental assertion-set maintenance

Taking section 4.2 literally implies that dataspaces convey entire sets of assertions back and forth every time some assertion changes. While wholesale transmission is a convenient illusion, it is intractable as an implementation strategy. Because the *change* in state from one moment to the next is usually small, actors and dataspaces transmit redundant information with each action and event. In short, the model needs an incremental semantics. Relatedly, while many actors find natural expression in terms of whole sets of assertions, some are best expressed in terms of reactions to changes in state. Supporting a change-oriented interface between leaf actors and their dataspaces simplifies the programmer's task in these cases.

Starting from the definitions of section 4.1, we replace assertion-set state-change notification events with *patches*. Patches allow incremental maintenance of the shared dataspace without materially changing the semantics in other respects. When extended to code in leaf actors, they permit incremental computation in response to changes. We will call the syntax and semantics already presented the *monolithic* dataspace model, and the altered syntax and semantics introduced in this section the *incremental* dataspace model.

The required changes to program syntax are small. We replace assertion sets $\pi$ with patches $\Delta$ in the syntax of events and actions:

$$\begin{aligned} \text{Events } e \in \textbf{Evt} &::= \langle c \rangle \mid \Delta \\ \text{Actions } a \in \textbf{Act} &::= \langle c \rangle \mid \Delta \mid P \\ \text{Patches } \Delta \in \textbf{Patch} &::= \frac{\pi_{in}}{\pi_{out}} \text{ where } \pi_{in} \cap \pi_{out} = \emptyset \end{aligned}$$

All other definitions from figures 12 and 14 remain the same. The configuration syntax is as before, except that queued events and actions now use patches instead of assertion sets. Behavior functions, too, exchange patches with their callers.

Patches denote changes in assertion sets. They are intended to be applied to some existing set of assertions. The notation is chosen to resemble a substitution, with elements to be added to the set written above the line and those to be removed below. We require that a patch's two sets be disjoint.[16]

**Definition 4.43** (Rule patch)**.** To match the exchange of patches for assertion sets, we replace the newtable reduction rule (definition 4.11 and figure 15) with a rule for applying patches:

$$\begin{aligned} &\langle \vec{e} \triangleright [\vec{q}(k,\Delta); R \;;\; \overrightarrow{A_Q}\;] \triangleright \vec{a} \rangle \qquad \text{(patch)} \\ \longrightarrow\; &\langle \vec{e} \triangleright [\vec{q} \;;\; R \oplus (k,\Delta'); \overrightarrow{\mathsf{bc}_\Delta\ k\ \Delta'\ R\ A_Q}] \triangleright (\mathsf{out}\ k\ \Delta'\ R)\vec{a} \rangle \end{aligned}$$

where $\Delta = \dfrac{\pi_{in}}{\pi_{out}}$ and $\Delta' = \dfrac{\pi_{in} - \{c \mid (k,c) \in R\}}{\pi_{out} \cap \{c \mid (k,c) \in R\}}$.

*Remark.* The effect of the definition of $\Delta'$ is to render harmless any attempt by $k$ to add an assertion it has already added or retract an assertion that is not asserted.

**Definition 4.44** (Dataspace patching)**.** The $\oplus$ operator, defined above for wholesale assertion-set updates (definition 4.10), is straightforwardly adapted to patches:

$$R \oplus (k, \frac{\pi_{in}}{\pi_{out}}) = R \cup \{(k,c) \mid c \in \pi_{in}\} - \{(k,c) \mid c \in \pi_{out}\}$$

16 Disjointness of $\pi_{in}$ and $\pi_{out}$ ensures that a patch can be applied either $\pi_{in}$-first or $\pi_{out}$-first without affecting the result.

**Definition 4.45** (Inbound patch transformation)**.** The inp metafunction is likewise easily adjusted:

$$\mathsf{inp}\ \frac{\pi_{in}}{\pi_{out}} = \frac{\{\upharpoonleft c \mid c \in \pi_{in}\ \}}{\{\upharpoonleft c \mid c \in \pi_{out}\}}$$

**Definition 4.46** (Outbound patch transformation)**.** It is the out metafunction that requires deep surgery. We must take care not only to correctly relabel assertions in the resulting patch but to signal only true changes to the aggregate set of assertions of the entire dataspace:[17]

$$\mathsf{out}\ \ell\ \frac{\pi_{in}}{\pi_{out}}\ R = \frac{\{c \mid\ \downharpoonleft c \in (\pi_{in}\ - \pi')\} \cup \{?c \mid ?\upharpoonleft c \in (\pi_{in}\ - \pi')\}}{\{c \mid\ \downharpoonleft c \in (\pi_{out} - \pi')\} \cup \{?c \mid ?\upharpoonleft c \in (\pi_{out} - \pi')\}}$$
$$\text{where } \pi' = \{c \mid (j, c) \in R,\ j \neq \ell\}$$

**Definition 4.47** (Patch event broadcast)**.** The metafunction $bc_\Delta$, used in the patch rule, constructs a state change notification patch event tailored to the interests of actor $\ell$. The notification describes the net change to the shared dataspace caused by actor $k$'s patch action—as far as that change is relevant to the interests of $\ell$.

$$bc_\Delta : \mathbf{ID} \times \mathbf{Patch} \times \mathbf{Space} \times \mathbf{Actor_Q} \to \mathbf{Actor}$$

$$bc_\Delta\ k\ \frac{\pi_{in}}{\pi_{out}}\ R_{old}\ \ (\ell \mapsto \langle \vec{e} \triangleright B \triangleright \cdot\rangle) = \begin{cases} \ell \mapsto \langle \Delta_{fb}\ \ \vec{e} \triangleright B \triangleright \cdot\rangle & \text{if } \ell = k \text{ and } \Delta_{fb} \neq \frac{\emptyset}{\emptyset} \\ \ell \mapsto \langle \Delta_{other}\ \vec{e} \triangleright B \triangleright \cdot\rangle & \text{if } \ell \neq k \text{ and } \Delta_{other} \neq \frac{\emptyset}{\emptyset} \\ \ell \mapsto \langle\ \ \vec{e} \triangleright B \triangleright \cdot\rangle & \text{otherwise} \end{cases}$$

$$\begin{aligned} \text{where } R_{new} &= R_{old} \oplus (k, \frac{\pi_{in}}{\pi_{out}}) \\ \pi^{\circ} &= \{c \mid (j, c) \in R_{old}\ \} \\ \pi^{\bullet} &= \{c \mid (j, c) \in R_{old}\ , j \neq k\} \\ \pi^{\bullet}_{in} &= \pi_{in} - \pi^{\bullet} \\ \pi^{\bullet}_{out} &= \pi_{out} - \pi^{\bullet} \\ \Delta_{other} &= \frac{\{c \mid c \in \pi^{\bullet}_{in}\ , (\ell, ?c) \in R_{old}\ \}}{\{c \mid c \in \pi^{\bullet}_{out}, (\ell, ?c) \in R_{old}\ \}} \\ \Delta_{fb} &= \frac{\{c \mid c \in \pi^{\bullet}_{in}\ , (\ell, ?c) \in R_{new}\} \cup \{c \mid c \in (\pi^{\circ} \cup \pi^{\bullet}_{in}\ - \pi^{\bullet}_{out}), ?c \in \pi_{in}\ \}}{\{c \mid c \in \pi^{\bullet}_{out}, (\ell, ?c) \in R_{old}\ \} \cup \{c \mid c \in \pi^{\circ}\ \ , ?c \in \pi_{out}\}} \end{aligned}$$

The patch $\Delta_{fb}$ that $bc_\Delta$ constructs as feedback when $\ell = k$ differs from the patch $\Delta_{other}$ delivered to $k$'s peers. While assertions made by $k$'s peers do not change during the reduction, $k$'s assertions do. Not only must new assertions in $\pi_{in}$ be considered as potentially worthy of inclusion, but new subscriptions in $\pi_{in}$ must be given the opportunity to examine the entirety of the aggregate state. Similar considerations arise for $\pi_{out}$.

17 The definition of $\pi'$ here is analogous to that of $\pi^{\bullet}$ in the definition of $bc_\Delta$, which also filters $R$ to compute a mask applied to the patch.

The final changes adjust the quit and spawn rules to produce patches instead of assertion set state change notifications in case of process termination and startup.

**Definition 4.48** (Incremental quit rule)**.** The quit rule becomes

$$\langle \overrightarrow{e}\, e_0 \triangleright \mathsf{pack}\,\langle \tau, (f_{beh}, u)\rangle \triangleright \overrightarrow{a}\rangle \longrightarrow \langle \overrightarrow{e} \triangleright \mathsf{pack}\,\langle \mathbf{1}, (\mathsf{noop}, ())\rangle \triangleright \frac{\emptyset}{\mathbf{Val}}\, \overrightarrow{a}'\, \overrightarrow{a}\rangle$$

when $f_{beh}(e_0, u) = \mathsf{exit}(\overrightarrow{a}')$. The sole change from definition 4.6 is use of $\frac{\emptyset}{\mathbf{Val}}$ in place of $\emptyset$.

**Definition 4.49** (Incremental spawn rule)**.** The spawn rule becomes

$$\langle \overrightarrow{e} \triangleright [\overrightarrow{q}(k, P); R; \overrightarrow{A_Q}] \triangleright \overrightarrow{a}\rangle \longrightarrow \langle \overrightarrow{e} \triangleright [\overrightarrow{q}(\ell, \frac{\pi}{\emptyset}); R; \overrightarrow{A_Q}(\ell \mapsto \Sigma)] \triangleright \overrightarrow{a}\rangle$$

where $\ell$ is chosen as in definition 4.15 and where $(\Sigma, \pi) = \mathsf{boot}\ P$. The only change from definition 4.15 is use of $\frac{\pi}{\emptyset}$ in place of $\pi$.

EQUIVALENCE BETWEEN MONOLITHIC AND INCREMENTAL MODELS. Programs using the incremental protocol and semantics are not directly comparable to those using the monolithic semantics. Each variation uses a unique language for communication between dataspaces and actors. However, any two assertion sets $\pi_1$ and $\pi_2$ can be equivalently represented by $\pi_1$ and a patch $\frac{\pi_2 - \pi_1}{\pi_1 - \pi_2}$, because $\pi_2 = \pi_1 \cup (\pi_2 - \pi_1) - (\pi_1 - \pi_2)$ and $(\pi_2 - \pi_1) \cap (\pi_1 - \pi_2) = \emptyset$.

This idea suggests a technique for embedding an actor communicating via the monolithic protocol into a dataspace that uses the incremental protocol.[18] Specifically, the actor *integrates* the series of incoming patches to obtain knowledge about the state of the world, and *differentiates* its outgoing assertion sets with respect to previous assertion sets.

Every monolithic leaf actor can be translated into an equivalent incremental actor by composing its behavior function with a wrapper that performs this on-the-fly integration and differentiation. The reduction rules ensure that, if every monolithic leaf actor in a program is translated into an incremental actor in this way, each underlying monolithic-protocol behavior function receives events and emits actions *identical* to those seen in the run of the unmodified program using the monolithic semantics.

**Definition 4.50.** We write $[\![P_M]\!]$ to denote the translation of a monolithic-protocol program into the incremental-protocol language using this wrapping technique, and use `M` and `I` subscripts for monolithic and incremental constructs generally.

The translation maintains additional state with each leaf actor in order to compute patches from assertion sets and vice versa and to expose information required for judging equivalence between the two kinds of machine state. Where a leaf actor has private state $u$ in an untranslated program, it has state $(u, \pi_i, \pi_o)$ in the translated program. The new registers $\pi_i$ and $\pi_o$ are the actor's most recently delivered and produced assertion sets, respectively.

**Definition 4.51.** We write $\Sigma_M \approx \Sigma_I$ to denote equivalence between monolithic and incremental actor states. To see what this means, let us imagine hierarchical configurations as trees like the

18 The symmetry of translation between patches and assertion sets also makes it possible to embed incremental-protocol actors in a monolithic-protocol environment.

one in figure 13. Each actor and each dataspace becomes a node, and each edge represents the pair of queues connecting an actor to its container. For a monolithic-protocol configuration to be equivalent to an incremental-protocol configuration, it must have the same tree shape and equivalent leaf actors with identical private states. Furthermore, at each internal monolithic node (i.e., at each dataspace), the assertion store must be identical to that in the corresponding incremental node. Finally, events and actions queued along a given edge on the monolithic side must have the same *effects* as those queued on the corresponding incremental edge.

The effects of monolithic and incremental action queues are the same when corresponding slots in the queues contain either identical message-send actions, spawn actions that result in equivalent actors, or state change notifications that have the same effect on the assertion store in the containing dataspace. Comparing event queues is similar, except that instead of requiring state change notifications to have identical effects on the shared dataspace, we require that they instead identically modify the *perspective* on the shared dataspace that the actor they are destined for has been accumulating.

If the conditions for establishing $\Sigma_M \approx \Sigma_I$ are satisfied, then reduction of $\Sigma_M$ proceeds in lockstep with reduction of the equivalent $\Sigma_I$, and equivalence is preserved at each step.

**Theorem 4.52.** *For every monolithic program* $P_M$*, let* $(\Sigma_M^0, \pi_M^0) = \mathsf{boot}(P_M)$ *and* $(\Sigma_I^0, \pi_I^0) = \mathsf{boot}([\![P_M]\!])$*. Then,*

1. $\pi_M^0 = \pi_I^0$.
2. *If there exists* $\Sigma_M$ *such that* $\Sigma_M^0 \longrightarrow_M^n \Sigma_M$ *for some* $n \in \mathbb{N}$*, then there exists a unique* $\Sigma_I$ *such that* $\Sigma_I^0 \longrightarrow_I^n \Sigma_I$ *and* $\Sigma_M \approx \Sigma_I$*.*

*Proof (Sketch).* Conclusion 1 follows trivially from the definition of boot and the fact that the translation process does not alter an actor's initial assertion set. The bulk of the proof is devoted to establishing conclusion 2. We first define $(\!|P_M|\!)$ to mean augmentation of the monolithic program with the same additional registers as provided by $[\![P_M]\!]$. Second, we define an equivalence $\approx_{MM}$ between $(\!|\cdot|\!)$-translated and untranslated monolithic machine states that ignores the extra registers, and prove that reduction respects $\approx_{MM}$. Third, we prove that $(\!|P_M|\!)$ and $[\![P_M]\!]$ reduce in lockstep, and that an equivalence $\approx_{MI}$ between translated monolithic and incremental states is preserved by reduction. Finally, we prove that the two notions of equivalence $\approx_{MM}$ and $\approx_{MI}$ together imply the desired equivalence $\approx$. The full proof takes the form of a Coq script. □

## 4.7 PROGRAMMING WITH THE INCREMENTAL PROTOCOL

The incremental protocol occasionally simplifies programs for leaf actors. This applies not only to examples in Dataspace ISWIM, but also to large programs written for the Racket or JavaScript dataspace model implementations. Occasional simplification is not the only advantage of incrementality: the incremental protocol often improves the efficiency of programs. Theorem 4.52 allows programmers to choose on an actor-by-actor basis which protocol is most appropriate for a given task.

| | Short-lived observables (i.e. messages) | Long-lived observables (i.e. assertions) |
|---|---|---|
| Short-lived interest | — | Query-like behavior |
| Long-lived interest | Publish-subscribe | State replication, streaming queries |

Figure 16: Behavior resulting from variation of subscription lifetime and fact lifetime

For example, the demand-matcher example (numbered 4.3 above) can be implemented in a *locally-stateless* manner using patch-based state change notifications. It is no longer forced to maintain a record of the most recent set of active conversations, and thus no set subtraction is required. Instead, it can rely upon the added and removed sets in patch events it receives from its dataspace. The revised *demandMatcher* behavior function takes $()$ as its actor-private state value, since each event it receives conveys all the information it needs:

$$\mathit{demandMatcher}\ (\frac{\pi_{in}}{\pi_{out}}, ()) = \mathtt{continue}([\mathit{mkWorker}\ x \mid (\mathtt{hello}, x) \in \pi_{in}], ())$$

More generally, theorem 4.53 can free actors written using the incremental protocol from maintaining sets of assertions they have "seen before"; they may rely on the dataspace to unambiguously signal (dis)appearance of assertions.

**Theorem 4.53** (Concision)**.** *For all pairs of events* $e = \frac{\pi_1}{\pi_2}$ *and* $e' = \frac{\pi_3}{\pi_4}$ *delivered to an actor,* $c \in \pi_1 \cap \pi_3$ *only if some event* $\frac{\pi_5}{\pi_6}$ *was delivered between* $e$ *and* $e'$*, where* $c \in \pi_6$*. Symmetrically,* $c$ *cannot be retracted twice without being asserted in the interim.*

*Proof (Sketch).* The patch rule prunes patch actions against $R$ to ensure that only real changes are passed on in events. $R$ itself is then updated to incorporate the patch so that subsequent patches can be accurately pruned in turn. □

## 4.8 STYLES OF INTERACTION

The dataspace model offers a selection of different styles of interaction. In order for communication to occur at all, some actors must assert items of knowledge $c$, and others must simultaneously assert interest in such knowledge, $?c$. (Here, we may treat message-sending $\langle c \rangle$ as fleeting assertions of $c$, as discussed in section 4.4.) Varying the *lifetimes* of assertions placed in the dataspace gives rise to patterns of information exchange reminiscent of publish/subscribe messaging, state replication, streaming queries, and instantaneous queries.

Figure 16 summarizes the situation. There are four regions of interest shown. Only three yield interesting patterns of interaction: if both assertions of interest and assertions of knowledge are very short-lived, no communication can occur. There is no moment when the two kinds of assertion exist simultaneously.

When assertions of interest tend to be long-lived and assertions of the items of interest themselves tend to be brief in duration, a *publish/subscribe* pattern of interaction results. The as-sertions of interest can be thought of as *subscriptions* in this case. Publish/subscribe communication is naturally multi-party; however, point-to-point, channel-like messaging is readily available via a convention for assignment and use of channel names.

As the lifetimes of assertions representing knowledge increase, the pattern of interaction takes on a different character. It begins to resemble a *streaming query* style of knowledge transfer, where long-lived queries over a changing set of rows yield incrementally-maintained result sets. The resemblance is particularly strong when cast in terms of the incremental patch actions $\Delta$ introduced in section 4.6. Seen from a different perspective, this pattern of interaction appears similar to *state replication*, where spatially-distinct replicas of a set of information are maintained by exchange of messages. The monolithic state change notifications $\pi$ first introduced in section 4.1 most clearly capture the intuition backing this perspective.

Finally, if we consider long-lived assertions of knowledge in combination with briefer and briefer assertions of interest in this knowledge, the style of interaction approaches that of clients making `SELECT` queries against a shared SQL database. Here, the assertions of interest can usefully be thought of as *queries*. An important consideration in this style of interaction is the length of time that each query is maintained.[19]

There is no general answer to the question of how long an assertion of interest should be maintained in order to effectively act as a query over matching assertions. It varies from protocol to protocol. In some protocols, it is certain that the assertions of interest will be present at the moment the query is established, in which case an immediate retraction of interest is sound. In other protocols, queries must be held in place for some time to allow them to be detected and responded to. The specific duration depends on the mechanism by which such responses are to be produced: a local actor may be able to compute a result in one round-trip of control transfer, on demand; an actor communicating with a remote system over a network link may require queries to be held for a certain number of seconds.

An actor maintaining an assertion of interest for any non-trivial length of time at all runs the risk of the result set changing during the lifetime of the query. The longer the query is maintained, the more the style of interaction begins to resemble a streaming query and the less it has in common with SQL-style point-in-time queries of a snapshot of system state.

19 Abstractly, of course, time is measured in number of reduction steps rather than any real-world measure.

# 5

# *Computational Model II:* Syndicate

With the dataspace model, we have a flexible facility for communicating changes in conversational state among a group of actors. We are able to express styles of interaction ranging from unicast, multicast and broadcast messaging through streaming queries and state replication to shared-database-like protocols. The model's emphasis on structured exchange of public aspects of component state allows us to express a wide range of effects including service presence, fate sharing, and demand matching. These effects in turn serve as mechanisms by which a range of resource-allocation, -management, and -release policies may be expressed.

The dataspace model brings actors together into a conversational group, but says nothing about the internal structure of each leaf actor. Such actors are not only stateful, but *internally* concurrent. Each leaf actor is frequently engaged in more than one simultaneous conversation. Ordinary programming languages offer no assistance to the programmer for managing intra-actor control and state, even when (like Dataspace ISWIM) extended with dataspace-model-specific data types and functions. However, to simply discard such languages would be a mistake: practicality demands interoperability. If we design a new language specifically for leaf actor programming, we forfeit the benefits of the enormous quantity of useful software written in already-existing languages.[1] Instead, we seek tools for integrating the dataspace model not only with existing programs and libraries but with existing ways of thinking.

We will need new *control* structures reflecting the conversation-related concepts the dataspace model introduces. Programmers are asked to think in terms of asynchronous (nested sub-)conversations, but given ordinary sequential control flow. They are asked to maintain connections between actor-private state and published assertions in a shared space, but given ordinary variables and heaps. They are asked to respond to conversational implicatures expressing peers' needs, but offered no support for turning such demands into manageable units of programming. Conversely, they are asked to respond to signals indicating abandonment of a conversation by releasing local related resources, but given no means of precisely delimiting such resources. Finally, when a local control decision is made to end an interaction, programmers are left to manually ensure that this is communicated to affected peers.[2]

1 The idea of such a new language is nonetheless interesting, worthy of future exploration.

2 Actor languages face some of the same issues, especially as they relate to (de)multiplexing of conversations. Erlang, for example, is like the unadorned dataspace model in funneling all communication for an actor through a single behavior function. The E strategy of allocating a new object (what E terms a facet) to handle a given sub-conversation is an interesting approach that takes advantage of E's ability to offer peers different perspectives on shared state in a single vat. E facets thus overlap in intent with Syndicate facets at least in part.

The second part of the SYNDICATE design therefore builds on the dataspace model by proposing new language features to address these challenges. The new features are intended for incorporation into base languages used to express leaf actor behaviors. The central novelty is an explicit representation of a (sub-)conversation named a *facet*. Facets nest, forming a tree that mirrors the nested conversational structure of the actor's interactions. Each actor's private state is held in *fields*; each field is associated with a particular facet. Special declarations called *endpoints* allow the programmer to connect assertions in the dataspace with values held in local fields in a bidirectional manner. Endpoints describing *interest* in assertions—that is, endpoints that publish assertions of the form $?c$ into the dataspace—offer a convenient syntactic location for the specification of responses to the appearance and disappearance of matching assertions.

Facets, fields, and endpoints together allow the programmer to write programs in terms of conversations, conversational state, and conversational interactions. They connect local to shared state. They offer a unit of resource management that can come and go with changes in expressed demand. Finally, because the connection between a facet and the surrounding dataspace is bidirectional, adding or removing a facet automatically adds or removes its endpoints' assertions, allowing peers to detect and respond to the change. In the extreme case of an actor crash, all its facets are removed, automatically (if abruptly) ending all of its conversations.

SYNDICATE/λ. Chapter 4 used an informal quasi-language, Dataspace ISWIM, to illustrate the formal system underpinning the dataspace model. Here, we take a slightly different tack, illustrating new language features by presenting them as part of an otherwise-minimal, mathematical, λ-calculus-inspired base language with just enough structure to act as a backdrop. We call this language SYNDICATE/λ, by analogy with the full prototype implementations SYNDICATE/RKT and SYNDICATE/JS. In our formal presentation, we abstract away from concrete details of base value types and specific built-in operations; where needed for examples, we reuse the notation and concepts sketched for Dataspace ISWIM.

## 5.1 ABSTRACT SYNDICATE/λ SYNTAX AND INFORMAL SEMANTICS

Figure 17 displays the syntax of SYNDICATE/λ. It is stratified into expressions $e \in \mathbf{Expr}$ and reactive, imperative *programs* $\mathsf{Pr} \in \mathbf{Pr}$. Expressions are both terminating and pure up to exceptions caused by partial primitive functions. Programs describe the interesting features of the language. While expressions yield values, programs are evaluated solely for their side effects.

The empty or inert program is written $0$. A semicolon is used to denote a form of sequential composition, $\mathsf{Pr}_1;\mathsf{Pr}_2$. The inert program $0$ is both a left and a right identity for this form of composition. In this chapter, we identify terms up to arbitrary composition with $0$. This avoids spurious nondeterminism in reduction.

The usual λ-calculus syntax for application, $e_1\ e_2$, is only available to programs, because the language includes only *procedure* values $\lambda\,[(\mathsf{P}.\mathsf{Pr})\ \ldots]$ instead of the function values familiar from λ-calculus. Each $(\mathsf{P}.\mathsf{Pr})$ in a procedure value is a branch of a pattern-match construct. When the procedure is called, the supplied argument is tested against each $\mathsf{P}$ in left-to-right order, and the entire call reduces to the corresponding $\mathsf{Pr}$, substituted appropriately.

$$
\begin{array}{rcll}
\text{Programs } Pr \in \mathbf{Pr} & := & 0 & \text{inert} \\
& | & Pr; Pr & \text{composition} \\
& | & e\ e & \text{procedure call} \\
& | & \mathsf{let}\ x\ =\ e\ \mathsf{in}\ Pr & \text{bind immutable variable} \\
& | & \mathsf{let}\ x\ :=\ e\ \mathsf{in}\ Pr & \text{allocate mutable field} \\
& | & x \leftarrow e & \text{update mutable field} \\
& | & \mathsf{send}\ e & \text{send message via dataspace} \\
& | & \mathsf{spawn}\ Pr & \text{spawn actor} \\
& | & \mathsf{dataspace}\ Pr & \text{spawn dataspace} \\
& | & x\,[A\ \ (D\ Pr)\ \ldots] & \text{start facet} \\
& | & \mathsf{stop}\ x\ Pr & \text{stop facet}
\end{array}
$$

$$
\begin{array}{rcl}
\text{Expressions } e \in \mathbf{Expr} & := & b \mid (e, \ldots) \mid p\ e\ \ldots \mid x \mid \lambda\,[(P.Pr)\ \ldots] \\
\text{Local values } v \in \mathbf{Val}^{\lambda} & := & b \mid (v, \ldots) \mid \qquad\qquad\quad \lambda\,[(P.Pr)\ \ldots] \\
\text{Assertions } c \in \mathbf{Val} & := & b \mid (c, \ldots) \\
\text{Assertion sets } \pi \in \mathbf{ASet} & = & \mathcal{P}(\mathbf{Val}) \\
\text{Names } x \in \mathbf{Var} & & \text{(used to denote variables, fields, facets)}
\end{array}
$$

$$
\begin{array}{rcl}
\text{Event patterns } D \in \mathbf{EPat} & := & \mathsf{asserted}\ P \\
& | & \mathsf{retracted}\ P \\
& | & \mathsf{message}\ \langle P \rangle \\
& | & \mathsf{start} \\
& | & \mathsf{stop}
\end{array}
$$

$$
\begin{array}{rcl}
\text{Dataspace events } \epsilon \in \mathbf{Evt} & := & \langle c \rangle \mid \Delta \\
\text{Local events } \epsilon^{+} \in \mathbf{Evt}^{+} & := & \langle c \rangle \mid \Delta \mid \mathsf{start} \mid \mathsf{stop} \\
\text{Dataspace actions } a \in \mathbf{Act} & := & \langle c \rangle \mid \Delta \mid \mathsf{actor}\ g\ \pi
\end{array}
$$

$$
\begin{array}{rcl}
\text{Patterns } P \in \mathbf{Pat} & := & \star \mid b \mid (P, \ldots) \mid p\ e\ \ldots \mid x \mid \$x \\
\text{Assertion templates } k \in \mathbf{Tmpl} & := & \star \mid b \mid (k, \ldots) \mid p\ e\ \ldots \mid x \\
\text{Pattern values } \mathcal{I} \in \mathbf{PVal} & := & \star \mid b \mid (k, \ldots) \mid \qquad\qquad \$x \\
\text{Assertion endpoints } A \in \mathbf{Tmpls} & := & \emptyset \mid k \cup A \\
\text{Base values } b \in \mathbf{BVal} & = & \text{Atoms, incl. strings, symbols, numbers, etc.} \\
\text{Primitive functions } p \in \mathbf{Prim} & &
\end{array}
$$

Figure 17: Syntax of Syndicate/λ programs

It is not only SYNDICATE/λ syntax that is stratified. SYNDICATE/λ bindings come in three flavors: immutable variables ("variables"), mutable fields ("fields"), and names for facets ("facet names"). The first two are introduced by the two forms of let, and the third is introduced as an automatic consequence of creating a facet. Variables may include values containing procedures, but fields must not. While not strictly required, this restriction captures some of the spirit of programming in SYNDICATE; recall from section 2.6 the desire to eschew sharing of higher-order data. Field update, $x \leftarrow e$, naturally applies only to fields, not variables,[3] and the value to be stored in the field must not directly or indirectly contain a procedure value.

The command send $e$ emits a dataspace model action of the form $\langle c \rangle$, where $c$ is the result of evaluating $e$. Similarly, the command spawn $Pr$ spawns a sibling actor in the dataspace, and dataspace $Pr$ spawns a sibling *dataspace* initially containing a lone actor with behavior $Pr$. Spawned programs $Pr$ may refer to arbitrary variables and fields of their spawning actor; at the moment of the spawn, the store is effectively duplicated, meaning that mutations to fields subsequently performed affect only the actor performing them.

The final two syntactic forms create and destroy facets. The form $x\,[A\ \ (D\ Pr)\ \ldots]$ specifies a facet template which is instantiated at the moment the form is interpreted. Once instantiated, the new facet's endpoints—the assertion endpoint $A$ and the event-handling endpoints $(D\ Pr)$—become active and contribute assertions to the aggregate of assertions published by the actor as a whole.

Each assertion endpoint $A$ is written using syntax chosen to connote set construction. The meaning of such an endpoint is exactly a set of assertions, the union of the sets denoted by the assertion templates $k$ embedded in the syntax of the assertion endpoint. Changing a field that is referred to by an assertion endpoint *automatically* changes the assertions published by that endpoint. In this way, SYNDICATE/λ programs are able to publish assertions that track changes in local state.

Similarly, event-handling endpoints $(D\ Pr)$ contribute assertions of interest derived from the event pattern $D$ into the dataspace, as well as specifying a subprogram $Pr$ to run when any event relating to $D$ is delivered. Event patterns $D$ may select the appearance (asserted $P$) or disappearance (retracted $P$) of assertions matching some pattern, the arrival of a message (message $\langle P \rangle$) matching some pattern, or the *synthetic* events start and stop which relate to facet lifecycle.[4] Patterns that contain binders $\$x$ capture portions of assertions in matching events, making $x$ available in subprograms $Pr$. As with assertion endpoints, every pattern $P$ automatically tracks changes in fields it refers to.

The form stop $x$ $Pr$, only legal when surrounded by a facet named $x$, causes that facet—and all its nested subfacets—to terminate cleanly, executing any stop event handlers they might have. Once a terminating facet becomes inert, after its stop handlers have completed their tasks, its assertions are removed from the shared dataspace and the facet itself is then deleted. The program $Pr$ in stop $x$ $Pr$ is then scheduled to execute *alongside* the terminating facet, so that any facets that $Pr$ creates will exist in the actor's facet tree as *siblings* of the just-stopped facet $x$.

3 The "well-formedness" judgment of section 5.5 enforces this requirement, among others.

4 The start and stop events are purely internal, having no connection to any dataspace-level events or actions. They are used for structuring the ordering of side-effects within a SYNDICATE/λ actor.

Despite being layered atop the dataspace model, the events and actions of that model are not directly exposed to the Syndicate/λ programmer the way that they are in Dataspace ISWIM. Instead of yielding values describing actions to perform in a functional style, programs perform side-effecting operations like send and spawn. Instead of functional state transduction, programs imperatively update fields. Instead of describing changes to published assertion sets, programs create facets with embedded endpoints. Finally, instead of manually directing control flow by analyzing and interpreting received events, programs declare event-handling endpoints, which are activated as appropriate.

**Example 5.1.** For our first example, let us revisit the shared mutable reference cell actors of example 4.2. First, we spawn an actor implementing the cell itself:[5]

$$\mathsf{spawn}\ (\mathsf{let}\ v\ :=\ 0\ \mathsf{in}\ \mathit{box}\ \big[\emptyset \cup (\mathtt{value}, v)\ \ \big(\mathsf{message}\ \langle(\mathtt{set}, \$v')\rangle\ (v \leftarrow v')\big)\big])$$

This actor first creates a new field $v$, initialized to zero. It then creates a single facet named *box*, which has an assertion endpoint that places the assertion $(\mathtt{value}, v)$ into the shared dataspace. The semantics of Syndicate/λ automatically update this published assertion as the value of $v$ changes in response to subsequent events. The *box* facet also has a single event-handling endpoint. In response to an incoming set message, the endpoint updates the field $v$ to contain the new value $v'$ specified in the received message.

The client actor from example 4.2 can be written as follows:

$$\mathsf{spawn}\ \mathit{boxClient}\ [\emptyset\ \ (\mathsf{asserted}\ (\mathtt{value}, \$v)\ (\mathsf{send}\ (\mathtt{set}, v+1)))]$$

This actor is stateless, having no fields. It creates a single facet, *boxClient*, which makes no assertions but contains a single event-handling endpoint which responds to patch events. If such a patch event describes the appearance of an assertion matching the pattern $(\mathtt{value}, \$v)$, the endpoint sends a message $\langle(\mathtt{set}, v+1)\rangle$ via the dataspace. (We imagine here that **Prim** includes functions for arithmetic and assume a convenient infix syntax.) Of course, the *box* actor responds to such messages by updating its value assertion, which triggers *boxClient* again. This cycle repeats *ad infinitum*. ◊

**Example 5.2.** Next, we translate the demand-matcher from example 4.3 to Syndicate/λ:

$$\mathsf{let}\ \mathit{worker}\ =\ \lambda\,[(\$x.w\,[\emptyset\ \ (\mathsf{retracted}\ (\mathtt{hello}, x)\ \ (\mathsf{stop}\ w\ 0))\ \ (\mathsf{start}\ \ldots)])]\ \ \mathsf{in}$$
$$\mathsf{spawn}\ \mathit{demandMatcher}\,[\emptyset\ \ (\mathsf{asserted}\ (\mathtt{hello}, \$x)\ \ (\mathsf{spawn}\ (\mathit{worker}\ x)))]$$

The single event-handling endpoint in facet *demandMatcher* responds to each asserted hello tuple by spawning a new actor, which begins its existence by calling the procedure *worker*, passing it the $x$ from the assertion that led to its creation. In turn, *worker* creates a facet $w$ which monitors *retraction* of $(\mathtt{hello}, x)$ in addition to performing whichever startup actions a worker should perform. When the last peer to assert $(\mathtt{hello}, x)$ retracts its assertion, the worker terminates itself by performing a stop command on its top-level facet, supplying $0$ to replace it.

The concision of Syndicate/λ has allowed us to show how a worker terminates itself once demand for its existence disappears. The Dataspace ISWIM version of example 4.3 omits this functionality: it is possible but verbose to express in Dataspace ISWIM. ◊

5 We have dispensed here with the *id* field of example 4.2.

$$
\begin{array}{rll}
\text{Facet trees } S, T \in \textbf{Tree} := & \mathrm{Pr} & \text{unreduced code} \\
| & \spadesuit & \text{exception} \\
| & S; S & \text{composition} \\
| & x\,[A\ \ (D\ \mathrm{Pr})\ \ldots].S & \text{running facet} \\
| & x\,[A\ D\ \ldots]\dagger S & \text{stopping facet} \\
| & \%\,[S] & \text{termination boundary}
\end{array}
$$

$$\text{Inert facet trees } S_I, T_I \in \textbf{Tree}_\mathbf{I} := 0 \mid S_I; S_I \mid x\,[A\ \ (D\ \mathrm{Pr})\ \ldots].S_I$$

$$\text{Contexts } E, F \in \textbf{Ctxt} := \square \mid E; S \mid S_I; E \mid x\,[A\ \ (D\ \mathrm{Pr})\ \ldots].E \mid x\,[A\ D\ \ldots]\dagger E \mid \%\,[E]$$

$$
\begin{array}{rl}
\text{Field stores } \sigma \in \textbf{Store} = & \textbf{Var} \rightharpoonup \textbf{Val} \\
\text{Machine states } M \in \textbf{M} := & \langle \sigma, \pi, \pi, \overrightarrow{a}, S \rangle \\
\text{Inert machine states } M_I \in \textbf{M}_\mathbf{I} := & \langle \sigma, \pi, \pi, \cdot, S_I \rangle
\end{array}
$$

Figure 18: Evaluation Syntax, Contexts and Machine States

## 5.2 FORMAL SEMANTICS OF SYNDICATE/λ

Figure 18 introduces syntax for describing evaluation states of **Tree**s of facets, as well as a specification of what it means for such a tree to be inert, a definition of evaluation contexts (**Ctxt**), field **Store**s, and reducible and inert machine states (**M** and $\mathbf{M_I}$).

A tree of facets may include unreduced commands drawn from **Pr**. Reduction interprets these commands, applying any side effects they entail to the machine state. A tree may also include an exception marker, ♠, which arises as a result of various run-time error conditions and leads to abrupt actor termination. The composition operator on facet trees loses much of the flavor of sequentiality that it enjoys in programs, and acts instead primarily to separate (and order) adjacent sibling facets in the tree. However, evaluation contexts prefer redexes in the left-hand side of a composition to those in the right-hand side, thus preserving the intuitive ordering of effects.

The form $x\,[A\ \ (D\ \mathrm{Pr})\ \ldots].S$ describes an instantiated, running facet, with active endpoints. It serves as an interior node in a facet tree. Any facets contained in $S$ are considered nested children of $x$. If $x$ is later stopped, all facets in $S$ are stopped as well.

The final two syntactic forms describing facet trees relate to shutdown of facets. First, $x\,[A\ D\ \ldots]\dagger S$ describes a facet that is marked as *terminating*. The facet cannot be deleted until $S$ has reached inertness, but it will no longer react to incoming events, as can be seen from the lack of $\mathrm{Pr}$ event handlers associated with each $D$. Second, $\%\,[S]$ marks a *contour* within the tree. Contained facets and subfacets of $S$ will transition to terminating state as soon as they become inert. An explicit contour is necessary because a facet may create a sibling or child facet as a response to being terminated, and such "hail mary" facets must not be allowed to escape termination.[6]

The reduction relation $M \longrightarrow M'$ operates on a machine state $\langle \sigma, \pi_i, \pi_o, \overrightarrow{a}, S \rangle$ containing five registers:

1. $\sigma$ is the store, mapping field identifiers in **Var** to field values in **Val**. Higher-order values such as procedures may not be placed in the store.
2. $\pi_i$ is the actor's record of the assertions it has learned from the dataspace. As patch events arrive from the dataspace, $\pi_i$ is updated.
3. $\pi_o$ is the actor's record of the assertions it has placed into the dataspace. As fields are updated and facets are created and destroyed, the actor issues patch actions and updates $\pi_o$ to account for the changes.
4. $\overrightarrow{a}$ is an accumulator of dataspace model actions produced. As messages are sent, actors are spawned, and changes are made to published assertions, actions are appended to this register.
5. $S$ is the tree of facets, the actor's behavior and control state. Reduction drives this tree of facets toward inertness.

EVALUATION OF EXPRESSIONS AND PATTERNS. The semantics of SYNDICATE/λ depends on evaluation of expressions in a number of places. Evaluation of expressions is straightforward, since no function or procedure calls (other than to primitives) are allowed. In addition, because SYNDICATE/λ *patterns* include calls to primitive functions and references to field values, the semantics requires a means of "evaluating" a pattern.

**Definition 5.3** (Evaluation of expressions)**.** The partial metafunction $eval^\lambda$ evaluates an **Expr** to a $\mathbf{Val}^\lambda$, resolving field references using a **Store**.[7]

$$
\begin{aligned}
eval^\lambda &: \mathbf{Store} \times \mathbf{Expr} \rightharpoonup \mathbf{Val}^\lambda \\
eval^\lambda\ \sigma\ b &= b \\
eval^\lambda\ \sigma\ (e, \dots) &= (eval^\lambda\ \sigma\ e, \dots) \\
eval^\lambda\ \sigma\ (p\ e\ \dots) &= delta^\lambda\ p\ \overrightarrow{eval^\lambda\ \sigma\ e} \\
eval^\lambda\ \sigma\ x &= \sigma[x] \\
eval^\lambda\ \sigma\ \lambda\,[(P.Pr)\ \dots] &= \lambda\,[(P.Pr)\ \dots]
\end{aligned}
$$

$$
\begin{aligned}
eval &: \mathbf{Store} \times \mathbf{Expr} \rightharpoonup \mathbf{Val} \\
eval\ \sigma\ e &= v \text{ if } v = eval^\lambda\ \sigma\ e \in \mathbf{Val}
\end{aligned}
$$

6 Note, however, that stop $x$ $Pr$ explicitly hoists $Pr$ out of any termination boundary associated with facet $x$.

7 Field references are not resolved under $\lambda$ (per the last line of the definition of $eval^\lambda$), because to do so would be premature: updates to the store between the use of $eval^\lambda$ and subsequent invocation of the procedure would be lost.

The metafunction *eval* is like $eval^\lambda$, but with domain **Val** instead of $\mathbf{Val}^\lambda$. It is used in contexts where procedure values are forbidden, such as values used to initialize or update a field, or values serving as the body of a message to be transmitted. Both $eval^\lambda$ and *eval* are undefined in cases where they depend on a use of $delta^\lambda$ that is in turn undefined.

**Definition 5.4** (Primitive functions)**.** The partial metafunction $delta^\lambda$ interprets applications of primitive functions $p \in \mathbf{Prim}$, and *delta* is to $delta^\lambda$ as *eval* is to $eval^\lambda$. We do not specify a fixed **Prim** here, and so escape the need to fix $delta^\lambda$ in any detail.

$$delta^\lambda : \mathbf{Prim} \times \overrightarrow{\mathbf{Val}^\lambda} \rightharpoonup \mathbf{Val}^\lambda$$

$$delta : \mathbf{Prim} \times \overrightarrow{\mathbf{Val}^\lambda} \rightharpoonup \mathbf{Val}$$

**Definition 5.5** ("Evaluation" of patterns)**.** The metafunction *snapshot* "evaluates" a pattern by computing the results of any embedded calls to primitive operations or references to field values from the store. This "evaluation" process may fail with an exception; however, if it succeeds, the resulting pattern does not include any primitive operations or field references, and therefore is guaranteed not to signal an exception when used.

$$snapshot : \mathbf{Store} \times \mathbf{Pat} \to \mathbf{PVal}_\spadesuit$$

$$snapshot\ \sigma\ \star = \star$$

$$snapshot\ \sigma\ b = b$$

$$snapshot\ \sigma\ () = ()$$

$$snapshot\ \sigma\ (P_1, P_2, \ldots) = \begin{cases} (P_1', P_2', \ldots) & \text{if } P_1' = snapshot\ \sigma\ P_1 \\ & \text{and } (P_2', \ldots) = snapshot\ \sigma\ (P_2, \ldots) \\ \spadesuit & \text{otherwise} \end{cases}$$

$$snapshot\ \sigma\ (p\ e\ \ldots) = \begin{cases} v & \text{if } v = delta\ p\ \overrightarrow{eval^\lambda\ \sigma\ e} \\ \spadesuit & \text{otherwise} \end{cases}$$

$$snapshot\ \sigma\ x = \sigma[x]$$

$$snapshot\ \sigma\ \$x = \$x$$

THE ACTIVE ASSERTION SET. As facets come and go and fields change their values, the set of assertions to be placed into the surrounding dataspace by a SYNDICATE/λ actor changes. The set must be tracked and, as it changes, corresponding patch actions must be computed and emitted.

**Definition 5.6.** The metafunction *assertions*, which extracts the current set of assertions from a tree of facets, is defined in figure 19.[8] It is a pedestrian structural traversal of syntax except when processing an event pattern D. In that case, it specially adds the assertion-of-interest constructor ?· to each assertion arising from the pattern inside D.

8 For simplicity of presentation, *assertions* is given as a partial function; it is undefined where *snapshot* yields ♠.

$$\textit{assertions} : \textbf{Store} \times \textbf{Tree} \rightharpoonup \textbf{ASet}$$

$$\textit{assertions}\ \sigma\ S = \begin{cases} \textit{assertions}\ \sigma\ T; \textit{assertions}\ \sigma\ T' & \text{if } S = T; T' \\ \textit{assertions}\ \sigma\ A \cup \textit{assertions}\ \sigma\ D \cup \cdots \cup \textit{assertions}\ \sigma\ T & \text{if } S = x\,[A\ \ (D\ Pr)\ \ \ldots].T \\ \textit{assertions}\ \sigma\ A \cup \textit{assertions}\ \sigma\ D \cup \cdots \cup \textit{assertions}\ \sigma\ T & \text{if } S = x\,[A\ D\ \ \ldots] \dagger T \\ \emptyset & \text{otherwise} \end{cases}$$

$$\textit{assertions} : \textbf{Store} \times \textbf{EPat} \rightharpoonup \textbf{ASet}$$

$$\textit{assertions}\ \sigma\ D = \begin{cases} \emptyset & \text{if } D = \mathsf{start} \text{ or } D = \mathsf{stop} \\ \{?c \mid c \in \pi\} & \text{if } D = \mathsf{asserted}\ P,\ D = \mathsf{retracted}\ P \text{ or } D = \mathsf{message}\ \langle P \rangle \\ & \text{where } \pi = \textit{assertions}'\ \ (\textit{snapshot}\ \sigma\ P) \end{cases}$$

$$\textit{assertions}' : \textbf{PVal} \to \textbf{ASet}$$

$$\textit{assertions}'\ P = \begin{cases} \textbf{Val} & \text{if } P = \star \text{ or } P = \$x \\ \{b\} & \text{if } P = b \\ \{()\} & \text{if } P = () \\ \{v \times v' \mid v \in \textit{assertions}'\ P', v' \in \textit{assertions}'\ (P'', \ldots)\} & \text{if } P = (P', P'', \ldots) \end{cases}$$

$$\textit{assertions} : \textbf{Store} \times \textbf{Tmpls} \rightharpoonup \textbf{ASet}$$

$$\textit{assertions}\ \sigma\ A = \begin{cases} \emptyset & \text{if } A = \emptyset \\ \textit{assertions}\ \sigma\ k \cup \textit{assertions}\ \sigma\ A' & \text{if } A = k \cup A' \end{cases}$$

$$\textit{assertions} : \textbf{Store} \times \textbf{Tmpl} \rightharpoonup \textbf{ASet}$$

$$\textit{assertions}\ \sigma\ k = \textit{assertions}'\ \ (\textit{snapshot}\ \sigma\ k)$$

Figure 19: The (overloaded) *assertions* metafunction

**Definition 5.7.** In situations where an actor's assertion set may have changed, the metafunction *patch* is used to compute an updated $\pi_o$ register as well as a patch to be appended to the pending action accumulator.

$$\mathit{patch} : \mathbf{Store} \times \mathbf{ASet} \times \mathbf{Tree} \to (\mathbf{ASet} \times \mathbf{Patch})_{\spadesuit}$$

$$\mathit{patch}\ \sigma\ \pi_o\ S = \begin{cases} \left(\pi'_o, \frac{(\pi'_o - \pi_o)}{(\pi_o - \pi'_o)}\right) & \text{if } \pi'_o = \mathit{assertions}\ \sigma\ S \\ \spadesuit & \text{otherwise} \end{cases}$$

**Definition 5.8.** The metafunction *emit* takes care of combining a patch action (often resulting from *patch*) with an existing action queue. Any adjacent enqueued patch actions are *coalesced* using a patch composition operator. By contrast, no such coalescing is desired (or possible) when enqueueing message or actor-creation actions.

$$\mathit{emit} : \overrightarrow{\mathbf{Act}} \times \mathbf{Patch} \to \overrightarrow{\mathbf{Act}}$$

$$\mathit{emit}\ \cdot\ \Delta = \Delta$$

$$\mathit{emit}\ (\overrightarrow{a}\ a')\ \Delta = \begin{cases} \overrightarrow{a}\ (\Delta \circ \Delta') & \text{if } a' = \Delta' \\ \overrightarrow{a}\ a'\ \Delta & \text{otherwise} \end{cases}$$

**Definition 5.9** (Patch composition)**.** The patch composition operator is defined as follows:

$$\cdot \circ \cdot : \mathbf{Patch} \times \mathbf{Patch} \to \mathbf{Patch}$$

$$\frac{\pi'_{in}}{\pi'_{out}} \circ \frac{\pi_{in}}{\pi_{out}} = \frac{\pi_{in} \cup \pi'_{in} - \pi'_{out}}{\pi_{out} - \pi'_{in} \cup \pi'_{out}}$$

PATTERN MATCHING. The SYNDICATE/$\lambda$ semantics also makes use of pattern matching in a number of places. Occasionally, a suite of patterns with matching continuations must be searched for the first match for some value; at other times, matching of a single pattern with a single value is required.

**Definition 5.10.** The metafunction *matchInOrder* searches a collection of $(P.Pr)$ branches, often extracted from a procedure value, to find the first that matches the argument value given. If none of the branches match, an exception is signaled.

$$\mathit{matchInOrder} : \mathbf{Store} \times \mathbf{Val}^{\lambda} \times \overrightarrow{(\mathbf{Pat} \times \mathbf{Pr})} \to \mathbf{Pr}_{\spadesuit}$$

$$\mathit{matchInOrder}\ \sigma\ \nu\ \cdot = \spadesuit$$

$$\mathit{matchInOrder}\ \sigma\ \nu\ ((P, Pr)\ \overrightarrow{(P', Pr')}) = \begin{cases} \spadesuit & \text{if } \mathit{snapshot}\ \sigma\ P = \spadesuit \\ Pr'' & \text{if } \mathit{match}\ (\mathit{snapshot}\ \sigma\ P)\ \nu\ Pr = Pr'' \\ \mathit{matchInOrder}\ \sigma\ \nu\ \overrightarrow{(P', Pr')} & \text{if } \mathit{match}\ (\mathit{snapshot}\ \sigma\ P)\ \nu\ Pr \text{ is undefined} \end{cases}$$

**Definition 5.11.** The partial metafunction *match* is defined when the given $\mathbf{Val}^\lambda$ matches the given **PVal**, and is otherwise undefined. The result of *match* is a program fragment that, when interpreted, uses let to bind pattern variables before continuing with the **Pr** given to *match*.[9]

$$\begin{aligned} & \mathit{match} : \mathbf{PVal} \times \mathbf{Val}^\lambda \times \mathbf{Pr} \rightharpoonup \mathbf{Pr} \\ \mathit{match}\ \star\ v\ \mathrm{Pr} &= \mathrm{Pr} \\ \mathit{match}\ b\ b\ \mathrm{Pr} &= \mathrm{Pr} \\ \mathit{match}\ ()\ ()\ \mathrm{Pr} &= \mathrm{Pr} \\ \mathit{match}\ (P, P', \ldots)\ (v, v', \ldots)\ \mathrm{Pr} &= \mathit{match}\ P\ v\ (\mathit{match}\ (P', \ldots)\ (v', \ldots)\ \mathrm{Pr}) \\ \mathit{match}\ \$x\ v\ \mathrm{Pr} &= \mathsf{let}\ x\ =\ v\ \mathsf{in}\ \mathrm{Pr} \end{aligned}$$

REDUCTION RELATION. The reduction relation is defined by fourteen rules,[10] shown in full in figures 20 and 21. The call rule implements procedure call, and rule let allows introduction of immutable variables. The new-field and set-field rules manipulate fields, while rules send, spawn and dataspace produce actions for interpretation by an actor's surrounding dataspace. The remainder of the rules relate to facet startup and shutdown: boot-facet instantiates a facet, while the two stop-facet rules, three stop-child rules, and burial rule combine to handle the process of facet termination.

**Definition 5.12** (Rule call)**.** The call rule interprets procedure calls $e_1\ e_2$:

$$\left\langle \sigma, \pi_i, \pi_o, \overrightarrow{a}, E[e_1\ e_2] \right\rangle \longrightarrow \left\langle \sigma, \pi_i, \pi_o, \overrightarrow{a}, E\,[S] \right\rangle$$

It first attempts to evaluate both $e_1$ and $e_2$ to elements of $\mathbf{Val}^\lambda$ via the metafunction $\mathit{eval}^\lambda$. If both $\mathit{eval}^\lambda\ \sigma\ e_1 = \lambda\,[(P.\mathrm{Pr})\ \ldots] \in \mathbf{Val}^\lambda$ and $\mathit{eval}^\lambda\ \sigma\ e_2 = v \in \mathbf{Val}^\lambda$, then $S = \mathit{matchInOrder}\ \sigma\ v\ \overrightarrow{(P, \mathrm{Pr})}$ on the right hand side of the relation; otherwise, $S = \spadesuit$.

**Definition 5.13** (Rule let)**.** The first kind of let construct allows programmers to give names to values drawn from $\mathbf{Val}^\lambda$. Machine states do not include an environment, and so our presentation makes use of hygienic substitution[11] to replace references to a bound variable $x$ with its let-computed value while respecting the notion of free names captured by the metafunction fv (figure 22).

$$\left\langle \sigma, \pi_i, \pi_o, \overrightarrow{a}, E[\mathsf{let}\ x\ =\ e\ \mathsf{in}\ \mathrm{Pr}] \right\rangle \longrightarrow \left\langle \sigma, \pi_i, \pi_o, \overrightarrow{a}, E[S] \right\rangle$$

If $\mathit{eval}^\lambda\ \sigma\ e = v \in \mathbf{Val}^\lambda$, then $S = \{v/x\}\,\mathrm{Pr}$ on the right hand side; otherwise, $S = \spadesuit$.

**Definition 5.14** (Rule new-field)**.** The second kind of let construct creates a new field, allocating a fresh name $y$ for the field and substituting $y$ for references to the field in the body of the

9 As written, *match* admits repeated pattern variables, allowing later uses of a binder to shadow earlier uses. Implementations of the SYNDICATE design may reasonably vary in their responses to this situation, depending on the idioms of the base language.

10 The development of the reduction rules was informed by discussions with Sam Caldwell.

11 See Barendregt (1984, ch. 2). Our notation $\{v/x\}\,\mathrm{Pr}$ reads "replace $x$ with $v$ in Pr".

$$\left\langle \sigma, \pi_i, \pi_o, \overrightarrow{a}, E[e_1\ e_2] \right\rangle \longrightarrow \left\langle \sigma, \pi_i, \pi_o, \overrightarrow{a}, E\,[S] \right\rangle \qquad \text{(call)}$$

$$\text{where } S = \begin{cases} \textit{matchInOrder}\ \sigma\ v\ \overrightarrow{(P, Pr)} & \text{if } \lambda\,[(P.Pr)\ \ldots] = \textit{eval}^{\lambda}\ \sigma\ e_1 \\ & \text{and } v = \textit{eval}^{\lambda}\ \sigma\ e_2 \\ \spadesuit & \text{otherwise} \end{cases}$$

$$\left\langle \sigma, \pi_i, \pi_o, \overrightarrow{a}, E[\mathsf{let}\ x\ =\ e\ \mathsf{in}\ Pr] \right\rangle \longrightarrow \left\langle \sigma, \pi_i, \pi_o, \overrightarrow{a}, E[S] \right\rangle \qquad \text{(let)}$$

$$\text{where } S = \begin{cases} \{v/x\}\,Pr & \text{if } v = \textit{eval}^{\lambda}\ \sigma\ e \\ \spadesuit & \text{otherwise} \end{cases}$$

$$\left\langle \sigma, \pi_i, \pi_o, \overrightarrow{a}, E[\mathsf{let}\ x\ :=\ e\ \mathsf{in}\ Pr] \right\rangle \longrightarrow \left\langle \sigma', \pi_i, \pi_o, \overrightarrow{a}, E[S] \right\rangle \qquad \text{(new-field)}$$

$$\text{where } y \text{ fresh and } \left(\sigma', S\right) = \begin{cases} (\sigma[y \mapsto v], \{y/x\}\,Pr) & \text{if } v = \textit{eval}\ \sigma\ e \\ (\sigma, \spadesuit) & \text{otherwise} \end{cases}$$

$$\left\langle \sigma, \pi_i, \pi_o, \overrightarrow{a}, E[x \leftarrow e] \right\rangle \longrightarrow \left\langle \sigma', \pi_i, \pi_o', \textit{emit}\ \overrightarrow{a}\ \Delta, E[S] \right\rangle \qquad \text{(set-field)}$$

$$\text{where } x \in \mathsf{dom}(\sigma)$$

$$\left(\sigma', S, \pi_o', \Delta\right) = \begin{cases} (\sigma[x \mapsto v], 0, \pi_o', \Delta) & \text{if } v = \textit{eval}\ \sigma\ e \\ & \text{and } (\pi_o', \Delta) = \textit{patch}\ \sigma\ \pi_o\ E[0] \\ \left(\sigma, \spadesuit, \pi_o, \frac{\emptyset}{\emptyset}\right) & \text{otherwise} \end{cases}$$

$$\left\langle \sigma, \pi_i, \pi_o, \overrightarrow{a}, E[\mathsf{send}\ e] \right\rangle \longrightarrow \left\langle \sigma, \pi_i, \pi_o, \overrightarrow{a}\ \overrightarrow{a}', E[S] \right\rangle \qquad \text{(send)}$$

$$\text{where } \left(\overrightarrow{a}', S\right) = \begin{cases} (\langle v \rangle, 0) & \text{if } v = \textit{eval}\ \sigma\ e \\ (\cdot, \spadesuit) & \text{otherwise} \end{cases}$$

$$\left\langle \sigma, \pi_i, \pi_o, \overrightarrow{a}, E[\mathsf{spawn}\ Pr] \right\rangle \longrightarrow \left\langle \sigma, \pi_i, \pi_o, \overrightarrow{a}\ (\mathsf{actor}\ (\textit{setup}\ (\sigma, Pr))\ \emptyset), E[0] \right\rangle \qquad \text{(spawn)}$$

$$\left\langle \sigma, \pi_i, \pi_o, \overrightarrow{a}, E[\mathsf{dataspace}\ Pr] \right\rangle \longrightarrow \left\langle \sigma, \pi_i, \pi_o, \overrightarrow{a}\ a', E[0] \right\rangle \qquad \text{(dataspace)}$$

$$\text{where } a' = \mathsf{dataspace}\ (\mathsf{actor}\ (\textit{setup}\ (\sigma, Pr))\ \emptyset)$$

Figure 20: SYNDICATE/λ reduction rules (procedure call, variables, fields, actions)

$$\left\langle \sigma, \pi_i, \pi_o, \overrightarrow{a}, E[x\,[A\ \ (D\ Pr)\ \ldots]] \right\rangle \longrightarrow \left\langle \sigma, \pi_i, \pi_o', \textit{emit}\ \overrightarrow{a}\ \Delta, E\left[S'\right] \right\rangle \qquad \text{(boot-facet)}$$

$$\text{where } S = (y\,[A\ \ (D\ \ (\{y/x\}\,Pr))\ \ \ldots]\,.T)$$

$$(S', \pi_o', \Delta) = \begin{cases} (S, \pi_o', \Delta) & \text{if } (\pi_o', \Delta) = \textit{patch}\ \sigma\ \pi_o\ E\,[S] \\ \left(♠, \pi_o, \frac{\emptyset}{\emptyset}\right) & \text{otherwise} \end{cases}$$

$$T_{\mathsf{start}} = \textit{handle}\ \emptyset\ \pi_i\ \sigma\ \mathsf{start}\ \overrightarrow{(D, Pr)}$$

$$T_{\mathsf{asserted}} = \textit{handle}\ \emptyset\ \pi_i\ \sigma\ \frac{\pi_i}{\emptyset}\ \overrightarrow{(D, Pr)}$$

$$T = \{y/x\}\,(T_{\mathsf{start}}; T_{\mathsf{asserted}})$$

$$y \text{ fresh}$$

$$\left\langle \sigma, \pi_i, \pi_o, \overrightarrow{a}, E[x\,[A\ \ (D\ Pr)\ \ \ldots]\,.F[\mathsf{stop}\ x\ Pr']] \right\rangle \longrightarrow \left\langle \sigma, \pi_i, \pi_o, \overrightarrow{a}, E[\%\,[x\,[A\ \ (D\ Pr)\ \ \ldots]\,.F[0]]\,; Pr'] \right\rangle \qquad \text{(stop-facet-1)}$$

$$\text{where } x \notin \mathsf{bv}(F)$$

$$\left\langle \sigma, \pi_i, \pi_o, \overrightarrow{a}, E[x\,[A\ D\ \ \ldots]\dagger F[\mathsf{stop}\ x\ Pr']] \right\rangle \longrightarrow \left\langle \sigma, \pi_i, \pi_o, \overrightarrow{a}, E[x\,[A\ D\ \ \ldots]\dagger F[0]; Pr'] \right\rangle \qquad \text{(stop-facet-2)}$$

$$\text{where } x \notin \mathsf{bv}(F)$$

$$\left\langle \sigma, \pi_i, \pi_o, \overrightarrow{a}, E[\%\,[0]] \right\rangle \longrightarrow \left\langle \sigma, \pi_i, \pi_o, \overrightarrow{a}, E[0] \right\rangle \qquad \text{(stop-child-1)}$$

$$\left\langle \sigma, \pi_i, \pi_o, \overrightarrow{a}, E[\%\,[S_I; T_I]] \right\rangle \longrightarrow \left\langle \sigma, \pi_i, \pi_o, \overrightarrow{a}, E[\%\,[S_I]\,; \%\,[T_I]] \right\rangle \qquad \text{(stop-child-2)}$$

$$\left\langle \sigma, \pi_i, \pi_o, \overrightarrow{a}, E[\%\,[x\,[A\ \ (D\ Pr)\ \ \ldots]\,.S_I]] \right\rangle \longrightarrow \left\langle \sigma, \pi_i, \pi_o, \overrightarrow{a}, E[x\,[A\ D\ \ \ldots]\dagger\%\,[S_I; T]] \right\rangle \qquad \text{(stop-child-3)}$$

$$\text{where } T = \textit{handle}\ \pi_i\ \pi_i\ \sigma\ \mathsf{stop}\ \overrightarrow{(D, Pr)}$$

$$\left\langle \sigma, \pi_i, \pi_o, \overrightarrow{a}, E[x\,[A\ D\ \ \ldots]\dagger 0] \right\rangle \longrightarrow \left\langle \sigma, \pi_i, \pi_o', \textit{emit}\ \overrightarrow{a}\ \Delta, E[S'] \right\rangle \qquad \text{(burial)}$$

$$\text{where } (S', \pi_o', \Delta) = \begin{cases} (0, \pi_o', \Delta) & \text{if } (\pi_o', \Delta) = \textit{patch}\ \sigma\ \pi_o\ E[0] \\ \left(♠, \pi_o, \frac{\emptyset}{\emptyset}\right) & \text{otherwise} \end{cases}$$

Figure 21: SYNDICATE/λ reduction rules (facet startup and shutdown)

$$
\begin{aligned}
\mathsf{fv} &: \textbf{Tree} \to \mathcal{P}(\textbf{Var})\\
\mathsf{fv}(0) &= \emptyset\\
\mathsf{fv}(S;T) &= \mathsf{fv}(S) \cup \mathsf{fv}(T)\\
\mathsf{fv}(e_1\ e_2) &= \mathsf{fv}(e_1) \cup \mathsf{fv}(e_2)\\
\mathsf{fv}(\mathsf{let}\ x\ :=\ e\ \mathsf{in}\ Pr) &= \mathsf{fv}(e) \cup (\mathsf{fv}(Pr) - \{x\})\\
\mathsf{fv}(\mathsf{let}\ x\ =\ e\ \mathsf{in}\ Pr) &= \mathsf{fv}(e) \cup (\mathsf{fv}(Pr) - \{x\})\\
\mathsf{fv}(x \leftarrow e) &= \{x\} \cup \mathsf{fv}(e)\\
\mathsf{fv}(\mathsf{spawn}\ Pr) &= \mathsf{fv}(Pr)\\
\mathsf{fv}(\mathsf{dataspace}\ Pr) &= \mathsf{fv}(Pr)\\
\mathsf{fv}(\mathsf{stop}\ x\ Pr) &= \{x\} \cup \mathsf{fv}(Pr)\\
\mathsf{fv}(x\,[A\ \ (D\ Pr)\ \ldots]) &= (\mathsf{fv}(A) \cup (\mathsf{fv}(D) \cup (\mathsf{fv}(Pr) - \mathit{formals}(D))) \cup \ldots) - \{x\}\\
\mathsf{fv}(x\,[A\ \ (D\ Pr)\ \ldots].T) &= (\mathsf{fv}(A) \cup (\mathsf{fv}(D) \cup (\mathsf{fv}(Pr) - \mathit{formals}(D))) \cup \cdots \cup \mathsf{fv}(T)) - \{x\}\\
\mathsf{fv}(x\,[A\ D\ \ldots] \dagger T) &= (\mathsf{fv}(A) \cup \mathsf{fv}(D) \cup \cdots \cup \mathsf{fv}(T)) - \{x\}\\
\mathsf{fv}(\%\,[S]) &= \mathsf{fv}(S)
\end{aligned}
$$

$$
\begin{aligned}
\mathsf{bv} &: \textbf{Ctxt} \to \mathcal{P}(\textbf{Var})\\
\mathsf{bv}(\square) &= \emptyset\\
\mathsf{bv}(E;S) &= \mathsf{bv}(E)\\
\mathsf{bv}(S_I;E) &= \mathsf{bv}(E)\\
\mathsf{bv}(x\,[A\ \ (D\ Pr)\ \ldots].E) &= \{x\} \cup \mathsf{bv}(E)\\
\mathsf{bv}(x\,[A\ D\ \ldots] \dagger E) &= \{x\} \cup \mathsf{bv}(E)\\
\mathsf{bv}(\%\,[E]) &= \mathsf{bv}(E)
\end{aligned}
$$

$$
\begin{aligned}
\mathsf{fv} &: \textbf{Tmpls} \to \mathcal{P}(\textbf{Var})\\
\mathsf{fv}(\emptyset) &= \emptyset\\
\mathsf{fv}(k \cup A) &= \mathsf{fv}(k) \cup \mathsf{fv}(A)
\end{aligned}
$$

$$
\begin{aligned}
\mathsf{fv} &: \textbf{Expr} \to \mathcal{P}(\textbf{Var})\\
\mathsf{fv}(b) &= \emptyset\\
\mathsf{fv}((e, \ldots)) &= \mathsf{fv}(e) \cup \ldots\\
\mathsf{fv}(p\ e\ \ldots) &= \mathsf{fv}(e) \cup \ldots\\
\mathsf{fv}(x) &= \{x\}\\
\mathsf{fv}(\lambda\,[(P.Pr)\ \ldots]) &= (\mathsf{fv}(P) \cup (\mathsf{fv}(Pr) - \mathit{formals}(P))) \cup \ldots
\end{aligned}
$$

$$
\begin{aligned}
\mathsf{fv} &: \textbf{Tmpl} \to \mathcal{P}(\textbf{Var})\\
\mathsf{fv}(\star) &= \emptyset\\
\mathsf{fv}(b) &= \emptyset\\
\mathsf{fv}((k, \ldots)) &= \mathsf{fv}(e) \cup \ldots\\
\mathsf{fv}(p\ e\ \ldots) &= \mathsf{fv}(e) \cup \ldots\\
\mathsf{fv}(x) &= \{x\}
\end{aligned}
$$

$$
\begin{aligned}
\mathsf{fv} &: \textbf{EPat} \to \mathcal{P}(\textbf{Var})\\
\mathsf{fv}(\mathsf{asserted}\ P) &= \mathsf{fv}(P)\\
\mathsf{fv}(\mathsf{retracted}\ P) &= \mathsf{fv}(P)\\
\mathsf{fv}(\mathsf{message}\ \langle P\rangle) &= \mathsf{fv}(P)\\
\mathsf{fv}(\mathsf{start}) = \mathsf{fv}(\mathsf{stop}) &= \emptyset
\end{aligned}
$$

$$
\begin{aligned}
\mathit{formals} &: \textbf{EPat} \to \mathcal{P}(\textbf{Var})\\
\mathit{formals}(\mathsf{asserted}\ P) &= \mathit{formals}(P)\\
\mathit{formals}(\mathsf{retracted}\ P) &= \mathit{formals}(P)\\
\mathit{formals}(\mathsf{message}\ \langle P\rangle) &= \mathit{formals}(P)\\
\mathit{formals}(\mathsf{start}) = \mathit{formals}(\mathsf{stop}) &= \emptyset
\end{aligned}
$$

$$
\begin{aligned}
\mathsf{fv} &: \textbf{Pat} \to \mathcal{P}(\textbf{Var})\\
\mathsf{fv}(\star) &= \emptyset\\
\mathsf{fv}(b) &= \emptyset\\
\mathsf{fv}((P, \ldots)) &= \mathsf{fv}(P) \cup \ldots\\
\mathsf{fv}(p\ e\ \ldots) &= \mathsf{fv}(e) \cup \ldots\\
\mathsf{fv}(x) &= \{x\}\\
\mathsf{fv}(\$x) &= \emptyset
\end{aligned}
$$

$$
\begin{aligned}
\mathit{formals} &: \textbf{Pat} \to \mathcal{P}(\textbf{Var})\\
\mathit{formals}(\star) &= \emptyset\\
\mathit{formals}(b) &= \emptyset\\
\mathit{formals}((P, \ldots)) &= \mathit{formals}(P) \cup \ldots\\
\mathit{formals}(p\ e\ \ldots) &= \emptyset\\
\mathit{formals}(x) &= \emptyset\\
\mathit{formals}(\$x) &= \{x\}
\end{aligned}
$$

Figure 22: Free and bound names

let. The store $\sigma$ in the machine state is updated with the initial value of the field, which is constrained to be drawn from **Val**.

$$\left\langle \sigma, \pi_i, \pi_o, \vec{a}, E[\mathsf{let}\ x\ :=\ e\ \mathsf{in}\ Pr] \right\rangle \longrightarrow \left\langle \sigma', \pi_i, \pi_o, \vec{a}, E[S] \right\rangle$$

If *eval* $\sigma\ e = v \in$ **Val** then $\sigma' = \sigma[y \mapsto v]$ and $S = Pr$ on the right hand side; otherwise, $\sigma' = \sigma$ and $S = \spadesuit$.

**Definition 5.15** (Rule set-field)**.** In rule set-field, we see the first production of an action for transmission to the surrounding dataspace. Updating a field affects any assertions depending on the field, and a patch action must be issued to communicate any changed assertions to the actor's peers.

$$\left\langle \sigma, \pi_i, \pi_o, \vec{a}, E[x \leftarrow e] \right\rangle \longrightarrow \left\langle \sigma', \pi_i, \pi'_o, \mathit{emit}\ \vec{a}\ \Delta, E[S] \right\rangle \text{ where } x \in \mathsf{dom}(\sigma)$$

If *eval* $\sigma\ e = v \in$ **Val**, then $\sigma' = \sigma[x \mapsto v]$, $S = 0$, and $(\pi'_o, \Delta) = \mathit{patch}\ \sigma\ \pi_o\ E[0]$. Otherwise, $\sigma' = \sigma$, $S = \spadesuit$, and $(\pi'_o, \Delta) = (\pi_o, \frac{\emptyset}{\emptyset})$.

**Definition 5.16** (Rule send)**.** The send rule is entirely straightforward:

$$\left\langle \sigma, \pi_i, \pi_o, \vec{a}, E[\mathsf{send}\ e] \right\rangle \longrightarrow \left\langle \sigma, \pi_i, \pi_o, \vec{a}\ \vec{a}', E[S] \right\rangle$$

If *eval* $\sigma\ e = v \in$ **Val**, then $S = 0$ and $\vec{a}' = \langle v \rangle$; otherwise, $S = \spadesuit$ and $\vec{a}'$ is the empty sequence.

**Definition 5.17** (Rules spawn and dataspace)**.** The spawn and dataspace rules are also uncomplicated, but depend on the *setup* metafunction, which we will not discuss until section 5.4.

$$\left\langle \sigma, \pi_i, \pi_o, \vec{a}, E[\mathsf{spawn}\ Pr] \right\rangle \longrightarrow \left\langle \sigma, \pi_i, \pi_o, \vec{a}\ (\mathsf{actor}\ (\mathit{setup}\ (\sigma, Pr))\ \emptyset), E[0] \right\rangle$$

$$\left\langle \sigma, \pi_i, \pi_o, \vec{a}, E[\mathsf{dataspace}\ Pr] \right\rangle \longrightarrow \left\langle \sigma, \pi_i, \pi_o, \vec{a}\ (\mathsf{dataspace}\ (\mathsf{actor}\ (\mathit{setup}\ (\sigma, Pr))\ \emptyset)), E[0] \right\rangle$$

The remaining reduction rules (figure 21) all relate to various stages of a facet's lifecycle.

**Definition 5.18.** Rule boot-facet interprets a facet template $x\,[A\ \ (D\ Pr)\ \ldots]$, renaming it, transforming it to an interior node in the facet tree and delivering two *synthetic events* to it.

$$\left\langle \sigma, \pi_i, \pi_o, \vec{a}, E[x\,[A\ \ (D\ Pr)\ \ldots]] \right\rangle \longrightarrow \left\langle \sigma, \pi_i, \pi'_o, \mathit{emit}\ \vec{a}\ \Delta, E\left[S'\right] \right\rangle$$

$$\text{where } S = (y\,[A\ \ (D\ \ (\{y/x\}\,Pr))\ \ \ldots]\,.T)$$

$$(S', \pi'_o, \Delta) = \begin{cases} (S, \pi'_o, \Delta) & \text{if } (\pi'_o, \Delta) = \mathit{patch}\ \sigma\ \pi_o\ E\,[S] \\ \left(\spadesuit, \pi_o, \frac{\emptyset}{\emptyset}\right) & \text{otherwise} \end{cases}$$

$$T_{\mathsf{start}} = \mathit{handle}\ \emptyset\ \pi_i\ \sigma\ \mathsf{start}\ \overrightarrow{(D, Pr)}$$

$$T_{\mathsf{asserted}} = \mathit{handle}\ \emptyset\ \pi_i\ \sigma\ \frac{\pi_i}{\emptyset}\ \overrightarrow{(D, Pr)}$$

$$T = \{y/x\}\,(T_{\mathsf{start}}; T_{\mathsf{asserted}})$$

$$y \text{ fresh}$$

First, a start event allows the facet to execute any startup actions necessary following the establishment of its assertions and endpoints by the action $\Delta$. Second, a synthetic *patch* $\frac{\pi_i}{\emptyset}$ is delivered to the new facet, intended to "catch it up" on events preceding its instantiation. The patch conveys to the facet the sum total of the assertions that the actor has already learned from its dataspace. This latter event is necessary because otherwise any event-handlers in the new facet do not have a chance to react to existing assertions; the dataspace is economical with its events, never repeating itself unnecessarily, as shown by theorem 4.53. The final effect of boot-facet is to update $\pi_o$ and issue a patch $\Delta$ to account for the assertions of the new facet.

**Definition 5.19** (The stop-facet rules)**.** Rules stop-facet-1 and stop-facet-2 handle explicit facet termination requests:

$$\left\langle \sigma, \pi_i, \pi_o, \vec{a}, E[x\,[A\ \ (D\ Pr)\ \ldots]\,.F[\mathsf{stop}\ x\ Pr']]\right\rangle \longrightarrow \left\langle \sigma, \pi_i, \pi_o, \vec{a}, E[\%\,[x\,[A\ \ (D\ Pr)\ \ldots]\,.F[0]]\,; Pr']\right\rangle$$
$$\left\langle \sigma, \pi_i, \pi_o, \vec{a}, E[x\,[A\ D\ \ldots]\dagger F[\mathsf{stop}\ x\ Pr']]\right\rangle \longrightarrow \left\langle \sigma, \pi_i, \pi_o, \vec{a}, E[x\,[A\ D\ \ldots]\dagger F[0]; Pr']\right\rangle$$

The nested context F is used to connect the containing facet named $x$ with the redex requesting its termination, stop $x$ $Pr'$. A side-condition $x \notin \mathsf{bv}(F)$ applies (see figure 22); it ensures that the facet name $x$ is not captured by any node in F sitting between the identified facet $x$ and the termination request. In the first of the two rules, stop-facet-1, facet $x$ is an ordinary running facet that has not yet begun its termination process. The rule encloses it in $\%\,[\cdot]$ to trigger termination. In the second, stop-facet-2, facet $x$ is an already-terminated facet that is awaiting final tear-down, and no additional $\%\,[\cdot]$ is required. In each case, the $Pr'$ is hoisted to a position adjacent to facet $x$, just inside the outer context E, but outside the scope of the $\%\,[\cdot]$ termination contour corresponding to $x$.

**Definition 5.20** (The stop-child rules)**.** Termination boundaries $\%\,[\cdot]$ are moved leafward through a facet tree by rules stop-child-1, stop-child-2, and stop-child-3.

$$\left\langle \sigma, \pi_i, \pi_o, \vec{a}, E[\%\,[0]]\right\rangle \longrightarrow \left\langle \sigma, \pi_i, \pi_o, \vec{a}, E[0]\right\rangle$$
$$\left\langle \sigma, \pi_i, \pi_o, \vec{a}, E[\%\,[S_I; T_I]]\right\rangle \longrightarrow \left\langle \sigma, \pi_i, \pi_o, \vec{a}, E[\%\,[S_I]\,; \%\,[T_I]]\right\rangle$$
$$\left\langle \sigma, \pi_i, \pi_o, \vec{a}, E[\%\,[x\,[A\ \ (D\ Pr)\ \ldots]\,.S_I]]\right\rangle \longrightarrow \left\langle \sigma, \pi_i, \pi_o, \vec{a}, E[x\,[A\ D\ \ldots]\dagger \%\,[S_I; T]]\right\rangle$$
$$\text{where } T = \textit{handle}\ \pi_i\ \pi_i\ \sigma\ \mathsf{stop}\ \overrightarrow{(D, Pr)}$$

The first two of the three are simple structural rules. It is stop-child-3 where a termination boundary and a running facet interact. The rule applies only when the facet is inert; that is, where any previously-triggered event handlers have run their course. As the termination boundary passes by the facet's node, the node is converted from the form $x\,[A\ \ (D\ Pr)\ \ldots]\,.S$ to the form $x\,[A\ D\ \ldots]\dagger S$ and a stop event is synthesized and delivered to the facet's event-handling endpoints. Any resulting commands are inserted adjacent to the existing (inert) children, but remain *inside* the termination contour.

**Definition 5.21** (Rule burial)**.** The final tear-down of a terminated facet does not take place until all of its children have not only become inert but have actually reduced to a literal 0. The burial rule takes care of this case. It is here that we finally see a patch action issued to remove the assertions of the terminating facet from the actor's aggregate assertion set.

$$\left\langle \sigma, \pi_i, \pi_o, \overrightarrow{a}, E[x\,[A\ D\ \ldots]\dagger 0]\right\rangle \longrightarrow \left\langle \sigma, \pi_i, \pi'_o, emit\ \overrightarrow{a}\ \Delta, E[S']\right\rangle$$

If *patch* $\sigma\ \pi_o\ E[0]$ yields a pair $(\pi'_o, \Delta)$, then $S' = 0$; otherwise, *patch* yields ♠ and we set $S' = ♠$, $\pi'_o = \pi_o$ and $\Delta = \frac{\emptyset}{\emptyset}$.

## 5.3 INTERPRETATION OF EVENTS

Several of the reduction rules appeal to a metafunction *handle* to compute the consequences of a reaction to an event by a collection of event-handling endpoints. As we will see in section 5.4, the same metafunction is used to distribute events arriving from the containing dataspace among the facets in an actor's facet tree.

**Definition 5.22.** The *handle* function itself is straightforward:

$$handle : \mathbf{ASet} \times \mathbf{ASet} \times \mathbf{Store} \times \mathbf{Evt}^+ \times \overrightarrow{(\mathbf{EPat} \times \mathbf{Pr})} \to \mathbf{Tree}$$
$$handle\ \pi_i\ \pi'_i\ \sigma\ \epsilon^+\ \cdot = 0$$
$$handle\ \pi_i\ \pi'_i\ \sigma\ \epsilon^+\ ((D, Pr)\ \overrightarrow{(D', Pr')}) = S; handle\ \pi_i\ \pi'_i\ \sigma\ \epsilon^+\ \overrightarrow{(D', Pr')}$$

where S in the second clause is defined by cases:

- if $D = \mathsf{asserted}\ P$ and $\epsilon^+ = \frac{\pi_{in}}{\pi_{out}}$, then $S = project\ \pi_i\ \pi'_i\ \sigma\ \pi_{in}\ P\ Pr$; otherwise,
- if $D = \mathsf{retracted}\ P$ and $\epsilon^+ = \frac{\pi_{in}}{\pi_{out}}$, then $S = project\ \pi_i\ \pi'_i\ \sigma\ \pi_{out}\ P\ Pr$; otherwise,
- if $D = \mathsf{message}\ \langle P\rangle$ and $\epsilon^+ = \langle c\rangle$, then $S = matchInOrder\ \sigma\ c\ ((P, Pr)\ (\star, 0))$; otherwise,
- if $D = \mathsf{start}$ and $\epsilon^+ = \mathsf{start}$, then $S = Pr$; otherwise,
- if $D = \mathsf{stop}$ and $\epsilon^+ = \mathsf{stop}$, then $S = Pr$; otherwise,
- $S = 0$.

The sequence of event-handling endpoints becomes a composition of programs. Each endpoint becomes 0 if the given event does not apply. Patch events apply to asserted and retracted endpoints; message events to message endpoints; and start and stop events to start and stop endpoints. The interesting cases are message delivery and patch handling. Message delivery delegates to *matchInOrder* with the event-handler's pattern P and continuation Pr augmented with a catch-all 0 clause to handle the case where the incoming message does not match P. Patch processing delegates to the metafunction *project*.

**Definition 5.23.** The *project* metafunction extracts a *finite sequence* of assertions matching pattern P from an assertion set carried in a patch. Each relevant assertion should generate one instance of the event handler program Pr. It is clearly an error to attempt to iterate over an infinite set;

therefore, *project* yields an exception in cases where the assertion set $\pi$ being projected contains an infinite number of individual assertions that happen to match the pattern $P$.

$$\textit{project} : \textbf{ASet} \times \textbf{ASet} \times \textbf{Store} \times \textbf{ASet} \times \textbf{Pat} \times \textbf{Pr} \to \textbf{Tree}$$

$$\textit{project}\ \pi_i\ \pi_i'\ \sigma\ \pi\ P\ Pr = \begin{cases} \spadesuit & \text{if } P' = \spadesuit \\ \textit{unroll}\ m & \text{if } |m| \in \mathbb{N} \text{ (i.e., } m \text{ is finite)} \\ \spadesuit & \text{otherwise} \end{cases}$$

where

$$\begin{aligned} P' &= \textit{snapshot}\ \sigma\ P \\ m &= \left\{\textit{match}\ P'\ \mathcal{I}\ Pr \mid \mathcal{I} \in \{\textit{inst}\ P'\ v \mid v \in \pi\}, \textit{known}(\mathcal{I}, \pi_i) \neq \textit{known}(\mathcal{I}, \pi_i')\right\} \\ \textit{known}(\mathcal{I}, \pi'') &= 1, \text{ if some } c \in \pi'' \text{ exists s.t. } \textit{match}\ \mathcal{I}\ c\ 0 \text{ is defined; } 0, \text{ otherwise} \\ \textit{unroll}\ \left\{S, S', \ldots\right\} &= S; S'; \ldots; 0 \end{aligned}$$

The first step in *project*'s operation is to filter the set $\pi$ using metafunction *inst*, retaining only those assertions that match $P'$.

**Definition 5.24.** The partial function *inst* is similar to *match* (definition 5.11), in that it is defined only where the structure of the pattern matches the assertion; however, it is different in that it yields a **PVal** as a result that includes detail only where it is *relevant* to the supplied pattern.

$$\textit{inst} : \textbf{PVal} \times \textbf{Val} \rightharpoonup \textbf{PVal}$$

$$\begin{aligned} \textit{inst}\ \star\ c &= \star \\ \textit{inst}\ b\ b &= b \\ \textit{inst}\ (P, \ldots)\ (v, \ldots) &= (\textit{inst}\ P\ v, \ldots) \\ \textit{inst}\ \$x\ v &= v \end{aligned}$$

Where the pattern is $\star$, meaning "any value is acceptable", the precise value that was given is *obscured* in the output of *inst*. This causes irrelevant detail to be eliminated from consideration. By gathering together results from *inst*, *project* collapses together assertions from $\pi$ that are identical up to "uninteresting" positions in the syntax of $P$.

Returning to the operation of *project*, the next step after filtering and partial transformation of the input set $\pi$ is to take each $\mathcal{I} \in$ **PVal** drawn from the set of *inst* results and use the arguments $\pi_i$ and $\pi_i'$ given to *project* to decide whether $\mathcal{I}$ is *novel* or not.

The set $\pi_i$ denotes the set of *known assertions* just prior to the arrival of the event that *project* is processing. The set $\pi_i'$ denotes the result of updating $\pi_i$ with the contents of the arriving event. That is, $\pi_i$ is "what the actor knew before", and $\pi_i'$ is "what the actor knows now."

If a particular $\mathcal{I}$ corresponds to some assertion in $\pi_i$, but not to any in $\pi_i'$, or conversely corresponds to some assertion in $\pi_i'$ but none in $\pi_i$, then the actor has learned something new, and the handler program $Pr$ should be instantiated for this $\mathcal{I}$. However, if $\mathcal{I}$ corresponds to some assertion in both or neither of $\pi_i$ and $\pi_i'$, then nothing relevant has changed for the actor, and $Pr$ should not be instantiated.

**Example 5.25.** Consider the presence-management portion of a chat service with multiple rooms. Assertions $(\mathit{userName}, \texttt{in}, \mathit{roomName})$ denote presence of the named user in the named room. Rooms are said to "exist" only when inhabited by at least one user. Users joining the system are presented with a list of currently-extant rooms to choose from. A program for calculating this list might be written (assuming suitable data structures and primitive operations for sets):

$$\mathsf{let}\ \mathit{rooms}\ :=\ \emptyset\ \mathsf{in}$$
$$\mathit{track}\begin{bmatrix} \emptyset & \\ (\mathsf{asserted}\ (\star, \texttt{in}, \$r) & (\mathit{rooms} \leftarrow \mathit{rooms} \cup \{r\})) \\ (\mathsf{retracted}\ (\star, \texttt{in}, \$r) & (\mathit{rooms} \leftarrow \mathit{rooms} - \{r\})) \end{bmatrix}$$

Imagine now that two users, `Alice` and `Bob`, arrive and join the room `Lobby` *simultaneously*. This results in delivery of a patch event $\Delta = \frac{\pi^+}{\emptyset}$ where $\pi^+ = \{(\texttt{Alice}, \texttt{in}, \texttt{Lobby}), (\texttt{Bob}, \texttt{in}, \texttt{Lobby})\}$ to our list-management actor. Ultimately, a call to *handle* takes place:

$$\mathit{handle}\ \emptyset\ \pi^+\ \sigma\ \Delta\ ((\mathsf{asserted}\ (\star, \texttt{in}, \$r), (\mathit{rooms} \leftarrow \mathit{rooms} \cup \{r\}))\ (\mathsf{retracted}\ (\star, \texttt{in}, \$r), (\mathit{rooms} \leftarrow \mathit{rooms} - \{r\})))$$

For the retracted endpoint, *handle* delegates to *project*:

$$\mathit{project}\ \emptyset\ \pi^+\ \sigma\ \emptyset\ (\star, \texttt{in}, \$r)\ (\mathit{rooms} \leftarrow \mathit{rooms} - \{r\})$$

which yields $0$. The situation for the asserted endpoint is more interesting:

$$\mathit{project}\ \emptyset\ \pi^+\ \sigma\ \pi^+\ (\star, \texttt{in}, \$r)\ (\mathit{rooms} \leftarrow \mathit{rooms} \cup \{r\})$$

Because the pattern ignores the first component of matching triples, we have that

$$\{(\star, \texttt{in}, \texttt{Lobby})\} = \{\mathit{inst}\ (\star, \texttt{in}, \$r)\ v \mid v \in \pi^+\}$$

Now, $\mathit{known}((\star, \texttt{in}, \texttt{Lobby}), \emptyset) \neq \mathit{known}((\star, \texttt{in}, \texttt{Lobby}), \pi^+)$, so *match* is invoked and the actor processes the new knowledge of the room `Lobby`. ◊

**Example 5.26.** Imagine now that `Alice` leaves the room, while `Bob` stays on. This results in a patch event $\Delta = \frac{\emptyset}{\pi^-}$ where $\pi^- = \{(\texttt{Alice}, \texttt{in}, \texttt{Lobby})\}$. At the time of the event, the total knowledge of the actor is $\pi_i = \{(\texttt{Alice}, \texttt{in}, \texttt{Lobby}), (\texttt{Bob}, \texttt{in}, \texttt{Lobby})\}$. Updating $\pi_i$ with the patch yields $\pi_i' = \{(\texttt{Bob}, \texttt{in}, \texttt{Lobby})\}$. This time, the asserted endpoint has nothing to do, but the retracted endpoint triggers:

$$\mathit{project}\ \pi_i\ \pi_i'\ \sigma\ \pi^-\ (\star, \texttt{in}, \$r)\ (\mathit{rooms} \leftarrow \mathit{rooms} - \{r\})$$

Again, the pattern ignores the first component of matching triples in $\pi^-$, so

$$\{(\star, \texttt{in}, \texttt{Lobby})\} = \{\mathit{inst}\ (\star, \texttt{in}, \$r)\ v \mid v \in \pi^-\}$$

However, this time, $\mathit{known}((\star, \texttt{in}, \texttt{Lobby}), \pi_i) = \mathit{known}((\star, \texttt{in}, \texttt{Lobby}), \pi_i')$ since in each case *some* assertion matching the pattern is contained in the assertion set. Therefore, this event does *not* lead to our list-tracking actor updating its *rooms* field. This is what we want: `Bob` is still present in `Lobby`. Even though `Alice` left, the room itself has not vanished yet. ◊

**Example 5.27.** Finally, Bob leaves the room. The patch event is $\Delta = \frac{\emptyset}{\pi^-}$ again but with $\pi^- = \{(\texttt{Bob}, \texttt{in}, \texttt{Lobby})\}$ this time. At the time of the event, $\pi_i = \{(\texttt{Bob}, \texttt{in}, \texttt{Lobby})\}$, and so $\pi_i' = \emptyset$. The retracted endpoint triggers again, as before; and, as before, *inst* leaves us a single value for $\mathcal{I}$, namely $(\star, \texttt{in}, \texttt{Lobby})$. This time, however, $\mathit{known}((\star, \texttt{in}, \texttt{Lobby}), \pi_i) \neq \mathit{known}((\star, \texttt{in}, \texttt{Lobby}), \pi_i')$ because $\pi_i'$ is empty, and so $(\mathit{rooms} \leftarrow \mathit{rooms} - \{r\})$ is instantiated with $r = \texttt{Lobby}$, and the actor removes Lobby from *rooms*. ♢

## 5.4 INTERFACING SYNDICATE/λ TO THE DATASPACE MODEL

Thus far, we have discussed the internal operation of SYNDICATE/λ actors, but have not addressed the question of their interface to the wider world. The path to an answer begins with the way SYNDICATE/λ constructs actor actions. To start an actor with store $\sigma$ and code $\mathit{Pr}$, SYNDICATE/λ issues the dataspace model action actor $(\mathit{setup}\ (\sigma, \mathit{Pr}))\ \emptyset$. This term appears in rules spawn and dataspace, as well.

**Definition 5.28.** The function *setup* produces a boot function of type **Boot** (figure 12) which in turn describes the behavior function and initial state of a new actor. Every SYNDICATE/λ actor has behavior function *interp* and a state value drawn from set $\mathbf{M_I}$ (fig. 18).

$$\mathit{setup} : \mathbf{Store} \times \mathbf{Pr} \to \mathbf{Boot}$$

$$\mathit{setup}\ (\sigma, \mathit{Pr}) = \lambda().\begin{cases} \mathsf{init}(\overrightarrow{a}, \mathsf{pack}\ \langle \mathbf{M_I}, (\mathit{interp}, \langle \sigma', \emptyset, \pi_o, \cdot, S\rangle)\rangle) & \text{if } S \in \mathbf{Tree_I} \text{ and } S \neq 0 \\ \mathsf{exit}(\overrightarrow{a}) & \text{otherwise} \end{cases}$$

$$\text{where } \langle \sigma, \emptyset, \emptyset, \cdot, \mathit{Pr}\rangle \longrightarrow^* \langle \sigma', \emptyset, \pi_o, \overrightarrow{a}, S\rangle \not\longrightarrow$$

The initial state value contains information extracted from a use of the reduction relation, starting from $\sigma$ and $\mathit{Pr}$. If reduction stops in an exception-signaling configuration or fails to generate at least one running facet, *setup* instructs the dataspace to terminate the nascent actor.

**Definition 5.29.** The operator $\pm$ incorporates changes described by an incoming event to a previous record of the contents of the surrounding dataspace. When given a patch event, it updates the assertion set. By contrast, a message event is treated as an infinitesimally-brief assertion of its carried value, as discussed in section 4.4, and the assertion set remains unchanged.

$$\cdot \pm \cdot : \mathbf{ASet} \times \mathbf{Evt} \to \mathbf{ASet}$$

$$\pi \pm \frac{\pi_{in}}{\pi_{out}} = \pi \cup \pi_{in} - \pi_{out}$$

$$\pi \pm \langle c\rangle = \pi$$

**Definition 5.30.** The *inject* function traverses a facet tree, using *handle* to deliver an incoming event to the event-handler endpoints of every running facet.

$$\mathit{inject} : \mathbf{ASet} \times \mathbf{ASet} \times \mathbf{Store} \times \mathbf{Evt} \times \mathbf{Tree_I} \to \mathbf{Tree}$$

$$\mathit{inject}\ \pi_i\ \pi_i'\ \sigma\ e\ 0 = 0$$

$$\mathit{inject}\ \pi_i\ \pi_i'\ \sigma\ e\ (S_I; T_I) = \mathit{inject}\ \pi_i\ \pi_i'\ \sigma\ e\ S_I; \mathit{inject}\ \pi_i\ \pi_i'\ \sigma\ e\ T_I$$

$$\mathit{inject}\ \pi_i\ \pi_i'\ \sigma\ e\ x\,[A\ \ (D\ \mathit{Pr})\ \ldots].S_I = x\,[A\ \ (D\ \mathit{Pr})\ \ldots].(\mathit{inject}\ \pi_i\ \pi_i'\ \sigma\ e\ S_I; \mathit{handle}\ \pi_i\ \pi_i'\ \sigma\ e\ \overrightarrow{(D, \mathit{Pr})})$$

**Definition 5.31.** The behavior function *interp* integrates an event arriving from the dataspace with the machine state held in the actor's private state value, reduces the result, and returns. If the actor terminates all its facets or if reduction yields an exception, *interp* instructs the dataspace to terminate the actor.

$$\mathit{interp} : \mathcal{F}_{\mathbf{M_I}}$$

$$\mathit{interp}\ (\epsilon, \langle \sigma, \pi_i, \pi_o, \cdot, S_I \rangle) = \begin{cases} \mathsf{continue}(\mathit{emit}\ \vec{a}\ \Delta,\ \langle \sigma', \pi_i', \pi_o'', \cdot, S'' \rangle) & \text{if } S'' \in \mathbf{Tree_I} \text{ and } S'' \neq 0 \\ \mathsf{exit}(\vec{a}) & \text{otherwise} \end{cases}$$

$$\text{where } \pi_i' = \pi_i \pm \epsilon$$

$$\langle \sigma, \pi_i', \pi_o, \cdot, \mathit{inject}\ \pi_i\ \pi_i'\ \sigma\ \epsilon\ S_I \rangle \longrightarrow^* \langle \sigma', \pi_i', \pi_o', \vec{a}, S' \rangle \not\longrightarrow$$

$$(S'', \pi_o'', \Delta) = \begin{cases} (S', \pi_o'', \Delta) & \text{if } (\pi_o'', \Delta) = \mathit{patch}\ \sigma'\ \pi_o'\ S' \\ \left(\spadesuit, \pi_o', \frac{\emptyset}{\emptyset}\right) & \text{otherwise} \end{cases}$$

*Remark.* SYNDICATE/$\lambda$ is an untyped language, and can express nontermination:

$$\begin{aligned} & \lambda\,[(\$x.\,(x\ x))]\ \ \lambda\,[(\$x.\,(x\ x))] \\ \longrightarrow\ & \mathsf{let}\ x\ =\ \lambda\,[(\$x.\,(x\ x))]\ \ \mathsf{in}\ \ (x\ x) \\ \longrightarrow\ & \lambda\,[(\$x.\,(x\ x))]\ \ \lambda\,[(\$x.\,(x\ x))] \\ \longrightarrow\ & \cdots \end{aligned}$$

Despite this, we have equipped it with the behavior function *interp* for interfacing it with the dataspace model, even though, strictly speaking, the dataspace model demands a terminating leaf actor language.[12] SYNDICATE/$\lambda$ thus shares with its extant implementations the flaw that programmers must take care to ensure their programs terminate.

## 5.5 WELL-FORMEDNESS AND ERRORS

Reduction of SYNDICATE/$\lambda$ programs can stop for many reasons. First of all, as in practically all interesting uses of $\lambda$-calculus-like machinery, certain primitive operations may be partial functions. The classic example is arithmetic division, undefined at a zero denominator. This partiality manifests via $\mathit{delta}^\lambda$ and *delta* yielding no answer. In turn, this affects most of the other core metafunctions as well as the lion's share of the reduction rules.

More interesting are type errors. Certain errors, such as attempts to call a non-procedure or invoke an arithmetic primitive with a non-numeric value, may be prevented by developing a conventional type system (Pierce 2002). Standard techniques also exist for enforcing exhaustive pattern-matching in procedures. Other errors are peculiar to SYNDICATE/$\lambda$. Figures 23 and 24 sketch a "well-formedness" judgment $\Gamma \vdash Pr$ wf intended to catch three kinds of scope error: reference to an unbound variable, field, or facet; update to a name that is non-existent or not a

12 Ongoing collaborative work includes the development of a type system which ensures termination of SYNDICATE/$\lambda$ programs, among other benefits (Caldwell, Garnock-Jones and Felleisen 2017).

$$\tau ::= \mathsf{var} \mid \mathsf{field} \mid \mathsf{facet}$$
$$\Gamma ::= \cdot \mid \Gamma, x : \tau$$

$$\begin{aligned}
\mathit{prune}(\cdot) &= \cdot \\
\mathit{prune}(\Gamma, x : \mathsf{var}) &= \mathit{prune}(\Gamma), x : \mathsf{var} \\
\mathit{prune}(\Gamma, x : \mathsf{field}) &= \mathit{prune}(\Gamma), x : \mathsf{field} \\
\mathit{prune}(\Gamma, x : \mathsf{facet}) &= \mathit{prune}(\Gamma) \\
\\
\mathit{extend}(\Gamma, \{\vec{x}\}) &= \Gamma, \overrightarrow{x : \mathsf{var}}
\end{aligned}$$

$$\begin{aligned}
\mathit{pruneUpTo}(x, \cdot) &= \cdot \\
\mathit{pruneUpTo}(x,\ \Gamma, z : \mathsf{var}) &= \mathit{pruneUpTo}(x,\ \Gamma), z : \mathsf{var} \\
\mathit{pruneUpTo}(x,\ \Gamma, z : \mathsf{field}) &= \mathit{pruneUpTo}(x,\ \Gamma), z : \mathsf{field} \\
\mathit{pruneUpTo}(x,\ \Gamma, z : \mathsf{facet}) &= \mathit{pruneUpTo}(x,\ \Gamma) \qquad (x \neq z) \\
\mathit{pruneUpTo}(x,\ \Gamma, x : \mathsf{facet}) &= \Gamma
\end{aligned}$$

Figure 23: "Types", type environments, and their metafunctions

field; and inappropriate use of a facet name in a stop command. For an example of the latter, consider the two programs

$$x\,[\emptyset\ \ (\mathsf{start}\ \ (\mathsf{stop}\ x\ \ (\mathsf{stop}\ x\ 0)))]$$
$$x\,[\emptyset\ \ (\mathsf{start}\ y\,[\emptyset\ \ (\mathsf{start}\ \ (\mathsf{stop}\ x\ \ (\mathsf{stop}\ y\ 0)))]])]$$

In the first, the outer stop terminates the facet $x$, effectively replacing it with stop $x$ $0$, which is stuck because it is not contained in an $x\,[\cdots].\square$ context. Similarly, in the second, the outer stop terminates $x$ but also all its child facets, including $y$. Ultimately, reduction becomes stuck at stop $y$ $0$ for lack of a $y\,[\cdots].\square$ context.[13] The well-formedness judgment aims to prevent such errors by removing all facet names from the type environment when checking the bodies of spawn and dataspace commands and by removing facet names *at* or *below* a certain name when checking the continuation of each stop command.

Going beyond simple scope errors, SYNDICATE/$\lambda$ programs can fail in two important ways relating to the assertions they exchange with peers via the shared dataspace. First, programs may make simple data-type errors in their assertions and subscriptions. For example, a particular protocol may require that peers interact by asserting and expressing interest in tuples $(\mathsf{square}, n, m)$, where $n, m \in \mathbb{N}$ and $m = n^2$. It is an error, then, for a program to assert $(\mathsf{square}, \text{"a"}, \text{"aa"})$, to misspell square, or to assert a tuple such as $(\mathsf{square}, 10, 1000)$. Second, the metafunction *project* signals an exception when the set of relevant matches to a given pattern is infinite. Consider the following program, which computes and asserts squares in response to detected interest:[14]

$$\mathsf{spawn}\ \mathit{squareServer} \left[\emptyset \left(\mathsf{asserted}\ ?\,(\mathsf{square}, \$x, \star)\ \ \mathit{ans} \left[\begin{array}{l} \emptyset \cup (\mathsf{square}, x, x \times x) \\ (\mathsf{retracted}\ ?\,(\mathsf{square}, x, \star)\ \ (\mathsf{stop}\ \mathit{ans}\ 0)) \end{array}\right]\right)\right]$$

13 As a matter of practicality, the SYNDICATE prototypes, both untyped, ignore this error, treating it as a no-op.
14 In section 5.7 we will introduce a more elegant approach to programming such services.

$$\boxed{\Gamma \vdash \mathsf{Pr}\ \mathsf{wf}} \qquad \frac{}{\Gamma \vdash 0\ \mathsf{wf}} \qquad \frac{\Gamma \vdash \mathsf{Pr}_1\ \mathsf{wf} \quad \Gamma \vdash \mathsf{Pr}_2\ \mathsf{wf}}{\Gamma \vdash \mathsf{Pr}_1 ; \mathsf{Pr}_2\ \mathsf{wf}} \qquad \frac{\Gamma \vdash e_1\ \mathsf{wf} \quad \Gamma \vdash e_2\ \mathsf{wf}}{\Gamma \vdash e_1\ e_2\ \mathsf{wf}}$$

$$\frac{\Gamma \vdash e\ \mathsf{wf} \quad \Gamma, x : \mathsf{var} \vdash \mathsf{Pr}\ \mathsf{wf}}{\Gamma \vdash \mathsf{let}\ x\ =\ e\ \mathsf{in}\ \mathsf{Pr}\ \mathsf{wf}} \qquad \frac{\Gamma \vdash e\ \mathsf{wf} \quad \Gamma, x : \mathsf{field} \vdash \mathsf{Pr}\ \mathsf{wf}}{\Gamma \vdash \mathsf{let}\ x\ :=\ e\ \mathsf{in}\ \mathsf{Pr}\ \mathsf{wf}} \qquad \frac{\Gamma(x) = \mathsf{field} \quad \Gamma \vdash e\ \mathsf{wf}}{\Gamma \vdash x \leftarrow e\ \mathsf{wf}}$$

$$\frac{\Gamma \vdash e\ \mathsf{wf}}{\Gamma \vdash \mathsf{send}\ e\ \mathsf{wf}} \qquad \frac{\mathit{prune}(\Gamma) \vdash \mathsf{Pr}\ \mathsf{wf}}{\Gamma \vdash \mathsf{spawn}\ \mathsf{Pr}\ \mathsf{wf}} \qquad \frac{\mathit{prune}(\Gamma) \vdash \mathsf{Pr}\ \mathsf{wf}}{\Gamma \vdash \mathsf{dataspace}\ \mathsf{Pr}\ \mathsf{wf}}$$

$$\frac{\Gamma' = \Gamma, x : \mathsf{facet} \quad \Gamma' \vdash A\ \mathsf{wf} \quad (\Gamma' \vdash D\ \mathsf{wf} \wedge \mathit{extend}(\Gamma', \mathit{formals}(D)) \vdash \mathsf{Pr}\ \mathsf{wf}) \quad \cdots}{\Gamma \vdash x\,[A\ \ (D\ \mathsf{Pr})\ \ldots]\ \ \mathsf{wf}}$$

$$\frac{\Gamma(x) = \mathsf{facet} \quad \mathit{pruneUpTo}(x, \Gamma) \vdash \mathsf{Pr}\ \mathsf{wf}}{\Gamma \vdash \mathsf{stop}\ x\ \mathsf{Pr}\ \mathsf{wf}}$$

$$\boxed{\Gamma \vdash e\ \mathsf{wf}} \qquad \frac{}{\Gamma \vdash b\ \mathsf{wf}} \qquad \frac{\Gamma \vdash e\ \mathsf{wf} \quad \cdots}{\Gamma \vdash (e, \ldots)\ \mathsf{wf}} \qquad \frac{\Gamma \vdash e\ \mathsf{wf} \quad \cdots}{\Gamma \vdash p\ e\ \ldots\ \mathsf{wf}} \qquad \frac{\Gamma(x) = \mathsf{var}\ \mathsf{or}\ \Gamma(x) = \mathsf{field}}{\Gamma \vdash x\ \mathsf{wf}}$$

$$\frac{(\Gamma \vdash P\ \mathsf{wf} \wedge \mathit{extend}(\Gamma, \mathit{formals}(P)) \vdash \mathsf{Pr}\ \mathsf{wf}) \quad \cdots}{\Gamma \vdash \lambda\,[(P.\mathsf{Pr})\ \ldots]\ \ \mathsf{wf}}$$

$$\boxed{\Gamma \vdash A\ \mathsf{wf}} \qquad \frac{}{\Gamma \vdash \emptyset\ \mathsf{wf}} \qquad \frac{\Gamma \vdash k\ \mathsf{wf} \quad \Gamma \vdash A\ \mathsf{wf}}{\Gamma \vdash k \cup A\ \mathsf{wf}}$$

$$\boxed{\Gamma \vdash P\ \mathsf{wf}} \qquad \frac{}{\Gamma \vdash \star\ \mathsf{wf}} \qquad \frac{}{\Gamma \vdash b\ \mathsf{wf}} \qquad \frac{\Gamma \vdash P\ \mathsf{wf} \quad \cdots}{\Gamma \vdash (P, \ldots)\ \mathsf{wf}} \qquad \frac{\Gamma \vdash e\ \mathsf{wf} \quad \cdots}{\Gamma \vdash p\ e\ \ldots\ \mathsf{wf}}$$

$$\frac{\Gamma(x) = \mathsf{var}\ \mathsf{or}\ \Gamma(x) = \mathsf{field}}{\Gamma \vdash x\ \mathsf{wf}} \qquad \frac{}{\Gamma \vdash \$x\ \mathsf{wf}}$$

$$\boxed{\Gamma \vdash k\ \mathsf{wf}} \qquad \text{(like } \Gamma \vdash P\ \mathsf{wf} \text{ but without the case for } \$x\text{)}$$

$$\boxed{\Gamma \vdash D\ \mathsf{wf}} \qquad \frac{}{\Gamma \vdash \mathsf{start}\ \mathsf{wf}} \qquad \frac{}{\Gamma \vdash \mathsf{stop}\ \mathsf{wf}}$$

$$\frac{\Gamma \vdash P\ \mathsf{wf}}{\Gamma \vdash \mathsf{asserted}\ P\ \mathsf{wf}} \qquad \frac{\Gamma \vdash P\ \mathsf{wf}}{\Gamma \vdash \mathsf{retracted}\ P\ \mathsf{wf}} \qquad \frac{\Gamma \vdash P\ \mathsf{wf}}{\Gamma \vdash \mathsf{message}\ \langle P \rangle\ \mathsf{wf}}$$

$$\boxed{\Gamma \vdash S\ \mathsf{wf}} \qquad \frac{\Gamma \vdash S\ \mathsf{wf}}{\Gamma \vdash \%\,[S]\ \ \mathsf{wf}} \qquad \frac{\Gamma \vdash S_1\ \mathsf{wf} \quad \Gamma \vdash S_2\ \mathsf{wf}}{\Gamma \vdash S_1 ; S_2\ \mathsf{wf}}$$

$$\frac{\Gamma' = \Gamma, x : \mathsf{facet} \quad \Gamma' \vdash A\ \mathsf{wf} \quad (\Gamma' \vdash D\ \mathsf{wf} \wedge \mathit{extend}(\Gamma', \mathit{formals}(D)) \vdash \mathsf{Pr}\ \mathsf{wf}) \quad \cdots \quad \Gamma' \vdash S\ \mathsf{wf}}{\Gamma \vdash x\,[A\ \ (D\ \mathsf{Pr})\ \ldots].S\ \mathsf{wf}}$$

$$\frac{\Gamma' = \Gamma, x : \mathsf{facet} \quad \Gamma' \vdash A\ \mathsf{wf} \quad \Gamma' \vdash D\ \mathsf{wf} \quad \cdots \quad \Gamma' \vdash S\ \mathsf{wf}}{\Gamma \vdash x\,[A\ D\ \ldots]\dagger S\ \mathsf{wf}}$$

Figure 24: Well-formedness judgments

$$\begin{array}{ccccccc}
\langle \cdot, \emptyset, \emptyset, \cdot, \mathrm{Pr}\rangle & \longrightarrow^* & \left\langle \sigma, \emptyset, \pi_o, \vec{a}, S_I \right\rangle & & \left\langle \sigma, \pi_i, \pi_o', \cdot, S \right\rangle & \longrightarrow^* & \left\langle \sigma', \pi_i, \pi_o'', \vec{a}', S_I' \right\rangle \\
& & \downarrow & & \uparrow & & \downarrow \\
& & \vec{a}\Delta, \langle \sigma, \emptyset, \pi_o', \cdot, S_I \rangle & \dashrightarrow & \epsilon, \langle \sigma, \emptyset, \pi_o', \cdot, S_I \rangle & & \vec{a}'\Delta', \left\langle \sigma', \pi_i, \pi_o''', \cdot, S_I' \right\rangle \quad \dashrightarrow
\end{array}$$

Figure 25: Internal reduction and external interaction

All is well if some peer includes an endpoint $(\mathsf{asserted}\ (\mathsf{square}, 3, \$\mathit{nine})\ \mathrm{Pr})$. But if a programmer makes an error, violating our square-computing protocol by attempting to enumerate all squares using an endpoint $(\mathsf{asserted}\ (\mathsf{square}, \star, \$v)\ \mathrm{Pr})$, the resulting assertion of interest, $?\,(\mathsf{square}, \star, \star)$, causes *squareServer* to signal an exception as it *project*s that infinite set against the pattern $?\,(\mathsf{square}, \$x, \star)$. Even though the client was at fault, the server is the component which crashes, since the server is the component relying upon the finiteness of a certain subspace of assertions. Ongoing research investigates type-system-based approaches to ruling out these forms of assertion-set-related error (Caldwell, Garnock-Jones and Felleisen 2017).

In order to use our well-formedness judgment to work towards a statement of overall soundness, we need to account for the way an actor transmits actions to its environment and receives events in reply. Figure 25 illustrates the alternation between the reduction relation explored in this chapter (upper row) and the exchange of information with an actor's surrounding dataspace, as explored in chapter 4 (lower row). Setting aside cases where an actor exits because of a signaled exception or termination of all of its facets, we can abbreviate the excursions to the lower row of the figure as a pseudo-reduction-rule. From the actor's perspective, it is as if an oracle supplies a fresh (relevant) event $\epsilon$ at just the right moment, as the actor achieves an inert configuration:

**Definition 5.32** (Pseudo-rule interact)**.**

$$\begin{aligned}
\left\langle \sigma, \pi_i, \pi_o, \vec{a}, S_I \right\rangle &\longrightarrow \left\langle \sigma, \pi_i', \pi_o', \cdot, \mathit{inject}\ \pi_i\ \pi_i'\ \sigma\ \epsilon\ S_I \right\rangle && (\mathsf{interact}) \\
\text{where } (\pi_o', \Delta) &= \mathit{patch}\ \sigma\ \pi_o\ S_I \\
\pi_i' &= \pi_i \pm \epsilon
\end{aligned}$$

The uses of $\pm$ (definition 5.29), *patch* (definition 5.7) and *inject* (definition 5.30) here show that the rule is effectively an "inlining" of *interp* (definition 5.31).

**Definition 5.33** (Extended reduction relation)**.** We will write $\longrightarrow_{\mathsf{IO}}$ to mean the reduction relation $\longrightarrow$ extended with the interact pseudo-rule.

**Conjecture 5.34.** *If* $\cdot \vdash \mathrm{Pr}\ \mathsf{wf}$ *and* $\langle \cdot, \emptyset, \emptyset, \cdot, \mathrm{Pr}\rangle \longrightarrow_{\mathsf{IO}}^* \left\langle \sigma, \pi_i, \pi_o, \vec{a}, S \right\rangle$, *then either*

1. $S \in \mathbf{Tree}_I$*; or*
2. $S = E\,[\spadesuit]$ *for some* $E$*; or*
3. $S \notin \mathbf{Tree}_I$ *and there exists a unique* $M'$ *such that* $\left\langle \sigma, \pi_i, \pi_o, \vec{a}, S \right\rangle \longrightarrow M'$.

That is, at every step in a reduction chain, one of three conditions holds. First, the facet tree $S$ may be inert, in which case the actor terminates ($S = 0$) or yields to its dataspace ($S \neq 0$).

Second, the facet tree may have a signaled exception as its selected redex, in which case it is terminated abruptly. Third, the tree may be neither inert nor in an exception state, in which case there is always another non-interact reduction step that may be taken.

Examination of the reduction rules and metafunctions shows that ♠ is signaled in two situations: when use of a primitive function yields no result, either due to intrinsic partiality or a type error, and when *project* encounters an infinite set of matching assertions. The well-formedness judgment rules out stuckness from misuse of names. If a program makes a simple data-type error in the content of an assertion, a number of consequences may unfold: the actor may simply sit inert forever, having failed to solicit events from peers; the actor may later receive events containing information it is not prepared to handle, resulting in an exception; or the actor may unintentionally trigger crashes in its peers, having supplied them with incoherent information.

## 5.6 ATOMICITY AND ISOLATION

With fields, we have introduced mutable state, opening the door to potential unpredictability. SYNDICATE mitigates this unpredictability by limiting the scope of mutability to individual actors. In addition, SYNDICATE's facet model enforces three kinds of atomicity that together help the programmer in reasoning about field updates. First, an actor's behavior function is never preempted. As a result, events internal to an actor occur "infinitely quickly" from the perspective of the surrounding dataspace. This yields "synchrony" similar to that of languages such as Esterel (Berry and Gonthier 1992) and Céu (Sant'Anna, Ierusalimschy and Rodriguez 2015). Second, each actor's private state is not only isolated but threaded through its behavior function in a linear fashion. This yields a natural boundary within which private state may safely be updated via mutation. Third, exceptions during event processing tear down the entire actor at once, including its private state. The same happens for deliberate termination of an actor. Termination is again instantaneous, and damaged private state cannot affect peers. This yields a form of "fail-stop" programming (Schlichting and Schneider 1983). Together, these forms of atomicity allow facet fields to be mutable, while events continue to be handled with sequential code, resolving all questions of internal consistency in the face of imperative updates.

By coalescing adjacent patch actions with *emit* during reduction, the SYNDICATE/λ semantics hides momentary "glitches" from observers. This allows actors to stop one facet and start another publishing the same assertion(s) without observers ever detecting the change, and without being forced to explicitly indicate that a smooth changeover is desired. Contrast this automatic ability to seamlessly delegate responsibility to a new facet with the equivalent ability for glitch-free handover of assertions to a new *actor*. In the latter case, programmers must explicitly make use of the "initial assertion set" field in the actor action describing the new actor as described in section 4.2.

Events are dispatched to facets all at once: the metafunction *inject* matches each event against all endpoints of an actor's facets simultaneously. Actors thus make all decisions about which event-handlers are to run before any particular event-handler begins execution. This separation of a planning phase from an execution phase helps reduce dependence on order-of-operations (cf. section 2.6) by ensuring no field updates can occur during planning.[15]

## 5.7 DERIVED FORMS: during AND select

Examples 5.2 and 5.25 highlight two common idioms in SYNDICATE programming worthy of promotion to language feature. In example 5.2, we saw a scenario in which appearance of an assertion led to creation of a resource—in this case, a separate actor—and disappearance of the same assertion led to the resource's release. In example 5.25, we saw a scenario in which an actor aggregated specific information from a set of assertions into a local set data structure held in a field. SYNDICATE offers support for the former scenario via a new form of endpoint called during, and support for the latter via a family of forms called select.[16]

"during" ENDPOINTS. In a facet template $x\,[A\ (D\ Pr)\ \ldots]$, each $(D\ Pr)$ declares a single event-handling endpoint. We add during to the language by extending the class of event patterns:

$$\text{Event patterns } D \in \textbf{EPat} := \ldots \mid \mathsf{during}\ P$$

We interpret the new construct in terms of existing syntax. An endpoint $(\mathsf{during}\ P\ Pr)$ is interpreted as if the programmer had written

$$(\mathsf{asserted}\ P\ x\,\big[\emptyset\ (\mathsf{start}\ Pr)\ \big(\mathsf{retracted}\ P'\ (\mathsf{stop}\ x\ 0)\big)\big])$$

where $x$ is fresh and $P'$ is $P$ with each binder $\$z$ rewritten to $z$, a reference to the specific value bound at that position during the firing of the asserted event pattern. As an example, a program that asserts $(\mathtt{room}, \mathit{roomName})$ whenever some assertion $(\mathit{userName}, \mathtt{in}, \mathit{roomName})$ exists in the dataspace might be written

$$\mathit{listRooms}\,[\emptyset\ (\mathsf{during}\ (\star, \mathtt{in}, \$r)\ \mathit{entry}\,[\emptyset \cup (\mathtt{room}, r)])]$$

Concrete syntax aside, during is reminiscent of a form of *logical implication*

$$\forall u, r.\,(u, \mathtt{in}, r) \implies (\mathtt{room}, r)$$

where assertions are interpreted as ground facts.

A related derived event pattern, during P spawn, is able to help us with our demand-matcher example 5.2, where during is not directly applicable. In contrast to during P, the event pattern during P spawn does not create a facet but instead spawns an entire sibling actor to handle each assertion matching P. The critical difference concerns failure. While a failure in a during P endpoint tears down the current actor, a failure in a during P spawn endpoint terminates the

15 Both SYNDICATE/RKT and SYNDICATE/JS fastidiously maintain this phase distinction. However, as each integrates SYNDICATE features with an imperative host language, a loophole remains where field updates during computation of *pattern* expressions evaluated during planning may affect later stages. In practice, this seems not to occur.

16 The new forms during and select can be compared to similar features of the fact space model, namely its *rule-based sub-language* and its *reactive context-aware collections* (Mostinckx et al. 2007; Mostinckx, Lombide Carreton and De Meuter 2008).

separate actor, leaving its siblings and parent intact. Equipped with this new construct, we may reformulate example 5.2 as just

$$\text{spawn } \textit{demandMatcher}\,[\emptyset\ (\text{during}\ (\texttt{hello}, \$x)\ \text{spawn}\ \texttt{...})]$$

"select" EXPRESSIONS. The ability of facets to automatically update published assertions in response to changes in fields provides a unidirectional link from the local state of an actor to the shared state held in its dataspace. To establish a bidirectional link, we require a construct describing a local data structure to maintain in response to changes in observed assertions:

$$\text{Programs } \mathrm{Pr} \in \mathbf{Pr} := \ldots \mid \text{select } P \text{ into } x := \{e\} \text{ in } \mathrm{Pr}$$

Like during, the new select construct is interpreted in terms of existing syntax. A program select $P$ into $x := \{e\}$ in $\mathrm{Pr}$ is interpreted as if it were written

$$\text{let } x\ :=\ \emptyset \text{ in } \left( y \begin{bmatrix} \emptyset & \\ (\text{asserted } P & (x \leftarrow x \cup \{e\})) \\ (\text{retracted } P & (x \leftarrow x - \{e\})) \end{bmatrix} ; \mathrm{Pr} \right)$$

where $y$ is fresh. The expression $e$ may refer to bindings introduced by pattern $P$.

The new construct allows us to recast example 5.25 as just

$$\text{select } (\star, \texttt{in}, \$r) \text{ into } \textit{rooms} := \{r\} \text{ in } 0$$

We may usefully generalize select from maintenance of fields containing sets to fields containing hash-tables, counts of matching rows, sums of matching rows, and other forms of aggregate summary of a set of assertions:

$$\begin{aligned} \text{Hash table: } & \text{select } P \text{ into } x := \{e \mapsto e\} \text{ in } \mathrm{Pr} \\ \text{Count: } & \text{select } P \text{ into } x := \text{count}\,(e) \text{ in } \mathrm{Pr} \\ \text{Sum: } & \text{select } P \text{ into } x := \text{sum}\,(e) \text{ in } \mathrm{Pr} \\ & \vdots \end{aligned}$$

The interpretations of these forms in terms of asserted and retracted endpoints should follow that of the form for sets, *mutatis mutandis*.

Finally, while event-handling endpoints in SYNDICATE/λ allow a program to react to changes in shared assertions, there is no general symmetric ability in this minimal language for a program to directly react to changes in local fields. The only such reaction available is the automatic republication of assertions depending on field values. We will see in chapter 6 a more general form of reaction to field changes that allows the programmer to express dataflow-style dependencies on and among fields. In our example here, this ability might find use in reacting to changes in the set-valued field *rooms*, updating a graphical display of the room list.

## 5.8 PROPERTIES

While Syndicate/λ is a mathematical fiction used to explain a language design, it highlights a number of properties that a Syndicate implementation must enjoy in order to satisfy a programmer's expectations. First of all, when extending a host language with Syndicate features, care must be taken to reconcile the host language's own soundness property with the invariants demanded by constructs such as facet creation and termination, field allocation, reference and update, and so on. If the errors discussed in section 5.5 cannot be ruled out statically, they should be checked for dynamically. Of particular importance is the check for a finite result set in *project*; experience writing Syndicate programs thus far suggests that programmer errors of this kind are not uncommon while designing a new protocol.

Second, with the introduction of mutable state and sequencing comes the obligation to offer programmers a comprehensible model of order-of-evaluation and of visibility of intra-actor side-effects from the perspective of an actor's peers. As section 5.6 explains, Syndicate/λ supports reasoning about various kinds of atomicity preserved during evaluation. Whichever guarantees about order-of-evaluation and transactionality a host language offers should be extended to Syndicate features so as to preserve these forms of atomicity.

Finally, programmers rely on theorem 4.35's promise of the conversational cooperation of the dataspace connecting a group of actors. In the same way, they rely on an extension of this promise to include Syndicate/λ's endpoints. An implementation of Syndicate must ensure that the event-handling code associated with an endpoint runs *only* for relevant events, for *every* relevant event, and never *redundantly* for the same piece of knowledge. In particular, the notion of *necessity* developed in lemma 4.42 must be adapted in the setting of facets and endpoints to account for the way in which *inst* elides irrelevant detail from incoming patch events.

# Part III

# PRACTICE

# *Overview*

In order to evaluate the Syndicate design, we must be able to write programs using it. In order to write programs, we need three things: algorithms, data structures and implementation techniques allowing us to realize the language design; a concrete instance of integration of the design with a host language; and a number of illustrative examples.

Chapter 6 builds on the formal models of chapters 4 and 5, presenting Syndicate/rkt, an extension of the Racket programming language with Syndicate features.

Chapter 7 then discusses general issues related to Syndicate implementation. First, it presents a new data structure, the *assertion trie*, important for efficient representation and manipulation of the sets of assertions ubiquitous to the dataspace model. Second, it turns to techniques for implementation and integration of Syndicate with a host language. Third, it describes some experimental tools for visualization and debugging of Syndicate programs.

Finally, chapter 8 presents a large number of examples demonstrating Syndicate idioms.

# 6

# SYNDICATE/RKT *Tutorial*

Now that we have explored the details of the SYNDICATE design in the abstract, it is time to apply the design ideas to a concrete language. This chapter introduces SYNDICATE/RKT, a language which extends Racket (Flatt and PLT 2010) with SYNDICATE language features. The aim of the chapter is to explain SYNDICATE/RKT in enough detail to allow the reader to appreciate the implementation ideas of chapter 7 and engage with the examples of chapter 8.[1]

## 6.1 INSTALLATION AND BRIEF EXAMPLE

The Racket-based SYNDICATE implementation is supplied as a Racket package.[2] After installing Racket itself, use the command-line or DrRacket-based interactive package management tool to install the `syndicate` package. For example, on a Unix-like system, run the command

```
raco pkg install syndicate
```

The implementation uses Racket's `#lang` facility to provide a custom language dialect with SYNDICATE language features built-in. A Racket source file starting with

```
#lang syndicate
```

declares itself to be a SYNDICATE program. Before we examine details of the language, a brief example demonstrates the big picture.

**Example 6.1.** Figure 26 shows a complete SYNDICATE/RKT program analogous to the box-and-client programs shown in previous chapters (examples 4.2 and 5.1). Typing it into a file and loading it into the DrRacket IDE or running it from the command line produces an unbounded stream of output that begins

```
client: learned that box's value is now 0
box: taking on new-value 1
client: learned that box's value is now 1
box: taking on new-value 2
...
```

1 A brief overview of SYNDICATE/JS is given in appendix A.
2 It is also available for download separately. See http://syndicate-lang.org/ for details.

```
1  #lang syndicate

2  (message-struct set-box (new-value))
3  (assertion-struct box-state (value))

4  (spawn (field [current-value 0])
5         (assert (box-state (current-value)))
6         (on (message (set-box $new-value))
7             (printf "box: taking on new-value ~v\n" new-value)
8             (current-value new-value)))

9  (spawn (on (asserted (box-state $v))
10            (printf "client: learned that box's value is now ~v\n" v)
11            (send! (set-box (+ v 1)))))
```

Figure 26: Syndicate/rkt box-and-client example

Line 1 declares that the module is written using Syndicate/rkt. Lines 2 and 3 declare Racket structures: `set-box` is declared as a structure to be used as a *message*, and `box-state` to be used as an *assertion*. Lines 4–8 and 9–11 start two actors together in the same dataspace. The first actor provides a mutable reference cell service, and the second accesses the cell.

The cell initially contains the value `0` (line 4). It publishes its value as a `box-state` record in the shared dataspace (line 5). When it hears a `set-box` message (line 6), it prints a message to the console and updates its `current-value` field. This leads to automatic update of the assertion of line 5. The cell actor is the only party able to alter the `box-state` assertion in the dataspace: peers may submit *requests* to change the assertion, but cannot themselves change the value.

The client actor begins its life waiting to hear about the assertion of `box-state` records. When it learns that a new record has been `asserted`, it prints a message to the console and sends a `set-box` message, which causes the cell to update itself, closing the loop. ◊

## 6.2 The structure of a running program: ground dataspace, driver actors

Figure 27 shows a schematic of a running Syndicate/rkt program. The main thread drives execution of the Syndicate world, dispatching events to actors and collecting and interpreting the resulting actions. All actors and dataspaces are gathered into a single, special *ground dataspace* which connects Syndicate to the "outside world" of plain Racket and hence, indirectly, to the underlying operating system.

The figure shows actors relating to the program itself—some of which are running in a nested dataspace—as well as actors supplying services offered by library modules. Two *driver actors* are shown alongside these. The role of a driver actor in Syndicate/rkt is to offer an assertion- and message-based Syndicate perspective on some external service—the "hardware" to the actor's "driver". Such driver actors call ordinary Racket library code, spawning Racket-level threads to perform long-lived, CPU-intensive or I/O-heavy tasks. Such threads may inject events—"hardware interrupts"—to the ground dataspace as if they were peers in some surrounding dataspace.

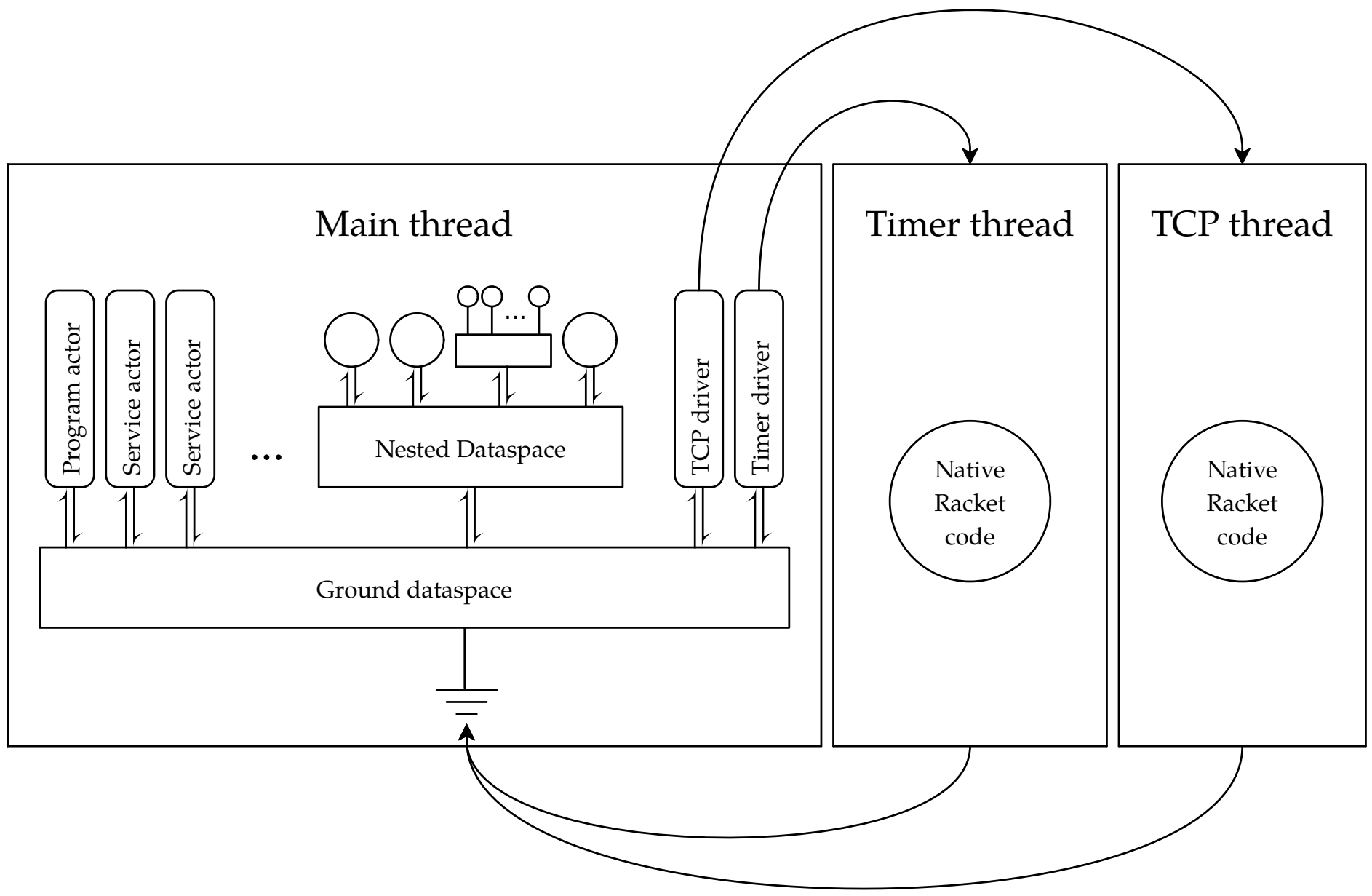

Figure 27: The structure of a running Syndicate/rkt program

For example, the Syndicate/rkt TCP driver interacts with peers via a protocol of assertions and messages describing TCP/IP sockets and transmitted and received TCP segments. When a socket is requested, the driver spawns not only a Syndicate/rkt actor but also a Racket-level thread. The thread uses Racket's native event libraries to wait for activity on the TCP/IP socket, and sends Syndicate messages describing received packets, errors, or changes in socket state. These messages are delivered to the Syndicate/rkt actor corresponding to the socket, which translates them and forwards them on to the driver's peers.

Similarly, the Timer driver responds to Syndicate/rkt messages requesting its services by updating a priority-queue of pending timers which it shares with a Racket-level thread. The thread interfaces with Racket's native timer mechanisms. Each time it is signaled by Racket, it delivers an appropriate event to the ground dataspace, which is picked up by the Timer driver and forwarded to the original requesting actor.

Syndicate/rkt's driver model is inspired by Erlang's "ports" model for I/O (*Erlang/OTP Design Principles* 2012). The layer of indirection that a driver actor introduces between a user program and some external facility serves not only to isolate the external facility from user program failures and vice versa but also to separate concerns. The driver actor's responsibility is to implement the access protocol for the external service, no matter how complex and stateful, exposing its features in terms of a Syndicate protocol. The user program may thus concentrate on its own responsibilities, delegating management of the external service to the driver. If either party should fail, the other may gracefully shut down or take some compensating action.

## 6.3 EXPRESSIONS, VALUES, MUTABILITY, AND DATA TYPES

Expressions in SYNDICATE/RKT are ordinary Racket expressions. While SYNDICATE/λ maintains a strict separation between commands and expressions, SYNDICATE/RKT inherits Racket's expression-oriented approach. Racket's functions replace SYNDICATE/λ's procedures. Ordinary Racket side-effects are available, and Racket's sequencing and order-of-evaluation are used unchanged.

Values in SYNDICATE/RKT are ordinary Racket values. This includes values used as assertions and message bodies. While SYNDICATE/λ forbids higher-order and mutable values in fields and assertions, SYNDICATE/RKT makes no such restriction, trusting the programmer to avoid problematic situations.[3] Actors may exchange mutable data or use Racket's mutable variables as required, though programmers are encouraged to design protocols that honor the spirit of SYNDICATE by eschewing mutable structures.

The SYNDICATE/RKT implementation of the dataspace model must be able to inspect the elements of compound data types such as lists, vectors and records in order to fulfill its pattern-matching obligations. Racket's `struct` record facility defaults to creation of "opaque" records which cannot be inspected in the necessary way. While SYNDICATE/RKT does not forbid use of such `struct` definitions—in fact, their opacity is beneficial in certain circumstances (see section 7.2.1)—it is almost always better to use Racket's "prefab" structures, which allow the right kind of introspection.

The special dataspace model observation constructor $?\cdot$ and the cross-layer constructors $\downharpoonleft \cdot$ and $\upharpoonleft \cdot$ are represented in SYNDICATE/RKT as instances of `struct`s named `observe`, `outbound` and `inbound`, respectively.

| Mathematical notation (figure 12) | $?c$ | $\downharpoonleft c$ | $\upharpoonleft c$ |
|---|---|---|---|
| SYNDICATE/RKT notation | `(observe c)` | `(outbound c)` | `(inbound c)` |

## 6.4 CORE FORMS

Each of the constructs of the formal model in chapter 5 maps to a feature of the implementation. In some cases, a built-in Racket language feature corresponds well to a SYNDICATE feature, and is used directly. In others, a feature is provided by way of a Racket library exposing new functions and data structures. In yet others, new syntax is required, and Racket's `syntax-parse` facility (Culpepper and Felleisen 2010) is brought to bear. Figure 28 summarizes the core forms added to Racket to yield SYNDICATE/RKT; figure 29 sketches a rough rubric allowing interpretation of the syntax of SYNDICATE/λ in terms of SYNDICATE/RKT.

PROGRAMS AND MODULES. A module written in the `syndicate` dialect not only provides constants, functions, and structure type definitions to its clients, as ordinary Racket modules do, but also offers *services* in the form of actors to be started when the module is *activated*. Thus, each module does double duty, serving as either or both of a unit of program composition

3 In this, SYNDICATE/RKT follows many implementations of the actor model for previously-existing languages.

```
module-level-form := ...
                   | (require/activate require-spec ...)
                   | struct-declaration
                   | spawn

struct-declaration := ...
                    | (message-struct name (field ...))
                    | (assertion-struct name (field ...))

             spawn := (spawn {#:name expr} facet-setup-expr ...)
                    | (spawn* {#:name expr} script-expr ...)
                    | (dataspace {#:name expr} script-expr ...)

  facet-setup-expr := expr
                    | field-declaration
                    | endpoint-expr

 field-declaration := (field [field-name initial-value] ...)

              expr := ...
                    | (current-facet-id)
                    | (observe expr)
                    | (outbound expr)
                    | (inbound expr)

       script-expr := expr
                    | (react facet-setup-expr ...)
                    | (stop-facet expr script-expr ...)
                    | (stop-current-facet)
                    | field-declaration
                    | spawn
                    | (send! script-expr)

     endpoint-expr := (assert {#:when test-expr} pattern)
                    | (on-start script-expr ...)
                    | (on-stop script-expr ...)
                    | (on {#:when test-expr} event-pattern script-expr ...)

     event-pattern := (asserted pattern)
                    | (retracted pattern)
                    | (message pattern)
```

Figure 28: Core SYNDICATE/RKT forms

| | |
|---|---|
| $0$ | `(void)` |
| $Pr_1; \cdots ; Pr_n$ | `(begin` $Pr_1 \cdots Pr_n$`)` |
| $e_1\ e_2$ | `(`$e_1\ e_2$`)` |
| let $x\ =\ e$ in $Pr$ | `(let ((`$x\ e$`))` $Pr$`)` |
| let $x\ :=\ e$ in $Pr$ | `(begin (field [`$x\ e$`])` $Pr$`)` |
| $x \leftarrow e$ | `(`$x\ e$`)` |
| send $e$ | `(send!` $e$`)` |
| spawn $Pr$ | `(spawn*` $Pr$`)` |
| dataspace $Pr$ | `(dataspace` $CPr$`)` |
| $x\,[A$ (start $Pr_{start}$) (stop $Pr_{stop}$) (D $Pr$) $\cdots]$ | `(react (define` $x$ `(current-facet-id))`<br>`(assert` $A$`)`<br>`(on-start` $Pr_{start}$`)`<br>`(on-stop` $Pr_{stop}$`)`<br>`(on` D $Pr$`)` $\cdots$`)` |
| stop $x$ $Pr$ | `(stop-facet` $x$ $Pr$`)` |

Figure 29: Approximate translation from SYNDICATE/λ syntax to SYNDICATE/RKT syntax

and a unit of *system* composition. In order to start a SYNDICATE/RKT program running, a user specifies a module to serve as the entry point. That module is activated in a fresh, empty dataspace, along with any actors created during activation of service modules it depends upon.

The nonterminal *module-level-form* in figure 28 specifies the SYNDICATE/RKT extensions to Racket's module-level language. A client may `require` a module, as usual, or may require and activate it by using the `require/activate` form. Activation is idempotent within a given program: a particular module's services are only started once.

The module-level language is also extended with two new structure-definition forms (nonterminal *struct-declaration*), `message-struct` and `assertion-struct`. The former is intended to declare structures for use in *messages*, while the latter declares structures for *assertions*.[4] Each is a thin veneer over Racket's "prefab" structure definition facility.

ABSTRACTION FACILITIES. In order to remain minimal, SYNDICATE/λ includes little in the way of abstraction facilities. However, in SYNDICATE/RKT, we wish to permit abstraction over field declaration, assertion- and event-handling endpoint installation, and facet creation and tear-down, as well as the usual forms of abstraction common to Racket programming. Therefore, we make abstraction facilities like `define`, `let`, `define-syntax` and `let-syntax` available throughout the SYNDICATE/RKT language.

However, not all SYNDICATE constructs make sense in all contexts. For example, it is nonsensical to attempt to declare an endpoint outside a facet. SYNDICATE/λ includes special syntactic po-

4 The current implementation does not enforce the distinction: in fact, the definitions of `message-struct` and `assertion-struct` are identical. They are both equivalent to `(struct` *name* (*field* ...) `#:prefab)`.

sitions for declaration of endpoints, keeping them clearly (and statically) distinct from positions for specifying commands. This approach conflicts with the desire to reuse Racket's abstraction facilities in all such syntactic positions. SYNDICATE/RKT therefore brings most SYNDICATE constructs into a single syntactic class—that of expressions—and relies on a dynamic mechanism to rule out inappropriate usage of SYNDICATE constructs. An internal flag keeps track of whether the program is in "script" or "facet setup" context.

Figure 28 reflects this dynamic context in its use of nonterminals *script-expr* and *facet-setup-expr*. Script expressions may only be used within event-handlers and in ordinary straight-line Racket code. They include expressions which perform side effects such as spawning other actors or sending messages. Facet setup expressions may only be used in contexts where a new facet is being configured. They include expressions which construct both assertion and event-handling endpoints.[5]

SENDING MESSAGES. The `send!` form broadcasts its argument to peers via the dataspace. That is, a message action is enqueued for transmission to the dataspace when the actor's behavior function eventually returns. Message sending, like all other actions, is thus *asynchronous* from the perspective of the SYNDICATE/RKT programmer.

SPAWNING ACTORS AND DATASPACES. The nonterminal *spawn* in figure 28 is available not only at the module level but also anywhere a *script-expr* is permitted within a running actor. The three forms `spawn`, `spawn*`, and `dataspace` correspond to the SYNDICATE/λ commands `spawn` and `dataspace`. Each of the first two, like SYNDICATE/λ's `spawn`, constructs a sibling actor in the current dataspace; the third, a nested dataspace whose initial actor runs the *script-expr*s specified. The two variations `spawn` and `spawn*` relate to each other as follows:

$$\texttt{(spawn}\ \textit{facet-setup-expr}\ \ldots\texttt{)} \triangleq \texttt{(spawn* (react}\ \textit{facet-setup-expr}\ \ldots\texttt{))}$$

Initially, SYNDICATE/RKT included only `spawn*` (written, at the time, "`spawn`"); a survey of programs showed that the overwhelming majority of uses of `spawn*` were of the form that the current `spawn` abbreviates, namely an actor with a single initial facet.

If a `spawn`, `spawn*`, or `dataspace` is supplied with a `#:name` clause, the result of the corresponding *expr* is attached to the created actor as its name for debugging and tracing purposes. The name is never made available to peers via assertions or messages in the dataspace.

FACET CREATION AND TERMINATION. The `react` form causes addition of a new facet to the currently-running actor, nested beneath the currently-active facet, or as the root of the actor's facet tree if used immediately within `spawn*`. The body of the `react` form is in "facet setup" context, and declares the new facet's endpoints. Unlike in SYNDICATE/λ, the facet's name is not manifest in the syntax. Instead, a facet may retrieve its system-generated name with a call to the procedure `current-facet-id`. Facet IDs may be freely stored in variables, passed as procedure arguments, and so on.

5 The script/facet distinction is reminiscent of, and partially inspired by, the "step/process" distinction of Hancock's FLOGO II language (Hancock 2003, chapter 5).

A facet ID must be supplied as the first argument to a `stop-facet` form, which is the Syndicate/rkt analogue of Syndicate/λ's stop. For example, the following program starts an actor whose root facet immediately terminates itself:

```
(spawn (on-start (stop-facet (current-facet-id))))
```

The program is analogous to the Syndicate/λ program

$$\mathsf{spawn}\ \mathit{root}\ [\emptyset\ \ (\mathsf{start}\ \ (\mathsf{stop}\ \mathit{root}\ 0))]$$

The shortcut

(`stop-current-facet` *script-expr* ...) $\triangleq$ (`stop-facet (current-facet-id)` *script-expr* ...)

captures a common form of use of `stop-facet` and `current-facet-id`.

A `stop-facet` form includes an optional sequence of *script-expr*s. These are executed *outside* the stopping facet, once its subscriptions and assertions have been completely withdrawn, in the context of the facet's own containing facet (if any). That is, an expression such as

```
(react (on (message 'x) (stop-facet (current-facet-id) (react ...))))
```

(1) creates a facet, which upon receipt of a message `'x` (2) terminates itself, and (3) effectively *replaces* itself with another facet, whose body is shown as an ellipsis. Compare to

```
(react (on (message 'x) (react ...)))
```

which upon receipt of each `'x` creates an additional, nested subfacet; and

```
(react (on (message 'x) (stop-current-facet) (react ...)))
```

which not only terminates itself when it receives an `'x` message but also creates a nested subfacet, which is shortly thereafter destroyed as a consequence of the termination of its parent.

Field declaration, access and update. Syndicate/rkt allows declaration of fields in both "script" and "facet setup" contexts. The `field` form borrows its syntactic shape from Racket's support for object-oriented programming; it acts as a `define` of the supplied *field-name*s, initializing each with its *initial-value*.

Fields are represented as *procedures* in Syndicate/rkt. When called with no arguments, a field procedure returns the field's current value; when called with a single argument, the field procedure updates the field's value to the given argument value.

Endpoint declaration. In facet setup context, the forms `assert`, `on-start`, `on-stop`, and `on` are available for creation of assertion and event-handling endpoints.

The `assert` form allows a facet to place assertions into the shared dataspace. For example, given the following structure definitions,

```
(assertion-struct user-present (user-name))
(message-struct say-to (user-name utterance))
```

the endpoint

```
(assert (user-present 'Alice))
```

asserts a `user-present` record, keeping it there until the endpoint's facet is terminated.

The `on` form allows facets to react to events described by the *event-pattern* nonterminal. Each possibility corresponds to the analogous event-pattern in Syndicate/λ. For example,

```
(on (asserted (user-present $name))
    (send! (say-to name "Hello!")))
```

reacts to the assertion of `(user-present 'Alice)` with a message, `(say-to 'Alice "Hello!")`.

Both `assert` and `on` forms take a *pattern*, within which most Racket expressions are permitted. Use of the discard operator (`_`) in a pattern corresponds to Syndicate/λ's ⋆; that is, to a *wildcard* denoting the universe of assertion values. In an `on` form, it acts to accept and ignore arbitrary structure in matched assertions or messages. Variables `x` introduced via binders `$x` embedded in an `on` form's pattern are available in the form's *script-expr*s; it is an error to include a binder in a pattern in an `assert` form. Patterns may make reference to current field values, and any fields accessed are called the endpoint's *dependent fields*.

A pattern is evaluated to yield a set of assertions, both initially and every time a dependent field is updated. For `on` forms, an `observe` structure constructor is added to each assertion in the denoted set. This process of assertion-set extraction is analogous to Syndicate/λ's use of the *assertions* metafunction of figure 19.

Whenever an endpoint's pattern is re-evaluated, the resulting assertions are placed in the surrounding dataspace by way of a state-change notification action. However, if a `#:when` clause is present in the `assert` or `on` form, the corresponding *test-expr* is evaluated just before actually issuing the action. If *test-expr* yields a false value, no action is produced. This allows *conditional* assertion and *conditional* subscription. In particular, a `#:when` clause *test-expr* may depend on field values; if it does, the fields are considered part of the dependent fields.

The `on-start` and `on-stop` forms introduce facet-startup and -shutdown event handlers. The former are executed once the block of *facet-setup-expr*s has finished configuring the endpoints of the facet and after the facet's new assertions (including its subscriptions) have been sent to the surrounding dataspace. The latter are executed just prior to withdrawal of the facet's endpoint subscriptions during the facet shutdown process.

An `on-start` form may be used to send a message in a context where a corresponding reply-listener is guaranteed to be active and listening; for example, in

```
(react (on-start (send! 'request))
       (on (message 'reply ...)))
```

the `'request` is guaranteed to be sent only after the subscription to the `'reply` has been established, ensuring that the requesting party will receive the reply even if the replying party responds immediately.[6]

6 In fact, `on-start` is the only way to send a message or spawn an actor during facet startup. Both `send!` and `spawn` are *script-expr*s, not *facet-setup-expr*s.

An on-stop form may be used to perform cleanup actions just prior to the end of the conversational context modeled by a facet; for example, in

```
(react (on (retracted 'connection) (stop-current-facet))
       (on-stop (send! 'goodbye)))
```

the 'goodbye message is guaranteed to be sent before the subscription to 'connection assertions is withdrawn. Any number of on-start and on-stop forms may be added during facet setup.

## 6.5 Derived and additional forms

Figure 30 summarizes derived forms that build upon the core forms to allow concise expression of frequently-employed concepts.

**Facet termination.** A common idiom is to terminate a facet in response to an event. The abbreviation stop-when is intended for this case:

(stop-when *P E* ...) ≜ (on *P* (stop-facet (current-facet-id) *E* ...))

The *script-expr*s (*E* ...) are placed inside the stop-facet command, and so are executed outside the stopping facet. The example from above could be written

```
(react (stop-when (message 'x) (react ...)))
```

This style of use of stop-when gives something of the flavor of a state-transition in a state machine, since the *script-expr*s are in a kind of "tail position" with respect to the stopping facet.

**Sub-conversations and subfacets.** A second, even more common idiom is that of Syndicate/λ's during (section 5.7), which introduces a nested subfacet to an actor for the duration of each assertion matching a given pattern. The triggering assertion acts as a *conversational frame*, delimiting a sub-conversation. The Syndicate/rkt during form corresponds to a stereotypical usage of core forms:

(during P E ...) ≜ (on (asserted P) (react (stop-when (retracted P′)) E ...))

where P′ is derived from P by replacing every binder $x in P with the corresponding x.

Just as Syndicate/λ's during had a spawning variant, so Syndicate/rkt has during/spawn. The variant form spawns an actor in the dataspace instead of creating a subfacet, confining the scope of failure to an individual sub-conversation rather than allowing a crashing sub-conversation to terminate the actor as a whole.

**Streaming queries.** The assert form allows actors to construct shared assertions from values of local fields. To operate in the other direction, updating a field based on some aggregate function over a set of assertions, Syndicate/rkt offers a suite of define/query-* forms. For example, given a structure definition

```
(assertion-struct mood (user-name description))
```

```
         script-expr := ...
                      | blocking-facet-expr

    facet-setup-expr := ...
                      | derived-endpoint-expr
                      | dataflow-expr

derived-endpoint-expr := (stop-when {#:when expr} event-pattern script-expr ...)
                      | (stop-when-true expr script-expr ...)
                      | (during pattern facet-setup-expr ...)
                      | (during/spawn pattern {#:name expr} script-expr ...)
                      | query-endpoint-expr

  query-endpoint-expr :=
          (define/query-value field-name expr pattern script-expr add/remove)
        | (define/query-set field-name pattern script-expr add/remove)
        | (define/query-hash field-name pattern script-expr script-expr add/remove)
        | (define/query-hash-set field-name pattern script-expr script-expr add/remove)
        | (define/query-count field-name pattern script-expr add/remove)

          add/remove := {#:on-add script-expr} {#:on-remove script-expr}

  blocking-facet-expr := (react/suspend (id) facet-setup-expr ...)
                      | (until event-pattern facet-setup-expr ...)
                      | (flush!)
                      | immediate-query

       dataflow-expr := (begin/dataflow script-expr ...)
                      | (define/dataflow field-name script-expr {#:default expr})

     immediate-query := (immediate-query query-spec ...)

          query-spec := (query-value expr pattern script-expr)
                      | (query-set pattern script-expr)
                      | (query-hash pattern script-expr script-expr)
                      | (query-hash-set pattern script-expr script-expr)
                      | (query-count pattern script-expr)
```

Figure 30: Derived and additional Syndicate/rkt forms

representing a user's mood, we may declare a field that tracks the set of all user names that have an associated mood via

```
(define/query-set moody-users (mood $n _) n)
```

or a field that collects all available mood descriptions into a local hash table via

```
(define/query-hash all-moods (mood $n $m) n m)
```

The resulting fields contain ordinary Racket sets and hash tables. In cases where only a single assertion matching a given pattern is expected, the `define/query-value` form extracts that single value, falling back on a default during periods when no matching assertion exists in the shared dataspace:

```
(define/query-value alice-mood 'unknown (mood 'Alice $m) m)
```

If an `#:on-add` or `#:on-remove` clause is supplied to a `define/query-*` form, the corresponding expressions are evaluated immediately prior to updating the form's associated field upon receiving a relevant change notification.

**general-purpose field dependencies.** The Syndicate/rkt implementation uses a simple dependency-tracking mechanism to determine which endpoint patterns depend on which fields, and exposes this mechanism to the programmer. Each `begin/dataflow` form in a facet setup context introduces a block of code that may depend on zero or more fields. Like an endpoint's pattern, it is executed once at startup and then every time one of its dependent fields is changed. For example, the following prints a message each time Alice's mood changes in the dataspace:

```
(react (define/query-value alice-mood 'unknown (mood 'Alice $m) m)
       (begin/dataflow
         (printf "Alice's mood is now: ~a\n" (alice-mood))))
```

Of course, for a simple example like this there are many alternative approaches, including use of `during` with an `on-start` handler:

```
(react (during (mood 'Alice $m)
         (on-start (printf "Alice's mood is now: ~a\n" (alice-mood)))))
```

or use of an `#:on-add` clause in `define/query-value`:

```
(react (define/query-value alice-mood 'unknown (mood 'Alice $m) m
         #:on-add (printf "Alice's mood is now: ~a\n" (alice-mood))))
```

An important difference between the latter two and the first variation is that only the first variation prints a message if *some other* piece of code updates the `alice-mood` field. It is this ability to react to field changes *specifically*, rather than to dataspace assertion changes generally, that makes `begin/dataflow` useful.

The form `define/dataflow` is an abbreviation for definition of a field whose value directly depends on the values of other fields and that should be updated whenever its dependents change:

(define/dataflow *F E*) ≜ (begin (field [*F* #f]) (begin/dataflow (*F E*)))

and the form stop-when-true is an abbreviation useful for terminating a facet in response to a predicate over a collection of fields becoming true:

(stop-when-true *test-expr E ...*) ≜
(begin/dataflow (when *test-expr* (stop-facet (current-facet-id) *E ...*)))

**BLOCKING EXPRESSIONS.** The uniformly event-driven style of SYNDICATE can make it difficult to express certain patterns of control-flow involving *blocking requests*. Absent special support, the programmer must fall back on manual conversion to continuation-passing style followed by defunctionalization of the resulting continuations, yielding an event-driven state machine (Felleisen et al. 1988). Fortunately, Racket includes a rich library supporting delimited continuations (Felleisen 1988; Flatt et al. 2007). SYNDICATE/RKT allows event-handlers to capture their continuation up to invocation of the event-handler body and no further, replacing it by a nested subfacet. The subfacet may, in response to a later event, restore the suspended continuation, resuming its computation. This allows the programmer to use a blocking style to describe remote-procedure-call-like interactions with an actor's peers, reminiscent of the style made available by a similar use of continuations in the context of web applications (Queinnec 2000; Graunke et al. 2001).

The react/suspend form suspends the active event-handler, binds the suspended continuation to identifier *id*, and adds a new subfacet configured with the form's *facet-setup-expr*s. If one of the new subfacet's endpoints later invokes the continuation in *id*, the subfacet is terminated, and the arguments to *id* are returned as the result of the react/suspend form. For example,

```
(printf "Received: ~a" (react/suspend (k) (on (message (say-to 'Alice $v) (k v))))
```

waits for the next message sent to 'Alice, and when one arrives, prints it out.

The until form is built on react/suspend, and allows temporary establishment of a set of endpoints until some event occurs. It is defined as

(until *E F ...*) ≜ (react/suspend (k) (stop-when *E* (k (void))) *F ...*)

As an example of its use, the following SYNDICATE/RKT library procedure interacts with a timer service and causes a running event-handler to pause for a set number of seconds:

```
(define (sleep sec)
  (define timer-id (gensym 'sleep))
  (until (message (timer-expired timer-id _))
         (on-start (send! (set-timer timer-id (* sec 1000.0) 'relative)))))
```

An important consideration when programming with react/suspend and its derivatives is that the world may change during the time that an event-handler is "blocked". For example, the following actor has no guarantee that the two messages it prints display the same value:

```
(message-struct request-printout ())
(message-struct increment-counter ())
(spawn (field [counter 0])
       (on (message (increment-counter))
           (counter (+ (counter) 1))))
       (on (message (request-printout))
           (printf "Counter (before sleep): ~a\n" (counter))
           (sleep 1)
           (printf "Counter (after sleep): ~a\n" (counter))))
```

**point-in-time queries.** The `define/query-*` forms allow an actor to reflect a set of assertions into a local field on an ongoing basis. However, some programs call for sampling of the set of assertions present at a given moment in time, rather than establishment of a long-running streaming query. For these programs, Syndicate/rkt offers the `immediate-query` form, built atop a construct called `flush!`. The latter is approximately defined as

```
(flush!) ≜ (let ((x (gensym))) (until (message x) (on-start (send! x))))
```

and acts to force all preceding actions to the dataspace and allow them to take effect before proceeding. In particular, if new endpoints have established subscriptions to some set of assertions, then `flush!` allows the dataspace a chance to transmit matching assertions to those endpoints before execution of the continuation of `flush!` proceeds.[7]

The `immediate-query` form makes use of `flush!` by establishing a temporary subfacet using `react/suspend`, creating temporary fields tracking the requested information, and performing a single `flush!` round in an `on-start` handler before releasing its suspended continuation. For example,

```
(immediate-query [query-value 'unknown (mood 'Alice $m) m])
= (react/suspend (k)
    (define/query-value v 'unknown (mood 'Alice $m) m)
    (on-start (flush!) (k (v))))
```

retrieves the *current* mood of `'Alice` without setting up a facet to track it over the long term. Likewise,

```
(immediate-query [query-set (mood $n _) n])
= (react/suspend (k)
    (define/query-set v (mood $n _) n)
    (on-start (flush!) (k (v))))
```

7 The notion can usefully be generalized to perform a round-trip to some *other* distinct locus of state than the dataspace. For example, consider an actor managing a connection to an external SQL database. A form of "flush" that performed a round-trip communication with that actor would assure the caller that all previous commands for the actor had been seen and (presumably) interpreted. Further examples that aim at connected but distinct loci of state can be seen in 80x86 `MFENCE` instruction, the `fflush(3)`, `fsync(2)` and `sync(2)` POSIX library routines, and the "force unit access" variations on SATA write commands (IEEE 2009; INCITS T13 Committee 2006).

retrieves the names of all users with some mood recorded at the time of the query.

Users of `immediate-query` should remain aware that the world may change during the time the query is executing, since it is based on `react/suspend`. Because `immediate-query` yields to the dataspace, other events may arrive between the moment the query is issued and the moment it completes.

**Example 6.2.** Figure 31 demonstrates a number of the language features just introduced. The program starts four actors: a `printer` (line 6), which displays messages on the standard output; a `flip-flop` (line 9), which transitions back and forth between active and inactive states in response to a `toggle` message, and which asserts an `active` record only when active; a `monitor-flip-flop` actor (line 18), which displays a message (via the `printer`) every time the `flip-flop` changes state; and a `periodic-toggle` actor (line 21), which interfaces to the system timer driver and arranges for delivery of a `toggle` message every second. Figure 32 shows the structure of the running program.

The `flip-flop` actor makes use of SYNDICATE/RKT's abstraction facility to define two local procedures, `active-state` and `inactive-state` (lines 10 and 14, respectively). When called, `active-state` constructs a facet that asserts the `active` record (line 11) and waits for a `toggle` message (line 12). When such a message arrives, the facet terminates itself using `stop-when` and performs the action of calling the `inactive-state` procedure. In effect, when toggled, an active facet *replaces* itself with an inactive facet, demonstrating the state-transition-like nature of `stop-when` endpoints. The `inactive-state` procedure is similar, omitting only the assertion of the `active` record.

Despite its lack of explicit fields, the `flip-flop` actor is stateful. Its state is implicit in its facet structure. Each time it transitions from inactive to active state, or vice versa, the facet tree that forms part of the actor's implicit *control* state is updated.

Only one of the four actors, `periodic-toggle`, maintains explicit state. Its `next-toggle-time` field keeps track of the next moment that a `toggle` message should be transmitted. Each time the field is updated (line 25), SYNDICATE/RKT's change-propagation mechanism ensures that the assertion resulting from the subscription of line 23, namely

```
(observe (later-than (next-toggle-time)))
```

is updated in the dataspace. The service actor started by the timestate driver,[8] activated by the `require/activate` form on line 2, is observing *observers* of `later-than` records, and coordinates with the underlying Racket timer mechanism to ensure that an appropriate record is asserted once the moment of interest has passed. ◊

## 6.6 AD-HOC ASSERTIONS

From time to time, it is convenient to augment the set of assertions currently expressed by an actor without constructing an endpoint or even a whole facet to maintain the new assertions.

8 See the implementation of the timestate protocol in example 8.12 in section 8.4.

```
#lang syndicate

(require/activate syndicate/drivers/timestate)

(assertion-struct active ())
(message-struct toggle ())
(message-struct stdout-message (body))

(spawn #:name 'printer
       (on (message (stdout-message $body))
           (displayln body)))

(spawn* #:name 'flip-flop
        (define (active-state)
          (react (assert (active))
                 (stop-when (message (toggle))
                    (inactive-state))))
        (define (inactive-state)
          (react (stop-when (message (toggle))
                    (active-state))))
        (inactive-state))

(spawn #:name 'monitor-flip-flop
       (on (asserted (active)) (send! (stdout-message "Flip-flop is active")))
       (on (retracted (active)) (send! (stdout-message "Flip-flop is inactive"))))

(spawn #:name 'periodic-toggle
       (field [next-toggle-time (current-inexact-milliseconds)])
       (on (asserted (later-than (next-toggle-time)))
           (send! (toggle))
           (next-toggle-time (+ (next-toggle-time) 1000))))
```

Figure 31: Flip-flop example

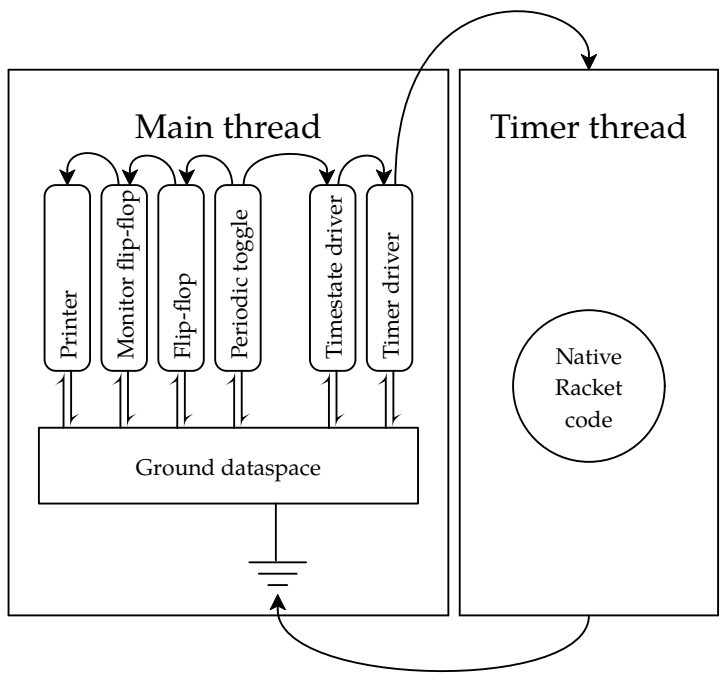

Figure 32: The structure of the running flip-flop example

```
1 (assertion-struct file (name content))
2 (message-struct save (name content))
3 (message-struct delete (name))

4 (spawn (field [files (hash)])
5       (during (observe (file $name _))
6              (assert (file name (hash-ref (files) name #f))))
7       (on (message (save $name $content))
8          (files (hash-set (files) name content)))
9       (on (message (delete $name))
10         (files (hash-remove (files) name))))
```

Figure 33: "File system" using `during`

SYNDICATE/RKT provides two imperative commands, `assert!` and `retract!` which allow ad-hoc addition and removal of assertions. While most programs will never use these commands, they occasionally greatly simplify certain tasks.

Consider figure 33, which implements a simple "file system" abstraction using the protocol structures defined on lines 1–3. Clients assert interest in `file` records, and in response the server examines its database (the hash table held in its `files` field) and supplies the record of interest. Because the server uses `during` (line 5), each distinct file name results in a distinct *sub-conversation* responsible for maintaining a single assertion (line 6). Subsequent `save` or `delete` requests (lines 7–10) update the `files` table, which automatically causes recomputation of the assertions resulting from different instances of line 6.

An alternative approach is shown in figure 34. Here, conversational state is explicitly maintained in a new field, `monitored`, which holds a set of the file names known to be of current interest. In response to a newly-appeared assertion of interest (line 2), the server updates `monitored` (line 4) but also uses `assert!` to publish an initial response record in the dataspace. Subsequent `save` or `delete` requests (lines 8–17) *replace* this record by using `retract!` followed by `assert!`, but only do so if the modified file is known to be of interest.

A second example of the use of `assert!` and `retract!` is shown in figure 35. The program implements something like a counting semaphore. Here, `assert!` and `retract!` are used to maintain up to `total-available-leases` separate `lease-assignment` records describing parties requesting use of the semaphore. The first-in, first-out nature of the lease assignment process does not naturally correspond to a nested facet structure; no obvious solution using `during` springs to mind. A remarkable aspect of this program is the use of `retract!` on line 15, in response to a withdrawn lease request. The party withdrawing its request may *or may not* currently hold one of the available resources. If it does, the `retract!` corresponds to a previous `assert!` (on either line 8 or line 21) and so results in a patch transmitted to the dataspace and a corresponding triggering of the endpoint on line 16; if it does not, then the `retract!` has no effect on the actor's assertion set, since set subtraction is idempotent, and therefore the lease request vanishes without disturbing the rest of the state of the system.

```
(spawn (field [files (hash)] [monitored (set)])
       (on (asserted (observe (file $name _)))
           (assert! (file name (hash-ref (files) name #f)))
           (monitored (set-add (monitored) name)))
       (on (retracted (observe (file $name _)))
           (retract! (file name (hash-ref (files) name #f)))
           (monitored (set-remove (monitored) name)))
       (on (message (save $name $content))
           (when (set-member? (monitored) name)
             (retract! (file name (hash-ref (files) name #f)))
             (assert! (file name content)))
           (files (hash-set (files) name content)))
       (on (message (delete $name))
           (when (set-member? (monitored) name)
             (retract! (file name (hash-ref (files) name #f)))
             (assert! (file name #f)))
           (files (hash-remove (files) name))))
```

Figure 34: "File system" using `assert!` and `retract!`

The `assert!` and `retract!` commands manipulate a "virtual" endpoint which is considered to exist alongside the real endpoints within the actor's facets. The effects of the commands are therefore only visible to the extent they do not interfere with assertions made by other facets in the actor. For example, if an actor has an existing facet that has an endpoint `(assert 1)`, and it subsequently performs `(assert! 1)` and `(assert! 2)` in response to some event, the patch action sent to the dataspace contains only `2`, since `1` was previously asserted by the assertion endpoint. If, later still, it performs `(retract! 1)` and `(retract! 2)`, the resulting patch action will again only mention `2`, since `1` remains asserted by the assertion endpoint.

Very few programs make use of this feature; it is not implemented at all in Syndicate/js. Usually, given freedom to design a protocol appropriate for Syndicate, pervasive use of assertions over messages allows `during` and nested facets in general to be used instead of `assert!` and `retract!`. Setting aside the unusual case of the "semaphore" of figure 35, there are two general areas in which the two commands are helpful:

1. They can be used to react to the *absence* of particular assertions in the dataspace (as seen in the "loading indicator" example discussed in section 9.6).

2. They are useful in cases where Syndicate implements a protocol not designed with dataspaces in mind, where *messages* in the *foreign* protocol update conversational state that corresponds to *assertions* on the Syndicate side.

An example of this latter is the IRC client driver.[9] When a user joins an IRC channel, the IRC protocol requires that the server send a list of existing members of the channel to the new user

9 Source code file `racket/syndicate/drivers/irc.rkt` in the Syndicate repository.

```
(assertion-struct lease-request (resource-id request-id))
(assertion-struct lease-assignment (resource-id request-id))

(define (spawn-resource resource-id total-available-leases)
  (spawn (field [waiters (make-queue)]
                [free-lease-count total-available-leases])

         (on (asserted (lease-request resource-id $w))
             (cond [(positive? (free-lease-count))
                    (assert! (lease-assignment resource-id w))
                    (free-lease-count (- (free-lease-count) 1))]
                   [else
                    (waiters (enqueue (waiters) w))]))

         (on (retracted (lease-request resource-id $w))
             (waiters (queue-filter (lambda (x) (not (equal? w x)))
                                    (waiters)))
             (retract! (lease-assignment resource-id w)))

         (on (retracted (lease-assignment resource-id $w))
             (cond [(queue-empty? (waiters))
                    (free-lease-count (+ (free-lease-count) 1))]
                   [else
                    (define-values (w remainder) (dequeue (waiters)))
                    (assert! (lease-assignment resource-id w))
                    (waiters remainder)]))))
```

Figure 35: “Semaphore” protocol, suitable for implementing Dining Philosophers

using special syntax—zero or more `353` messages containing nicknames, followed by a `366` message to signal the end of the list—before transitioning into an incremental group-maintenance phase, where separate `JOIN` and `PART` messages signal the appearance and disappearance of channel members. The Syndicate protocol for participating in an IRC channel, however, maintains assertions describing the membership of each channel. The mismatch between the IRC protocol's use of messages and the Syndicate protocol's use of assertions is addressed by trusting the server to send appropriate messages, and reflecting each `353`, `JOIN`, and `PART` into appropriate `assert!` and `retract!` actions.

# 7

# *Implementation*

So far, our discussion of Syndicate has been abstract in order to concentrate on the essence of the programming model, with just enough concrete detail provided to allow us to examine small, realistic examples. Now it is time to turn to techniques for making implementations of Syndicate suitable for exploration of the model using practical programs.

The notion of sets of assertions is at the heart of the dataspace model. The formal models presented in chapters 4 and 5 rely on representations of potentially-infinite assertion sets that are amenable to efficient implementations of set union, intersection, complement, and subtraction. These operations are the foundations of the set comprehensions used extensively in the metafunctions used throughout chapter 4. For these reasons, I will begin by examining a trie-based data structure I have developed, that I call an *assertion trie* (section 7.1), which represents and allows efficient operation on assertion sets. After presenting the data structure in the abstract, I will turn to concrete details of its implementation.

With a suitable data structure in hand, producing an implementation of the dataspace model can be as simple as following the formal model of chapter 4. Section 7.2 gives an overview of the core of an implementation of the dataspace model.

In addition, Syndicate offers not only a means of structuring interactions among groups of components, but also features for structuring state and reactions within a single component, as presented in chapter 5. Section 7.3 describes implementation of the full design.

Finally, by directly representing structures from the formal model in an implementation, we unlock the possibility of reflecting on running programs in terms of the concepts of the formal model. Section 7.4 describes initial steps toward programming tools that exploit this connection to assist in visualization and debugging of Syndicate programs.

While it is both intuitively apparent and borne out by experience that assertion tries are able to faithfully reflect the semantics of a certain kind of sets, and that the prototype Syndicate implementations as a whole live up to the formal models of previous chapters, this chapter will not go beyond making informal claims about these issues. The level of rigor of the implementation work thus far is the usual informal connection to what is proposed to be an underlying formal model. To get to the level of rigor of, say, VLISP (Oliva, Ramsdell and Wand 1995), we would have to formalize the claims and prove them; to get to the level of CompCert (Leroy 2009), we would have to mechanize the claims and proofs.[1]

1 The formal claims and proofs from previous chapters do not depend on anything in this chapter. Instead, the proofs there use the semantics of sets directly, and remain agnostic to concrete representations of the sets concerned.

## 7.1 REPRESENTING ASSERTION SETS

As the model shows, evaluation of SYNDICATE programs involves frequent and complex manipulation of sets of assertions. While the grammar and reduction rules of the calculus itself depend on only a few elements of the syntax of assertions, namely the unary constructors $\downharpoonleft$, $\upharpoonleft$, and ?, much of the power of the system comes from the ability to use a wildcard $\star$ in user programs specifying sets of assertions to be placed in a dataspace. The wildcard symbol is interpreted as *the infinite set of all possible assertions*. This leads us to a central challenge: the choice of representation for assertion sets in implementations of SYNDICATE.

There are a number of requirements that must be satisfied by our representation.

EFFICIENT COMPUTATION OF METAFUNCTIONS. Metafunctions such as $\mathsf{bc}_\Delta$, $\mathsf{inp}$, and $\mathsf{out}$ (section 4.6) must have efficient and accurate implementations in order to deliver correct SCN events to the correct subset of actors in a dataspace in response to a SCN action, without wasting time examining assertions made by actors that are not interested in the change represented by the SCN action.

EFFICIENT MESSAGE ROUTING. Some SYNDICATE programs make heavy use of message sending. For these programs, it is important to be able to quickly discover the set of actors interested in a given message.

COMPACTNESS. The representation of the dataspace must not grow unreasonably large as time goes by. In particular, many SYNDICATE programs assert and retract the same assertions over and over as they proceed, and it is important that the representation of the dataspace not grow without bound in this case.

GENERALITY. Assertions are the data structures of SYNDICATE programming. The representation of assertion sets must handle structures rich enough to support the protocols designed by SYNDICATE programmers. Likewise, it must support embedding wildcards in assertions in a general enough way that common uses of wildcard are not ruled out.

In this section, I will present *Assertion Tries*. An Assertion Trie is a data structure based on a *trie* that satisfies these requirements, offering support for semi-structured *S-expression* assertions with embedded wildcards, sufficient for the examples and case studies presented in this dissertation. In an implementation of SYNDICATE, these tries are used in many different contexts. First and foremost, they are used to index dataspaces, mapping assertions to actor identities, but they are also used to represent sets of assertions alone, mapping assertions to the unit value, and therefore to represent both monolithic and incremental SCNs (patches).

### 7.1.1 *Background*

A *trie* (de la Briandais 1959; Fredkin 1960), also known as a *radix tree* or *prefix tree*, is a kind of search tree keyed by sequences of symbols. Each edge in the trie is labeled with a particular symbol, and searches proceed from the root down toward a leaf, following each symbol in turn from the sequence comprising the key being searched for. Tries are commonly used in applications such as phone switches for routing telephone calls, and in publish/subscribe messaging middleware (Ionescu 2010, 2011) for routing messages to subscribers. In the former case, the phone number is used as a key, and each digit is one symbol; in the latter, some property of the message itself, commonly its "topic", serves as the key. In both cases, tries are a good fit because they permit rapid discarding of irrelevant portions of the search space. Standard data structures texts give a good overview of the basics (for example, Cormen et al. 2009, section 2-12).

Each trie node logically has a finite number of edges emanating from it, making tries directly applicable to tasks such as phone call routing, where the set of symbols at each step is finite and small. They also work well for cases of message routing where, while the set of possible symbols at each step is not finite, each *subscription* involves a specific sequence of symbols without wildcards.

Given two tries, interpreting each as the set of keys that it matches, it is straightforward and efficient to compute the trie corresponding to the set union, set intersection, or set difference of the inputs. However, set *complement* poses a problem: tries cannot represent *cofinite* sets of edges emanating from a node. This poses a difficulty for supporting wildcards, since a wildcard is supposed to correspond to the set of all possible symbols, a special case of a cofinite set.

Finally, tries work well where edges are labeled with *unstructured* symbols. Tries cannot easily represent patterns over semi-structured data such as S-expressions.

The data structure must be adapted in order to properly handle both semi-structured keys and wildcards.

### 7.1.2 *Semi-structured assertions & wildcards*

While a trie must use sequences of tokens as keys, we wish to key on trees. Hence, we must map our tree-shaped assertions, which have both hierarchical and linear structure, to sequences of tokens that encode both forms of structure.[2] To this end, we reinterpret assertions as sequences of tokens by reading them left to right and inserting a distinct *tuple-marker* token $\ll_n$ labeled with the *arity* of the tuple it introduces to mark entry to a nested subsequence.

Let us limit our attention to S-expressions over atoms, sometimes extended with a wildcard marker, $\star \notin$ **Atom**:

$$\begin{aligned} \text{Atoms } x, y, z \in \mathbf{Atom} &= \mathbb{Z} \cup \mathbf{String} \cup \mathbf{Symbol} \cup \dots \\ \text{S-expressions } v, w \in \mathbf{Sexp} &:= x \mid (v, w, \dots) \\ \text{Wild S-expressions } v^+, w^+ \in \mathbf{Sexp}^+ &:= x \mid (v^+, w^+, \dots) \mid \star \\ \text{Sets of S-expressions } V \in \mathcal{P}(\mathbf{Sexp}) & \end{aligned}$$

Wildcards are not the only option for matching multiple values. In principle, any (decidable) predicate can be used, so long as it can be indexed suitably efficiently. As examples of something narrower than wildcard, but more general than matching specific values, consider *range queries* over integers and *type predicates*. Range queries such as $\lambda x. (10 \leqslant x \wedge x < 20)$ can

2 This approach is inspired by Alur and Madhusudan's work on *nested-word automata* (Alur and Madhusudan 2009).

be used in protocols involving contiguous *message identifiers* for bulk acknowledgement as well as for flow control. Type predicates such as Racket's `number?` and `string?` can extend our language of assertion patterns with something reminiscent of *occurrence typing* (Tobin-Hochstadt and Felleisen 2008).

**Definition 7.1** (Meaning of wild S-expressions)**.** Each element of $\mathbf{Sexp}^+$ has a straightforward interpretation as a set of **Sexp**s:

$$
\begin{aligned}
\mathit{meaning} &: \mathbf{Sexp}^+ \to \mathcal{P}(\mathbf{Sexp}) \\
\mathit{meaning}(x) &= \{x\} \\
\mathit{meaning}((v^+, w^+, \dots)) &= \{(v', w', \dots) \mid v' \in \mathit{meaning}(v^+), w' \in \mathit{meaning}(w^+), \dots)\\
\mathit{meaning}(\star) &= \mathbf{Sexp}
\end{aligned}
$$

The alphabet of tokens we will use, **Tok**, consists of the atoms, plus a family of tuple-markers (not themselves **Atom**s), each subscripted with its arity: $\ll_0$ is the token introducing a 0-ary tuple, $\ll_1$ a unary tuple, $\ll_2$ a pair, and so on. Matching $\gg$ "pop" tokens are not included: they follow implicitly from arities used on tuple-markers in sequences of tokens. We will write $\#t$ for the arity of a given token: for all atoms, $\#x = 0$; for all tuple-markers, $\# \ll_n = n$. We will sometimes want to extend the set of tokens with our wildcard marker; we will write $\mathbf{Tok}^+$ for this set.

$$
\begin{aligned}
\text{Tokens } s, t \in \mathbf{Tok} &= \mathbf{Atom} \cup \{\ll_0, \ll_1, \ll_2, \dots\} \\
\text{Wild tokens } s^+, t^+ \in \mathbf{Tok}^+ &= \mathbf{Tok} \cup \{\star\}
\end{aligned}
$$

**Definition 7.2** (Serialization of S-expressions)**.** We now have the ingredients we need to read S-expressions as sequences of tokens using the following definition, and the analogous $[\![\cdot]\!]^+$ extended to wild S-expressions and wild tokens by $[\![\star]\!]^+ = \star$:

$$
\begin{aligned}
[\![\cdot]\!] &: \mathbf{Sexp} \to \overrightarrow{\mathbf{Tok}} \\
[\![x]\!] &= x \\
[\![(v, w, \dots)]\!] &= \ll_n \; [\![v]\!] \; [\![w]\!] \; \dots \text{ where } n = |(v, w, ...)|
\end{aligned}
$$

**Example 7.3.** The S-expression $(\texttt{sale}, \texttt{milk}, (1, \texttt{pt}), (1.17, \texttt{usd}))$ translates to the following token sequence:

$$
\begin{aligned}
&[\![(\texttt{sale}, \texttt{milk}, (1, \texttt{pt}), (1.17, \texttt{usd}))]\!] \\
= &\ll_4 \; \texttt{sale milk} \; \ll_2 \; 1 \; \texttt{pt} \; \ll_2 \; 1.17 \; \texttt{usd}
\end{aligned}
$$

◊

The correctness of some of our operations on assertion tries depends on the idea of a *well-formed* sequence of tokens; namely, one that corresponds to some **Sexp**.

**Definition 7.4** (Parsing of token sequences)**.** We define the (partial) function $(\!|\cdot|\!)$ (and its obvious extension $(\!|\cdot|\!)^+$, $(\!|\star \; t \dots|\!)^+ = (\star, t \dots))$ to parse a sequence of tokens into a **Sexp** and an unconsumed portion of the input:

$$\begin{aligned}
(\!|\cdot|\!) :\ & \overrightarrow{\textbf{Tok}} \rightharpoonup \textbf{Sexp} \times \overrightarrow{\textbf{Tok}} \\
(\!|x\ t \ldots|\!) &= (x, t \ldots) \\
(\!|\ll_n\ t_0 \ldots|\!) &= ((v_1, v_2, \ldots, v_n), t_n \ldots) \\
\text{where } (v_1, t_1 \ldots) &= (\!|t_0 \ldots|\!) \\
(v_2, t_2 \ldots) &= (\!|t_1 \ldots|\!) \\
&\vdots \\
(v_n, t_n \ldots) &= (\!|t_{n-1} \ldots|\!)
\end{aligned}$$

We extend $(\!|\cdot|\!)$ and $(\!|\cdot|\!)^+$ to sequences of tokens representing an $n$-tuple of **Sexp**s with $(\!|t \ldots|\!)_n = (\!|\ll_n\ t \ldots|\!)$ and $(\!|t \ldots|\!)^+_n = (\!|\ll_n\ t \ldots|\!)^+$.

**Definition 7.5** (Well-formed token sequences)**.** Exactly those token sequences for which $(\!|\cdot|\!)_n$ is defined and yields an empty remainder are the $n$-well-formed token sequences:

$$\begin{aligned}
\mathcal{WF}_n(t \ldots) &\iff \exists v.\ (\!|t \ldots|\!)_n = (v, \cdot) \\
\mathcal{WF}^+_n(t^+ \ldots) &\iff \exists v^+.\ (\!|t^+ \ldots|\!)^+_n = (v^+, \cdot)
\end{aligned}$$

We write $\mathcal{WF}(t \ldots)$ to mean $\mathcal{WF}_1(t \ldots)$ and $\mathcal{WF}^+(t^+ \ldots)$ to mean $\mathcal{WF}^+_1(t^+ \ldots)$.

**Proposition 7.6.** *For all $v$, $(\!|[\![v]\!]|\!) = (v, \cdot)$ and likewise for $v^+$,* mutatis mutandis.

*Proof.* By induction on $v$ (respectively $v^+$). ☐

**Proposition 7.7.** *For all $(t \ldots)$, if $(\!|t \ldots|\!) = (v, \cdot)$ then $[\![v]\!] = (t \ldots)$, and likewise for $(t^+ \ldots)$,* mutatis mutandis.

*Proof.* By induction on $(\!|t \ldots|\!)$ (respectively $(\!|t^+ \ldots|\!)^+$). ☐

### 7.1.3 *Assertion trie syntax*

Tries $\mathsf{T}$ themselves are polymorphic in the values carried at their leaves, and consist of the following recursive definitions:

$$\begin{aligned}
\text{Tries } T, W \in \textbf{Trie}_\mathbf{A} &:= \mathsf{mt} \mid \mathsf{ok}(\alpha) \mid \mathsf{br}(W, M) \qquad \text{where } \alpha \in \mathbf{A} \\
\text{Branch nodes } M \in \textbf{Node}_\mathbf{A} &= \textbf{Tok} \to_{\textit{finite}} \textbf{Trie}_\mathbf{A}
\end{aligned}$$

There are three types of node: $\mathsf{mt}$ denotes an empty trie, a failing match; $\mathsf{ok}(\alpha)$ denotes a leaf node carrying a value, a (potentially) successful match; and $\mathsf{br}(W, M)$ an internal node with a *default* branch $W$ and a finite collection of *token-labeled* branches $(s \mapsto T) \in M$. Key to the interpretation of this syntax is that the wildcard branch $W$ represents the trie to be associated with any token $s'$ not mentioned, $s' \notin \mathsf{dom}(M)$. A sequence of tokens stretching from the root of a trie to one of its leaves represents an assertion, if every followed edge is token-labeled, or a set of assertions, if any default branches are taken.

$$
\begin{aligned}
\mathit{meaning} &: \mathcal{WF}_n(\mathsf{T}) \implies n : \mathbb{N} \times \mathsf{T} : \mathbf{Trie_A} \to \mathcal{P}(\mathbf{Sexp} \times \mathbf{A}) \\
\mathit{meaning}\,(n, \mathsf{T}) &= \mathit{collect}\,(n, \cdot, \emptyset, \mathsf{T}) \\
\mathit{collect} &: \mathbb{N} \times \overrightarrow{\mathbf{Tok}^+} \times \mathcal{P}(\mathbf{Sexp}) \times \mathbf{Trie_A} \to \mathcal{P}(\mathbf{Sexp} \times \mathbf{A}) \\
\mathit{collect}\left(n, t^+ \ldots, V, \mathsf{mt}\right) &= \emptyset \\
\mathit{collect}\left(n, t^+ \ldots, V, \mathsf{ok}(\alpha)\right) &= (\mathit{meaning}(v^+) - V) \times \{\alpha\} \text{ where } (v^+, \cdot) = (\!|t^+ \ldots|\!)^+_n \\
\mathit{collect}\left(n, t^+ \ldots, V, \mathsf{br}(W, M)\right) &= \mathit{collect}\left(n, t^+ \cdots \star, \left(V \cup \mathit{prefixes}(n, t^+ \ldots, M)\right), W\right) \cup \bigcup_{s \in \mathsf{dom}(M)} \mathit{collect}\left(n, t^+ \ldots s, V, M(s)\right) \\
\mathit{prefixes} &: \mathbb{N} \times \overrightarrow{\mathbf{Tok}^+} \times \mathbf{Node_A} \to \mathcal{P}(\mathbf{Sexp}) \\
\mathit{prefixes}(n, t^+ \ldots, M) &= \left\{ v \mid v \in \mathit{meaning}(v^+), (v^+, \cdot) = (\!|t^+ \ldots s\ s' \ldots|\!)^+_n, s \in \mathsf{dom}(M) \right\}
\end{aligned}
$$

Figure 36: Interpretation of assertion tries

The notion of well-formedness developed for token sequences generalizes to assertion tries by reading token sequences along the edges in a trie stretching from the root to each leaf. The intuition is that if a path in an $n$-well-formed trie ends at an $\mathsf{ok}()$ node, then the tokens labeling that path denote exactly $n$ $\mathbf{Sexp}^+$s. In addition, all paths in an $n$-well-formed trie should be no longer than necessary. That is, if we traverse $n$ $\mathbf{Sexp}^+$s' worth of edges from the root of an $n$-well-formed trie, we will either arrive "early" at an $\mathsf{mt}$ node, or "exactly on time" at an $\mathsf{ok}()$ or an $\mathsf{mt}$ node, but will never end up at a $\mathsf{br}$ node.

**Definition 7.8** (Well-formed tries)**.** We reuse the notation $\mathcal{WF}_n$ for tries. We will again write $\mathcal{WF}(\mathsf{T})$ for $\mathcal{WF}_1(\mathsf{T})$. The predicate $\mathcal{WF}_n(\mathsf{T})$ is defined by structural induction on $\mathsf{T}$ by three cases:

1. $\mathcal{WF}_n(\mathsf{mt})$; that is, $\mathsf{mt}$ is $n$-well-formed for all $n$.
2. $\mathcal{WF}_0(\mathsf{ok}(\alpha))$; that is, an $\mathsf{ok}(\alpha)$ trie is only ever $0$-well-formed.
3. $\mathcal{WF}_{n+1}(\mathsf{br}(W, M))$ if both $\mathcal{WF}_n(W)$ and $\forall s \in \mathsf{dom}(M). \mathcal{WF}_{n+\#s}(M(s))$; that is, $\mathsf{br}(W, M)$ is $(n+1)$-well-formed if $W$ is $n$-well-formed and every $M(s)$ is $(n+\#s)$-well-formed for every $s$-labeled edge leading away from the $\mathsf{br}$ node.

This definition deserves an illustration. Following our intuition, the trie

$$\mathsf{br}(\mathsf{mt}, \{\ll_2 \mapsto \mathsf{br}(\mathsf{mt}, \{\mathsf{X} \mapsto \mathsf{br}(\mathsf{mt}, \{\mathsf{Y} \mapsto \mathsf{ok}(\alpha)\})\})\})$$

should be $\mathcal{WF}_1$ because the token sequence $\ll_2$ $\mathsf{X}$ $\mathsf{Y}$ along the path leading to $\mathsf{ok}(\alpha)$ denotes one $\mathbf{Sexp}$, $(\mathsf{X}, \mathsf{Y})$. Following the definition of $\mathcal{WF}_n$, however, it is $\mathcal{WF}_1$ because $\mathcal{WF}_0(\mathsf{mt})$ and $\mathcal{WF}_2(\mathsf{br}(\mathsf{mt}, \{\mathsf{X} \mapsto \mathsf{br}(\mathsf{mt}, \{\mathsf{Y} \mapsto \mathsf{ok}(\alpha)\})\}))$. As we traversed the $\ll_2$ edge, we added $\# \ll_2 = 2$ to $n$, taking into account the nesting structure implied by the tuple-marker.

**Definition 7.9** (Meaning of $\mathcal{WF}$ tries)**.** Each element $\mathsf{T}$ of $\mathbf{Trie_A}$ such that $\mathcal{WF}_n(\mathsf{T})$ has an interpretation as a set of pairs of $n$-tuples of $\mathbf{Sexp}$s and elements of $\mathbf{A}$, $\mathit{meaning}\,(n, \mathsf{T})$, defined

$$
\begin{aligned}
pat_{\mathbf{A}}(\cdot,\cdot) &: \ \mathbf{A} \to \mathbf{Sexp}^{+} \to \mathbf{Trie}_{\mathbf{A}} \\
pat_{\mathbf{A}}(\alpha, v^{+}) &= pat'_{\mathbf{A}}(\mathsf{ok}(\alpha), [\![ v^{+} ]\!]^{+})
\end{aligned}
$$

$$
\begin{aligned}
pat'_{\mathbf{A}}(\cdot,\cdot) &: \ \mathbf{Trie}_{\mathbf{A}} \to \overrightarrow{\mathbf{Tok}^{+}} \to \mathbf{Trie}_{\mathbf{A}} \\
pat'_{\mathbf{A}}(\mathsf{T},\cdot) &= \mathsf{T} \\
pat'_{\mathbf{A}}(\mathsf{T}, \star\ \mathsf{t}^{+} \ldots) &= \mathsf{br}(pat'_{\mathbf{A}}(\mathsf{T}, \mathsf{t}^{+} \ldots), \{\}) \\
pat'_{\mathbf{A}}(\mathsf{T}, s\ \mathsf{t}^{+} \ldots) &= \mathsf{br}(\mathsf{mt}, \{s \mapsto pat'_{\mathbf{A}}(\mathsf{T}, \mathsf{t}^{+} \ldots)\})
\end{aligned}
$$

Figure 37: Compilation of wild S-expressions to tries

in figure 36. Intuitively, *collect* traverses the trie, accumulating not only token sequences along paths but also a set of **Sexp**s that are "handled elsewhere"; when it reaches an $\mathsf{ok}()$ node, it interprets the sequence, and then rejects any **Sexp**s in the "handled elsewhere" set. When $\mathbf{A} = \mathbf{1}$, a $\mathcal{WF}_n$ trie represents a set of $n$-tuples of **Sexp**s.

### 7.1.4 *Compiling patterns to tries*

Equipped with syntax for tries, we may define a function $pat_{\mathbf{A}}(\alpha, v^{+})$ which translates wild S-expressions to tries (figure 37).

*Claim* 7.10. If $\mathsf{T} = pat_{\mathbf{A}}(\alpha, v^{+})$, then $\mathcal{WF}(\mathsf{T})$ and $meaning\,(1, \mathsf{T}) = meaning(v^{+}) \times \{\alpha\}$.

**Example 7.11.** Consider the wild S-expression $(\texttt{sale}, \texttt{milk}, \star, \star)$, representing the infinite set of all 4-ary tuple S-expressions with first element `sale` and second element `milk`. To translate this into an equivalent trie, also representing a simple set of assertions, we choose to instantiate **Trie** with $\mathbf{A} = \mathbf{1}$:

$$
\begin{aligned}
pat_{\mathbf{1}}((), (\texttt{sale}, \texttt{milk}, \star, \star)) &= pat'_{\mathbf{1}}(\mathsf{ok}(()), \ll_4\ \texttt{sale}\ \texttt{milk}\ \star\ \star) \\
&= \mathsf{br}(\mathsf{mt}, \{\ll_4 \mapsto pat'_{\mathbf{1}}(\mathsf{ok}(()), \texttt{sale}\ \texttt{milk}\ \star\ \star)\}) \\
\cdots &= \mathsf{br}(\mathsf{mt}, \{\ll_4 \mapsto \mathsf{br}(\mathsf{mt}, \{\texttt{sale} \mapsto \mathsf{br}(\mathsf{mt}, \{\texttt{milk} \mapsto \mathsf{br}(\mathsf{br}(\mathsf{ok}(()), \{\}), \{\})\})\})\})
\end{aligned}
$$

◊

### 7.1.5 *Representing* SYNDICATE *data structures with assertion tries*

SYNDICATE implementations use assertion tries in two ways. The first application is to represent a set of assertions. We use $\mathbf{Trie}_{\mathbf{1}}$, where trie leaves are placeholders $\mathsf{ok}(())$, for this purpose. For example, such tries represent assertion sets in patch events and actions. The second application is to represent the contents of a dataspace; namely, a set of pairs of assertions and actor IDs, or equivalently a map from assertions to sets of actor IDs. Here, we use $\mathbf{Trie}_{\mathcal{P}(\mathbf{Loc})}$, and leaves carry sets of actor IDs.

A common operation in SYNDICATE implementations is *relabeling*, used among other things to convert back-and-forth between $\mathbf{Trie_1}$ and $\mathbf{Trie}_{\mathcal{P}(\mathbf{Loc})}$ instances:

$$
\begin{aligned}
\mathit{relabel} &: (\mathbf{A} \to \mathbf{B}) \to \mathbf{Trie_A} \to \mathbf{Trie_B} \\
\mathit{relabel}\ \mathsf{f}\ \mathsf{mt} &= \mathsf{mt} \\
\mathit{relabel}\ \mathsf{f}\ \mathsf{ok}(\alpha) &= \mathsf{ok}(\mathsf{f}\ \alpha) \\
\mathit{relabel}\ \mathsf{f}\ \mathsf{br}(\mathsf{T}, \{\mathsf{s} \mapsto \mathsf{T}', \ldots\}) &= \mathsf{br}(\mathit{relabel}\ \mathsf{f}\ \mathsf{T}, \{\mathsf{s} \mapsto \mathit{relabel}\ \mathsf{f}\ \mathsf{T}', \ldots\})
\end{aligned}
$$

*Claim* 7.12. If $\mathcal{WF}_n(T)$, then $T' = \mathit{relabel}\ f\ T$ implies $(\nu, \alpha) \in \mathit{meaning}(n, T)$ iff $(\nu, f\ \alpha) \in \mathit{meaning}(n, T')$.

### 7.1.6 *Searching*

A straightforward adaptation of the usual trie-searching algorithm to take wildcards into account (figure 38) allows discovery of the set of actors interested in receiving a copy of a given message. Given a candidate message and a trie representing a dataspace, we first convert the message to an equivalent sequence of tokens, and then walk the trie using the token-sequence as a key:

$$
\begin{aligned}
R &: \mathbf{Trie}_{\mathcal{P}(\mathbf{Loc})} \\
m &: \mathbf{Sexp} \\
\mathit{ids} &: \mathcal{P}(\mathbf{Loc}) \\
\mathit{ids} &= \begin{cases} \mathit{locs} & \text{if } \mathit{search}_{\mathcal{P}(\mathbf{Loc})}([\![m]\!], R) = \mathsf{found}(\mathit{locs}) \\ \emptyset & \text{if } \mathit{search}_{\mathcal{P}(\mathbf{Loc})}([\![m]\!], R) = \mathsf{notfound} \end{cases}
\end{aligned}
$$

The key distinction from the normal trie-searching algorithm is the case where a token is not found in the trie. Normally, the search would yield failure at this point. Instead, we fall back on the *wildcard* case, following that branch as if the sought token had been present all along.

*Claim* 7.13 (Sound searching). If $\mathcal{WF}(s\ldots)$, meaning that $(\!|s\ldots|\!) = (\nu, \cdot)$ for some $\nu$, and $\mathcal{WF}(T)$, then both (a) $\mathit{search}_{\mathbf{A}}(s\ldots, T) = \mathsf{notfound}$ iff there is no $\alpha$ such that $(\nu, \alpha) \in \mathit{meaning}(1, T)$, and (b) $\mathit{search}_{\mathbf{A}}(s\ldots, T) = \mathsf{found}(\alpha)$ iff some unique $\alpha$ exists such that $(\nu, \alpha) \in \mathit{meaning}(1, T)$.

The algorithm can be further adapted to support wildcards embedded in *messages*, representing simultaneous searching for a particular infinite set of keys. This extended version of *search* has signature $\overrightarrow{\mathbf{Tok}^{+}} \to \mathbf{Trie_A} \to (\mathbf{A} \to \mathbf{A} \to \mathbf{A}) \to \mathsf{notfound} + \mathsf{found}(\mathbf{A})$, not only allowing $\star$ in the input token sequence but also demanding a function used to combine **A**s when a wildcard input matches more than one branch of the trie. Allowing wildcards in messages gives a flavor of broadcast messaging dual to the normal type: where usually a pattern declaring interest in messages admits multiple possible messages, and each delivery includes a single message, here we may use a group of patterns each declaring interest in a single message, while each delivery includes *multiple* messages, by way of the wildcard mechanism.

Hybrids are also possible and useful. For example, consider an instant-messaging system where each connected user has asserted interest in the pair of their own name and the wildcard,

$$\begin{aligned}
search_{\mathbf{A}}(\cdot,\cdot) &: \overrightarrow{\mathbf{Tok}} \to \mathbf{Trie}_{\mathbf{A}} \to \mathsf{notfound} + \mathsf{found}(\mathbf{A}) \\
search_{\mathbf{A}}(\mathsf{s}\ldots,\mathsf{mt}) &= \mathsf{notfound} \\
search_{\mathbf{A}}(\cdot,\mathsf{ok}(\alpha)) &= \mathsf{found}(\alpha) \\
search_{\mathbf{A}}(\mathsf{s}\ \mathsf{t}\ldots,\mathsf{ok}(\alpha)) &= \mathsf{notfound} \\
search_{\mathbf{A}}(\cdot,\mathsf{br}(\mathsf{T},\mathsf{M})) &= \mathsf{notfound} \\
search_{\mathbf{A}}(\mathsf{s}\ \mathsf{t}\ldots,\mathsf{br}(\mathsf{T},\{\mathsf{s}'\mapsto\mathsf{T}',\ldots\})) &= \begin{cases} search_{\mathbf{A}}(\mathsf{t}\ldots,\mathsf{T}'') & \text{if } (\mathsf{s}\mapsto\mathsf{T}'')\in\{\mathsf{s}'\mapsto\mathsf{T}',\ldots\} \\ search_{\mathbf{A}}(\mathsf{t}\ldots,\mathit{makeTail}\ \#s\ \mathsf{T}) & \text{otherwise} \end{cases}
\end{aligned}$$

$$\mathit{makeTail}\ n\ \mathsf{T} = \underbrace{\mathsf{br}(\mathsf{br}(\ldots\mathsf{br}(\mathsf{T},\{\})\ldots,\{\}),\{\})}_{n \text{ layers of } \mathsf{br}}$$

Figure 38: Searching an assertion trie

for example (`Alice`, $\star$) and (`Bob`, $\star$). Sending a wildcard message $\langle\star,$"Hello!"$\rangle$ delivers a greeting to all connected users, and sending a specifically-addressed message $\langle$`Alice`, "Hello, Alice!"$\rangle$ delivers the greeting to a single participant.

The prototype Syndicate implementations use *search* in a few different ways. First, as discussed, to route a message to a set of actors. Here, the extension of *search* to wildcard-carrying messages is natural. Second, in Syndicate facets, *search* is used in the implementation of *project* (definition 5.23) to interrogate the actor's memory when a patch arrives, to see whether a given assertion of interest was "already known" or whether it is new to the actor concerned. Finally, *search* finds use in filtering of messages by "firewall" proxies; see section 11.3.

### 7.1.7 *Set operations*

Algorithms for computing set union, intersection, and difference on well-formed tries carrying various kinds of data in their ok() nodes can be formulated as specializations of a general *combine* function (figure 39).

We are careful to specify that *combine* may only be used with well-formed tries. This has an important consequence for the operation of the algorithm: during traversal of the two input tries, if one of the tries is an ok() node, then at the same moment, the other trie is either an ok() or an mt node. Since the function f is called only when one or both of the tries is ok(), we know that f need only handle ok() and mt inputs, leaving treatment of br nodes entirely to the *combine*/*foldKeys* functions. The effect of *combine* is to walk the interior nodes of the tries it is given, delegating processing of leaf nodes to the f passed in. In addition, *combine* itself produces a well-formed output, given a well-formed input and an f that answers only ok() or mt nodes.

$$\begin{aligned}
\textit{combine} : \ & \mathcal{WF}(\mathsf{T_L}) \implies \mathcal{WF}(\mathsf{T_r}) \implies (\mathcal{WF}(\mathsf{T}_{ans}) \land \\
& (\mathbf{Trie_L} \to \mathbf{Trie_R} \to \mathbf{Trie_A}) \\
& \to (\mathbf{Trie_L} \to \mathbf{Trie_A}) \\
& \to (\mathbf{Trie_R} \to \mathbf{Trie_A}) \\
& \to (\mathsf{T_L} : \mathbf{Trie_L}) \to (\mathsf{T_R} : \mathbf{Trie_R}) \to (\mathsf{T}_{ans} : \mathbf{Trie_A}))
\end{aligned}$$

$$\begin{aligned}
\textit{combine}\ \mathsf{f}\ \mathsf{e_L}\ \mathsf{e_R}\ \mathsf{T_L}\ \mathsf{T_R} &= \mathsf{g}\ \mathsf{T_L}\ \mathsf{T_R} \\
\text{where } \mathsf{g}\ \mathsf{ok}(\alpha)\ \mathsf{T_R} &= \mathsf{f}\ \mathsf{ok}(\alpha)\ \mathsf{T_R} \\
\mathsf{g}\ \mathsf{T_L}\ \mathsf{ok}(\alpha) &= \mathsf{f}\ \mathsf{T_L}\ \mathsf{ok}(\alpha) \\
\mathsf{g}\ \mathsf{mt}\ \mathsf{T_R} &= \textit{collapse}\ (\mathsf{e_R}\ \mathsf{T_R}) \\
\mathsf{g}\ \mathsf{T_L}\mathsf{mt} &= \textit{collapse}\ (\mathsf{e_L}\ \mathsf{T_L}) \\
\mathsf{g}\ \mathsf{br}(W_\mathsf{L}, M_\mathsf{L})\ \mathsf{br}(W_\mathsf{R}, M_\mathsf{R}) &= \textit{collapse}\ (\textit{foldKeys}\ \mathsf{g}\ \mathsf{br}(W_\mathsf{L}, M_\mathsf{L})\ \mathsf{br}(W_\mathsf{R}, M_\mathsf{R}))
\end{aligned}$$

$$\begin{aligned}
\textit{foldKeys} : \ & (\mathbf{Trie_L} \to \mathbf{Trie_R} \to \mathbf{Trie_A}) \\
& \to \mathbf{Trie_L} \to \mathbf{Trie_R} \to \mathbf{Trie_A}
\end{aligned}$$

$$\begin{aligned}
\textit{foldKeys}\ \mathsf{g}\ \mathsf{br}(W_\mathsf{L}, M_\mathsf{L})\ \mathsf{br}(W_\mathsf{R}, M_\mathsf{R}) &= \mathsf{br}(W, M) \\
\text{where } W &= \mathsf{g}\ W_\mathsf{L}\ W_\mathsf{R} \\
M &= \{s \mapsto h(s) \mid s \in \mathsf{dom}(M_\mathsf{L}) \cup \mathsf{dom}(M_\mathsf{R}), \\
& \qquad h(s) \neq \textit{makeTail}\ \#s\ W\} \\
h(s) &= \mathsf{g}\ (\textit{lookup}\ M_\mathsf{L}\ s\ W_\mathsf{L})\ (\textit{lookup}\ M_\mathsf{R}\ s\ W_\mathsf{R})
\end{aligned}$$

$$\textit{lookup}\ M\ s\ \mathsf{T} = \begin{cases} \mathsf{T}' & \text{if } (s \mapsto \mathsf{T}') \in M \\ \textit{makeTail}\ \#s\ \mathsf{T} & \text{otherwise} \end{cases}$$

$$\textit{collapse}\ \mathsf{T} = \begin{cases} \mathsf{mt} & \text{if } \mathsf{T} = \mathsf{br}(\mathsf{mt}, \{\}) \\ \mathsf{T} & \text{otherwise} \end{cases}$$

Figure 39: The *combine* function for performing set operations on assertion tries.

The three set operations on **Trie₁** instances are:

$$\begin{aligned} T_1 \cup T_2 &= \textit{combine}\ f_{un}\ \mathtt{id}\ \mathtt{id}\ T_1\ T_2 \\ T_1 \cap T_2 &= \textit{combine}\ f_{int}\ (\lambda x.\mathsf{mt})\ (\lambda x.\mathsf{mt})\ T_1\ T_2 \\ T_1 - T_2 &= \textit{combine}\ f_{sub}\ \mathtt{id}\ (\lambda x.\mathsf{mt})\ T_1\ T_2 \end{aligned}$$

$$\begin{aligned} f_{un}\ \mathsf{ok}(())\ \mathsf{ok}(()) &= \mathsf{ok}(()) \\ f_{un}\ \mathsf{mt}\ \mathsf{T} &= \mathsf{T} \\ f_{un}\ \mathsf{T}\ \mathsf{mt} &= \mathsf{T} \end{aligned}$$

$$\begin{aligned} f_{int}\ \mathsf{ok}(())\ \mathsf{ok}(()) &= \mathsf{ok}(()) \\ f_{int}\ \mathsf{mt}\ \mathsf{T} = f_{int}\ \mathsf{T}\ \mathsf{mt} &= \mathsf{mt} \end{aligned}$$

$$\begin{aligned} f_{sub}\ \mathsf{ok}(())\ \mathsf{ok}(()) &= \mathsf{mt} \\ f_{sub}\ \mathsf{mt}\ \mathsf{T} &= \mathsf{mt} \\ f_{sub}\ \mathsf{T}\ \mathsf{mt} &= \mathsf{T} \end{aligned}$$

The same operations have similar definitions for $\mathbf{Trie}_{\mathcal{P}(\mathbf{Loc})}$ instances used to represent dataspace contents, with f computing various functions over the sets carried in ok() nodes. It is also possible to use *combine* asymmetrically, operating on a **Trie₁** instance and a $\mathbf{Trie}_{\mathcal{P}(\mathbf{Loc})}$ instance in various ways.

*Claim* 7.14. For each $\mathsf{R} \in \{\cup, \cap, -\}$, $\textit{meaning}\,(1, T_1\ \mathsf{R}\ T_2) = (\textit{meaning}\,(1, T_1))\ \ \mathsf{R}\ \ (\textit{meaning}\,(1, T_2))$.

In addition, complements of sets represented as **Trie₁** can be computed by exchanging mt nodes for ok() nodes:

$$\begin{aligned} \textit{neg}(\cdot) :\ & \mathbf{Trie_1} \to \mathbf{Trie_1} \\ \textit{neg}(\mathsf{mt}) &= \mathsf{ok}(()) \\ \textit{neg}(\mathsf{ok}(())) &= \mathsf{mt} \\ \textit{neg}(\mathsf{br}(\mathsf{T}, \{s \mapsto \mathsf{T}', \ldots\})) &= \mathsf{br}(\textit{neg}(\mathsf{T}), \{s \mapsto \textit{neg}(\mathsf{T}'), \ldots\}) \end{aligned}$$

**Example 7.15.** Consider the set of all tuples *not* matching $(\star, 1)$—that is, any assertion that is either not a tuple, not a pair, or has something other than 1 as its second element (figure 40):

$$\begin{aligned} & \textit{neg}(\textit{pat}_1((), (\star, 1))) \\ &= \textit{neg}(\mathsf{br}(\mathsf{mt}, \{\ll_2 \mapsto \mathsf{br}(\mathsf{br}(\mathsf{mt}, \{1 \mapsto \mathsf{ok}(())\}), \{\})\})) \\ &= \mathsf{br}(\mathsf{ok}(()), \{\ll_2 \mapsto \mathsf{br}(\mathsf{br}(\mathsf{ok}(()), \{1 \mapsto \mathsf{mt}\}), \{\})\}) \end{aligned}$$

◊

*Claim* 7.16. If $\mathcal{WF}_n(\mathsf{T})$ and $\mathsf{T}' = \textit{neg}(\mathsf{T})$ then $\textit{meaning}\,(n, \mathsf{T}') = \mathbf{Sexp}^n - \textit{meaning}\,(n, \mathsf{T})$.

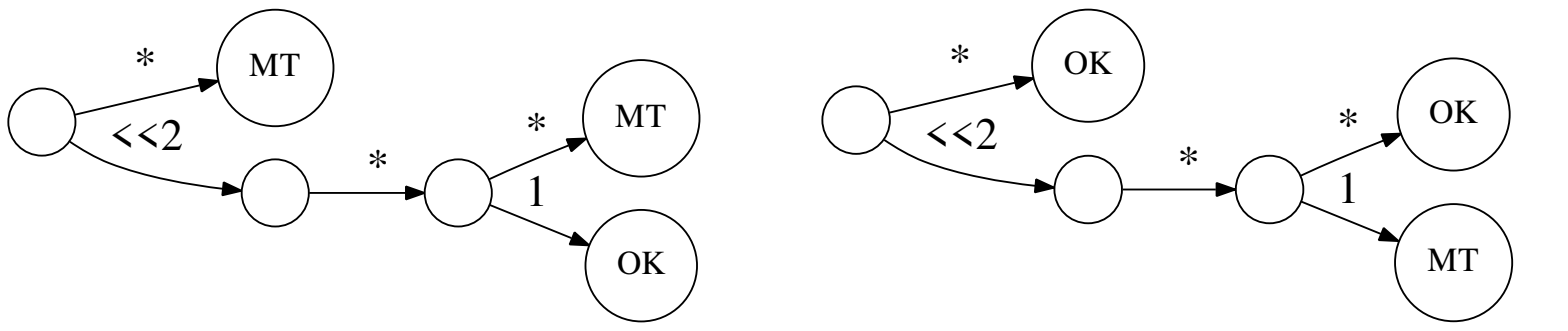

(a) Trie representing all tuples matching $(\star, 1)$.

(b) Trie representing the complement of (a).

Figure 40: A trie and its complement

$$\begin{aligned} project_{spec} &: \ \mathbf{Proj} \to \mathcal{P}(\mathbf{Sexp}) \to \mathcal{P}(\mathbf{Sexp}) \\ project_{spec}\ p\ \pi &= \{w \mid v \in \pi, w = match\ p\ v\} \end{aligned}$$

$$\begin{aligned} match &: \ \mathbf{Proj} \to \mathbf{Sexp}^+ \rightharpoonup \mathbf{Sexp}^+ \\ match\ x\ x &= () \\ match\ x\ \star &= () \\ match\ (p_0, \ldots, p_n)\ (v_0^+, \ldots, v_n^+) &= match\ p_0\ v_0^+ \times \cdots \times match\ p_n\ v_n^+ \\ match\ (p_0, \ldots, p_n)\ \star &= match\ p_0\ \star \times \cdots \times match\ p_n\ \star \\ match\ _\ v^+ &= () \\ match\ \$\ v^+ &= (v^+) \end{aligned}$$

Figure 41: Specification of *project* in terms of sets of **Sexp**s.

### 7.1.8 *Projection*

A key operation on assertion sets is *projection,* guided by a pattern with embedded binders. Projection is relevant both for the raw dataspace model and SYNDICATE's proposed language extensions. Projection is to pattern-matching as sets are to elements of sets, and allows programs to specify and extract relevant portions of assertion sets for later processing.

We call the patterns used in projection *specifications*. Projection specifications over $\mathbf{Sexp}^+$s include *capture marks,* \$, and a *discard* operator, _, both analogous to wildcard:

$$\text{Projection specifications } p, q \in \mathbf{Proj} = x \mid (p, q, \ldots) \mid _ \mid \$$$

A projection specification both *filters* and *reshapes* a given assertion set: it discards entire assertions if they fail to match its structure, and retains only the portions of assertions corresponding to its embedded capture marks.

Figure 41 specifies the desired behavior of projection as a function $project_{spec}$, in terms of mathematical sets and a partial function *match* that performs both filtering and reshaping. It is

$$
\begin{aligned}
\textit{walk} :\ & [\mathbf{Proj}] \to \mathbf{Trie_1} \to (\mathbf{Trie_1} \to \mathbf{Trie_1}) \to \mathbf{Trie_1} \\
\textit{walk}\ []\ \mathsf{T}\ \mathsf{k} &= \mathsf{k}\ \mathsf{T} \\
\textit{walk}\ [p, \ldots]\ \mathsf{mt}\ \mathsf{k} &= \mathsf{mt} \\
\textit{walk}\ [p, \ldots]\ \mathsf{ok}(())\ \mathsf{k} &= \mathsf{mt} \\
\textit{walk}\ [x, p, \ldots]\ \mathsf{br}(\mathsf{T}, \mathsf{M})\ \mathsf{k} &= \textit{walk}\ (\textit{lookup}\ \mathsf{M}\ x\ \mathsf{T})\ [p, \ldots]\ \mathsf{k} \\
\textit{walk}\ [(q_0, \ldots, q_n), p, \ldots]\ \mathsf{br}(\mathsf{T}, \mathsf{M})\ \mathsf{k} &= \textit{walk}\ (\textit{lookup}\ \mathsf{M}\ \ll_n\ \mathsf{T})\ [q_0, \ldots, q_n, p, \ldots]\ \mathsf{k} \\
\textit{walk}\ [\sqcup, p, \ldots]\ \mathsf{br}(\mathsf{T}, \mathsf{M})\ \mathsf{k} &= \textit{walk}\ [p, \ldots]\ \mathsf{T}\ \mathsf{k}\ \cup \bigcup_{s \in \mathsf{dom}(\mathsf{M})} \textit{walk}\ [\underbrace{\sqcup, _, \ldots,}_{\#s\ \text{discards}} p, \ldots]\ (\textit{lookup}\ \mathsf{M}\ s\ \mathsf{T})\ \mathsf{k} \\
\textit{walk}\ [\$, p, \ldots]\ \mathsf{br}(\mathsf{T}, \mathsf{M})\ \mathsf{k} &= \textit{capture}\ 1\ \mathsf{br}(\mathsf{T}, \mathsf{M})\ (\lambda \mathsf{T}'.\ \textit{walk}\ [p, \ldots]\ \mathsf{T}'\ \mathsf{k})
\end{aligned}
$$

Figure 42: Skipping of unwanted structure during projection

defined only for $\mathbf{Sexp}^+$s that have the specified shape, and yields a tuple with an element for each capture mark. For example,

$$
\begin{aligned}
\textit{match}\ (\mathtt{present}, \$)\ (\mathtt{present}, a) &= (a) \\
\textit{match}\ (\mathtt{present}, \$)\ (\mathtt{says}, a, \text{"hello"}) &\ \text{is not defined} \\
\textit{match}\ (\mathtt{says}, \$, \$)\ (\mathtt{present}, a) &\ \text{is not defined} \\
\textit{match}\ (\mathtt{says}, \$, \$)\ (\mathtt{says}, a, \text{"hello"}) &= (a, \text{"hello"})
\end{aligned}
$$

$$
\begin{aligned}
\textit{project}_{spec}\ (\mathtt{says}, \$, \$)\ \{(\mathtt{present}, a), (\mathtt{present}, b), (\mathtt{says}, a, \text{"hello"})\} &= \{(a, \text{"hello"})\} \\
\textit{project}_{spec}\ (\mathtt{present}, \$)\ \{(\mathtt{present}, a), (\mathtt{present}, b), (\mathtt{says}, a, \text{"hello"})\} &= \{(a), (b)\}
\end{aligned}
$$

The implementation of projection is not quite so succinct as the specification. While the main function is simple, its helpers *walk* and *capture* (figures 42 and 43) are more complex:

$$
\begin{aligned}
\textit{project} :\ & \mathcal{WF}(\mathsf{T}) \implies \mathbf{Proj} \to \mathsf{T} : \mathbf{Trie_1} \to \mathbf{Trie_1} \\
\textit{project}\ p\ \mathsf{T} &= \textit{walk}\ [p]\ \mathsf{T}\ (\lambda \mathsf{T}'.\ \mathsf{T}')
\end{aligned}
$$

A precondition to *project* is that the input trie be $\mathcal{WF}$; however, the trie that results from the projection is $\mathcal{WF}_n$ where $n$ is the number of capture marks in the projection specification given to *project*. The helper function *walk* follows the structure of the projection specifications in its first argument, discarding portions of the input trie that do not match. When it encounters a capture mark, it transitions to the *capture* helper function, which copies one $\mathbf{Sexp}^+$'s worth of structure from the input trie to the result. Both functions terminate early in cases of mismatch.

*Claim* 7.17. If $\mathcal{WF}(\mathsf{T})$ and $\mathsf{T}' = \textit{project}\ p\ \mathsf{T}$, then $\textit{meaning}\,(n, \mathsf{T}') = \textit{project}_{spec}\ p\ \ (\textit{meaning}\,(1, \mathsf{T}))$ and $\mathcal{WF}_n(\mathsf{T}')$, where $n$ is the number of capture marks in $p$.

The implementations in Syndicate/rkt and Syndicate/js extend the algorithm in two ways: first, they support $\mathbf{Trie_A}$ for any $\mathbf{A}$ rather than just $\mathbf{Trie_1}$, by allowing customization of the

$$\begin{aligned}
\mathit{capture} :\ & \mathbb{N} \to \mathbf{Trie_1} \to (\mathbf{Trie_1} \to \mathbf{Trie_1}) \to \mathbf{Trie_1} \\
\mathit{capture}\ 0\ \mathsf{T}\ \mathsf{k} &= \mathsf{k}\ \mathsf{T} \\
\mathit{capture}\ (n+1)\ \mathsf{mt}\ \mathsf{k} &= \mathsf{mt} \\
\mathit{capture}\ (n+1)\ \mathsf{ok}(())\ \mathsf{k} &= \mathsf{mt} \\
\mathit{capture}\ (n+1)\ \mathsf{br}(\mathsf{T}, M)\ \mathsf{k} &= \mathit{collapse}\ \mathsf{br}(\mathsf{T}', \{s \mapsto h(s) \mid s \in \mathsf{dom}(M), h(s) \neq \mathit{makeTail}\ \#s\ \mathsf{T}'\}) \\
&\quad \text{where } \mathsf{T}' = \mathit{capture}\ n\ \mathsf{T}\ \mathsf{k} \\
&\quad\quad\quad h(s) = \mathit{capture}\ (n + \#s)\ (\mathit{lookup}\ M\ s\ \mathsf{T})\ \mathsf{k}
\end{aligned}$$

Figure 43: Capturing of structure during projection

union-function used in the discard case of *walk*; and second, they incorporate "and-patterns" in projection specifications, thus allowing the placement of structural conditions on the fragments of assertions to be captured by a capture mark. This latter mainly affects the structure of *capture*, making it more similar to *walk*.

### 7.1.9 *Iteration*

We often want to examine the assertions in the set represented by some $\mathcal{WF}_n$ assertion trie, one at a time, accumulating some result as we go. This is only possible when the set is finite, corresponding to a *structurally finite* trie:

**Definition 7.18** (Structurally finite tries)**.** A trie with every $\mathsf{br}$ node of the form $\mathsf{br}(\mathsf{mt}, M)$ for some $M$ is called *structurally finite*.

*Claim* 7.19. If a trie $\mathsf{T}$ is structurally finite and $\mathcal{WF}_n(\mathsf{T})$, then *meaning* $(n, \mathsf{T})$ is a finite set.

The partial function *keySet* shown in figure 44 traverses a $\mathcal{WF}$ trie, reconstructing **Sexp**s from the tokens laid out along paths in the trie. Tuple-marker tokens cause construction of a nested **Sexp** tuple in the output. The function is defined only for structurally finite input tries. The well-formedness of the input ensures that *take* and $k_0$ are exhaustively defined despite appearances:

- $\mathsf{ok}(\alpha)$ cannot appear as second argument to *take* except when the first argument is $0$, because that would imply that the trie was "short": that paths from the root to the $\mathsf{ok}(\alpha)$ node were not long enough for the original trie to be $\mathcal{WF}$.
- no $\mathsf{br}$ node can appear as argument to $k_0$, because that would imply that the trie was "long": that paths from the root included too many tokens for the original trie to be $\mathcal{WF}$.

### 7.1.10 *Implementation considerations*

JUST-IN-TIME TOKENIZATION. As presented, *search* requires any sought **Sexp** to have been converted to a token sequence up-front. The implementations perform this conversion just-in-

$$
\begin{aligned}
\mathit{keySet} : \ \mathcal{WF}(T) \implies T : \mathbf{Trie_A} &\rightharpoonup \mathcal{P}([\mathbf{Sexp}]) \\
\mathit{keySet}\ T &= \mathit{take}\ 1\ T\ []\ k_0 \\
&\quad \text{where } k_0\ [v, \ldots]\ \mathsf{mt} = \emptyset \\
&\qquad\quad k_0\ [v, \ldots]\ \mathsf{ok}(\alpha) = [v, \ldots]
\end{aligned}
$$

$$
\begin{aligned}
\mathit{take} : \ \mathbb{N} \to \mathbf{Trie_A} \to [\mathbf{Sexp}] & \\
\to ([\mathbf{Sexp}] \to \mathbf{Trie_A} \rightharpoonup \mathcal{P}([\mathbf{Sexp}])) & \\
\rightharpoonup \mathcal{P}([\mathbf{Sexp}]) & \\
\mathit{take}\ 0\ T\ vs[v, \ldots]\ k &= k\ [v, \ldots]\ T \\
\mathit{take}\ (n+1)\ \mathsf{mt}\ [v, \ldots]\ k &= \emptyset \\
\mathit{take}\ (n+1)\ \mathsf{br}(\mathsf{mt}, M)\ [v, \ldots]\ k &= \bigcup_{(s \mapsto T) \in M} h(s, T) \\
&\quad \text{where } h(x, T) = \mathit{take}\ n\ T\ [v, \ldots, x]\ k \\
&\qquad\quad h(\ll_m, T) = \mathit{take}\ m\ T\ []\ k' \\
&\qquad\quad k'([w, \ldots], T') = \mathit{take}\ n\ T'\ [v, \ldots, (w, \ldots)]\ k
\end{aligned}
$$

Figure 44: Conversion of finite, $\mathcal{WF}$ tries to sets of (lists of) **Sexp**s.

time, thereby avoiding the need to examine uninteresting portions of input **Sexp**s and the need to embed a trie in variable numbers of br wrappers in the case when a tuple-marker is not explicitly catered for in a br node.

EXAMINATION OF ONLY THE SMALLER INPUT. The version of *foldKeys* shown in figure 39 captures the essence of the algorithm, but suffers from an inefficiency that is both avoidable and critically important to an efficient implementation of SYNDICATE. Consider a situation where a SYNDICATE program encodes actor-like point-to-point message delivery semantics, where each actor is addressed by a unique integer, and expresses interest in messages addressed to it by asserting $\{?(\mathit{id}, \star)\}$. In situations with a large number $n$ of running actors, the resulting tree is wide but shallow (figure 45a). Imagine now spawning a new actor with $\mathit{id} = n + 1$. The new actor asserts $\{?(n+1, \star)\}$ by issuing a patch containing the assertion trie

$$\mathit{pat}_{\mathcal{P}(\mathbf{Loc})}(\{n+1\}, (\text{"?"}, (n+1, \star)))$$

shown in figure 45b.

Computing the union of these tries in order to update the containing dataspace involves consideration of *every* edge leading away from the node following the "?" edge in figure 45a—a total of $O(n)$ work. However, nothing along any of the existing branches changes. An efficient SYNDICATE implementation demands that it be possible to combine a smaller with a larger trie doing only an amount of work proportional to the size of the smaller trie.

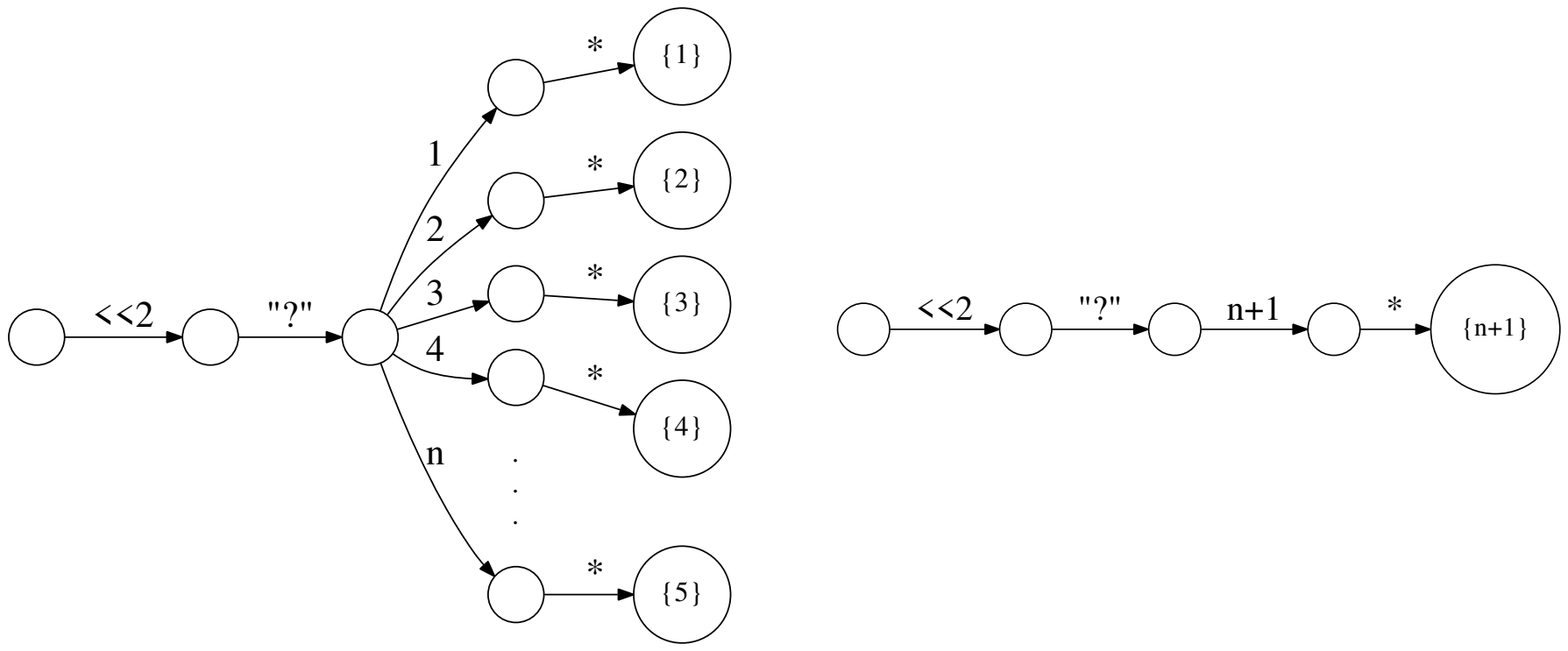

(a) A wide but shallow routing trie (b) Assertion trie produced by a new actor

Figure 45: Wide, shallow assertion tries

The key is to alter *foldKeys* so that it treats the larger of its two arguments as a base against which the smaller of the two is applied. Accordingly, *combine* is modified to accept an additional pair of functions which, given the larger trie, determine the starting point for *foldKeys*. After these changes, set operations on tries take time at each step proportional to the number of edges leading away from the *smaller* of the two given br nodes.

One consequence of this requirement is that it must be possible to efficiently count the number of edges leading away from a node. Not all hash-table or dictionary-like data structures offered by programming languages satisfy this requirement; some care must be taken in these cases.

CANONICAL CONSTRUCTORS. We use canonicalizing constructors extensively to enforce invariants that would otherwise be distributed throughout the codebase. For example, the uses of *collapse* in *combine* are implicit in our functions for constructing and extending br instances.

HASH-CONSING FOR CHEAP EQUALITY TESTING. Naively implemented, the side condition $\mathtt{h}(\mathtt{s}) \neq \mathit{makeTail}\ \#s\ W$ in the definition of *foldKeys* may examine a large amount of the structure of the tries on each side of the inequality. The time to decide this inequality is unacceptable, because the test is on the "hot" path of every set operation on assertion tries. Implementations must provide a cheap yet accurate way of testing equality between tries. In SYNDICATE/RKT, we hash-cons (Ershov 1958; Goubault 1994; Filliâtre and Conchon 2006) to force pointer-equality (`eq?`) to hold exactly when set equality (`equal?`) holds for our tries.[3]

Unfortunately, however, SYNDICATE/JS cannot currently provide this optimization. If we implemented hash-consing in JavaScript, we would forfeit proper garbage-collection behavior because JavaScript lacks suitable hooks into the garbage-collection subsystem. The `WeakMap` and `WeakSet` objects provided by ECMAScript 6 are unsuited to the task, since they are keyed by object identity, not object structure. Therefore, SYNDICATE/JS simply uses naive recursive

3 According to Baker (1992), "Hash consing was invented by Ershov (1958) for the detection of common subexpressions in a compiler and popularized by Goto (1976) for use in a Lisp-based symbolic algebra systems."

structural comparison of tries in *foldKeys*. The SYNDICATE/JS programs we have written to date have performed well enough to be usable, despite the performance penalty.

EFFICIENTLY CANONICALIZABLE DICTIONARIES. Even when hash-consing can ensure that the results of `eq?` and `equal?` are identical, care must be taken in choosing a representation for `br` nodes. Initially, we used Racket's built-in hash tables. However, for nodes with many edges, the hash table itself grew large, resulting in a large amount of time spent in `canonicalize`. To avoid this problem we need a representation of collections of edges that can be efficiently searched, efficiently counted, and efficiently updated, while also admitting a canonical representation suitable for hash-consing. An ideal data structure for the situation where tokens can be represented as bit strings is the *crit-bit tree* (Bernstein 2004; Finch 2016); however, rather than force repeated conversion back-and-forth between Racket values and bit strings, we chose instead to use *treaps* (Seidel and Aragon 1996; Cormen et al. 2009, problem 13-4). Treaps are trees which augment each node with a randomly-chosen *priority*, used to ensure a balanced tree. However, we require deterministic, canonical tree shapes for each unique set of key-value pairs. A deterministically-chosen priority can easily lead to unbalanced trees. The compromise that we have settled on is to use a fragment of a strong hash function to compute a deterministic pseudo-random priority from the key associated with each tree node. Experimental results (chapter 10) show that the results are acceptable, though questions remain as to whether this deterministic pseudo-random function leads to well-balanced trees in general.[4]

EFFICIENTLY CANONICALIZABLE SETS. Care must also be taken to ensure that the sets of actor IDs used in `ok()` nodes when representing values from $\mathbf{Trie}_{\mathcal{P}(\mathbf{Loc})}$ are efficiently canonicalizable. The implementation reuses canonicalized treaps (mapping keys to `#t`) for this purpose.

COMPOUND DATA STRUCTURES. Our **Sexp**s include $n$-tuples as the only compound, represented with special $\ll_n$ tokens when converted to token sequences. By contrast, both Racket and JavaScript enjoy a rich variety of compound data structures. Racket offers the programmer *structures*, *lists*, and *vectors*, while JavaScript offers *arrays* and *objects*.

Racket's structures may be interrogated to determine their *struct type*, which in turn can be examined to determine its arity. This suggests replacing generic tuple-markers $\ll_n$ with a tuple-marker for each struct type. For example, a structure type `present` with a single field would appear in an assertion trie as a tuple-marker $\mathit{present}_1$; and a structure type `says` with two fields would appear as $\mathit{says}_2$. Lists and vectors are represented with tuple-markers $\mathit{list}_n$ and $\mathit{vector}_n$, for arbitrary $n$, respectively. Improper lists are disallowed: an alternative is to support *pairs* natively, and then to represent lists as nested pairs.

JavaScript arrays are treated roughly as Racket's vectors, and for programmer convenience, SYNDICATE/JS includes a crude `struct`-like facility, as well, allowing rough parity and reasonably smooth interoperability with SYNDICATE/RKT's assertions. JavaScript objects present a

4 Sundar and Tarjan (1989) discuss the *unique representation problem* for binary search trees; Andersson and Ottmann (1995) improve on Sundar and Tarjan's solution. The property of canonicity in our setting is also known as *history-independence*. Pugh's *skip lists* built with a deterministic hash function offer a potential alternative to our pseudo-randomized treaps (Pugh 1990; Golovin 2010), though their pure-functional implementation may be challenging.

```
;; A Question is a
;;     (question DomainName QueryType QueryClass QuestionContext)
;; representing a DNS question: "What are the RRs for the given name,
;; type and class?" as well as a possible parent question that the
;; answer to this question is to contribute to the answer to.
```

(a)

```
(struct: (TName TType TClass TContext)
         question-repr
         ([name : TName] [type : TType] [class : TClass] [context : TContext])
         #:transparent)
(pseudo-substruct: (question-repr DomainName
                                  QueryType
                                  QueryClass
                                  QuestionContext)
                   Question question question?)
(pseudo-substruct: (question-repr (U Wild DomainName)
                                  (U Wild QueryType)
                                  (U Wild QueryClass)
                                  (U Wild QuestionContext))
                   QuestionPattern question-pattern question-pattern?)
```

(b)

```
(struct question (name type class context) #:transparent)
```

(c)

Figure 46: Typed Racket and Racket code describing a structure type, `Question`, used in Syndicate messages and assertions. (a) The comment remained the same in both implementations. (b) The Typed Racket implementation. (c) The untyped Racket implementation.

problem, however. There is no "natural" interpretation of an object with an embedded wildcard as a pattern over assertions: should fields not mentioned in the "pattern" be ignored for the purposes of matching, or should they cause a mismatch? There is no clear "best" design option; for now, inclusion of assertions containing objects is forbidden. Similar problems occur in Racket's own built-in pattern-matcher, `racket/match`, when it comes to hash tables; patterns over hash tables come in many varieties. It may be possible to support objects and object patterns in an ergonomic way in future by taking inspiration from the use of the "interleave" operator as seen in pattern languages for XML (Clark and Murata 2001).

REPRESENTING WILDCARD. In languages like Typed Racket (Tobin-Hochstadt and Felleisen 2008), the types of values that may appear in fields of structures are precisely specified. Our trick of representing patterns over structures by embedding a special marker value does not work in this setting. Early experiments with a Typed Racket implementation of SYNDICATE required painstaking work to specify

- the structure type itself with type parameters for all fields that could potentially carry a wildcard; plus
- an auxiliary type definition that instantiated the basic type with concrete types, for use in value contexts; and
- another that instantiated it again, with concrete types *plus a* `Wild` *type*, for use in pattern contexts.

The result proved awkward, verbose and brittle, as demonstrated by the example shown in figure 46. The "hack" of representing wildcard as an ordinary value does not work well for typed languages; instead, I suspect that deeper integration of wildcards with the type system is indicated.

Despite the poor ergonomics of the experimental approach explored in Typed Racket, the ability of the system to *forbid* wildcard from appearing in certain positions was useful. Future work on type systems for SYNDICATE should support this feature: it allows static encoding of restrictions on the kinds of patterns that may be placed into the shared dataspace. For example, one application is to forbid actors from asserting $?(\star,\star)$ in situations where they should only be allowed to subscribe to messages explicitly addressed to them, $?(id,\star)$.

### 7.1.11 *Evaluation of assertion tries*

At the beginning of the subsection, we listed a handful of requirements that a worthy assertion set representation must satisfy. Let us revisit them in light of the presented design:

EFFICIENT COMPUTATION OF METAFUNCTIONS. The set operations needed by the core metafunctions of SYNDICATE can be effectively implemented in terms of assertion trie operations. By careful choice of data structure and implementation technique (section 7.1.10), we can efficiently update our dataspace structures as changes are made; that is, without having to traverse the entirety of the dataspace.

EFFICIENT MESSAGE ROUTING. Assertion tries can efficiently route messages to sets of actors (section 7.1.6).

COMPACTNESS. Use of hash-consing and elimination of redundant branches in trie br nodes ensures that dataspace representations stay compact (section 7.1.10).

GENERALITY. Assertion tries support semi-structured data well, including local variations such as *structs* (Racket) and *arrays* (JavaScript). Support for hash-tables and objects remains future work, along with improved techniques for specifying allowable wildcard positions in assertions in typed languages.

### 7.1.12 *Work related to assertion tries*

The routing problem faced by SYNDICATE is a recurring challenge in networking, distributed systems, and coordination languages. Tries matching prefixes of *flat* data find frequent application in IP datagram routing (Sklower 1991) and are also used for topic-matching in industrial publish-subscribe middleware (Eugster et al. 2003; Baldoni, Querzoni and Virgillito 2005). I do not know of any other uses of tries exploiting visibly-pushdown languages (Alur and Madhusudan 2009; Alur 2007) (VPLs) for simultaneously evaluating multiple patterns over semi-structured data (such as the language of our assertions), though Mozafari et al. (Mozafari, Zeng and Zaniolo 2012) compile single XPath queries into NFAs using VPLs in a complex event processing setting. A cousin to the technique described in this section is YFilter (Diao et al. 2003), which uses tries to aggregate multiple XPath queries into a single NFA for routing XML documents to collections of subscriptions. Depth in their tries corresponds to depth in the XML document; depth in ours, to *position* in the input tree.

More closely related to our tries are the tries of Hinze (2000), keyed by type-directed preorder readings of tree-shaped values. Hinze's tries, like those presented here, have implicit "pop" tokens; however, they rely on types, where our tries rely merely on arity, which may be computed either dynamically (as we do) or statically, and they furthermore lack wildcards in any form.

Ionescu (2010) presents a trie including a form of wildcard, and compiles it to a DFA via an equivalent NFA that corresponds directly to the trie. He reports that *backtracking* is the chief disadvantage of the naive trie representation they chose, and that compilation to DFA avoided this problem. However, the DFA representation is no panacea: he writes that "it occupies significantly more memory than the trie; there is a significant cost for adding new bindings, since the entire DFA has to be dropped and rebuilt; and it is more complex and therefore harder to implement and maintain." Furthermore, it is not clear how to extend it to more general forms of predicate over tokens, as sketched above for our tries.

In previously-published work (Garnock-Jones and Felleisen 2016), we introduced our trie structure, but used distinct "push" and "pop" tokens, $\ll$ and $\gg$, which were not labeled with the arity of the tuple nested between them. Here, we use a family of "push" tokens with an associated arity instead, leaving the "pop" implicit. While using an explicit "pop" token allows prefix-matching of sequences (via a special $\mathsf{tl}()$ trie constructor representing a arbitrary number of *balanced* tokens, followed by a "pop" token), there are a number of disadvantages that leaving "pop" tokens implicit ameliorates. Most importantly, SYNDICATE relies heavily on extracting sets of assertions labeled with constructors such as $\downharpoonleft$ or ? from larger assertion sets.

For example, it is common to wish to compute the set of assertions $\{c \mid \downharpoonleft c \in \pi\}$ to be relayed to an outer dataspace from some local set of assertions $\pi$; or to compute the set of current assertions in some dataspace $R$ that are relevant to some $j$-labeled actor, $\{c \mid (c, k) \in R, (?c, j) \in R\}$. To do so using explicit "pop" tokens, we must extract the portion of the trie *between* the "push" and "pop" tokens surrounding the structured terms $\downharpoonleft c$ and $?c$, as shown in figure 47. By omitting the "pop" token and instead labeling the "push" token with an arity, we are able to simply discard two tokens, $\ll_2$ and ?, thereby avoiding traversal of the remainder of the trie, as shown in figure 48. A secondary benefit is simplicity: the algorithm presented in our previous publication involved the tl() constructor mentioned above, while the presentation we choose here avoids this complication.

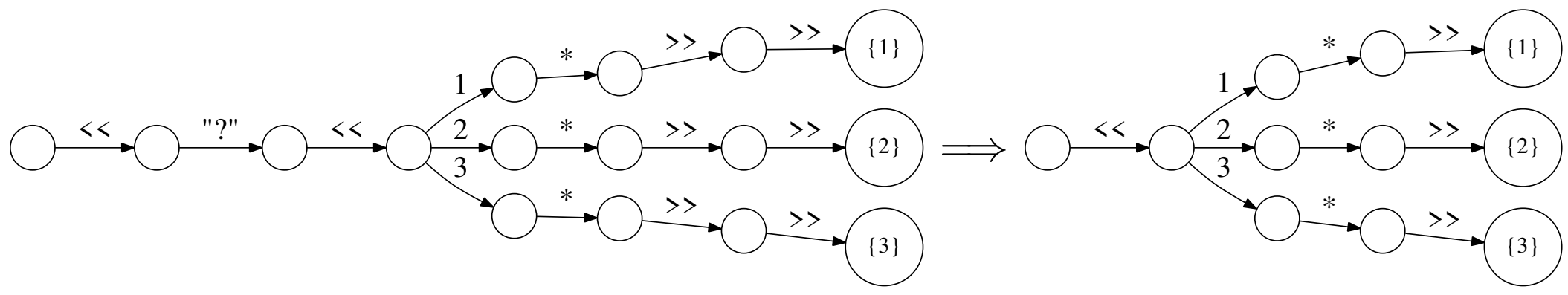

Figure 47: Extracting assertions labeled by some constructor, using explicit "pop" tokens

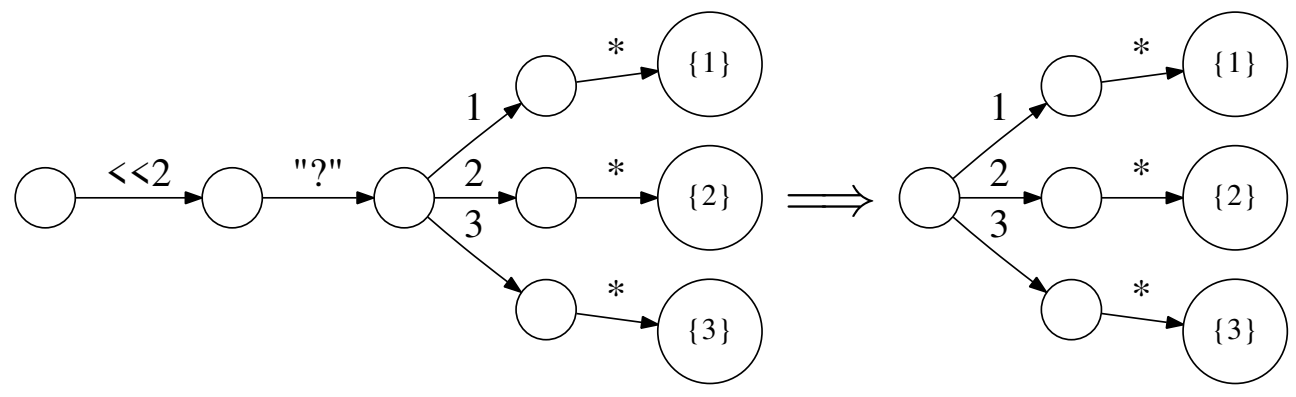

Figure 48: Extracting assertions labeled by some constructor, using implicit "pop" tokens and arity-labelled "push" tokens.

## 7.2 IMPLEMENTING THE DATASPACE MODEL

The prototype implementations of SYNDICATE closely follow the formal model described in chapter 4. SYNDICATE/RKT is written in a pure functional style, taking the signature of behavior functions as its central organizing principle. SYNDICATE/JS is written in a more imperative style, making use of object-oriented idioms appropriate to JavaScript programming.

There are two important differences between the model as described and as implemented. First, where the model treats dataspaces specially, the implementation treats them just like any other kind of actor. To do this, the interface to behavior functions is altered slightly. Second, the compactness of the model hides a number of useful abstractions, and the implementation benefits from explicitly recognizing these. In particular, the implementation separates representation of dataspace contents from the implementation of dataspace actors, and introduces a data structure that corresponds to the existential packages $\exists\tau.(\mathcal{F}_\tau \times \tau) \subset \mathbf{Beh}$ of the model (figure 14), precisely capturing the state of a running actor. The former allows reuse of the dataspace structure in other code, and the latter not only allows decomposition of the dataspace behavior into simpler components but also provides a useful interface to general reflective manipulation of actors.

The implementation is layered. The innermost layer (section 7.2.1) consists of the implementation of assertion tries along with utilities for hash-consing and implementations of (canonicalized) maps and sets. It is at this level that the mapping between host-language data structures and SYNDICATE assertions is made. The second layer (section 7.2.2) comprises two central datastructures. First, *patches* describe changes in assertion sets. Second, *multiplexors* or *muxes* form the central structure of each SYNDICATE dataspace; namely, the map between assertions and ac-tor IDs. The final layer (sections 7.2.3–7.2.5) contains the data-structures and behavior functions implementing the semantics of the dataspace model itself.

### 7.2.1 *Assertions*

An implementation of assertion tries in a given language must map that language's data structures onto the tokens **Tok** that label edges in br nodes. Each token must have an arity associated with it. It must also be possible to map *backwards* from a sequence of tokens to an implementation-language value. This means making choices about the representation of containers such as pairs, lists, vectors, sets, hash tables, `struct`s, reference cells, and objects, as well as about non-container data such as numbers, strings, symbols, and procedures.

The prototype implementations include non-container data directly in **Tok**, each with arity 0. The reverse mapping from such tokens to host-language data is immediate. *Objects* in Racket (Flatt, Findler and Felleisen 2006) are also treated as non-compound in order to sidestep difficulties generically analyzing such objects as well as generically reconstructing them from token sequences. Racket's procedures are also treated as atomic data. Likewise, Racket's "boxes", mutable reference cells, are treated as opaque atoms, following Baker's `egal` design (Baker 1993). For simplicity, SYNDICATE/JS restricts the range of assertions able to be placed within its dataspaces, limiting them to atoms (including procedures, as for Racket), arrays, and "structs". In particular, JavaScript objects (dictionaries) are forbidden entirely; to see why, consider the many different possible patterns one might wish to write over unordered key-value dictionaries, and the demands that each places on our trie data structure.[5]

Lists are handled with a family of tokens $\{\mathsf{list}_0, \mathsf{list}_1, \ldots\}$ for marking the beginning of a container in a token sequence. Arrays and vectors are similar. The arity of $\mathsf{list}_n$ is just $n$.[6]

Sets and hash tables pose a problem. The relevant equivalences for sets and tables do not coincide with the natural notion of equivalence for sequences of tokens. Therefore, the implementations treat sets and hash tables as opaque atoms when part of an assertion.

Racket's `struct`s are the primary means by which programmers extend the data types of the language, and as such are prominent in SYNDICATE/RKT protocols. JavaScript does not include a native `struct`-like facility, and so SYNDICATE/JS makes heavy use of a small support library providing a rough equivalent. In both languages, we may retrieve a *structure type* object describing the arity of a given structure instance, plus a sequence of the structure's field values. Furthermore, given a structure type and a sequence of field values, we may reconstruct a structure instance. We include these structure type objects in **Tok**. Each `struct` definition of the form

$$(\mathtt{struct}\ S\ (x_0\ x_1\ \ldots\ x_n))$$

(and its JavaScript equivalent) leads to inclusion of $S$ in **Tok** with arity $n$. The encoding of instances of $S$ is

$$[\![(S\ v_0\ v_1\ \ldots\ v_n)]\!] = S\ [\![v_0]\!]\ [\![v_1]\!]\ \ldots\ [\![v_n]\!]$$

5 It is left to future work to incorporate patterns over objects into SYNDICATE and its trie data structures.
6 Racket pairs that are not part of a proper list may not be used in assertions, since modern Racket style eschews non-list uses of pairs, and accommodating both list and non-list uses would significantly complicate matters.

If we are to strictly follow the Syndicate design principles laid out in section 2.6, then higher-order data such as procedures, objects and mutable data structures should be forbidden from appearing in assertions. However, given that Syndicate is not yet distributed and so does not suffer the associated restriction to first-order data, and that interoperability with some libraries demands trafficking in higher-order data, the implementations turn a blind eye to these cases.[7]

It is usually an advantage that Syndicate can see deeply into data structures: without such deep destructuring, subscriptions matching on fields of a compound datum are impossible to construct. However, from time to time, a protocol will involve a compound datum that *could* be destructured but that *should* be treated as an atomic value. It may be very large, taking an unreasonable amount of space and time to convert to a token sequence and back; or it may be able to be converted to a token sequence, but not back to a host-language value; or the relevant notion of equality for the value may not coincide with the notion of equality entailed by conversion to a token sequence, as already seen for sets and hash tables. For these cases, the implementation offers a simple remedy: a predefined standard `struct` type called `seal` with a single field:

```
(struct seal (contents))
```

Its equivalence predicate is pointer-equality and it is treated as completely opaque by the assertion trie code. Examples of its use include the transport of Racket *picts* (Felleisen et al. 2009) in the Syndicate assertions describing 2D graphics for display by Syndicate/rkt's OpenGL driver, and transport of HTML fragments in assertions describing portions of a web page for display by Syndicate/js's user-interface driver.

### 7.2.2 *Patches and multiplexors*

A *patch* represents a concrete change to be made to an assertion set or dataspace. As defined in section 4.6, each patch consists of a pair of assertion sets: one containing assertions to be removed, and the other assertions to be added. This becomes a structure or object with two members, each an instance of an assertion trie. Tries representing sets ($\mathbf{Trie_1}$) are used in most cases, but occasionally the implementation makes use of patches carrying $\mathbf{Trie}_{\mathcal{P}(\mathbf{Loc})}$ instances.

Most of the functions manipulating patches are straightforward, but there is one exception: the `compute-aggregate-patch` function (and its JavaScript cognate), which computes the *net effect* of a patch on an existing dataspace (sets $\pi^{\bullet}_{in}$ and $\pi^{\bullet}_{out}$ in metafunction $\mathtt{bc}_{\Delta}$, definition 4.47). If some actor labeled $\ell$ has produced a patch action $\Delta$, the changes to the dataspace it carries may be of interest to $\ell$ itself or to its peers in the dataspace. However, a newly-added assertion is only relayed on if no other actor has already made the same assertion, and a newly-retracted assertion likewise has no visible effect if some other actor happens to be making the same assertion at the time of retraction. The `compute-aggregate-patch` function makes use of various preconditions to optimize its calculation of the maximum visible net change to the dataspace, given the collection of assertions made by the dataspace's group of actors as a whole.

7 Even though JavaScript array values are pervasively mutable, they are (roughly speaking) *copied* into assertion tries. This effectively forces programmers to treat arrays as immutable when communicating them via a Syndicate dataspace.

The implementation frequently needs to discover the IDs of actors affected by a particular change, as well as the assertions currently being made by a particular actor. These correspond to reading off the dataspace structure, an instance of $\mathcal{P}(\mathbf{Sexp} \times \mathbf{Loc})$, in either a forwards or a reverse direction. A specialized object type, a *multiplexor* or *mux*, combines the necessary state and operations along with a source of fresh **Loc**s. A mux, then, effectively represents the core data structures of the dataspace itself along with an ID allocator. It is useful anywhere routing needs to be performed: to actors within dataspaces, as well as to facets within individual actors.

Each mux instance presents an interface involving a collection of named *streams*. The mux allocates stream names and allows addition, removal, and update of a set of assertions associated with each stream. It also offers convenient functions for computing events to be delivered to each stream in response to a given action or message. Within a dataspace, each actor is a stream; within an actor, each facet is a stream.

### 7.2.3 *Processes and behavior functions*

Recall the signature of behavior functions from figure 12:

$$\text{Behavior functions } f_{beh} \in \mathcal{F}_\tau = \mathbf{Evt} \times \tau \to \mathsf{continue}(\overrightarrow{\mathbf{Act}} \times \tau) + \mathsf{exit}(\overrightarrow{\mathbf{Act}})$$

The core of the implementation builds representations of the components of this signature. Events and actions are represented as structures; patches, in particular, make use of the patch and assertion-trie libraries described previously. An abstraction called a *transition* captures the type of the result from a behavior function. However, while the mathematical definition offers two possibilities, $\mathsf{continue}()$ or $\mathsf{exit}()$, the implementation offers *three*. A behavior function may yield a `transition` structure, corresponding to $\mathsf{continue}()$, bearing an updated state and a sequence of actions to perform. Alternatively it may produce a `quit` structure, corresponding to $\mathsf{exit}()$, instructing the dataspace to terminate the actor following the included sequence of final actions. The new third option is that a behavior function may return `#f`, signaling that the behavior is now inert and does not need to be *polled* until the next event arrives. This option is made available to ease implementation of dataspaces, and is described in the next subsection.

Around this representation of a behavior function, we introduce an abstraction called a *process*. A process is a pair of a behavior function and an associated private state. Processes correspond to the existential packages $\mathsf{pack}\,\langle\tau,(f_{beh},u)\rangle \in \exists\tau.(\mathcal{F}_\tau \times \tau)$ seen in the formalism of chapter 4. Making processes a first-class concept not only simplifies the implementation but allows for some reuse in situations calling for reflective representations of running actors. There are many examples: embedding SYNDICATE actors in Racket's `big-bang` framework; firewalling interactions between an actor and its dataspace; interfacing SYNDICATE actors to the rest of a running Racket system; simple approaches to supervision of actors; running individual SYNDICATE actors in separate Racket threads within a single dataspace; embedding dataspaces as ordinary actors within another dataspace, translating between assertions in the outer and the inner dataspaces appropriately; and of course embedding running actors within dataspaces themselves.

### 7.2.4 *Dataspaces*

While the formal model of chapter 4 treats dataspaces specially, in both SYNDICATE/RKT and SYNDICATE/JS they are implemented as ordinary actors like any other. The private state of a dataspace actor contains a *mux*; a queue of pending actions, each labeled with the ID of the actor that issued it; a set of "runnable IDs", used to manage the distinction between quiescent and inert actors; and a hash table mapping actor ID to `process` structures. Upon receipt of an event, the event is (trivially) translated into an action, labeled with a special ID—the symbol `'meta`—representing the containing context, and placed in the pending action queue. Then, the pending action queue is atomically exchanged for an empty queue, which will gather actions to be performed in the next event cycle, and the actions previously enqueued are interpreted. Any events that result from processing of an action are immediately delivered to the relevant actors during this stage, and the resulting transition structures both update the private state associated with the transitioning actor and enqueue actions for the next round of interpretation. Once all the queued actions from the current round have been processed, the dataspace *polls* any of its children that are marked as "not provably inert"; that is, those whose IDs are stored in the "runnable ID" set. An actor is considered not provably inert whenever its behavior function answers anything other than `#f` in response to a poll or a delivered event. This is critical for allowing some kind of approximation of fair scheduling: without such polling, the implementation would be forced to evaluate each actor, *including each dataspace*, to inertness every time a behavior function was invoked. This way, actors can choose to perform some small amount of work and to *yield* to any peers that may also wish to do work, thus interleaving stimuli from the outside world with internal reductions, while also allowing the system as a whole to become fully inert once there genuinely remains nothing to do.[8]

It is here that host-language exceptions raised by behavior functions are transmuted into synthetic `quit` transitions, leading to the termination of the faulting actor.

Some care must be taken to ensure that an actor that has issued a `quit` transition (or raised an exception) is *immediately* disabled. Its behavior function must not be called again, even though its final actions may remain to be interpreted. The dataspace cannot completely forget about a terminated actor until all its queued actions have been performed.

### 7.2.5 *Relays*

We have been claiming that dataspaces are implemented as ordinary actors like any other, but this is not quite accurate. A dataspace actor will, left to its own devices, never produce any actions. This is because it treats the connection to its containing context identically to the connections it maintains to its contained actors. Given that a contained actor will never receive a notification about an event it did not previously declare interest in, and that the same applies to the dataspace's treatment of the containing context, we see that the context will never "receive an event" (even though, syntactically, this is presented as the *dataspace* never *performing an action*).

8 A connection can be made here to the *parallel-or* construct of PCF (Plotkin 1977).

Furthermore, when one dataspace is embedded within another, we want to maintain a distinction between the two contained assertion sets. The protocol involving constructors $\downharpoonright$ ("outbound", "outgoing") and $\upharpoonright$ ("inbound", "incoming") should embody the connection between assertions in the two spaces.

The job of a *relay actor* is to solve these problems. Upon startup, a relay injects a synthetic *event* into its contained dataspace actor's behavior function, expressing interest in $\downharpoonright \star$ and in $? \upharpoonright \star$. When given an event, the relay's behavior function rewrites it, prepending $\upharpoonright$ to each contained assertion, before delivering it to its contained dataspace actor. Because the relay previously expressed interest in certain assertions, the dataspace will from time to time produce actions mentioning these assertions; the relay rewrites the actions it receives, translating not only $\downharpoonright c$ into $c$ but also $? \upharpoonright c$ into $?c$ before transmitting the action to its own surrounding context.[9]

The effect of the rewrite from $? \upharpoonright c$ into $?c$ is to allow expressed interest in *incoming* assertions to automatically result in an *outbound* expression of interest in those assertions. Without it, a contained actor would have to assert $\downharpoonright ?c$ as well as $? \upharpoonright c$. The problem compounds with multiple layers of dataspace nesting: with the approach taken by SYNDICATE relays, an actor needs only assert $? \upharpoonright\upharpoonright\upharpoonright c$ in order to be informed of $c$ three levels out; without it, it would be necessary to assert $? \upharpoonright\upharpoonright\upharpoonright c$, $\downharpoonright ? \upharpoonright\upharpoonright c$, $\downharpoonright\downharpoonright ? \upharpoonright c$, and $\downharpoonright\downharpoonright\downharpoonright ?c$.

By injecting a synthetic event into its contained dataspace, a relay kicks off the exchange of information between the outer and inner dataspaces, and by carefully relabeling assertions traveling in each direction, it maintains the correct distinction between "local" and "remote" assertions in the inner dataspace.

## 7.3 IMPLEMENTING THE FULL SYNDICATE DESIGN

The implementation of the SYNDICATE language atop the dataspace implementation has three main pieces: a runtime, which provides functions and data structures implementing facets, fields, endpoints, queries, and so forth; a syntax layer, which provides a pleasant domain-specific language for making use of the runtime; and a simple, imperative *dataflow* implementation, which tracks changes to fields and schedules re-evaluation of dependent computations. No special knowledge of intra-actor features such as fields, facets or endpoints has been added to the implementation of the dataspace model itself.[10]

### 7.3.1 *Runtime*

The runtime differs from the formal model of chapter 5 primarily in its support for efficient re-evaluation of the assertions of an actor as its fields are updated, but also in its approach to

9 As of this writing, SYNDICATE/JS lags the Racket implementation in that its dataspaces combine the functionality of Racket's dataspaces and relays, fused together. This is the original design; SYNDICATE/RKT was initially like this. It took some time before the benefits of separating the functions of "relay" and "dataspace" became clear.

10 It remains future work to explore potential performance advantages from making the dataspace implementation aware of the internal structure of SYNDICATE actors.

tracking facet state during facet shutdown. A simple recursive procedure is used, contrasting with the small-step approach of the model's stop-child rules.

An additional difference is the approach taken to representing non-inert actors: while the formal model embeds pending statements within the facet tree, the implementation maintains the facet tree as a data structure separate from a priority queue used to hold pending scripts. A *script* is a sequence of expressions to be evaluated in-order within the context of a given facet.

While the formalism of chapter 5 runs endpoint event handlers in the order they were written in the program, the implementations take a different approach. Endpoints are stored in an unordered hash-table; when ordering is relevant to an application, a system of endpoint *priorities* can be used. For example, the definitions of the `define/query-*` forms all involve two endpoints, one for responding to relevant assertions, and one for relevant retractions. Both endpoints are run at a priority level higher than the default, ensuring that the side-effects on the fields maintained by the queries are visible to ordinary endpoint event handlers. Furthermore, the retraction endpoint is placed at a slightly higher priority-level still, ensuring that *removal* of elements from sets, hash-tables, and so on is performed before addition of elements. For the specific case of hash-tables mapping each key to a single value, this is crucial: given a patch that simultaneously adds $(k, v_{new})$ and removes $(k, v_{old})$, processing addition of $(k, v_{new})$ before removal of $(k, v_{old})$ would result in an entirely absent entry for $k$. Priority levels *lower* than the default also have their uses: for example, if the synthetic endpoint corresponding to a `begin/dataflow` block is placed at a very low priority, then it will run *after* other code, toward the end of the actor's turn. This is a convenient time to check invariants among fields in the actor: a form of "actor contract" analogous to a class contract in an object-oriented language.

The SYNDICATE/JS runtime differs from the SYNDICATE/RKT runtime in its treatment of fields. Fields in SYNDICATE/JS are represented as properties on a special object used as the `this` object when running facet setup code and event handler code. They are thus "second-class" entities in the language, similar to SYNDICATE/λ but in contrast to SYNDICATE/RKT, where fields are values in their own right. Facets are nestable in SYNDICATE, and code in a given facet must be able to access not only the facet's own fields but those in any of its parents. In SYNDICATE/RKT, the "first-class" nature of fields makes it natural for there to be an actor-global collection of fields; in SYNDICATE/JS, the situation is different. A form of *inheritance* is required, with field objects of nested facets extending the field objects of their parents. The inheritance tree is ultimately rooted in an actor-global field object. SYNDICATE/JS provides this by way of JavaScript's own `prototype`-based object inheritance mechanism.

### 7.3.2 *Syntax*

The syntax layer adapts syntactic forms reminiscent of the formal model into calls to functions provided by the runtime. In the case of SYNDICATE/RKT, it makes use of Racket's syntactic extension system (Culpepper and Felleisen 2010), which greatly facilitates the addition of new constructs to a language. However, JavaScript lacks a built-in syntactic extension facility. Therefore, I developed a separate compiler based on Ohm (Warth, Dubroy and Garnock-Jones 2016) that translates the language extended with SYNDICATE syntax to core JavaScript. Appendix A describes the syntactic extensions.

### 7.3.3 *Dataflow*

Sophisticated pure-functional dataflow implementations such as that of Cooper and Krishnamurthi (2006) are well-suited to pure languages. However, idiomatic programs in the SYNDICATE design presented here make extensive use of mutation. Therefore, we chose a trivially simple imperative dataflow design with moderate efficiency and an easily-understood evaluation order and cost model.

Each SYNDICATE actor maintains a bipartite, directed *dataflow graph*: source nodes represent fields, target nodes represent endpoints, and edges represent dependencies of the endpoints on the fields. Each endpoint contains a procedure that is used to compute the set of assertions to be associated with the endpoint. By recording field dependencies during the execution of such procedures, the implementation learns which endpoints must have their assertion sets recomputed in response to a given field change. In addition, this dataflow facility is exposed to the programmer in the form of a special `begin/dataflow` form,[11] which creates a synthetic pseudo-endpoint whose assertion-set procedure always returns the empty assertion set but may perform arbitrary (side-effecting) computations. Commonly, these computations update a field with a computed expression depending on another field, potentially triggering further dataflow-induced recomputation.

Figure 49 sketches the interface to the Racket implementation of the dataflow library; full source code for the library is shown in appendix D. The JavaScript implementation is similar. The `current-dataflow-subject-id` parameter records the identity of the currently-evaluating endpoint. Whenever a field is read, the runtime invokes `dataflow-record-observation!` with the identity of the field, thus recording a connection between the executing endpoint and the observed field. Whenever a field is updated, the runtime calls `dataflow-record-damage!`. Later in the behavior function of the actor, the runtime calls `dataflow-repair-damage!` with a *repair* procedure which, given an endpoint, calls its assertion-set recomputation procedure, collecting the results into a patch action which updates the overall assertion set of the actor in the dataspace. The synthetic endpoints generated by `begin/dataflow` are simply a special case, where the side-effects of the assertion-set procedure are the interesting part of the computation.

As time goes by and fields change state, the precise set of fields that a given endpoint computation depends upon may change. The `dataflow-repair-damage!` procedure takes care to call `dataflow-forget-subject!` for each endpoint, just before invoking its repair procedure for that endpoint, in order to clear its previous memory of the endpoint's dependencies. The repair procedure, during its execution, records the currently-relevant set of dependencies for the endpoint. Finally, when an endpoint is removed from an actor as part of the facet shutdown process, `dataflow-forget-subject!` is used to remove obsolete dependency information for each removed endpoint.

---

11 The analogous SYNDICATE/JS syntax is a `dataflow { ... }` block.

| | |
|---|---|
| `current-dataflow-subject-id` : *Parameter*(*Endpoint*) | Used by `dataflow-record-observation!` to implicitly supply a depending endpoint. |
| `dataflow-record-observation!` : DFG × *Field* → 1 | Records a dependency of the implicit endpoint on the given field. |
| `dataflow-record-damage!` : DFG × *Field* → 1 | Marks the given field as "damaged". |
| `dataflow-forget-subject!` : DFG × *Endpoint* → 1 | Removes the given endpoint (and its edges) from the graph. |
| `dataflow-repair-damage!` : DFG × (*Endpoint* → 1) → 1 | Passes endpoints depending on damaged nodes to the given function one at a time, iterating until stability is reached. |

```
(begin/dataflow              ⟹   (add-endpoint! ...
  expr ...)                        (lambda ()
                                    (parameterize ((current-dataflow-subject-id ...))
                                       expr ...)))

(define/dataflow id expr)    ⟹   (begin
                                   (field [id #f])
                                   (begin/dataflow (field expr)))
```

Figure 49: Runtime- and programmer-level interfaces to imperative Racket dataflow library

The simple "dataflow" system described here is neither a "sibling" of nor a "cousin" to reactive programming in the sense of Bainomugisha et al. (2013), or even dataflow in the sense of Whiting and Pascoe (1994); rather, it is most similar to the simple *dependency tracking* approach to object-oriented reactive programming described by Salvaneschi and Mezini (2014, section 2.3), and was in fact directly inspired by the dependency tracking of JavaScript frameworks like Knockout[12] (Sanderson 2010) and Meteor.[13]

## 7.4 PROGRAMMING TOOLS

Because the prototype implementations of SYNDICATE are closely connected to the underlying formal models, the programmer is able to use concepts from the model in understanding the behavior of programs. Furthermore, points exist in the code implementing dataspace actors that correspond closely to the reduction rules given in chapter 4, and each invocation of a dataspace's actor behavior function itself corresponds roughly to use of the schedule rule. This gives us an opportunity to record *trace events* capturing the behavior of the program in terms of the formal model. In turn, these events enable visualization of program execution.

12 http://knockoutjs.com/
13 https://docs.meteor.com/api/tracker.html

```
(assertion-struct one-plus (n m))

(spawn #:name 'add1-server
       (during/spawn (observe (one-plus $n _))
         #:name (list 'solving 'one-plus n)
         (assert (one-plus n (+ n 1)))))

(spawn #:name 'client-process
       (stop-when (asserted (one-plus 3 $value))
         (printf "1 + 3 = ~a\n" value)))
```

Figure 50: Program generating the sequence diagram of figure 51

The lifecycle of an action can trigger multiple trace log entries from the moment of its production to the moment the dataspace events it causes are delivered:

1. an entry for the production of the action as a result from a behavior function;

2. an entry for the moment the action is enqueued in the dataspace's pending-actions queue;

3. an entry for its interpretation by the dataspace, which is the same moment that its effects are applied to the state of the dataspace, and the moment any resulting dataspace events are produced;

4. an entry for the moment such events are enqueued for delivery to an actor; and

5. an entry recording the final delivery of such events as input arguments to a behavior function.

Different Syndicate implementation strategies may combine some of these log entries together. For example, the prototype dataspace implementations combine entries 1 and 2 and entries 4 and 5. A hypothetical distributed implementation of Syndicate would likely maintain an observable distinction between all of the stages.

Thus far, I have implemented three consumers of generated trace log entries. The first is a console-based logging facility which simply displays each entry as colorized text on the standard error file descriptor. The remainder of this section is devoted to discussion of the other two: an offline renderer of sequence diagrams and a live display of program activity.

### 7.4.1 *Sequence diagrams*

Recorded trace events can be automatically rendered to a kind of *sequence diagram* displaying actor lifecycle events and causal connections between emitted actions, delivered events, and dataspace state. Any Syndicate/rkt program, if run with an environment variable `SYNDICATE_MSD` naming an output file name, fills the named file with recorded trace events as the program runs. Unix signals may be used to selectively enable and disable tracing during long executions. Once acquired, a trace file may be rendered with a command-line tool, `syndicate-render-msd`, able to display directly to the screen or produce PNG or PDF files.

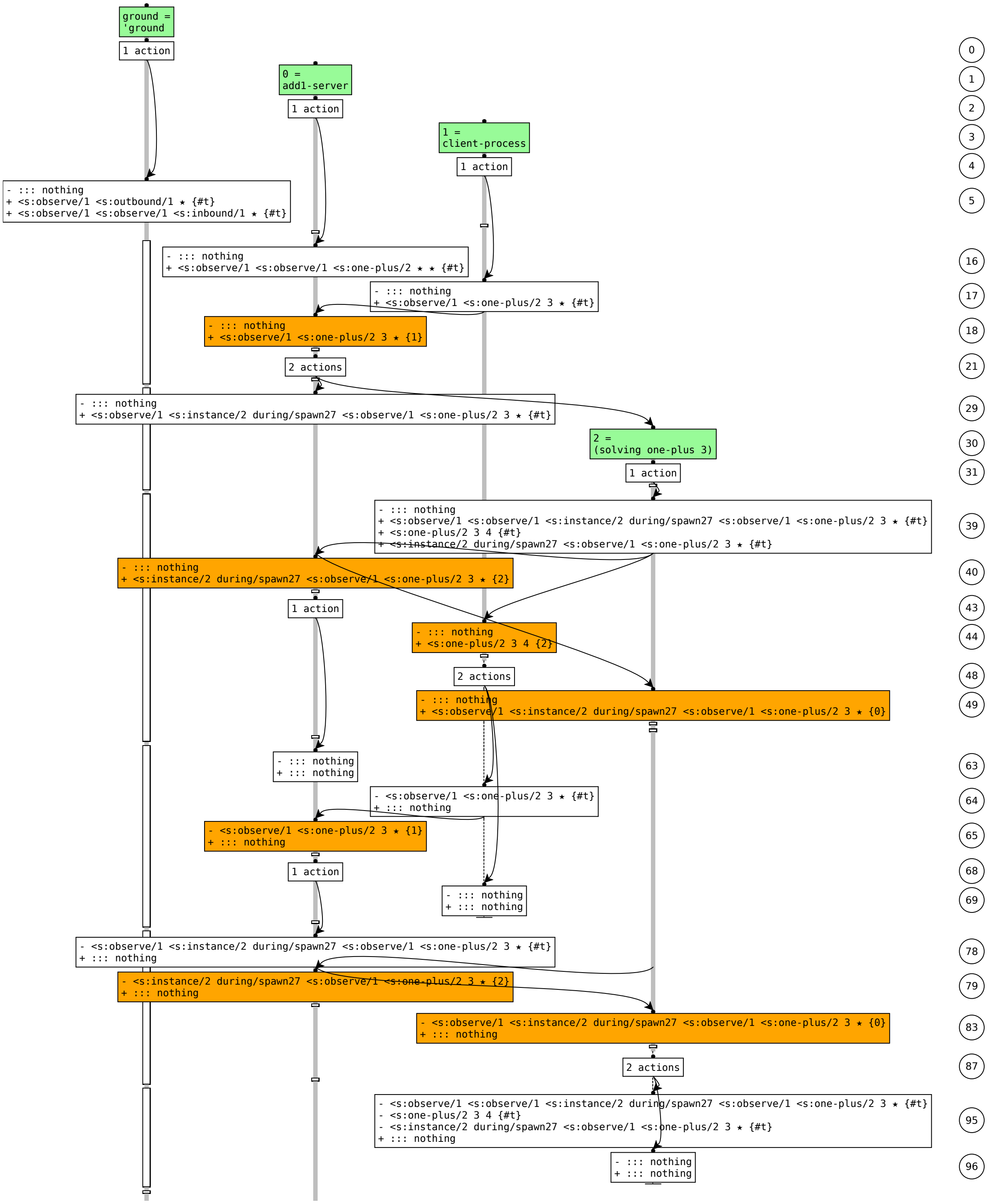

Figure 51: Sequence diagram of the program of figure 50

The rendered trace of the program of figure 50 is shown in figure 51. On the right of the diagram are the internal *step numbers* associated with displayed events.[14] Each "lifeline" corresponds to a single actor and is headed by a green rectangle containing the `#:name` of the actor, if any. In the example, we see from left to right swimlanes corresponding to the ground dataspace itself, the `add1-server` actor of lines 2–5 in the source code, the `client-process` of lines 6–8, and finally a server process named `(solving one-plus 3)` which is started in response to the `client-process`'s request. The vertical lines backing each lifeline are narrow and light gray when an actor is inactive, but are covered with empty white vertical rectangles when an actor's behavior function is executing. More than a single actor can be "executing" at once, because SYNDICATE/RKT is *functional* and a containing dataspace must be active in order for one of its children to be active. As a consequence, the ground dataspace in the leftmost lifeline is almost always executing; pauses in its execution correspond to moments when the system polled the outside world for any pending input. White rectangles on a lifeline correspond to *actions* performed by the actor, and orange rectangles correspond to *events* delivered to an actor.

The arrows overlaid on the diagram represent *causal influence*. They connect swimlanes of actors that *contributed to* or *caused* an event to the event's displayed rectangle. For example, at step 30, we see that the spawning of the `solving` actor is caused by one of the actions emitted by `add1-server` at step 21. This in turn is caused by the event of step 18, which contained information about assertions placed in the dataspace by `client-process` at step 17.

A more complex example of causal influence can be seen at step 39, where the `solving` actor emits a patch action asserting three groups of assertions:

1. `(observe (observe (instance 'during/spawn27 (observe (one-plus 3 _)))))`

2. `(one-plus 3 4)`

3. `(instance 'during/spawn27 (observe (one-plus 3 _)))`

The second in the list, `(one-plus 3 4)`, is the only one manifest in the source code (line 5). The others are assertions allowing `add1-server` to *supervise* the actors it spawns in its `during/spawn` form. The third in the list asserts an `instance` record that is interpreted by `add1-server` as "the child you spawned to handle the situation of `(observe (one-plus 3 _))` is alive." The first in the list allows the *child* to monitor the *parent*. The semantics of SYNDICATE requires that if a `during` or `during/spawn` endpoint disappears, all its subordinate facets *or actors* should also disappear; monitoring the parent arranges for this to happen.

The action of step 39 results in three events: step 40 for `add1-server`, letting it know its new child exists; step 44 for `client-process`, giving it the answer to the `one-plus` question it asked; and step 49, for the new `solving` actor itself. This latter event is a response to the child's expressed interest in the presence of its parent. The tail of the arrow connecting step 39 to step 49 is connected to `add1-server`, showing that some of the information in event 49—in this case, *all* the information—came from the set of assertions produced by `add1-server` at moment

14 They are non-contiguous because certain administrative events are not important for this form of visualization.

39. Looking back along the `add1-server` lifeline, we see that action 29, produced alongside the spawn action that created the `solving` actor, is the source of the assertion conveyed in event 49.

Event 44 causes the `client-process` to terminate, its task complete. Step 48 marks the transition: the lifeline is solid light gray above this point, but dashed black-and-white below this point, terminating in a crossbar just after step 69. Its final actions are implicitly computed as part of its termination, which must retract all its assertions; the synthetic action 64 does this. Action 64 influences `add1-server`, informing it that interest in `(one-plus 3 _)` no longer exists. In turn, this causes `add1-server` to terminate its internal facet responsible for expressing interest in the existence of the `solving` actor, leading to action 78. Action 78 causes two events: 79, removing the record of the `solving` actor's existence from `add1-server`, and 83, informing the `solving` actor that it is no longer needed. The `solving` actor terminates, producing its final actions at step 87 and being finally removed just after step 96. At the time the program ends, only the ground dataspace and the `add1-server` actor remain.

The sequence diagram renderer is a recent development, but has already been useful in my Syndicate programming, helping me find two interesting bugs. First, one program's accidental non-linear treatment of an accumulator led to duplicated spawn actions in response to an event. This mistake manifested itself on the trace as two identical new actors appearing as the result of one transaction. The fix was to treat the accumulator properly linearly.[15] Second, in a separate program, rapidly fluctuating assertions representing demand for a resource led to an actor outliving the demand that led to its creation. The problem was visible on the trace as a missing edge informing the new actor that its services were wanted. The fix was to ensure that the actor supplying the demanded resource began monitoring demand for its services as part of its initial assertions of interest (the $\pi$ in the syntax of `actor` actions described in figure 12).

### 7.4.2 *Live program display*

An experimental visualization based on trace information is shown in figure 52. Two screen captures are shown: on the left, only interactions between peers within a single dataspace are highlighted, while on the right, interactions between peers both within and *across* dataspace boundaries are shown. The diagrams are animated during the execution of the program whose structure they represent. The program depicted is a simple TCP/IP chat room service with four connected users, implemented with a nested dataspace isolating chat functionality from generic assertions and events relating to TCP, timers, and so forth. Figure 53 shows the nesting structure of the program.

Each of the circles in figure 52 represents an actor. The two larger circles correspond to the two dataspaces in the program; the smaller circles represent leaf actors. Edges connecting circles together represent recent *causal influence* between two actors. The thickness of an edge varies with the recent rolling-average rate of events exchanged between the edge's vertices; more recent events lead to thicker edges. As time goes by, interaction patterns among actors change, leading to changing patterns of connectivity in the visualization.

15 Such accidental reuse of "stale" values seems, in my experience, to be endemic in functional-programming simulations of mutable state. A monadic approach would have enforced the necessary invariants. An interesting alternative is to investigate whether some form of *contract* could help.

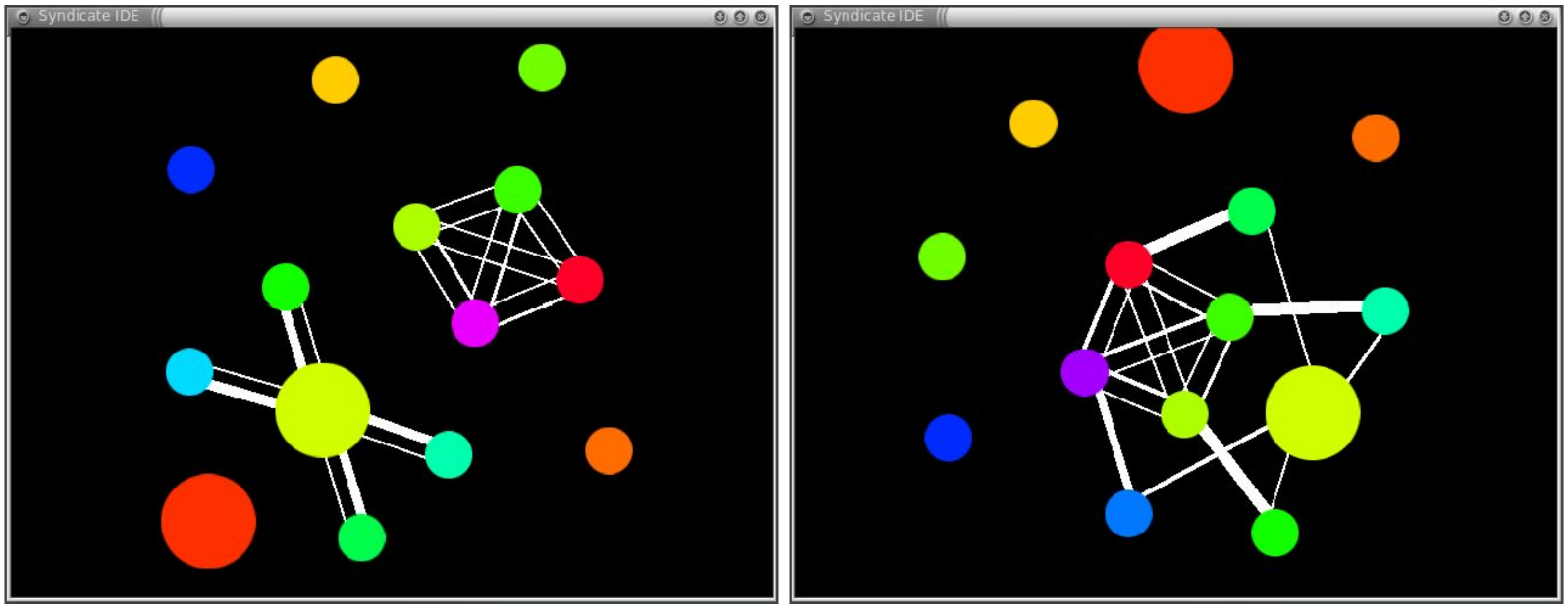

Figure 52: Two visualizations of a running chat server

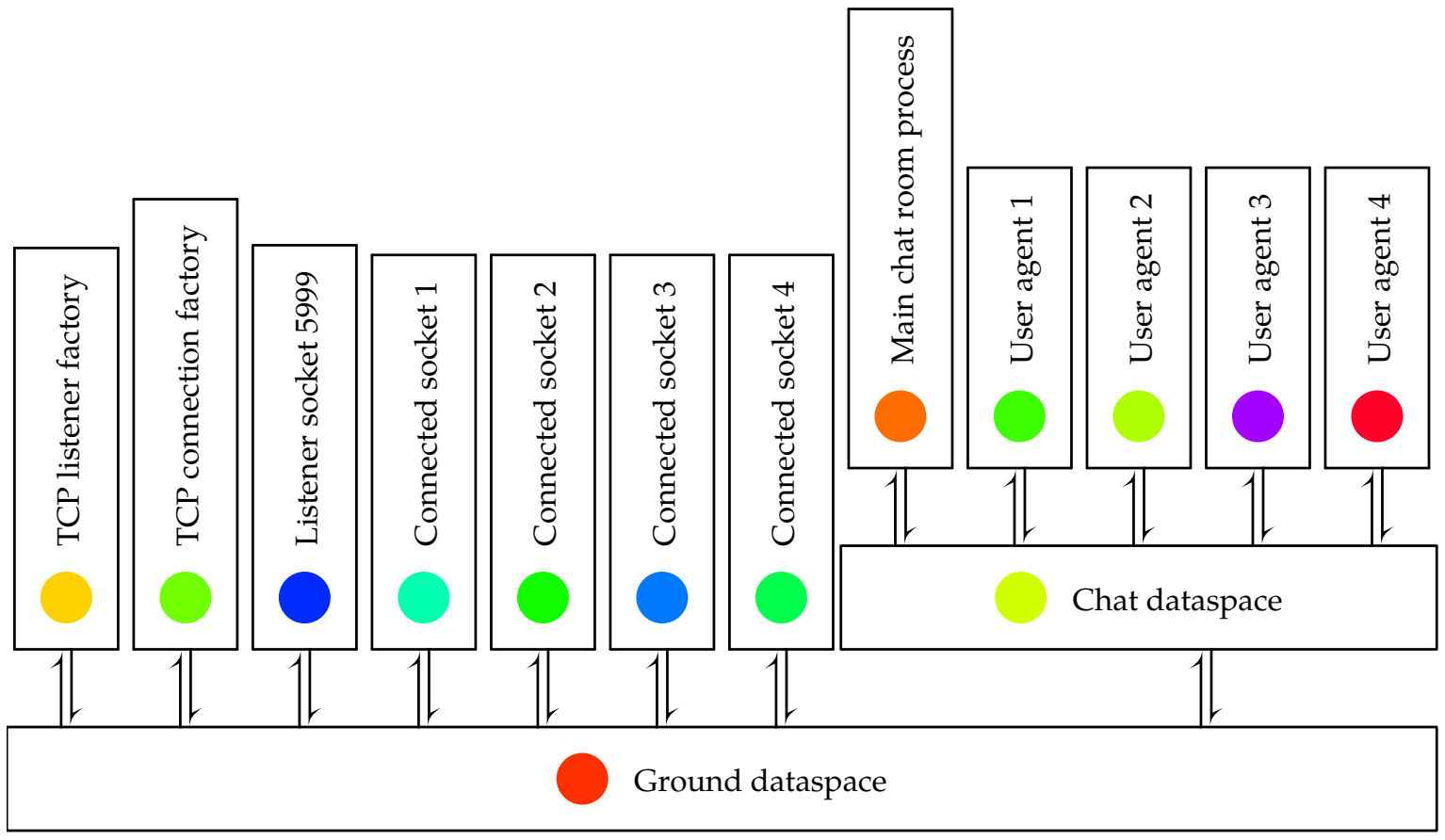

Figure 53: Actor structure of the displayed program

No nesting structure is represented. A simple spring-layout algorithm brings together interacting actors. Thicker edges lead to higher spring constants. The result is that groups of actors that interact with each other tend to move toward each other.

On the left of figure 52, we see two groups of interacting actors. The completely-connected group (toward the upper-right of the screenshot) is the four actors representing users in the inner dataspace exchanging chat messages. The other group (toward the lower-left) is the four TCP socket actors in the outer dataspace interacting with the inner dataspace actor itself in terms of TCP byte streams.

On the right, the same two groups are visible. However, the version on the right adds tracking of causal influence information across dataspace boundaries, allowing detection and display of the interactions between "Connected socket 1" and "User agent 1", and so on. The additional edges represent translation back and forth between chat messages and TCP byte stream events.

In both screenshots, we see five actors not interacting with any other. These are the ground dataspace, along with three actors directly running in the ground dataspace ("TCP listener factory", "TCP connection factory" and "Listener socket 5999") and one actor running in the inner dataspace ("Main chat room process").

This approach to visualization of a running program is still experimental and has not been integrated with the mainline implementation code. In future, exploration of ways of presenting nesting relationships among actors could prove useful.

# 8

# *Idiomatic* Syndicate

Having reviewed the theory of the dataspace model, the design of Syndicate's novel language features, and the fundamentals of programming with Syndicate/rkt, we are ready to explore practical aspects of the construction of Syndicate programs. In this chapter, we consider representative programs that illustrate idiomatic Syndicate programming techniques. We begin with the central concern in Syndicate programming: the design of Syndicate protocols.

## 8.1 protocols and protocol design

We have been calling the sum total of the related interactions among components a *protocol*, made up of *conversations* involving assertions and message transmissions. Each kind of conversation involves one or more actors playing *roles* within the conversation's context. Each role may include *responsibilities* and *obligations* that actors performing that role must live up to. The assertions and messages of each conversation form the shared *knowledge* exchanged among participants. The strong isolation afforded Syndicate actors dovetails with epistemic concerns about "who knows what" to force consideration of the *placement* of knowledge in a system. The notions of "schema", "role", "conversation" and so forth are, as yet, informal: they do not correspond either to Syndicate language features or to manifest aspects of the dataspace model. However, these latent ideas underpin each program that we examine in this chapter.

Designing a dataspace protocol is similar to designing an actor model program, but also has points in common with designing a relational database. Like the actor model, the focus is on knowledge exchanged between parties and the placement of the program's stateful components. Where the actor model focuses on exchange of domain messages, Syndicate concentrates on shared conversational state, represented as domain assertions in the shared dataspace. The structure and meaning of the assertions themselves are the primary point of similarity with relational database schema design, where interpretations of and relationships among rows in tables are carefully described. Every dataspace protocol has the rough analogue of a *schema* that describes its assertions and messages and their meanings. The schema is ontologically prior to other elements of a protocol; conversational exchanges take place within the framework provided by the schema. Consideration of the goals, abilities and needs of each participant in a conversation leads in turn to the notions of roles, responsibilities and obligations.

A second point of similarity between relational databases and the dataspace model is that both tend to construct rows from atomic data such as text, numbers, dates and domain-specific references to other rows. It is unusual to see a database include representations of programming-language concepts like thread IDs, exception values, mutable variables, or file handles. Likewise, in the dataspace model, it is rare to see such host-language implementation-level concepts communicated via the dataspace. This same point distinguishes dataspace programming from the actor model, which unavoidably communicates actor IDs as elements of message structures.[1]

**Protocol 8.1** (Toy "file system")**.** To demonstrate the pieces of a Syndicate protocol, we work through an example: a simple *file system* protocol as sketched in example 4.4, discussed in section 6.6, and implemented in figure 33.

*Schema.* Let us begin by examining the protocol's schema. Participants communicate primarily via assertions representing file contents:

```
(assertion-struct file (name content))
```

where `name` is a string denoting a file-system path and `content` is either `#f`, meaning that the file does not exist, or a string, the contents of the file. For example, asserting

```
(file "novel.txt" "Call me Ishmael.")
```

declares that the file named "`novel.txt`" currently contains the text "`Call me Ishmael.`" In principle, a `file` assertion could be maintained constantly for every file that exists, but in practice we allow an implementation to lazily manifest these assertions in response to detected demand. An endpoint like

```
(during (file "novel.txt" $text) ...)
```

results in an assertion of interest,

```
(observe (file "novel.txt" _))
```

and so our schema assigns an additional meaning to such assertions, beyond the intrinsic meaning of `observe` in expressing subscriptions. In this setting, these assertions of interest denote an active *demand* for production of a matching `file` record, not mere interest in any matching records that happen to exist.

Besides `file` assertions, our schema includes two message types:

```
(message-struct save (name content))
(message-struct delete (name))
```

where the `name` fields contain path strings, as for `file` records, but the `content` field must contain a string. The two messages denote requests to *update* or *delete* a named file, respectively.

*Roles.* There are three roles in our protocol: *Server*, *Reader*, and *Writer*.

1 The dataspace model is nameless, from the programmer's perspective; an actor label (section 4.2) is a purely dataspace-internal concept. Likewise, each facet name (section 5.1) is only meaningful to a single, specific actor.

- The Server is expected to be unique within a given protocol instance. It maintains the authoritative file store, reacts to demand by supplying file contents to Readers, and accepts file changes from Writers.
- Any number of Readers may exist within an instance. Readers observe file contents.
- Any number of Writers may exist within an instance. Writers save and delete files.

*Conversations.* There are three kinds of conversation in our protocol, each working toward satisfaction of the various *goals* that participants may have.

- *Reading* is an interaction between a Reader and the Server. The Server responds to a Reader's assertion of `(observe (file` *name* `_))` records. Each distinct *name* causes the Server to assert a `file` record with the current contents of the named file, if it exists, or with `#f` if it does not. A Reader asserts `(observe (file` *name* `_))` for some specific *name*, and responds to assertion of `(file` *name contents*`)` according to its needs.
- *Updating* is an interaction between a Writer and the Server. A Writer sends `(save` *name content*`)` to replace the content of the *name*d file with *content*, creating the file if it does not already exist. The Server responds to such messages by updating its store accordingly and updating any `(file` *name* `_)` assertions it has previously established to reference the new *content*.
- *Deleting* is also an interaction between a Writer and the Server. A Writer sends `(delete` *name*`)` to cause the deletion of the *name*d file. The Server responds to `(delete` *name*`)` messages by removing *name* from its records and updating any `(file` *name* `_)` assertions it has previously established to map *name* to `#f`.

---

An important part of the summary of a role is its expected *cardinality* within the dataspace. For example, in the example we imagine a unique file server; the protocol would require alteration to support multiple distinct file servers. Alternatively, if multiple *replica* servers were to be supported, the protocol would require changes to handle the necessary conversations among replicas. While we have described the server as unique within this protocol, we expect the protocol to support an arbitrary number of concurrent readers and writers.

The dataspace model allows wildcards to be placed freely within compound data structures, but not all SYNDICATE programs allow wildcards in all positions: families of assertions that a program expects to be able to iterate over must be finite in every position where a pattern variable exists (see section 5.5). Therefore, we must take care to specify which positions in assertions themselves and in *subscriptions* to such assertions may contain wildcards. In the description of the "Reading" conversation in our example, we see that the Server expects to be able to deduce distinct *name*s of files of interest; therefore, it is forbidden for any Reader to subscribe with a wildcard in the *name* position of a `file` assertion. In effect, we must be able to deduce the appropriate *finiteness* constraints on positions in assertions and messages (and subscriptions to those assertions and messages) from the protocol description.

Relatedly, certain positions in assertions may be required to be unique in the dataspace. In the file system example, a constraint exists that the server may not publish inconsistent information: for a given file `"a"`, only *one* `file` assertion of its contents may be placed in the dataspace at any given time.

## 8.2 BUILT-IN PROTOCOLS

Central to all dataspace protocols is an embedding of the protocol describing *interest* (?); without it, no communication takes place. Some programs also make use of *cross-layer relaying* (↿/⇂). These two protocols are special cases in that they are the *only* built-in protocols exposed to programmers.

**Protocol 8.2** (Interest)**.** This protocol is the fundamental unit of conversation in SYNDICATE; the smallest conversational frame that can exist. All other conversations and protocols are constructed from it.

*Schema.* A single family of assertions, `(observe x)`, describes interest in the assertion or assertions $x$. Asserting an `observe` record denotes *subscription* to matching assertions: not only to appearance and disappearance of assertions *per se*, but also to messages having a matching body. There are no inherent restrictions on wildcard use within `observe` records; indeed, wildcards are vital, as a wildcard used inside some $x$ in an `observe` record indicates a *range* of values of interest.

*Roles.* There are two overt, user-level roles, namely *subscriber* and *publisher*, but also a less apparent role: that of the dataspace itself acting as a *relay*.[2] Any actor asserting interest in $x$ is a subscriber to $x$; any actor asserting $x$ or sending $x$ as a message is a publisher of $x$. The dataspace, of course, is the unique relay in the scenario.

*Conversations.* Again, at an actor-to-actor level, only one kind of conversation exists: that between subscriber and publisher. The conversation between each actor and its dataspace is "at right angles" to, and facilitates, publisher-to-subscriber conversational interaction.

**Protocol 8.3** (Cross-layer relaying)**.** In every case where a dataspace is nested within another, the cross-layer relaying protocol exists to allow actors contained within the inner dataspace to access assertions and messages in the outer dataspace.

*Schema.* Two general-purpose unary records, `(inbound x)` and `(outbound x)` (corresponding to $\upharpoonleft x$ and $\downharpoonleft x$ in the formalism of section 4.1, respectively) are used for both assertions and messages $x$.

*Roles and Conversations.* This protocol is peculiar in that the relevant actors are, first, the actor $A$ asserting or sending an `outbound` assertion or message, and second, that actor's local dataspace, $D_1$. The dataspace, $D_1$, reacts to `(outbound x)` assertions or messages by relaying $x$ to its own containing dataspace, $D_2$.

2 A key difference between these two kinds of role is that the "relay" role performed by the dataspace has in some sense more to do with a *metalevel* protocol than any kind of domain-level protocol at all.

```
#lang syndicate
(assertion-struct greeting (text))

(spawn #:name "A" (assert (greeting "Hi from outer space!")))
(spawn #:name "B" (on (asserted (greeting $t))
                      (printf "Outer dataspace: ~a\n" t)))

(dataspace #:name "C"
  (spawn #:name "D" (assert (outbound (greeting "Hi from inner!"))))
  (spawn #:name "E" (on (asserted (inbound (greeting $t)))
                        (printf "Inner dataspace: ~a\n" t))))
```

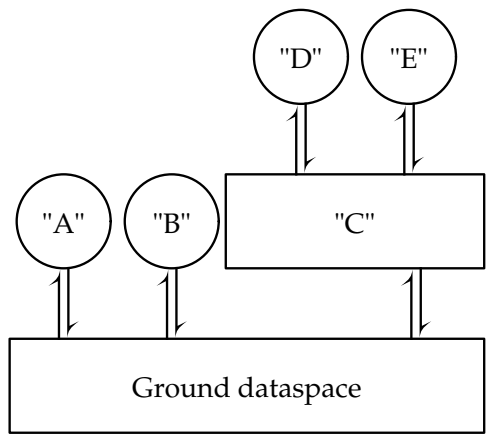

Figure 54: Cross-layer relaying example.

Asserting (outbound (observe x)) leads $D_1$ to assert (observe x) within the dataspace of $D_2$. This, of course, acts as a subscription to x, meaning that $D_1$ may receive assertions or messages x. In response to such events, $D_1$ wraps them in an inbound record and relays them on to its own dataspace. However, notice that actor A *never asserted interest in anything*. Actor A must assert (observe (inbound x)) in order to be notified when a relevant x-event in $D_2$'s dataspace takes place. For this reason, every dataspace interprets (observe (inbound x)) as if it *implied* (outbound (observe x)). Actors such as A need only assert the former to enjoy the effect of the latter.[3]

**Example 8.4** (Cross-layer relaying)**.** The program in figure 54 creates three actors (A, B and C) within the ground dataspace. Actor C is a dataspace itself. As C starts up, two further actors (D and E) are spawned within it. All interaction among actors A, B, D and E takes place via the ground dataspace; D and E communicate with the ground dataspace indirectly via C by using inbound and outbound constructors. Two greeting records end up being asserted within the ground dataspace, from A and D. As discussed above, E's assertion of interest in (inbound (greeting _)) assertions is automatically translated by C into an assertion of (observe (greeting _)) at the ground dataspace level. The matching records are relayed up into C's dataspace, appearing wrapped as (inbound (greeting ...)). At the time the program quiesces, the assertions in C's dataspace are:

- (outbound (greeting "Hi from inner!")), courtesy of actor D.
- (observe (inbound (greeting _))), from actor E.
- (inbound (greeting "Hi from outer space!")), from actor A, relayed up into C's dataspace by C itself.
- (inbound (greeting "Hi from inner!")), from actor D, relayed first down into the ground dataspace, where it matched C's own interest (on behalf of E) in greeting records and was relayed *back* again by C.

3 This is apparent from the specification of out (definition 4.14), where the assertion set relayed to a containing dataspace is given by $\{c \mid (j, \downharpoonleft c) \in R\} \cup \{?c \mid (j, ?\upharpoonright c) \in R\}$.

The assertions in the ground dataspace are:

- `(greeting "Hi from outer space!")`, from A.
- `(greeting "Hi from inner!")`, from C (on behalf of D).
- `(observe (greeting _))`, asserted by *both* C (on behalf of E) and B.

Figure 55 shows an execution trace of the program.[4] ◊

## 8.3 SHARED, MUTABLE STATE

Some protocols need assertions in the dataspace to be long-lived, outliving the actors that produced them and actors making use of them. In these situations, the dataspace takes on even more of the characteristics of a relational-style database. We have already seen two examples of this idiom: the toy "file system" of protocol 8.1, where `file` records persist until explicitly deleted, and the "box and client" program of example 6.1, where the `box-state` record persists indefinitely. Here, we present a *mutable cell* protocol and program that generalizes the latter.

**Protocol 8.5** (Mutable cell)**.** This protocol describes a mutable cell service, instantiable multiple times in a single dataspace. Cell IDs are auto-generated; a minor modification to this protocol yields a *key-value store*. The protocol is similar to so-called "CRUD" protocols (standing for Create, Read, Update and Delete). Here, creation and deletion of cells is explicit; an alternative could be to create cells implicitly at first mention of a hitherto-unseen ID.

*Schema.* One assertion describes the value of each cell, and three messages create, update, and destroy cells, respectively:

```
(assertion-struct cell (id value))
(message-struct create-cell (id value))
(message-struct update-cell (id value))
(message-struct delete-cell (id))
```

Cell IDs are arbitrary values, unique within one dataspace. At most one `cell` record is asserted for a given ID.

*Roles.* There are four roles: *CellFactory*, *Cell*, *Reader* and *Writer*. A unique CellFactory exists in the dataspace. A distinct Cell exists for each cell ID created. Any number of Readers or Writers may exist. Readers observe Cell contents; Writers request creation, deletion and update of Cells.

*Conversations.*

4 Unfortunately, the current tracing mechanism (section 7.4.1) does not capture the causal connection between `outbound` assertions and the assertions of the containing dataspace. The reader is left to deduce the connection between the assertions of actors D and E, and the subsequent actions of C.

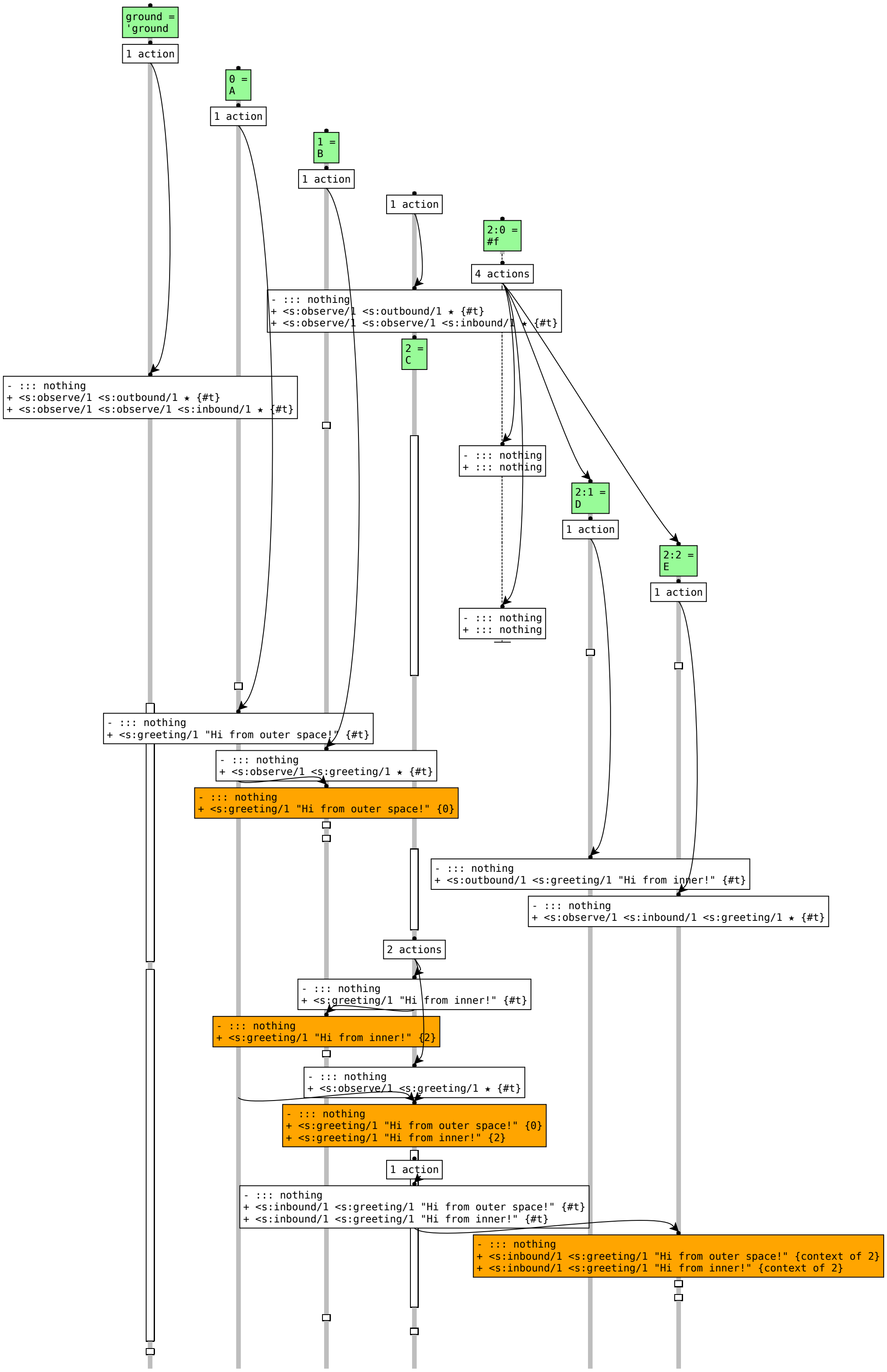

Figure 55: Execution trace of the cross-layer example 8.4

- *Creation* (Writer/CellFactory). The Writer chooses a dataspace-unique cell ID, and sends a `create-cell` message with the initial value to place in the new cell. In response, the CellFactory creates a new Cell with the given ID and value.

- *Reading* (Cell/Reader). The Cell is continuously publishing a `cell` assertion, which the Reader observes. The Cell updates the assertion as its value changes.

- *Updating* (Cell/Writer). The Writer sends an `update-cell` message; the Cell updates its value (and `cell` assertion) accordingly.

- *Deleting* (Cell/Writer). The Writer sends a `delete-cell` message; the Cell terminates in response.

*Programming interface.* A library routine, `spawn-cell`, allocates a fresh cell ID, sends a `create-cell` message, and returns the new ID:

```
(define (spawn-cell initial-value)
  (define id (gensym 'cell))
  (send! (create-cell id initial-value))
  id)
```

**Example 8.6** (Mutable cell)**.** The following listing shows an implementation of the Cell Factory and Cell roles, in context of the assertion- and message-structure definitions shown above:

```
1 (spawn #:name 'cell-factory
2   (on (message (create-cell $id $initial-value))
3       (spawn #:name (list 'cell id)
4         (field [value initial-value])
5         (assert (cell id (value)))
6         (on (message (update-cell id $new-value)) (value new-value))
7         (stop-when (message (delete-cell id))))))
```

The definition of an actor implementing Cell (lines 3–7) is embedded within the definition of the Cell Factory. Each time the Cell Factory receives a `create-cell` message (line 2), it spawns a new Cell instance (line 3), with a computed name incorporating the new cell's ID. Each Cell has a single `value` field (line 4), which is continuously published into the dataspace (line 5). Whenever an `update-cell` message is received, `value` is updated (line 6); recall that SYNDICATE/RKT fields are modeled as functions (see section 6.4). The Cell terminates itself when it receives an appropriately-addressed `delete-cell` message (line 7). ◊

Writers simply issue their requests by `send!`ing `update-cell` and `delete-cell` messages; Readers construct endpoints monitoring `cell` assertions.

**Example 8.7.** The following procedure spawns a simple actor that monitors the changing value of a cell:

```
1 (define (spawn-cell-monitor id)
2   (spawn #:name (list 'cell-monitor id)
3     (on (asserted (cell id $value))
4         (printf "Cell ~a updated to: ~a\n" id value))
5     (on (retracted (cell id _))
6         (printf "Cell ~a deleted\n" id))))
```

The endpoint of lines 3–4 monitors the *appearance* of each *distinct* ID/value combination for the ID given, thereby printing a message for each value the cell takes on. The endpoint of lines 5–6 monitors the *disappearance* of *all* `cell` assertions for the given ID, triggering only once: when no more assertions remain, namely when the cell's actor terminates itself. This is an example of the elision of irrelevant detail—here, the specific *value* in the `cell` record is irrelevant for this endpoint's purpose—performed by the metafunction *inst* (definition 5.24) as part of projection of incoming patch events. ◊

**Example 8.8.** Alternatively, a blocking `read-cell` routine can be constructed using `flush!` and `react/suspend` (section 6.5, page 119):

```
(define (read-cell id)
  (flush!)
  (react/suspend (k) (stop-when (asserted (cell id $value)) (k value))))
```

The use of `flush!` in `read-cell` deserves explanation. Recall that message sending is asynchronous. This means that if we `send!` an `update-cell` message, it is enqueued for transmission once the actor's behavior function returns control to the dataspace, and execution continues. If we omit the call to `flush!` before accessing the `cell` assertion, then programs calling `read-cell` multiple times in succession *reuse* the most-recently delivered information, without forcing queued actions (such as `update-cell` messages) out to the dataspace and waiting for new information. ◊

**Example 8.9.** Consider the following program, to be run alongside the definitions of protocol 8.5 and examples 8.6, 8.7 and 8.8:

```
(spawn* #:name 'main-actor
        (define id (spawn-cell 123))
        (spawn-cell-monitor id)
        (send! (update-cell id (+ (read-cell id) 1)))
        (send! (update-cell id (+ (read-cell id) 1)))
        (send! (update-cell id (+ (read-cell id) 1)))
        (send! (delete-cell id)))
```

With a `flush!` in `read-cell`, the output is

```
Cell cell27 updated to: 123
Cell cell27 updated to: 124
Cell cell27 updated to: 125
Cell cell27 updated to: 126
Cell cell27 deleted
```

Without a `flush!` in `read-cell`, the output is

```
Cell cell27 updated to: 123
Cell cell27 updated to: 124
Cell cell27 deleted
```

Figure 56 shows the reason why. The trace on the left includes `flush!`, while the trace on the right omits `flush!`. Recall that metafunction *emit* (section 5.2) coalesces adjacent patch actions produced by an actor. A chain of calls to a `flush!`less variation on `read-cell` results in repeated assertion and retraction of interest in `(cell id _)` assertions, which are then coalesced into *no-op, empty patches*. Observe the empty patches interleaved in the "8 actions" shown on the right in figure 56 as outputs of `main-actor` (center region of middle column). The version shown in example 8.8, however, breaks up the chain of patch actions with the message sent as part of the implementation of `flush!` that forces the round trip to the dataspace, leading to the (truncated) longer sequence of interactions shown on the left in figure 56. There, the retraction of interest in `(cell id _)` prior to the `flush!` causes the actor's cached record of the cell's value (stored in the $\pi_i$ register in Syndicate/λ's semantics, and an analogous location in the Syndicate/rkt implementation) to be evicted. ◊

## 8.4 i/o, time, timers and timeouts

Simple I/O in Syndicate/rkt programs can be performed as normal for Racket programs, via ordinary side-effecting function calls. If a particular I/O action could block—for example, a write to a buffered channel such as a TCP socket, or a read from a serial port—then an alternative strategy must be chosen to allow other conversations to proceed while the program waits. Generally speaking, identification of a blocking I/O facility results in design and construction of a driver actor. This includes pseudo-input operations such as waiting for a certain period of wall-clock time to elapse, exposed in Syndicate as a protocol like everything else.

**Protocol 8.10** (Timer Driver)**.** The *timer driver* implements this protocol.

*Module to activate.* `syndicate/drivers/timer`

*Schema.* The protocol involves two messages. The first is

```
(message-struct set-timer (label msecs kind))
```

where `label` is an arbitrary value and `msecs` is a count of milliseconds. If `kind` is `'absolute`, then `msecs` is interpreted as an absolute moment in time, counted in milliseconds from the machine's epoch; if `kind` is `'relative`, then `msecs` is interpreted as milliseconds in the future, counted from the moment the message is received by the timer driver implementation. The second message type is

```
(message-struct timer-expired (label msecs))
```

where `label` is the label from a previous `set-timer` message, and `msecs` is the absolute time that the message was sent from the timer driver implementation, counted in milliseconds from the machine's epoch.

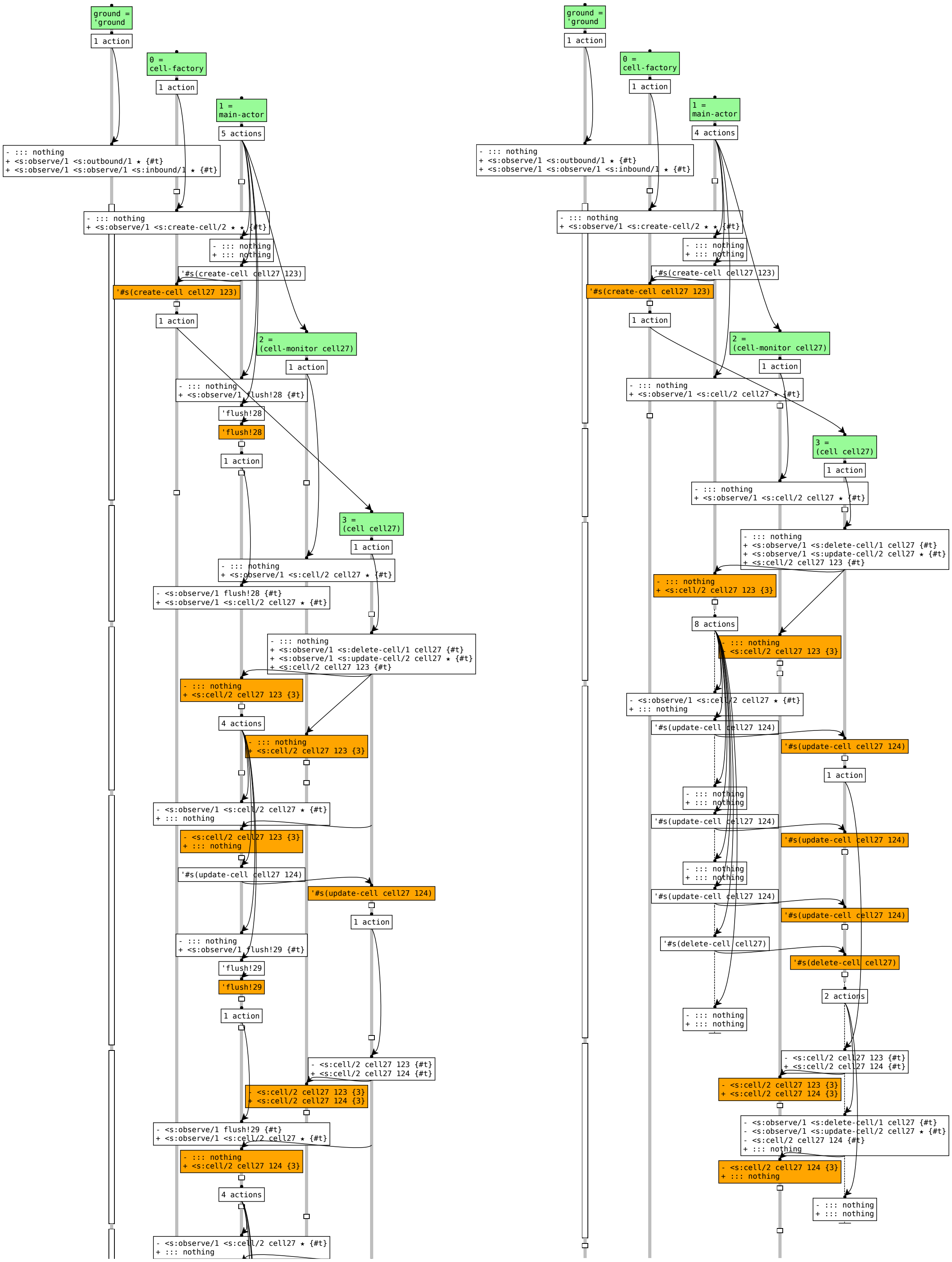

Figure 56: Execution trace of the mutable-cell example 8.9. On the left, a `flush!` call ensures the effects of `update-cell` messages are visible to `main-actor`; on the right, omitting `flush!` leads to reuse of cached knowledge.

*Roles.* The unique role of *Driver* is performed by the actor(s) started when a user program activates the driver's module. The roles of *AlarmSetter* and *AlarmReceiver* are performed by user code.

*Conversations.*

- *Setting* (AlarmSetter/Driver). The AlarmSetter sends `set-timer`. The Driver eventually responds with `timer-expired`.
- *Notifying* (Driver/AlarmReceiver). The Driver sends `timer-expired`, and the AlarmReceiver interprets it in an application-specific way.

Protocol 8.10 was the first to be implemented during the development of Syndicate and its predecessors, and so is in some ways anachronistic. To begin with, it uses *messages* in place of *assertions* for its functionality. This forces clients to take care with respect to ordering of operations. In particular, they must ensure their subscriptions to `timer-expired` messages are in place before the corresponding `set-timer` messages are sent, lest unfortunate scheduling cause them to miss their wake-up call. A so-called "timestate" driver provides an interface to timer functionality that is more idiomatic.

**Protocol 8.11** (Timestate)**.** The *timestate driver* is an ordinary Syndicate/rkt program that exposes this protocol to clients, using protocol 8.10 internally to implement its services.

*Module to activate.* `syndicate/drivers/timestate`

*Schema.* The protocol involves a single assertion, `(assertion-struct later-than (msecs))`, where `msecs` is an integer denoting an absolute moment in time counted in milliseconds from the machine's epoch. When asserted, it denotes a claim that wall-clock time is equal to or later than the moment mentioned.

*Roles.* The unique role of *Driver* is performed by the actor(s) started when a user program activates the driver's module. The role of *TimeObserver* is performed by user code.

*Conversations.*

- *Observing* (TimeObserver/Driver). The TimeObserver asserts interest in a particular `later-than` assertion. The Driver eventually responds by asserting it. Once the TimeObserver's interest is withdrawn, the Driver retracts it again.

**Example 8.12** (Timestate implementation)**.** The "driver" is extremely simple, as it is an ordinary program which reformulates protocol 8.10 into the more palatable form of protocol 8.11:

```
(spawn #:name 'drivers/timestate
       (during (observe (later-than $msecs))
         (define timer-id (gensym 'timestate))
         (on-start (send! (set-timer timer-id msecs 'absolute)))
         (on (message (timer-expired timer-id _))
             (react (assert (later-than msecs))))))
```

The use of `during` (lines 2–6) creates a facet whose lifetime is scoped to a particular conversation. Upon detection of interest in a particular `later-than` assertion, lines 3–6 run, creating the endpoints necessary for the conversation and kicking off the conversation between the timestate driver and the underlying timer driver. Line 3 uses Racket's `gensym` utility to generate a fresh symbol, unique within the running operating system process. This symbol is used on line 4 in a `set-timer` message. Recall from section 6.4 that `on-start` forms execute once the endpoints of a new facet are completely configured. This ensures that `set-timer` is transmitted in a context where a subscription to the corresponding `timer-expired` message has already been established (by lines 5–6). When triggered, that subscription creates a nested facet (line 6) which simply asserts the requested `later-than` record. When the TimeObserver that started this conversation retracts its interest in the `later-than` assertion, the entire `during` facet is terminated. Not only is the subscription to `timer-expired` then retracted, but the nested facet asserting `later-than` is also terminated. ◊

**Example 8.13** (Use of `later-than`)**.** The following program prints a message (line 2), and waits for five seconds (lines 4 and 5). Once the time has elapsed, the facet is terminated, triggering the message of line 3. Finally, the message of line 6 is printed.

```
1 (spawn #:name 'demo-later-than
2        (on-start (printf "Starting demo-later-than\n"))
3        (on-stop (printf "Stopping demo-later-than\n"))
4        (field [deadline (+ (current-inexact-milliseconds) 5000)])
5        (stop-when (asserted (later-than (deadline)))
6                   (printf "Deadline expired\n")))
```

◊

**Example 8.14** (Updating a deadline)**.** The following program prints out ten "Tick" messages (from line 5), waiting for one second between each, and then terminates. The endpoint of lines 4–7 is automatically withdrawn as soon as the value of the `counter` field exceeds 9, and is otherwise triggered every time the `deadline` is reached. After printing a "Tick" message, it increments its `counter` and adjusts the `deadline` forward by another second. Modifying `counter` causes reevaluation of the endpoint's `#:when` clause; modifying `deadline` causes reevaluation of the endpoint's subscription, and triggers the transmission of a patch action into the dataspace, which in turn informs the Timestate driver of the new state of affairs.

```
1 (spawn #:name 'demo-updating-later-than
2        (field [deadline (current-inexact-milliseconds)])
3        (field [counter 0])
4        (on #:when (< (counter) 10) (asserted (later-than (deadline)))
5            (printf "Tick ~v\n" (counter))
6            (counter (+ (counter) 1))
7            (deadline (+ (deadline) 1000))))
```

◊

Besides its primary purpose of simplifying interaction with the Timer driver, the Timestate driver offers a pair of utilities, `stop-when-timeout` and `sleep`, that capture frequently-occurring interaction patterns.

**Example 8.15** (Timeouts)**.** The `stop-when-timeout` macro offers a new kind of endpoint which terminates its facet after a certain number of milliseconds have elapsed. If the timeout occurs, the body expressions are executed; if the facet has already terminated for some other reason, the endpoint is withdrawn along with the other endpoints of the facet, and the body expressions are not executed.

```
(define-syntax-rule (stop-when-timeout relative-msecs body ...)
  (let ((timer-id (gensym 'timeout)))
    (on-start (send! (set-timer timer-id relative-msecs 'relative)))
    (stop-when (message (timer-expired timer-id _)) body ...)))
```

The macro expands into an expression to be executed in *facet-setup-expr* context. The expression creates an `on-start` endpoint which arms the timer, and a `stop-when` endpoint which reacts to the resulting `timer-expired` event by terminating the surrounding facet and executing the `body` forms. ◊

**Example 8.16** (Use of `stop-when-timeout`)**.** The following program terminates itself after three seconds have elapsed. During its execution, it prints the messages of lines 2, 3 and 4, in that order.

```
1 (spawn #:name 'demo-timeout
2        (on-start (printf "Starting demo-timeout\n"))
3        (on-stop (printf "Stopping demo-timeout\n"))
4        (stop-when-timeout 3000 (printf "Three second timeout fired\n")))
```

◊

**Example 8.17** (Use of `sleep`)**.** We have already seen the definition of `sleep` in section 6.5 on page 119. The following program uses `spawn*` to start a new actor in *script-expr* rather than *facet-setup-expr* context, allowing it to perform sequential actions such as sending a message, creating a facet, and so on. The program is a `sleep`-based reimplementation of example 8.14. Where that example was written in an event-based style, this is written in a threaded style (Li and Zdancewic 2007; Haller and Odersky 2009). It uses Racket's built-in looping construct, `for`, with a range of natural numbers (line 2). Example 8.14's `counter` field is replaced with an ordinary Racket variable, and `sleep` is used to cede control to neighboring actors until an appropriate wake-up event arrives, at which point the loop is resumed. The actor terminates once the loop finishes, since it contains no other facets.

```
1 (spawn* #:name 'demo-sleep
2         (for [(counter (in-range 10))]
3           (printf "Sleeping tick ~v\n" counter)
4           (sleep 1.0)))
```

◊

An interesting aspect of the Timestate protocol is that its purpose is to adapt *messages* to *assertions*. We will see the reverse case, adapting SYNDICATE assertions to *messages* sent over a non-SYNDICATE communications mechanism, in the context of a simple chat server (section 11.1).

The examples in this section so far have taken the approach of using a driver to perform blocking operations. An alternative, suitable for simple cases, is to make use of an *implicit* driver, that responds to interest in values yielded by Racket's CML-style events (Reppy 1999; Flatt and PLT 2010, version 6.2.1, section 11.2.1). The protocol is available at the (notional) dataspace surrounding the ground dataspace; that is, actors inhabiting the ground dataspace engage with the event driver via the cross-layer protocol (protocol 8.3).

**Protocol 8.18** (CML-style I/O Events)**.** Each of Racket's CML-style events yields zero or more values as its *synchronization result* when ready. For example, if `sock` is a Racket TCP server socket handle, then `(tcp-accept-evt sock)` is an event yielding two values, an input port and an output port, when ready. This protocol allows SYNDICATE/RKT programs to express interest in such synchronisation results.

*Schema.* A single assertion, `external-event`, pairs an event with its synchronisation results:

```
(assertion-struct external-event (descriptor values))
```

The `descriptor` is the event, and the `values` are a list containing the synchronisation results from the event.

*Roles.* The implicit, unique implementation of the *Driver* role exists just outside the ground dataspace. The *EventConsumer* role exists at or within the ground dataspace, and interacts with the *Driver* via the cross-layer protocol.

*Conversations.*

- *Subscription* (EventConsumer/Driver). Assertion of interest in `(external-event` $e$ `_)` for some particular Racket event value $e$ signals the Driver that $e$ should be added to its collection of active events. Retraction of interest withdraws $e$ from the same collection.

- *Delivery* (Driver/EventConsumer). Periodically, and whenever the ground dataspace as a whole is idle, the system will block, waiting for one of the $e$s in the collection of active events to become ready. The first to do so, yielding a list of results $r$, leads to a message `(external-event` $e$ $r$`)` being broadcast.

**Example 8.19** (Terminal I/O)**.** The program in figure 57 demonstrates the usage of protocol 8.18. Successive lines of text input appearing on standard input are, if they conform to the syntax for Racket numeric values, interpreted as such and added to an accumulator. Each time the accumulator changes, its new value is printed.

Line 2 `requires` the `read-bytes-line-evt` event constructor: when given an input port, the constructor yields an event whose synchronization result is either an end-of-file object or a byte-vector containing a single line's worth of text input. Line 3 constructs a single constant event that the program uses throughout. The field declaration on line 4 initializes the accumulator, and line 5 ensures that each time the `total` field is written to, its updated value is printed. Line 6 prints a message when the program terminates.

Line 7 is the point where the program interacts with the Driver role of protocol 8.18. The actor itself is running within the ground dataspace, but the driver is notionally one layer further out. Therefore, the actor subscribes to messages using an `inbound` constructor to signify that a cross-layer subscription should be established. Each time the event is selected and ready, it

```
1  #lang syndicate
2  (require (only-in racket/port read-bytes-line-evt))

3  (define e (read-bytes-line-evt (current-input-port) 'any))

4  (spawn (field [total 0])
5         (begin/dataflow (printf "The total is ~a.\n" (total)))
6         (on-stop (printf "Goodbye!\n"))
7         (on (message (inbound (external-event e (list $input))))
8             (cond
9               [(eof-object? input) (stop-current-facet)]
10              [(string->number (bytes->string/utf-8 input)) =>
11                 (lambda (n) (total (+ (total) n)))]
12              [else (void)])))
```

Figure 57: Terminal I/O "running total" program

yields either a line of text input or an end-of-file value, available as `input` in lines 8–12. On end-of-file, the program terminates itself (triggering line 6 in the process). If the input text, interpreted as UTF-8 text, can be converted to a number, that number is added to the current value of the `total` field. Otherwise, the input is ignored. ◊

Multiple subscriptions to such events may exist in a single running ground dataspace, from different drivers, actors, and protocols. Racket's underlying synchronization mechanism ensures fair (pseudo-random) selection from the set of ready events in case more than one is available at once.

Protocol 8.18 allows SYNDICATE/RKT programs to respond to Racket's CML-inspired I/O events. However, it is also possible to use Racket's event mechanism to transmit observations of the interior of a SYNDICATE/RKT program to other portions of a larger Racket program. For example, a Racket thread may run a SYNDICATE ground dataspace alongside other Racket-level threads. Within the dataspace, actors can respond to SYNDICATE events by sending non-SYNDICATE messages to those other Racket threads. In this way, the programmer may embed SYNDICATE/RKT subprograms within existing non-SYNDICATE code.

## 8.5 LOGIC, DEDUCTION, DATABASES, AND ELABORATION

We have seen that the SYNDICATE/λ notion of `during` is reminiscent of logical implication. The analogous SYNDICATE/RKT `during` construct is no different, and allows us to write SYNDICATE actors that perform *deductions* based on the assertions in the dataspace, expressed in a quasi-logical style. When relevant information is held elsewhere, such as an external SQL database, or the file system, actors may retrieve information from the external source on demand, presenting the results as assertions. In this way, multiple "proof strategies", including procedural knowledge, integrate smoothly with ordinary forward- and backward-chaining reasoning about assertions and demand for assertions.

```
1 parent(john, douglas).
2 parent(bob, john).
3 parent(ebbon, bob).
4 ancestor(A, C) :- parent(A, C).
5 ancestor(A, B) :- parent(A, C), ancestor(C, B).
```

Figure 58: Datalog "ancestor" program.

```
1 (assertion-struct parent (who of))
2 (assertion-struct ancestor (who of))

3 (spawn (assert (parent 'john 'douglas)))
4 (spawn (assert (parent 'bob 'john)))
5 (spawn (assert (parent 'ebbon 'bob)))

6 (spawn (during (parent $A $C)
7          (assert (ancestor A C))
8          (during (ancestor C $B)
9            (assert (ancestor A B)))))
```

Figure 59: Forward-chaining SYNDICATE "ancestor" program.

### 8.5.1 *Forward-chaining*

Writing a SYNDICATE program frequently feels similar to writing a Datalog program. Consider the "ancestor" Datalog predicate shown in figure 58. A SYNDICATE encoding of the predicate is shown in figure 59. Lines 1–2 declare the relations involved, implicit in the Datalog equivalent. Lines 3–5 assert ground terms describing a *parent* relation. Lines 6–9 define an *ancestor* relation in a form strongly reminiscent of a proposition involving implication:

$$\mathit{parent}(A, C) \implies (\mathit{ancestor}(A, C) \wedge (\mathit{ancestor}(C, B) \implies \mathit{ancestor}(A, B)))$$

Here, the program uses forward-chaining to prove all provable conclusions from the inputs given. The program reacts to *parent* assertions (line 6), immediately concluding the consequences of the base case of the inductive definition of *ancestor* (line 7; cf. figure 58 line 4) and enabling an additional reaction (lines 8–9; cf. figure 58 line 5) embodying the inductive step of the *ancestor* definition for a *specific* case. Line 8 reacts to assertion—interpreted as *proof*—of the inductive hypothesis for the specific case at hand, the specific binding of the variable `C`, and line 9 asserts a conclusion building upon that hypothesis.

### 8.5.2 *Backward-chaining and Hewitt's "Turing" Syllogism*

Carl Hewitt's paper describing PLANNER includes the following quote, which seems to anticipate the dataspace model (Hewitt 1971):

```
1 (assertion-struct human (who))
2 (assertion-struct fallible (who))

3 (spawn (assert (human 'turing)))

4 (spawn (during (observe (fallible $who))
5          (during (human who)
6            (assert (fallible who)))))

7 (spawn (during (fallible 'turing)
8          (on-start (printf "Turing: fallible\n"))
9          (on-stop (printf "Turing: infallible\n"))))
```

Figure 60: Hewitt's "Turing" syllogism

> ASSOCIATIVE MEMORY forms the basis for PLANNER'S data space which consists of directed graphs with labeled arcs. [...] Assertions are stored in buckets by their coordinates using the associative memory in order to provide efficient retrieval.

In the same paper, he offers a syllogistic proof of Turing's fallibility, which can be expressed in Syndicate as shown in figure 60. Where the example of figure 59 uses a forward-chaining strategy, our implementation of Hewitt's syllogism uses backwards-chaining. The key difference is the monitoring of interest in `fallible` assertions (line 4). When interest is detected, it is interpreted as a goal, and a small facet (lines 5–6) using a forward-chaining strategy is constructed to attempt to satisfy the goal.

### 8.5.3 *External knowledge sources: The file-system driver*

A particularly important external database for many applications is the file system provided by the underlying operating system.

**Protocol 8.20** (File system)**.** Syndicate/rkt's file system driver implements this protocol, which tracks and publishes the contents of files and directories.

*Module to activate.* `syndicate/drivers/filesystem`

*Schema.* The protocol involves a single assertion,

```
(assertion-struct file-content (name reader-proc content))
```

Each `file-content` structure represents a claim about the contents of the file system path `name`. When `content` is `#f`, the claim is that no file or directory exists at that path; otherwise, some file or directory exists at that name, and `content` is the output of `(reader-proc name)`.

*Roles.* The unique role of *Driver* is performed by the actor(s) started when a user program activates the driver's module. The role of *FileObserver* is performed by user code.

*Conversations.*

- *Observing* (FileObserver/Driver). The FileObserver asserts interest in a `file-content` assertion with a specific `name` string and specific `reader-proc`. The Driver will respond with a `file-content` assertion reporting the state of the `name`d file or directory in terms of `reader-proc`'s result. Each time the operating system reports a change to the file at `name`, the Driver re-executes `(reader-proc name)` and updates the assertion. Once interest is withdrawn, the Driver retracts the assertion and releases the operating-system-level resources associated with notifications about changes to the file.

This protocol is unusual in that it explicitly requires inclusion of a Racket procedure value in a field of an assertion, depending indirectly on Racket's primitive pointer-equality to compare such values. The reason is the large number and great variety of operations for reading or otherwise analyzing a file system resource. Supplying different `reader-proc` values allows the programmer to specify the nature of the information about the file that is of interest.

**Example 8.21** (Monitoring a file's contents)**.** Monitoring the actual contents of a file can be done using `file->bytes` as `reader-proc`,

```
(on (asserted (file-content "novel.txt" file->bytes $bytes)) ...)
```

In the event-handling code, `bytes` contains the raw bytes making up the file, or `#f` if the file does not exist or was deleted. ◊

**Example 8.22** (Monitoring a directory's entries)**.** Monitoring the list of files in a directory can be done with `directory-list` as `read-proc`,

```
(on (asserted (file-content "/tmp" directory-list $files)) ...)
```

The `files` variable contains a list of the names of the files in `/tmp`. Each time a file is added or removed, the `file-content` assertion is replaced. If the directory is deleted, `files` becomes `#f`. ◊

**Example 8.23** (Detecting whether a file's content has changed)**.** Given a small utility procedure `file->sha1` (shown in figure 61) to use for `read-proc`, we may track a secure hash of the file's contents with the endpoint

```
(on (asserted (file-content "novel.txt" file->sha1 $hash)) ...)
```

As usual, `hash` is `#f` if the file is not present; otherwise, it is a string containing a hexadecimal representation of the SHA-1 hash of the file's content. The properties of such secure hashes allow us to treat a changed `hash` as a change in the underlying file content, without having to relay the entirety of the contents via the dataspace. ◊

Commodity operating systems offer only the simplest of change-notification systems, essentially nothing more than a message signifying that *something* changed about a given path. This is akin to the monolithic SCN events of the dataspace model, which place the burden of determining the nature of a given change on the recipient of each event. The dataspace model's *incremental* patch SCN events convey the same information, but relieve actors of this burden. Augmenting operating systems with more fine-grained notifications would help in the same way, improving efficiency around reacting to changes in the file system.

```
1  #lang syndicate
2  (require/activate syndicate/drivers/filesystem)
3  (require racket/string racket/system file/sha1)

4  (define (file->sha1 p) (call-with-input-file p sha1))

5  (spawn (during (observe (file-content $name _ _))
6           (unless (string-suffix? name ".c")
7             (define name.c (string-append name ".c"))
8             (on (asserted (file-content name.c file->sha1 $hash)) ;; nb. $hash, not _
9                 (cond [(not hash) (printf "~a doesn't exist.\n" name.c)]
10                      [else
11                       (printf "~a has changed hash to ~a, recompiling\n" name.c hash)
12                       (system* (find-executable-path "cc") "-o" name name.c)])))))

13 (spawn (on (asserted (file-content "." directory-list $files))
14            (for [(name-path (in-list files))]
15              (match (path->string name-path)
16                [(pregexp #px"(.*)\\.c" (list _ name))
17                 (assert! (observe (file-content name file-exists? #t)))]
18                [_ (void)]))))
```

Figure 61: Automatic "Make"-like compiler

### 8.5.4 *Procedural knowledge and Elaboration: "Make"*

The Unix program make (IEEE 2009, "Shell and Utilities" volume) is a venerable tool for systematically producing conclusions (*target* files) from premises (*source* files) by way of procedural knowledge (*rules*). We may similarly combine deduction with procedural knowledge in SYNDICATE.

**Example 8.24** (Make-like compiler)**.** The program of figure 61 implements a pair of actors which, together, use the file system driver (protocol 8.20) to track the ".c" files in the current directory, compiling them to executables each time one changes.

The first actor (lines 5–12) interprets interest in a file named name to be a request that it should be compiled from name.c, if that file exists. Each time name.c is created or changes (cf. example 8.23), the actor shells out to cc(1) to compile the program.[5] There are two notable features of this portion of the program. The first is the use of unless on line 6 to *conditionally* add endpoints to the facet of the during of line 5. When interest is expressed in some file, we only attempt to build it from some corresponding C source file if the file of interest is not already a C source file. In the case that name ends with ".c", the facet created by the activation of during on line 5 is terminated automatically since it lacks endpoints entirely. The second interesting feature is the use of a binding, $hash, on line 8, where one might expect to see a

5 Use of a blocking call here is suboptimal: it indicates the need for a *subprocess* driver for starting, managing, and terminating subprocesses.

discard pattern, _. Recall that *inst* discards irrelevant structure from observed assertions during projection of incoming patch events. Had we used discard in place of `$hash`, we would have been declaring our lack of interest in such fine detail as the hash of the file being different, and would instead react only to the file having a hash *at all*. By using `$hash`, we convey that we care about specific values of `hash`, and thus that we should react every time the file's content changes.

The second actor (lines 13–18) monitors the files in the current directory. Every time it sees a file whose name ends in ".c", it strips that extension, and asserts interest in the existence of the resulting base filename. A more robust program would be able to *retract* interest in case such a file were erased (perhaps in turn leading to deletion of the corresponding build product); however, this program contents itself with an ever-growing collection of filenames of interest. It uses the ad-hoc assertion form, `assert!`, discussed in section 6.6. ◊

The form of the actor of lines 5–12 in figure 61 is essentially the same as that of lines 4–6 in our implementation of Hewitt's "Turing" syllogism (figure 60). This tells us that our "Make"-like program is also taking a backtracking strategy to goal satisfaction. The difference here is use of *procedural* knowledge as the local strategy for achieving some goal. In the "Make"-like program, we know that invoking the C compiler will achieve our goal, while in the "Turing" syllogism, the goal is an immediate logical consequence of the premise detected on line 5 of figure 60.

The notion of an *elaboration* of a formalism captures the idea of its modification to take into account new phenomena (McCarthy 1998). A simple example of this is the need to compute additional or derived information about a domain entity. The "Make" example can be seen as an instance of this, augmenting information about a source file with information about its compiled form. Such augmentation is promoted to a design pattern and given the name of *Content Enrichment* by Hohpe and Woolf (2004); the examples they present can be readily adapted to the idioms introduced in this section.

### 8.5.5 *Incremental truth-maintenance and Aggregation: All-pairs shortest paths*

In their recent paper, Conway et al. (2012) present a short program which solves the all-pairs shortest-paths problem, written in their distributed, Datalog-based language Bloom (Alvaro et al. 2011). An analogous Syndicate/rkt program is shown in figure 62. Each `link` assertion (lines 6–10) forms part of the program's input, describing a weighted edge in a directed graph. The program computes `path` assertions as it proceeds (lines 11–16), though these are an implementation detail; it is the `min-cost` assertions (lines 17–26) that are the outputs of the program. Each `min-cost` assertion describes a path between two nodes in the input graph, along with the computed minimum total cost for that path. As `link` edges are added and removed, the program reacts, converging on a solution and quiescing once it is achieved.

The program demonstrates two important Syndicate idioms. The first is the ability for programs expressed in this style to incrementally maintain outputs as inputs change. Altering the set of asserted `link` records leads to a corresponding update to the set of asserted `min-cost`

```
#lang syndicate

(require racket/set)

(assertion-struct link (from to cost))
(assertion-struct path (from to seen cost))
(assertion-struct min-cost (from to cost))

(spawn (assert (link 1 3 -2))
       (assert (link 2 1 4))
       (assert (link 2 3 3))
       (assert (link 3 4 2))
       (assert (link 4 2 -1)))

(spawn (during (link $from $to $cost)
               (assert (path from to (set from to) cost))))

(spawn (during (link $A $B $link-cost)
               (during (path B $C $seen $path-cost)
                 (assert #:when (not (set-member? seen A))
                         (path A C (set-add seen A) (+ link-cost path-cost))))))

(spawn (during (path $from $to _ _)
               (field [costs (set)] [least +inf.0])
               (assert (min-cost from to (least)))
               (on (asserted (path from to _ $cost))
                   (costs (set-add (costs) cost))
                   (least (min (least) cost)))
               (on (retracted (path from to _ $cost))
                   (define new-costs (set-remove (costs) cost))
                   (costs new-costs)
                   (least (for/fold [(least +inf.0)] [(x new-costs)] (min x least))))))
```

Figure 62: All-pairs shortest paths program. After Figure 1 of Conway et al. (2012), but modified with a path-seen set to ensure termination on input cycles.

records—though intermediate states become visible as the computation proceeds back toward consistency.[6]

The second idiom is aggregation of a set of assertions into a single *summary* of the set; here, this is seen in the actor of lines 17–26, which computes the *minimum* of a set of paths from A to B. The aggregation operator here is thus "set minimum". The pattern `(path $from $to _ _)` of line 17 scopes each computation to a particular source/sink node pair. Within this context, two fields are maintained: the first, `costs`, tracks all distinct path costs, while the second, `least`, contains the smallest cost in `(costs)`. As a new distinct cost appears (line 20), it is added to the set, and `least` is efficiently updated. However, when a distinct cost disappears (line 23), we must laboriously recompute `least` from an updated `costs`. A heap or ordered-set data structure would eliminate this problem.

This idiom of maintaining an order statistic could be abstracted into a new streaming query form, perhaps called `define/query-min` by analogy with the existing aggregate query forms introduced in section 6.5. In fact, the Bloom program of Conway et al. makes use of a built-in operator supporting a minimum-value calculation. Setting aside the explicit, non-library implementation of computing the minimum, our program is comparable in length to the Bloom program, showing that Bloom-like Datalog-style programming is achievable and useful in Syndicate, though of course Syndicate does not yet extend to distributed systems. An interesting question to examine is to what extent reasoning based on logical monotonicity, as introduced in the Bloom "CALM theorem" (Alvaro et al. 2011), translates well to Syndicate.

### 8.5.6 *Modal reasoning: Advertisement*

Earlier work on Network Calculus (NC) (Garnock-Jones, Tobin-Hochstadt and Felleisen 2014) included only a limited form of observable, replicated, shared state: the state of *subscriptions* to messages within each dataspace. The dataspace model generalizes this to allow observation of arbitrary shared assertions, and brings messages into the new setting by reinterpreting them as transient knowledge. However, NC included *two* forms of subscription. The traditional notion of "subscription" led to actors receiving messages produced by publishers, but a symmetric notion of "advertisement" led to publishers receiving *feedback* from subscribers. The dataspace model drops the idea of feedback, and with it the idea of a distinct publisher, leaving it to domain-specific protocols to include such notions as appropriate. Examining NC programs shows that the primary use of feedback and observation of "publisher" endpoints was to detect whether messages of a certain type *might potentially be produced in the near future*. Absence of a "publisher" was interpreted as meaning that there was no need to prepare to receive that publisher's communications; its presence, by contrast, suggested that it might begin speaking soon. Syndicate and the dataspace model captures this idea with an *advertisement* protocol.

**Protocol 8.25** (Advertisement)**.** The *advertisement* protocol decouples synchronization of conversational context from subsequent conversational interaction.

*Module to require.* `syndicate/protocol/advertise`

6 We discuss options for eliminating interference from intermediate states in section 8.8.

*Schema.* `(assertion-struct advertise (claim))`

An assertion of `(advertise c)` denotes the potential for *future* assertion of `c` itself, across some unspecified timescale. Other protocols will incorporate this protocol, as seen earlier in protocols 8.2 and 8.3. No particular obligations are placed on parties asserting `advertise` records, other than the loose notion that they may eventually produce an assertion of the underlying `claim`.

Advertisement allows us to explore an alternative factoring of protocol 8.5.

**Protocol 8.26** (Mutable cell, with advertisement)**.** In place of explicit command messages, we will create and destroy Cell actors in response to presence or absence of advertisements of potential `update-cell` messages and potential subscription to `cell` assertions.

*Schema.* As for protocol 8.5, omitting `create-cell` and `delete-cell`, and adding `(advertise (update-cell id _))` and `(advertise (observe (cell id _)))`.

*Roles.* As for protocol 8.5.

*Conversations.* As for protocol 8.5, but replacing Creation and Deleting as follows:

- *Creation* (Writer/Reader/CellFactory). In response to one of the forms of advertisement mentioned above, the CellFactory creates a new cell, initially with no value (and thus publishing no `cell` assertion).
- *Deleting* (Writer/Reader/Cell). Each Cell monitors both forms of advertisement (specific to its ID) mentioned above; once all advertisements have been retracted, it terminates itself.

**Example 8.27** (Mutable cell, with advertisement)**.**

```
1  (spawn #:name 'cell-factory
2    (assertion-struct cell-existence-demanded (id))
3    (during (advertise (update-cell $id _)) (assert (cell-existence-demanded id)))
4    (during (advertise (observe (cell $id _))) (assert (cell-existence-demanded id)))
5    (during/spawn (cell-existence-demanded $id)
6      (field [has-value? #f] [value (void)])
7      (assert #:when (has-value?) (cell id (value)))
8      (on (message (update-cell id $new-value))
9          (has-value? #t)
10         (value new-value))))
```

Line 2 declares an implementation-local structure type representing an intermediate piece of knowledge: that a cell with a particular ID should exist. Lines 3 and 4 deduce such facts from the two forms of relevant advertisement.[7] The consequences of a `cell-existence-demanded` assertion are spelled out on lines 5–10. Line 5 means that each distinct ID demanded results in a separate actor; once the demand for cell existence is retracted, by retraction of all corre-

7 In principle, we could imagine augmenting SYNDICATE's pattern language with an "or" construct that implemented this pattern automatically: `(advertise (or (update-cell $id _) (observe (cell $id _))))`. There is no fundamental obstacle to such a feature.

sponding advertisement assertions, the actor is automatically terminated. Lines 6–10 follow the implementation of our earlier mutable cell protocol closely. The main differences are a lack of a `stop-when` clause reacting to `delete-cell` messages, replaced by the action of `during/spawn` on line 5, and addition of the `has-value?` field, which accounts for the new protocol's cells lacking a value initially. Line 7 publishes a `cell` assertion only once a value is available. ◊

Syndicate/rkt's support for UDP communication makes use of protocol 8.25 to signal that the socket backing a request for service is ready.[8]

**Protocol 8.28** (UDP sockets)**.**

*Module to activate.* `syndicate/drivers/udp`

*Schema.* The core of the protocol is

`(assertion-struct udp-packet (source destination body))`

where `body` is a Racket byte-vector and `source` and `destination` are instances of

`(assertion-struct udp-remote-address (host port))` or

`(assertion-struct udp-listener (port))`

A `udp-packet` must either have a `udp-remote-address` in its `source` field, and a `udp-listener` in its `destination` field, or vice versa.

*Roles.* There are three roles: *SocketFactory*, which responds to demand for sockets; *Socket*, which mediates between local actors and a Racket socket resource; and *Client*, a local actor making use of UDP functionality.

*Conversations.*

- *Listening* (Client/SocketFactory/Socket). The Client chooses a port number `port` and asserts interest in `(udp-packet _ (udp-listener port) _)`.

  In response, the SocketFactory begins performing a corresponding Socket role (e.g. by delegating this responsibility to a new actor). The Socket asserts

  `(advertise (udp-packet _ (udp-listener port) _))`,

  which the Client may choose to observe to detect when the underlying UDP socket resource is ready to forward inbound packets.

  The Socket also expresses interest in `(udp-packet (udp-listener port) _ _)`, in order to receive packets intended to be relayed to remote parties; the Client may also make decisions based on the presence of such interest. Once a Socket actor is established and ready, *Reading* and *Writing* conversations take place.

8 This description covers only UDP *listener* sockets. Much other functionality including UDP multicast is available in the implementation.

- *Reading* (Client/Socket). When the underlying UDP socket receives a datagram from peer-host at peer-port with a certain body, it sends a

  `(udp-packet (udp-remote-address peer-host peer-port) (udp-listener port) body)`

  message. The Client, having previously declared interest in such messages, receives it.

- *Writing* (Client/Socket) The Client may send a

  `(udp-packet (udp-listener port) (udp-remote-address peer-host peer-port) body)`

  to deliver body to any peer-host and peer-port. The Socket will receive it and relay it via the Racket socket resource.

- *Closing* (Client/Socket) The Client may withdraw its interest in inbound udp-packets. The Socket detects this, closes the underlying UDP socket resource, and terminates, thus withdrawing its advertisement of readiness and its interest in outbound packets.

**Example 8.29** (UDP echo program)**.** The following actor listens for packets on port 5999, echoing each back to its sender as it is received; as soon as it knows packets may be forwarded to it, it prints a message saying so.

```
(spawn (on (message (udp-packet $peer (udp-listener 5999) $body))
           (send! (udp-packet (udp-listener 5999) peer body)))
       (on (asserted (advertise (udp-packet _ (udp-listener 5999) _)))
           (printf "Socket is ready and will forward datagrams.\n")))
```

◊

The UDP socket protocol was designed originally for our implementation of Network Calculus, which explains its awkward use of advertisement in place of a more straightforward udp-socket-ready assertion or similar. While the protocol of *interest* (protocol 8.2) is essential to the dataspace model, the protocol of *advertisement* appears to have much more limited applicability.

Despite this limited applicability, the general interpretation of the protocol remains of interest. Taking (advertise c) to mean "eventually c" or "possibly c" suggests a connection with the modal logic $\diamond$ operator (Manna and Pnueli 1991; van Ditmarsch, van der Hoek and Kooi 2017). We have seen that (during $P$ (assert $E$)) reads as $P \implies E$; perhaps it is, in truth, closer to some interpretation of $\Box(P \implies E)$. It remains future work to explore this connection further.

Finally, while advertisement has limited use within domain-specific protocols, it is of great benefit in the setting of publish/subscribe middleware, where it is used to optimize message routing overlays (Carzaniga, Rosenblum and Wolf 2000; Pietzuch and Bacon 2002; Jayaram and Eugster 2011; Martins and Duarte 2010; Eugster et al. 2003). Automatic, conservative overapproximation of the assertions an actor may produce could lead to efficiency gains in Syndicate implementations, which may become particularly useful in any attempt to scale the design to distributed systems.

```
(assertion-struct service-ready (name))

(spawn (assert (service-ready 'file-systems)) ...)

(spawn (stop-when (asserted (service-ready 'file-systems))
         (react (assert (service-ready 'database-service)) ...)))

(spawn (stop-when (asserted (service-ready 'file-systems))
         (react (assert (service-ready 'logging-service)) ...)))

(spawn (stop-when (asserted (service-ready 'database-service))
         (react (stop-when (asserted (service-ready 'logging-service))
                  (react (assert (service-ready 'web-application-server))
                         ...)))))
```

Figure 63: Service dependency resolution

## 8.6 Dependency Resolution and Lazy Startup: Service Presence

Unix systems start up their system service programs in an order which guarantees that the dependencies of each program are all ready before that program is started.[9] Many current Unix distributions manually schedule the system startup process. Because it is a complex process, such manually-arranged boot sequences tend to be strictly sequential. Other distributions are starting to use tools like `make` both to automatically compute a suitable startup ordering and to automatically parallelize system startup.

With Syndicate, we can both ensure correct ordering and automatically parallelize system startup where possible, by taking advantage of *service presence* information (Konieczny et al. 2009). Programs offer their services via endpoints; clients of these services interpret the presence of these endpoints as service availability and react, offering up their own services in turn when a service they depend upon becomes available.

Service availability must, at some level, be expressed in a concrete style, with endpoints interacting with their environment in terms of the actual messages of the protocols supported by the service. However, availability may also be expressed at a more abstract level. Consumers of a service may detect service presence by directly observing the presence of endpoints engaging in a protocol of interest, or by observing the presence of assertions describing the service more abstractly. The former corresponds to a kind of *structural* presence indicator, while the latter corresponds to a form of *nominal* service presence.

For example, a web application server may depend on a SQL database service as well as on the system logging service, which may in turn depend on the machine's file systems all being mounted. Figure 63 sketches a Syndicate realization of these service dependencies, with the actual implementations of each service replaced by ellipses. We may arbitrarily reorder the services in the file without changing the order in which they become available. The startup

9 At least, this is the ideal.

```
(define-syntax await-services
  (syntax-rules ()
    [(_ [] body ...)
     (begin body ...)]
    [(_ [service more ...] body ...)
     (stop-when (asserted (service-ready service))
       (react (await-services [more ...] body ...)))]))

(define-syntax spawn-service
  (syntax-rules (<-)
    [(_ target <- [service ...] body ...)
     (spawn (await-services [service ...]
              (assert (service-ready target))))]))

(spawn-service 'file-systems <- [] ...)
(spawn-service 'database-service <- ['file-systems] ...)
(spawn-service 'logging-service <- ['file-systems] ...)
(spawn-service 'web-application-server <- ['database-service
                                           'logging-service]
  ...)
```

Figure 64: Macros abstracting away details of the service dependency pattern.

procedures of the services in the sketch pause until they see the *names* of their dependencies asserted. An alternative would be to wait for assertions of interest in service requests to appear; concretely, the web application could wait until `(asserted (observe (log-message ...)))` rather than waiting for `(asserted 'logging-service-ready)`.

Examination of the sketch of figure 63 reveals a *design pattern*. Service actors start in a state awaiting their first dependency. When it appears, they transition to a state awaiting their second dependency; and so on, until all their dependencies are available, at which point the service configures its own offerings and asserts its own availability. A pair of simple macros allows us to abstract over this pattern; figure 64 shows an example. Having recognized and abstracted away details of this pattern, we may take further steps, such as to rearrange the implementation of the `await-services` macro to express interest in *all* dependencies at once rather than in one at a time.

The notion of service dependency can be readily extended to start services only when some demand for them exists. A service *factory* actor might observe `(observe (service-ready x))`, arranging for the program implementing service `x` to be run when such an interest appears in the dataspace. The protocol may also be enriched to allow a service to declare that it must run *before* some other service is started, rather than after. The combination of such forward and reverse dependencies, along with *milestones* such as "network configuration complete" masquerading as abstract services, yields a self-configuring system startup facility rivaling those available in many modern Unix distributions.

The notion of service startup applies not only at the level of a whole operating system, but also within specific applications in an arbitrarily fine-grained way. For example, a web server that depends on a database might wish to only start accepting TCP connections once (a) the database server itself is available, (b) a connection to the database server is established, and (c) the schema and contents of the database have been initialized.

## 8.7 transactions: rpc, streams, memoization

As we have discussed as far back as chapter 2, certain assertions serve as *framing knowledge* in a protocol, identifying and delimiting sub-conversations within an overarching interaction. Often, framing knowledge can be seen as statement of a goal; subsequent actions and interactions are then steps taken toward satisfaction of the goal. Establishment of a conversational frame is similar to establishment of a transaction boundary, and indeed various forms of transaction manifest themselves in Syndicate as conversational frames.

The simplest possible transaction is a one-way message. The message entails establishment and tear-down of a transactional context that lasts just long enough to process the message. Beyond this point, a vast design space opens up. Here we consider a few points in this space.

rpc. The simplest form of transaction involving feedback from recipient to sender is *request/response*. Such transactions allow us to encode *remote procedure call* (RPC); that is, *function calls* across Syndicate dataspaces.

**Example 8.30** (RPC, message/message)**.** At its simplest, a request message establishes context for the call at the same time as making a specific request, and a corresponding response message signals both completion of the request and discarding of the request's context.

```
(message-struct request (body))
(message-struct reply (body))
(spawn (on (message (request `(square ,$x)))
           (send! `(square-of ,x is ,(* x x)))))
```

Here, we know that computation of a square is *idempotent*, and so we may omit distinct request-identifiers. If the invoked action were non-idempotent, clients would have to allocate dataspace-unique request IDs, using them to tell otherwise-identical-seeming instances of the protocol apart. Similarly, here we see that the argument `x` can be used to correlate a response with a request, so that answers to simultaneous requests for `` `(square 3) `` and `` `(square 4) `` do not get mixed up. In cases where simple echoing of arguments does not suffice to correlate a response with its request, explicit correlation information should be supplied in a request and included in the response. ◊

**Example 8.31** (RPC, interest/assertion)**.** We may take a more relational approach to RPC by observing that a (pure) function is a relation between its argument and result. Interest in a subset of the elements of the relation can serve to establish the necessary context; assertion of the specific element produced by the function supplies the response; and retraction of interest signals the end of the conversation. In cases where requests are non-idempotent, and thus must be distinguished by some request ID, we may use a reply *message* instead of an assertion, since there is no risk of confusion in this case. Use of a reply message with idempotent request assertions would be an error, however: the dataspace model collapses multiple simultaneous assertions of the same request, risking a situation where a client is not supplied with an answer to its request.

```
(assertion-struct function (argument result))
(spawn (during (observe (function `(square ,$x) _))
         (assert (function `(square ,x) (* x x)))))
```

◊

**Example 8.32** (RPC, interest/assertion/error)**.** Error handling may be incorporated into our RPC protocols via *sum types*, as is traditional for pure functional languages. Alternatively, we may introduce a *nested conversational context* within which it is known to the *requestor* that processing of the request is ongoing. Closure of that nested context prior to assertion of a reply indicates abnormal termination. We may press protocol 8.25 into service as a convenient expression of this nested context, asserting our intention to *eventually* answer the request.

```
1 (spawn #:name 'division-server
2        (during/spawn (observe (function `(divide ,$n ,$d) _))
3          (assert (advertise (function `(divide ,n ,d) _)))
4          (on-start (flush!)
5                    (react (assert (function `(divide ,n ,d) (/ n d)))))))
```

On line 3, the service asserts its intention to reply.[10] The `flush!` call of line 4 is necessary to ensure that the patch SCN action resulting from line 3 reaches the dataspace safely before computation of the function begins.[11] Line 5 computes and publishes the answer. Once interest is retracted, the semantics of `during/spawn` ensures that the actor created for the specific request is terminated along with all its state and resources.

Naturally, a request entailing a division by zero causes a Racket exception to be signaled on line 5, terminating the request's actor (but not the service overall). We may take advantage of the careful separation of the advertisement of line 3 from the response of line 5 in order to make a *positive statement* of failure; a positive statement of an inference drawn from a *lack* of information:

```
(assertion-struct failed (argument))
(spawn #:name 'failure-detector
       (during/spawn (observe (function $req _))
         (on (retracted (advertise (function req _)))
           (react (assert (failed req))))))
```

10 An interesting generalization of this idea is to replace a simple `advertise` with a protocol for *progress reporting*; the service can then keep the client informed as a perhaps-complex request proceeds toward completion. This makes an RPC-like request into a kind of *stream*, discussed below.

11 An alternative to this use of `flush!` would be to use the *responsibility transfer* mechanism of the initial assertion set that is included with each actor-spawn action of the dataspace model, as discussed at the end of section 4.2. The `division-server`'s `during/spawn` would arrange for each spawned actor to be created *already asserting* its `advertise` record. That way, there would be zero risk of either a crash before the assertion of the `advertise` record, or accidentally beginning computation of the result before the `advertise` had safely made its way to the dataspace.

Every time a not-previously-asserted declaration of interest in a function result appears, an actor is spawned to monitor the situation (line 3). The `during/spawn` terminates the monitor as soon as interest in the function result is retracted. If a result—assertion of a `function` record—is transmitted, and the protocol is followed, the service maintains its assertion of `advertisement` until after our monitoring actor has terminated. However, if the service crashes before asserting its result, its `advertise` assertion is withdrawn, triggering lines 4 and 5 to report to interested parties that the overall request failed. Clients, then, pay attention to `failed` assertions, rather than observing retraction of the `advertisement` directly:

```
1 (spawn (define req `(divide 1 0))
2        (stop-when (asserted (failed req))
3          (printf "No answer was supplied!\n"))
4        (stop-when (asserted (function req $answer))
5          (printf "The answer is: ~a\n" answer)))
```

The endpoint of line 4 demands the answer to our division problem, triggering a computation in the division server. The endpoint of lines 2–3 causes the client to respond to failure `assertions`, should any appear. Alternatively, a normal answer from the server triggers the endpoint of lines 4–5. ♢

Recall the "eager" answer-production of the forward-chaining strategy of section 8.5.1 and the "lazy" nature of the backward-chaining strategy of section 8.5.2. Each example of RPC we have explored here combines *procedural knowledge* (section 8.5.4) with a *lazy* answer-production strategy. However, the decoupling of control from information flow that the dataspace model offers allows us to employ an *eager* strategy on a case-by-case basis, without altering any client code or protocol details. We may go further, offering *memoization* of computed results without altering callers.

**Example 8.33** (RPC, automatic memoization)**.** Here, a cache actor notices interest in answers to a request `req`, and "asks the same question" itself. This strategy exploits the way the dataspace model collapses identical assertions to *maintain* interest in answers to `req` for a certain length of time, presumably exceeding the duration of interest expressed by the original client.

```
(spawn (on (asserted (observe (function $req _)))
           (react (assert (observe (function req _)))
                  (stop-when (retracted (advertise (function req _))))
                  (stop-when-timeout 750))))
```

Because interest in a given answer is maintained without interruption, the service only performs its computation once. ♢

An alternative implementation of memoization might listen in on the answer from the service and take on responsibility for asserting that answer on its own. Then it may optionally coordinate with the server to relieve it of the burden of redundantly asserting the answer for the life time of the cache entry. By coordinating among different entries *within* a memoizing cache actor, making the life time of a cache entry depend on the life time of previously-demanded entries, we may achieve the effect of dynamic programming.

Finally, in a situation where one function-like service depends on another, we may wish to short-circuit the analogue of *tail calls*. Where request/reply correlation is done using the structure of the request, this can be difficult to achieve, but where explicit, arbitrary correlation identifiers exist distinct from request descriptions, an intermediary can reuse the identity of the request that triggered it, effectively *forwarding* the request to another service.

STREAMS. Moving beyond a single request and single response toward more long-lived transactions takes us toward *streams*. A stream is a conversational frame involving multiple interactions in either or both directions. Examples include the protocols SYNDICATE/RKT exposes as part of its TCP/IP socket, HTTP server, WebSocket, and IRC client drivers. We will examine these in more detail as part of our evaluation in chapter 9, focusing here on the example of the IRC driver protocol as it appears to clients.

---

**Protocol 8.34** (IRC client connection)**.** SYNDICATE/RKT may interact across the network via the IRC protocol (Oikarinen and Reed 1993; Kalt 2000), exposed by SYNDICATE/RKT's IRC client driver.

*Module to activate.* `syndicate/drivers/irc`

*Schema.* The IRC protocol allows participants to connect to a server and then to join zero or more separate named chat rooms, each known as a *channel*. Each connection is identified at the server by a server-unique nickname. A connection to an IRC server is represented by an `irc-connection` record,

```
(assertion-struct irc-connection (host port nick))
```

where `host` and `port` identify the server to connect to, and `nick` the nickname to associate with the connection.[12] The nicknames of connected users in a given channel are conveyed via `irc-presence` assertions,

```
(assertion-struct irc-presence (conn nick channel))
```

where `conn` is an `irc-connection` record, and `nick` and `channel` both strings. Messages from a given channel on the server appear as `irc-inbound` messages,

```
(message-struct irc-inbound (conn nick target body))
```

where `conn` is an `irc-connection` record, `body` is the message text, and `nick` and `target` identify the speaker and the channel, respectively. Messages traveling in the other direction, from the program to a given server channel, appear as `irc-outbound` messages,

```
(message-struct irc-outbound (conn target body))
```

with `conn` having its usual meaning, `body` being the message text, and `target` identifying the channel to which the IRC message should be directed.

*Roles.* The unique *ConnectionFactory* creates a *Connection* in response to user requests. In turn, each Connection interacts with a *User*.

*Conversations.*

- *Connecting* (User/ConnectionFactory). The User asserts an `irc-connection` record into the dataspace; the ConnectionFactory reacts to its appearance by creating a Connection. Alternatively, the User may simply assert interest in `irc-inbound` messages: the ConnectionFactory notices this, and asserts the `irc-connection` record carried in the `irc-inbound` subscription, thereby triggering the Connecting conversation automatically.
- *Joining* (User/Connection). The User asserts interest in `irc-inbound` messages for a specific, previously established connection and a specific channel name. In response, the Connection sends appropriate `JOIN` messages to the remote server. The Connection also commits to maintaining a local record of channel membership in terms of `irc-presence` assertions as the IRC server sends an initial bulk list of fellow channel members and subsequent incremental updates to this list. As a consequence, the User may use the `irc-presence` record indicating its *own* presence in the channel as an indication that the channel join operation is complete. When interest in `irc-inbound` messages is retracted, the Connection sends appropriate `PART` messages and retracts the channel-specific `irc-presence` assertions it has been maintaining.
- *Speaking* (User/Connection). The User sends `irc-outbound` messages, which the Connection relays on to the IRC server.
- *Listening* (User/Connection). Within the context of a joined channel, utterances from channel members are delivered by the Connection as `irc-inbound` messages to all listening Users.

**Example 8.35** (IRC bot)**.** Figure 65 shows a simple "bot" program which connects to the Freenode IRC network with nickname `syndicatebot`, joins channel `##syndicatelang`, and greets those in the channel as it joins. The driver notices the subscription of line 5, asserting `C`, the `irc-connection` record, in response. This triggers the actual creation of the connection. The endpoint of lines 5–8 reacts to incoming chat messages. The endpoint of lines 9–10 sends a greeting to the members of the channel once the connection has completed joining the channel. Finally, lines 11–13 react to changes in channel membership, including the connection's own membership and the members present at the time of channel join, by printing messages. ◊

In this example, the conversational context of membership in a particular IRC channel delimits two *streams* of messages. One of the two is the stream of `irc-inbound` messages from channel members; the other is the stream of `irc-outbound` messages from the local User to peers in the channel. The two streams interact: each `irc-outbound` message is reflected as an `irc-inbound` message, meaning that a connection "hears its own speech". Finally, these channel-specific streams are in fact *nested streams* (nested transactions) within the larger conversational context

12 The library is a drastically simplified prototype, not even supporting nick changes during a connection.

```
(define NICK "syndicatebot")
(define CHAN "##syndicatelang")
(define C (irc-connection "irc.freenode.net" 6667 NICK))

(spawn #:name 'irc-connection-example

       (on (message (irc-inbound C $who CHAN $body))
           (printf "~a says: ~a\n" who body)
           (when (not (equal? who NICK))
             (send! (irc-outbound C CHAN (format "Hey, ~a said '~a'" who body)))))

       (on (asserted (irc-presence C NICK CHAN))
           (send! (irc-outbound C CHAN "Hello, everybody!")))

       (during (irc-presence C $who CHAN)
         (on-start (printf "~a joins ~a\n" who CHAN))
         (on-stop (printf "~a leaves ~a\n" who CHAN))))
```

Figure 65: IRC bot

of the connection to the IRC server as a whole. Channel-specific sub-conversations come and go within a connection's context, interleaving arbitrarily.

**acknowledgement and flow control.** Within a single stream, it may be important to manage the sizes of various buffers. Assertions describing the amount of free available buffer space at a recipient act as *windowed flow control*. Assertions describing successfully-received messages act as *acknowledgements*. The former allow management of receive buffer space; the latter, management of send (retransmission) buffer space. Acknowledgements effectively "garbage-collect" slots in a sender's retransmission buffer. These ideas can be used to model TCP/IP-like sliding-window "reliable-delivery" transport protocols.

Waiting
−REQ/ − X ↑↓ +REQ/ + X
Transmitting
↓ +ACK/ − X
Complete

Figure 66: Flow control and acknowledgement

Consider the case of a single piece of information, to be transmitted from a sender to a receiver. In order to make effective use of bandwidth or other scarce resources, the sender might want to wait until the receiver is ready to listen before producing a message for its consumption. Likewise, if the medium or some relay in the communication path is unreliable, or if the receiver itself might fail at any time, the sender will keep trying to transfer until receipt (and/or processing) is confirmed.

Figure 66 depicts the lifecycle of the process from the sender's perspective. Starting in "Waiting" state, the sender learns that the receiver has `REQ`uested the item, transitioning to "Transmitting" state and asserting the item `X` itself. If the receiver crashes or changes its mind, the `REQ`uest is withdrawn, and the sender transitions back to "Waiting", retracting `X`.

If the receiver `ACK`nowledges the item, however, the sender transitions to state "Complete", retracting `X` and continuing about its business. By using assertions instead of messages, the Syndicate programmer has delegated to the dataspace the messy business of retries, timeouts and so forth, and can concentrate on the epistemic properties of the logic of the transfer itself.

**Example 8.36** (Acknowledgement and flow control)**.** In simple cases, the fact of an interest in a given assertion can be an implicature that the time is right to produce and to communicate it. As we saw above in cases such as example 8.31, when no explicit, positive indication of receipt is required, *retraction* of interest can serve as acknowledgement of receipt. However, this conflates an indication that the receiver has reneged on its previously-declared interest with an indication of successful delivery. When acknowledgement is important, we must make it explicit and separate from assertions of readiness to receive.

```
1 (assertion-struct envelope (payload))
2 (assertion-struct acknowledgement (payload))
3 ... (react/suspend (k)
4       (during (observe (envelope _))
5         (define item (compute-item))
6         (assert (envelope item))
7         (on (asserted (acknowledgement item)) (k)))) ...
```

On line 3, we enter the "Waiting" state of figure 66. Interest in our `envelope` assertion (line 4) constitutes a `REQ` signal from a recipient; a subfacet is created representing occupancy of the "Transmitting" state. The subfacet computes the item to transfer (line 5), asserts it (line 6) and awaits explicit, positive acknowledgement of receipt (line 7). Once acknowledgement is received, the call to `k` serves to terminate the facet opened on line 3, finishing at "Complete" state and releasing the continuation of the `react/suspend` form. Otherwise, if the `(observe (envelope _))` assertion is retracted before acknowledgement is received, the corresponding subfacet is destroyed and we return to "Waiting" state. ◊

## 8.8 DATAFLOW AND REACTIVE PROGRAMMING

Manna and Pnueli define a *reactive* program very generally as follows:

> A reactive program is a program whose role is to maintain an ongoing interaction with its environment rather than to compute some final value on termination. (Manna and Pnueli 1991)

This contrasts with the slightly more restrictive definition of Bainomugisha et al., who define reactive programming as "a programming paradigm that is built around the notion of continuous time-varying values and propagation of change" (Bainomugisha et al. 2013) that is in turn based on synchronous *dataflow* (Lee and Messerschmitt 1987). The dataspace model is clearly reactive in sense of Manna and Pnueli, and while it does not quite satisfy the "distinguishing features" of reactive languages given by Bainomugisha et al., it does enjoy similar strengths and suffer similar weaknesses to many of the reactive languages they describe.

Maintenance of a connection between a representation of a temperature in degrees Fahrenheit and in degrees Celsius is a classic challenge problem for dataflow languages (Ingalls et al.

```
(assertion-struct temperature (unit value))
(message-struct set-temperature (unit value))

(spawn #:name 'track-celsius
       (field [temp 0])
       (assert (temperature 'C (temp)))
       (on (message (set-temperature 'C $new-temp))
           (temp new-temp))
       (on (asserted (temperature 'F $other-temp))
           (temp (* (- other-temp 32) 5/9))))

(spawn #:name 'track-fahrenheit
       (field [temp 32])
       (assert (temperature 'F (temp)))
       (on (message (set-temperature 'F $new-temp))
           (temp new-temp))
       (on (asserted (temperature 'C $other-temp))
           (temp (+ (* other-temp 9/5) 32))))
```

Figure 67: Maintaining synchrony between two temperature scales

1988; Bainomugisha et al. 2013). The problem is to internally maintain a temperature value, presenting it to the user in both temperature scales and allowing the user to modify the value in terms of either temperature scale.[13] Figure 67 shows a SYNDICATE/RKT implementation of the problem. An actor exists for each of the two temperature scales, maintaining an appropriate assertion and responding to `set-temperature` messages by performing necessary conversions before updating internal state. Temperature displays (not shown) may monitor the `temperature` assertions maintained by each actor, and user interface controls allowing temperature update should issue `set-temperature` messages with appropriate `unit` and `value` fields.

Each of the two actors shown acts as a *unidirectional* propagator of changes. The difference between `temperature` assertions and `set-temperature` command messages suffices to rule out confusion: a `set-temperature` message is always the cause of a change to the temperature, while an update to a `temperature` assertion is never the cause of a change; rather, it simply reflects some previous change. As Radul (2009) observes, "multidirectional constraints are very easy to express in terms of unidirectional propagators", and indeed the combination of the two actors ensures a bidirectional connection between the Fahrenheit and Celsius `temperature` assertions. However, we must ask whether we have truly entered into the spirit of the problem: by allowing the Celsius actor to interpret events expressed in Fahrenheit, and vice versa, our solution lacks the modularity and extensibility of the multidirectional solutions available in true dataflow languages.

Figure 68 addresses the problem, separating the equations relating the Celsius and Fahrenheit representations from the actors maintaining the representations. Each time some *distinct* assertion of `temperature` appears, a `set-temperature` message is sent. Even though it seems

13 The specific presentation of this section is inspired by that of Ingalls et al. (1988).

```
(spawn #:name 'track-celsius
       (field [temp 0])
       (assert (temperature 'C (temp)))
       (on (message (set-temperature 'C $new-temp)) (temp new-temp)))

(spawn #:name 'track-fahrenheit
       (field [temp 32])
       (assert (temperature 'F (temp)))
       (on (message (set-temperature 'F $new-temp)) (temp new-temp)))

(spawn #:name 'convert-C-to-F
       (on (asserted (temperature 'C $other-temp))
           (send! (set-temperature 'F (+ (* other-temp 9/5) 32)))))

(spawn #:name 'convert-F-to-C
       (on (asserted (temperature 'F $other-temp))
           (send! (set-temperature 'C (* (- other-temp 32) 5/9)))))
```

Figure 68: Modular synchronization between two temperature scales

like this may lead to unbounded chains of updates, activity will eventually quiesce because the two equations are inverses. After a time, the interpretation of a `set-temperature` message will lead to no observable change in a corresponding `temperature` assertion.

While our solutions thus far enjoy multidirectionality, they exhibit observable *glitching* (Bainomugisha et al. 2013). For example, just after a `set-temperature` message expressed in degrees Celsius has been interpreted, a moment in the stream of events exists when the corresponding Celsius `temperature` assertion has been updated but the Fahrenheit assertion has not yet incorporated the change. In general, any computation that depends on events traveling through the dataspace to peers (and perhaps back again) involves unavoidable latency, which may manifest as a form of glitching in some protocols. One approach to resolution of the problem is to bring the mutually-dependent stateful entities into the same location; that is, publish both Celsius and Fahrenheit from a single actor. If we do so, we may use any number of off-the-shelf techniques for avoiding glitching, including reactive DSLs such as FrTime (Cooper and Krishnamurthi 2006). However, this approach shuffles the problem under the rug, as the domain-specific assertion protocol no longer embodies a dataflow system in any meaningful sense. An alternative approach is to extend the assertions in our protocol with *provenance* information (a.k.a "version" information or tracking of causality) to form a more complete picture of transient states in a system's evolution (Radul 2009; Shapiro et al. 2011). That way, while the assertions themselves are able to (and do) represent not-yet-consistent intermediate states, *under interpretation* the incomplete states are ignored. Provenance information allows us to reason *epistemically* about flows of information in our protocols.

Part IV

# REFLECTION

# *Overview*

Every design demands evaluation. For a language design, this takes the form of the investigation of properties such as well-definedness, usefulness, and performance. While the formal models of Syndicate include basic theorems that characterize evaluation in Syndicate/$\lambda$, this part presents an evaluation of the practical aspects of the design.

To begin, chapter 9 examines the usefulness of Syndicate, presenting a qualitative evaluation of the design in terms of its effect on *patterns* in program texts.

Performance is the focus of chapter 10, which develops a Syndicate-specific performance model. Programmers can rely on this model in the design and evaluation of their programs and in the understanding of the programs of others.

Chapter 11 places Syndicate within the concurrency design landscape introduced in chapter 3, analyzing it in terms of the criteria developed in section 3.1.

Finally, chapter 12 reflects on the thesis of this dissertation and outlines a handful of promising directions for future work.

# 9

# *Evaluation: Patterns*

> A programming language is low level when its programs require attention to the irrelevant.
>
> —Alan J. Perlis (1982)

The evaluation of programming models and language designs is a thorny topic. Where a design has been realized into a full language, and where mature implementations of that language exist, we may examine quantitative attributes such as performance on a suite of benchmarks. Where many large programs written in a language exist, we may plausibly look into quantitative attributes such as error rates or programmer productivity. However, programming models are mathematical constructs, and novel language designs are abstract. Quantitative measures are inappropriate.

We are left with the investigation of *qualitative* attributes of our models and designs. A key quality is the extent to which a model or design eliminates or simplifies *patterns* in program texts, because other attributes improve as a consequence. In evaluating Syndicate through the lens of design and programming patterns, I aim to show that the design is effective in concisely achieving the effects of several such patterns.

## 9.1 PATTERNS

I use the term "pattern" to cover two related concepts. The first is the idea of a *programming pattern* in the sense discussed by Felleisen in his work on expressiveness (Felleisen 1991), synonymous with an *encoding* of an otherwise-inexpressible concept. The second is the idea of a *design pattern* from the object-oriented programming literature (Beck and Cunningham 1987; Gamma et al. 1994). That is, a "pattern" appears in a text not only when a specific design pattern is mentioned, but also in any situation in which an encoding is applied.

An encoding is a precise, potentially-automatable program transformation for representing some linguistic feature that cannot be expressed directly. An example of an encoding is *store passing* in a functional language to achieve the effect of *mutable state*. The precision of an encoding makes it possible to develop tooling or language extensions to assist the programmer in working with it. Seen from another angle, however, this same precision makes working with an encoding by hand an exercise in "boilerplate" programming. For example, a program with a manual implementation of store passing has entirely routine and predictable placement and usage of the variable representing the store. Errors frequently arise in such programs.

Approaches to automation such as macros, code generation, and monadic style help reduce this boilerplate and rule out errors, but cannot usually ensure complete adherence to the abstraction the encoding represents. For example, a monadic state library hides the explicit store from the program text, but unless a type system rich enough to enforce the necessary invariants is available, it remains possible for the programmer to misapply the library and throw into doubt the guarantees offered by the abstraction. In other words, encodings generally yield leaky abstractions.

The notion of a design pattern originated in architecture (Alexander et al. 1977), but has been successfully transplanted to object-oriented programming (Beck and Cunningham 1987) and can also be applied to other programming paradigms such as asynchronous messaging (Hohpe and Woolf 2004; Hohpe 2017). A design pattern, in an object-oriented context, "names, abstracts, and identifies the key aspects of a common design structure that make it useful for creating a reusable object-oriented design" (Gamma et al. 1994). Unlike an encoding, a design pattern is often not precise enough to be captured as either a library implementation or a language feature, but like an encoding, its manual expression often involves boilerplate code and the problems that go with it. The lack of precision often makes it difficult to provide tooling for working with design patterns *per se.*

## 9.2 ELIMINATING AND SIMPLIFYING PATTERNS

In order to see what it might mean for a pattern to be eliminated or simplified, we must first understand how patterns manifest in programs. Broadly speaking, a pattern is characterized by realization of a program organization *goal* in terms of some *mechanism*, which frequently involves boilerplate code. We see *recursive* use of patterns: implementation of the mechanism for achieving some goal entails organizational requirements of its own, which in turn demand satisfaction by some means. This can lead to towers of patterns. A pattern is eliminated by a programming model or language feature if it is provided directly or made unnecessary. A pattern can be simplified in two ways: in the case that its implementation depends on a tower of patterns, some supporting layer of that tower can be eliminated; or its implementation may be made more obvious by some part of the model or language feature.

For example, consider the task of maintaining a consistent graphical view on a list of items as items are added to and removed from the list. Our ultimate goal is the *synchronization of state* between the state of the on-screen view and the state of the underlying list. We might choose to use the *observer pattern* to accomplish our synchronization task by processing signals from the list as it changes. In turn, the observer pattern might be implemented using *callbacks*, which ultimately depend on *function calls*. In the early days of computing, "function call" was a design pattern. It has since been *eliminated* from most programming languages; this has *simplified* not only the implementation of callbacks, but also the observer pattern and our original goal of state synchronization. Adding language-level support for the observer pattern to the language, as languages like C♯ have begun to explore, *eliminates* the need for callbacks in our pattern tower, *simplifying* the expression of our goal.

As another example, the addition of support for the actor model to a language makes obsolete many uses of shared memory for communication among components. In this sense, the pattern of a *shared store* has been eliminated not by being provided directly, but by being made irrelevant by a shift in perspective to a new way of thinking.

Turning our attention to design patterns in the sense of Gamma et al. *per se*, Norvig offers three "levels of implementation" for patterns: "informal", "formal" and "invisible" (Norvig 1996). An "informal" implementation of a pattern is expressed in program text as prose comments naming the pattern alongside a from-scratch, manual implementation of the required, stereotypical elements of the pattern at every site where the pattern is needed. A "formal" implementation allows reuse by providing the pattern as a kind of library or language extension, often in the form of a suite of macros, invoked for each separate use of the pattern. Finally, an "invisible" implementation is "so much a part of [the] language that you don't notice" its presence. This taxonomy gives us another approach to the topic of elimination and simplification of patterns: we may say that a pattern is *simplified* when it moves from "informal" to "formal", and *eliminated* when it is made entirely "invisible".

## 9.3 SIMPLIFICATION AS KEY QUALITY ATTRIBUTE

A language which eliminates or simplifies patterns in program texts, concisely and robustly achieving their effects without forcing the programmer to spell them out in full detail, is qualitatively better than one which does not. This claim is supported in several ways.

First, Felleisen's *Conciseness Conjecture* (Felleisen 1991) states that the more expressive a programming language is, the fewer programming patterns one tends to observe in texts written in that language. Felleisen argues that this is important because "pattern-oriented style is detrimental to the programming process," observing that "the most disturbing consequence of programming patterns is that they are an obstacle to an understanding of programs for both human readers and program-processing programs." For example, store-passing style requires a reader to analyze an entire program to learn whether the store has been properly propagated, accessed, and updated. Worse, certain encodings can have more than one interpretation, and determining which is intended requires analysis of fine detail of the text. Felleisen gives the example of continuation-passing style in a call-by-name language, which may encode either unusual control structure or a call-by-value protocol. Automated tooling suffers in a similar way: even with the precision offered by encodings, the global analyses required can be daunting. Tooling is also at a disadvantage compared to a human reader, since the human is able to read and understand comments conveying the *intent* behind a piece of code, while the tool is left to reason from the structure of the code alone. Turning to design patterns from encodings, we see that the problems of analysis are only made worse. The imprecision of design patterns forces humans and automated tools alike to make approximate guesses as to the intended design-pattern-level meaning of a particular piece of code.

Second, Felleisen's ideas surrounding expressiveness are formal reflections of more informal ideas of the quality of a given programming language, alluded to in the Perlis quote at the top of this chapter and discussed by researchers such as Brinch Hansen (Brinch Hansen 1993) and Hoare (Hoare 1974). Hoare, in an early and influential paper on programming-language design, writes that a programming language should give a programmer "the greatest assistance in the most difficult aspects of his art, namely program design, documentation, and debugging," and that "a necessary condition for the achievement of any of these objectives is the utmost simplicity in the design of the language" (Hoare 1974). Brinch Hansen, who frequently collaborated with Hoare, suggested that the primary contribution that a language makes toward achievement of this simplicity is "an abstract readable notation that makes the parts and structure of programs obvious to a reader," and goes on to say that "a programming language should be abstract":[1]

> An abstract programming language *suppresses machine detail* [...] [and] relies on *abstract concepts* [...] We shall also follow the crucial principle of language design suggested by Hoare: *The behavior of a program written in an abstract language should always be explainable in terms of the concepts of that language and should never require insight into the details of compilers and computers.* Otherwise, an abstract notation has no significant value in reducing complexity. (Brinch Hansen 1993, emphasis in original)

An abstract language, then, achieves "simplicity" in that the programmer's ideas find direct expression in terms of the language itself, rather than indirect expression in terms of "machine detail". This allows the programmer to reason in terms of the ideas rather than their representation. This is directly analogous to the relationship Felleisen remarks on between highly expressive languages and the programming patterns they suppress: a language able to avoid the need for programming patterns is abstract, i.e. good, in the sense of Brinch Hansen.

Third, the fields of Software Architecture and Software Engineering evaluate systems in terms of *quality attributes* (Bass, Clements and Kazman 1998; Clements, Kazman and Klein 2001), or so-called "-ilities", named for the common suffix of attributes such as maintainability, stability, portability, and so forth. While these attributes are, strictly speaking, only applicable to software architectures and not to programming models, they are not without value in our setting. Many "-ilities" benefit immediately from program pattern elimination. For example, *modifiability* depends on the programmer being able to understand the scope of a particular change: a global encoding of some pattern interferes with this aim. Likewise, *understandability* of a program hinges on concision and expressiveness, on the programmer's ability to say what they mean. In general, improvements in concision and expressiveness, and reduction of pattern boilerplate, should lead to improvements in terms of several frequently-discussed "-ilities". In the analysis to follow I illustrate specific points of connection between the SYNDICATE model and both general and scenario-specific "-ilities".

Finally, some small support for the claim of this section comes from previous analysis of design patterns in context of their implementation in various programming languages. Norvig reports on a study of the Design Patterns book of Gamma et al. in which 16 of the 23 patterns described in the book either find "qualitatively simpler implementation" or become en-

1 The full principle ends with "... and secure", later defined in terms that make it essentially a synonym of "abstract". This sense of "secure" was originally introduced by Hoare (1974).

tirely "invisible" when comparing implementations in Lisp or Dylan with implementations in C++ (Norvig 1996).

## 9.4 EVENT BROADCAST, THE OBSERVER PATTERN AND STATE REPLICATION

The *observer* pattern is a mainstay of object-oriented programming languages. Originating with Smalltalk,[2] its purpose is given by Gamma et al. (1994) as "a one-to-many dependency between objects so that when one object changes state, all its dependents are notified and updated automatically." Its intent is to communicate *state changes* from a *subject* to a set of *observers*.

The observer pattern frequently finds expression as part of a tower of patterns. Supporting it we find an *event broadcast* facility of some kind, and instances of the observer pattern in turn are often used to implement *state replication*. The three patterns differ in intent. State replication is used to synchronize disparate views on some stateful entity, while the observer pattern focuses on the fact of a *change* in a stateful entity, and event broadcasting is merely the vehicle by which some signal is delivered to a group of recipients. Roughly speaking, state replication is the integral to the observer pattern's differential, and event broadcast is a generic message transport mechanism. In particular, a state replica starts with an initial snapshot, while there is no such requirement for an observer requesting change-notifications from a subject.[3]

Event Broadcast<br>↓<br>Observer<br>↓<br>State Replication

Many popular programming languages include implementations of the observer pattern in their standard library, yielding what Norvig terms a "formal" implementation level. Others, however, make use of the pattern without a library implementation, yielding an "informal" implementation. One layer up our tower of patterns, state replication seldom is supported other than "informally", and one layer down, event broadcasting is often somewhere between "invisible" and informal-but-idiomatic.

A running example shows the patterns in action. The example involves a display of a set of names of users present in an online chat room. The display is to update itself as users arrive or depart the room, starting with the list of users present at the moment the display is initialized. The underlying set of users is the subject, and the display is an observer. The example embodies *state replication* in that the view, as it is created, interrogates the subject for its current members, and uses observer-pattern notifications as indications that it should incorporate some change that has just taken place.

2 ANSI Smalltalk does not include the "dependents" protocol because "there is nothing defined by the standard that requires any kind of dependency mechanism." (X3J20 Committee for NCITS 1997) However, inspection of a September 1986 source listing of Smalltalk-80 shows the dependents protocol in the form in which it survives in most Smalltalks today.

3 Hohpe (2017) has dubbed the distributed systems analogue of the observer pattern "the Subscribe-Notify conversational pattern". Surprisingly, "state replication" does not appear to be well-attested as a design pattern per se, either in the traditional OO or asynchronous messaging realms. The notion of event broadcasting appears in the asynchronous messaging literature simply as the role played by a message broker (Hohpe and Woolf 2004; Eugster, Guerraoui and Damm 2001).

```
1  nil subclass: Object [
2    Dependencies := nil.
3    Object class >> initialize [
4        self == Object ifFalse: [^self].
5        Dependencies := WeakKeyIdentityDictionary new.
6    ]
7    addDependent: anObject [
8        ^(Dependencies at: self ifAbsentPut: [OrderedCollection new]) add: anObject
9    ]
10   removeDependent: anObject [
11       | dependencies |
12       dependencies := Dependencies at: self ifAbsent: [^anObject].
13       dependencies remove: anObject ifAbsent: [].
14       dependencies size < 1 ifTrue: [Dependencies removeKey: self ifAbsent: []].
15       ^anObject
16   ]
17   changed: aParameter [
18       | dependencies |
19       dependencies := Dependencies at: self ifAbsent: [nil].
20       dependencies notNil ifTrue: [dependencies do: [:d | d update: aParameter]]
21   ]
22   changed [
23       self changed: self
24   ]
25   update: aParameter [
26       "Default behavior is to do nothing. Called by #changed and #changed:"
27   ]
28 ]
```

Figure 69: The GNU Smalltalk implementation of the classic Smalltalk dependents mechanism. (Excerpted from GNU Smalltalk version 3.2.91, ©1990–2015 Free Software Foundation, Inc.)

SMALLTALK. Figure 69 is the GNU Smalltalk library implementation of the classic Smalltalk "dependents" protocol. As part of the standard library, it fits Norvig's criterion for a "formal" instance of the observer pattern. Line 1 establishes a context in which we are supplying definitions for class `Object`. Line 2 declares a *class variable*, `Dependencies`, on class `Object`, initially with value `nil`. The constructor for class `Object` (lines 3–6) places a new `WeakKeyIdentityDictionary` in the class variable.

The class variable `Dependencies` contains a dictionary mapping each *subject* to an `OrderedCollection` of *observers*. The methods `addDependent:` and `removeDependent:` maintain the structure. Line 20 in the method `changed:` is the heart of the implementation. It sequentially visits each observer in turn, invoking the `update:` method on each with the given parameter value. Idiomatic Smalltalk code conventionally uses the nullary method `changed` (lines 22–24) to supply the subject itself as the parameter value of a change notification. The definition of `update:` within class `Object` ensures that *every* object in the system can act as an observer in this protocol.

```
1 Object subclass: UserList [
2   | users |
3   initialize      [ users := Set new. ]
4   userArrived: u  [ users add: u.    self changed. ]
5   userDeparted: u [ users remove: u. self changed. ]
6   users           [ ^users ]
7 ]

8 Object subclass: UserListDisplay [
9   | userList prevUsers |
10  userList: ul [
11    prevUsers := Set new.
12    userList := ul.
13    userList addDependent: self.
14    self update: userList.
15  ]
16  update: anObject [
17    | new old |
18    new := userList users - prevUsers.
19    old := prevUsers - userList users.
20    new do: [:u | Transcript nextPutAll: u, ' arrived.'; cr ].
21    old do: [:u | Transcript nextPutAll: u, ' departed.'; cr ].
22    prevUsers := userList users copy.
23  ]
24 ]
```

Figure 70: A GNU Smalltalk program making use of the "dependents" protocol.

State replication *per se* does not appear as a library pattern in Smalltalk. Instead, it appears as an informal pattern, implemented on a case-by-case basis, often making use of the "dependents" protocol. We see an almost-"invisible" instance of event broadcasting on line 20 of figure 69 in the loop that delivers a call to the `update:` method of each observer object.

Figure 70 shows a sketch of our example application. Class `UserList` implements the subject portion of the design, and class `UserListDisplay` implements the observer. The methods `userArrived:` and `userDeparted:` invoke the `UserList`'s `changed` method (lines 4 and 5) to notify observers that something has changed. In response, the `update:` method of `UserListDisplay` is run. Because the Smalltalk convention is to simply convey the *fact* of a change rather than any detail, `update:` must determine precisely what has changed in order to produce correct incremental output. The explicit call to `update:` on line 14 initializes the display with the set of users present at the time the display is created; if line 14 were omitted, no display updates would happen until the first time the `UserList` `changed`. Finally, the need for `copy` on line 22 is subtle. The `users` method of `UserList` returns a *reference* to the underlying set collection object, and collections in Smalltalk are imperatively updated. If `copy` on line 22 is omitted, the code fails to detect any changes after the initial update.

```
1  module Observable
2    def add_observer(observer, func=:update)
3      @observer_peers = {} unless defined? @observer_peers
4      @observer_peers[observer] = func
5    end
6    def delete_observer(observer)
7      @observer_peers.delete observer if defined? @observer_peers
8    end
9    def changed(state=true)
10     @observer_state = state
11   end
12   def notify_observers(*arg)
13     if defined? @observer_state and @observer_state
14       if defined? @observer_peers
15         @observer_peers.each do |k, v| k.send v, *arg end
16       end
17       @observer_state = false
18     end
19   end
20 end
```

Figure 71: The Ruby implementation of the observer pattern. (Excerpted from the Ruby standard library version 2.5.0, ©1998–2017 Yukihiro Matsumoto and other contributors.)

RUBY. Figure 71 is the Ruby standard library implementation of the observer pattern as a *mixin* module supplying the pattern's subject-side behavior. This, too, is a "formal" implementation of the pattern in Norvig's sense. The code relies on the Ruby idiom of dynamic addition of instance variables to individual objects, creating the `@observer_peers` collection if it does not exist at `add_observer` time (line 3). The implementation is more general than the Smalltalk implementation in that each registered observer may optionally specify a method name to invoke. By default, an observer's `:update` method will be called.

The implementation takes care to avoid unnecessary signals, delivering notifications from `notify_observers` only if the subject's `changed` method has been called since the previous `notify_observers` call. This can allow the programmer to *batch* multiple changes together, sending a single notification at an opportune time after a number of changes have taken place. While this technique is fragile unless the programmer is able to maintain tight control over the sequence of events at the subject, it can provide a form of *atomicity* for batched changes.

Like Smalltalk, Ruby does not formally support the state replication pattern as such. When it comes to event broadcast, line 15 is an almost-"invisible" implementation of event broadcast that is practically identical to the analogous Smalltalk idiom.

We omit a Ruby implementation of our `UserList` example, as it is substantially the same as the Smalltalk program, excepting the need for addition of a call to `notify_observers` after each call to `changed`.

```
public interface EventListener {}
public interface ChangeListener extends EventListener {
    void stateChanged(ChangeEvent e);
}
public class EventObject {
    protected transient Object source;
    public EventObject(Object source) { this.source = source; }
    public Object getSource() { return source; }
}
public class ChangeEvent extends EventObject {
    public ChangeEvent(Object source) { super(source); }
}
```

Figure 72: Abbreviated portion of an observer pattern from the Java Swing library. (Excerpted from the OpenJDK 7 source code, ©1996–2003 Oracle and/or its affiliates.)

JAVA. Java supplies a plethora of data classes and interfaces implementing variations on the observer pattern. Figure 72 shows one example from the standard Swing library. An observer is to implement the `ChangeListener` interface; its single method, `stateChanged`, is invoked by the subject when the relevant change occurs. The argument to `stateChanged` is a `ChangeEvent` bearing a single field, `source`, which by convention is a reference to the subject itself. Each variation on an `EventListener` in the Swing library comes with a corresponding subclass of `EventObject` carrying relevant details of a change, and may have more than one required handler method. For example, a `ListDataListener` implementation must respond to `intervalAdded`, `intervalRemoved` and `contentsChanged` events, each taking a `ListDataEvent`.

No public utility classes are made available to assist in the implementation of the subject role. This fact, along with the many variations and re-implementations of the pattern found both in the standard libraries and in third-party libraries and applications, leads us to the conclusion that Java offers only "informal" support for the observer pattern.

Just as in Smalltalk and Ruby, no formal support for state replication is on offer in Java. Inspection of the uses of the private `EventListenerList` class central to the Swing instances of the observer pattern shows that, again just as in Smalltalk and Ruby, idiomatic Java subjects iterate over a collection object in order to broadcast change notifications.

ERLANG/OTP. With Erlang/OTP we take a step away from shared-memory concurrency and move to a shared-nothing setting with strongly isolated processes. Here the distinction between the observer and state-replication patterns becomes more noticeable. Figure 73 shows the key portions of an actor implementing the `UserList` portion of our running example.

Erlang/OTP does not offer a library implementation of the observer pattern or of state replication,[4] making these patterns "informally" implemented in Norvig's terms. It does provide a "formal" library implementation of event broadcast called `gen_event`. The interface to `gen_event` requires a separate module for each kind of event callback; no standard callback imple-

4 In particular, the `observer` module is a graphical debugging tool, unrelated to the observer pattern.

```
1 -record(state, {users, listeners}).

2 init([]) ->
3     process_flag(trap_exit, true),
4     {ok, #state{users = [], listeners = []}}.

5 handle_info({arrive, Name}, State = #state{users = Users, listeners = Listeners}) ->
6     [L ! {arrived, self(), Name} || L <- Listeners],
7     {noreply, State#state{users = [Name | Users]}};
8 handle_info({depart, Name}, State = #state{users = Users, listeners = Listeners}) ->
9     [L ! {departed, self(), Name} || L <- Listeners],
10    {noreply, State#state{users = [N || N <- Users, N =/= Name]}};
11 handle_info({sub, Pid}, State = #state{users = Users, listeners = Listeners}) ->
12    link(Pid),
13    [Pid ! {arrived, self(), Name} || Name <- Users],
14    {noreply, State#state{listeners = [Pid | Listeners]}};
15 handle_info({unsub, Pid}, State = #state{listeners = Listeners}) ->
16    unlink(Pid),
17    {noreply, State#state{listeners = [P || P <- Listeners, P =/= Pid]}};
18 handle_info({'EXIT', Pid, _Reason}, State = #state{listeners = Listeners}) ->
19    {noreply, State#state{listeners = [P || P <- Listeners, P =/= Pid]}}.
```

Figure 73: An Erlang `UserList` program using the observer pattern.

mentation for the common case of sending an inter-actor message per event is provided. In addition, incorporating a `gen_event` broadcast mechanism as part of the behavior of a stateful actor is awkward, because each event source is implemented with a separate process. These additional processes must be managed carefully to avoid resource leaks, complicating what would ideally be a simple idiom. The `gen_event` module is seldom used outside of specialized situations, perhaps for these reasons. In cases where the observer pattern is appropriate, Erlang programmers generally prefer to roll their own broadcast mechanisms on a case-by-case basis.[5]

The code in figure 73 does just this. The figure shows part of a module implementing an Erlang/OTP `gen_server` service actor; the `init/1` function acts as constructor, `handle_info/2` handles messages delivered to the actor, and the `state` record declaration on line 1 specifies the structure used as the actor's private state value. The actor keeps track of a list representing a set of user names, as well as a list representing a set of observer process IDs.

The actor implements two protocols: one corresponding to the `UserList` protocol we saw on lines 4–5 in the Smalltalk example program (figure 70), and one corresponding to a use of the observer pattern to provide state replication. Lines 5–10 take care of the former, while lines 11–19 handle the latter.

Erlang/OTP `gen_server` actors, like Dataspace ISWIM actors, are functional event transducers. Each arriving message is passed to `handle_info/2` along with the actor's private state

5 Based on my first-hand experience of more than a decade of participation in the Erlang community.

value.[6] The actor is expected to return a functionally-updated state value along with an instruction regarding a possible reply. In response to an `{arrive, Name}` tuple message (line 5; analogous to the Smalltalk program's `userArrived:` method) our actor broadcasts a message to its current subscriber list (line 6), making use of a list comprehension containing an asynchronous message send instruction. It then returns an updated `state` record (line 7), placing the new `Name` at the head of the `users` list. Similarly, a `{depart, Name}` message results in a broadcast and an update removing the `Name` from the `users` list (lines 8–10).

Without line 13, this program would be closer to the observer pattern as described by Gamma et al. than to an instance of state replication. To explain, we must examine the entirety of the clause of lines 11–14. Line 11 matches a subscription request message, `{sub, Pid}`, carrying the process ID of an observing actor. Not only does our service add the new `Pid` to its `listeners` list, but it also supplies the new subscriber with a snapshot of the relevant portion of server state, i.e., the `users` list. It does so using the same protocol it uses for announcing subsequent incremental updates to the list.[7] There is an asymmetry here: if we announce the "arrival" of already-present users when a subscriber joins, we might expect it to be reasonable to announce the "departure" of those same users when a subscriber unsubscribes. However, no analogue of line 13 is present in the `unsub` clause (lines 15–17).

Part of the motivation for moving beyond the traditional scope of an observer pattern implementation, and toward a richer state replication design, is the strict "shared-nothing" isolation of Erlang processes. In shared-memory languages like Smalltalk, Ruby, and Java it makes sense for an observer to immediately interrogate the subject to access its current state; the two are co-located, and a simple call to a getter method suffices. In Erlang, however, analogous retrieval of the user list by an observer via RPC would not only be expensive, but could introduce concurrency bugs: the latency of the round-trip introduces an unavoidable lag during which further changes in the state of the processes in the system could take place. Conveying the relevant public aspects of the subject's state along with the change-notifications themselves elegantly solves this problem. It also obviates the need for anything like the call to `copy` we saw on line 22 of figure 70. Finally, this subtle shift in the implementation of the observer pattern in shared-nothing languages provides a clue that many uses of the pattern might be better thought of as mere mechanisms for state replication, rather than as ends in themselves.

We conclude our discussion of Erlang with investigation of lines 12, 16, and 18–19 of figure 73. These subscribe to, unsubscribe from, and react to notifications of process termination, respectively.[8] The call to `link` on line 12 ensures that if an observer terminates, either cleanly or with an exception, the subject receives a notification message called an "exit signal" in Erlang parlance. The call to `unlink` on line 16 cancels this subscription, and the clause of lines 18–19 treats receipt of an "exit signal" describing the termination of an observer as an implicit unsubscription. An understanding of this message-focused approach to error propagation allows us to see lines 6 and 9 in a new light. They are superficially similar to the one-line implementations of the event broadcast pattern seen in the Smalltalk (figure 69, line 20) and Ruby (figure 71,

6 I have used `handle_info/2` for simplicity. A real implementation would prefer `handle_call/3` and `handle_cast/2`.
7 An alternative implementation might use a special "initial snapshot" message format instead.
8 Line 3 is a standard incantation required by the language to ensure that exit signals are delivered as messages. Omitting line 3 would have the effect of causing our subject actor to crash if an observer process crashes.

```
1 (assertion-struct user-present (list-id user-name))

2 (define (spawn-user-list-display list-id)
3   (spawn (during (user-present list-id $user-name)
4            (on-start (printf "~a arrived.\n" user-name))
5            (on-stop  (printf "~a departed.\n" user-name)))))
```

Figure 74: A Syndicate/rkt UserList program.

```
(spawn (during/spawn (tcp-connection $id (tcp-listener 5999))
         (assert (tcp-accepted id))
         (on-start (send! (tcp-out id "What is your name? "))
                   (react (stop-when (message (tcp-in-line id $name))
                            (send! (tcp-out id (format "Hello, ~a!\n" name)))
                            (react (assert (user-present 'room1 name))))))))
```

Figure 75: Syndicate/rkt TCP service interacting with figure 74

line 15) library code. The key difference is that in Smalltalk and Ruby each notification is a synchronous method call without error handling. An exception from an observer will cause the remaining observers to miss their notification, and may damage the subject. Here, each notification is an asynchronous message send, which (in Erlang) never results in an exception. Error handling is separated into the code dealing with links and exit signals.

Syndicate. Finally, let us examine state replication, the observer pattern, and event broadcast in Syndicate. The design of Syndicate incorporates a number of lessons from the Erlang approach, but goes beyond it by placing state replication front and center in the programmer's mental model. The Syndicate design proceeds from the assumption that the *intent* of achieving state replication is more frequent than the intent of achieving the observer pattern, let alone a raw event broadcast. The language thus offers prominent, explicit linguistic support for sharing of public aspects of an actor's private state. Figure 74 implements the Syndicate/rkt equivalent of the Smalltalk program of figure 70.

An immediate difference is that class `UserList` is completely absent, appearing in vestigial form only in the declaration of the `user-present` record type (line 1). The function `spawn-user-list-display` is comparable to the Smalltalk class `UserListDisplay`. It observes public aspects of the state of the user list, reacting to appearance or disappearance of set elements with appropriate print commands. In the Smalltalk example, we imagined a component whose role was to call `userArrived:` and `userDeparted:` appropriately for each separate user. The Syndicate program cuts out this intermediary. Instead, an actor responsible for signaling presence of a particular user asserts a `user-present` record for the appropriate duration, and that record is directly communicated to observers by the dataspace. For example, adding the code in figure 75 to figure 74 causes the program to accept TCP/IP connections, ask for a name, send a greeting, and assert the connected user's presence in the user list until disconnection.

| | Smalltalk | Ruby | Java | Erlang/OTP | Syndicate |
|---|---|---|---|---|---|
| Event broadcasting | informal/invisible | informal/invisible | informal/invisible | informal/invisible | invisible |
| Observer pattern | formal | formal | informal | informal | invisible |
| State replication | informal | informal | informal | informal | invisible |

Figure 76: Levels of implementation for state replication, the observer pattern, and event broadcasting

The dataspace connecting actors to each other takes on the role that was played by class `UserList`, keeping observers up-to-date as relevant state changes, cleaning up subscriptions on exit, handling failures, and so on. The notion of a subject has become diffuse and domain-specific, rather than being tightly bound to the identity of a single object in the system. The state replication pattern has become "invisible" in Norvig's sense.

Object-oriented languages usually offer a notion of object *identity* that can be used as a marker for a specific topic of conversation. Syndicate does not offer anything like this. Instead, Syndicate encourages the programmer to take a relational view of shared state and demands explicit treatment of identity. The approach is similar to use of primary keys in relational databases. The programmer is free to choose a notion of identity appropriate to the domain.

Syndicate emphasizes state replication, but does not preclude use of the observer pattern. Not all uses of the observer pattern are intended to support state replication. The observer pattern is, like state replication, "invisible" in Syndicate. All that is required is for the subject to send change-notification messages with an appropriate structure,

`(send! (change-notification-record` *subject-id change-details*`))`

optionally also placing more aspects of its state into the dataspace as assertions or responding to RPC state queries. Observers express interest in such notifications in the usual way. Finally, the event broadcast pattern is also completely "invisible", as it is provided directly by the mechanics of the dataspace model.

Recall the asymmetry remarked upon earlier in the Erlang program of figure 73. When a new observer subscribes, the subject synthesizes "arrival" messages describing the users already present in the room, but sends no analogous "departure" messages to an unsubscribing observer. As we saw in section 4.5, the dataspace ensures that each actor is sent events describing assertions added to or removed from the intersection of the group's assertion set and the specific interests of the actor itself. This set changes *either* when a peer makes or removes assertions, *or* when the actor asserts or retracts interests. The dataspace makes no distinction, relaying changes in the relevant set no matter the cause. Thus, unusually, Syndicate is symmetric in the exact way that we observed that our Erlang subject actor is not. When a subscriber retracts interest in a set of assertions, the dataspace issues a state change event that correspondingly removes any extant matching assertions from that actor's view of the world.

ANALYSIS. Implementations of the observer pattern vary widely between languages and scenarios within a language. While figure 76 summarizes the situation in terms of Norvig's

"levels of implementation," we must step beyond this and consider practical concerns, which make the following questions relevant to the programmer.

1. What information is conveyed as part of the signal delivered from a subject to an observer?
   a) Is the entity which changed identified?
   b) Is the aspect of that entity which changed identified?
   c) Is there room for a detailed description of the particular change?
2. How does the implementation interact with garbage collection?
3. How does the implementation interact with errors?

The Smalltalk implementation allows the subject to send a single object to observers. This is conventionally a reference to the subject itself, but may be any object. A strong secondary convention, when multiple aspects of a subject may change, is to use a selector[9] as the notification payload. This has clear weaknesses: it no longer reliably identifies the subject, making it potentially challenging for a single observer to observe *multiple* subjects at once, and it is a simple atom with no room for additional detail.

The Ruby implementation is more flexible. Firstly, and most importantly, no matter the notification payload transmitted by the subject, each observer is given the opportunity to direct notifications on a per-subscription basis to specific entry points by passing an optional second argument ("`func`") to the `add_observer` method. Secondly, the subject may invoke `notify_observers` with any number of any type of arguments. These are passed on as arguments to each observer's chosen handler method. The Erlang implementation, "informal" as it is, is similarly flexible. No particular notification format or payload is required.

The SYNDICATE implementation is likewise flexible, but for a different reason. The specifics of any information communicated from a subject to observers is part of the ordinary SYNDICATE protocol design for the group. If the identity or nature of the entity which changed is relevant to the protocol, some value denoting it will be included in each assertion and message; likewise for the aspect which changed and any specific details of a given change.

The interactions between the observer pattern and garbage collection are straightforward to explain, but can be difficult to address in realistic programs. Consider the Ruby implementation of the `Observable` module (figure 71). Its use of an ordinary dictionary object establishes strong references to its observers. In cases where an observer becomes otherwise unreachable, the burden is on the programmer to explicitly break the connection between the two to avoid resources leaks or unwanted notifications. The situation is identical in the Java Swing `EventListenerList` subject implementation and is similar in Smalltalk, where a *global* dictionary with weakly-held keys achieves the same effect as Ruby's per-subject instance variable. In all three cases, care must be taken by the programmer to avoid accidental reference cycles and to develop a rigorous understanding of the lifecycles of all the objects involved.

9 Smalltalk's "selectors" are Lisp's "symbols".

Erlang and Syndicate, however, take a different approach. Actor lifetimes in both languages are under explicit programmer control. Despite this, there are no problems with dangling references. In the case of Erlang, such references are cleaned up as part of a subject's reaction to exit signals from terminated observers. However, solicitation of and responses to exit signals must be explicitly specified by the programmer. In Syndicate, fine-grained conversational *frames* associated syntactically with *facets* allow subjects and observers to precisely and automatically delimit the scope and duration of relationships. In the example, observers of `user-present` assertions may see them retracted due to an explicit *or* implicit retraction, as a conversation comes to an end, or a facet or entire actor terminates. Both Erlang and Syndicate are *symmetric* in that not only may subjects monitor their observers' lifecycles, but observers may also attend to the presence of observed subjects. In Erlang, this is achieved with links; in Syndicate, by the guaranteed availability and visibility of assertions of interest alongside other assertions. An observer may express interest in `(user-present` *id* `_)` assertions; a subject may express interest in `(observe (user-present` *id* `_))` assertions.

Finally, in every implementation of event broadcasting we have seen in object-oriented languages, the same error handling problems arise. An exception signaled by an observer's callback method will by default "spill over" into the context of the *subject*, potentially damaging it, even though it is the *observer* at fault. Worse, if the failing observer is in the middle of the subscriber list, entries in the list following the failure will not receive the notification. Error propagation with stack discipline in a situation where different *segments* of the stack belong to different *components* in a group is inappropriate.

Erlang and Syndicate both do better. Erlang's links and exit signals allow non-linear propagation of failure signals along graphs of components. Syndicate generalizes the idea of Erlang's links, observing that the "liveness" attribute of an actor is just another piece of public state, representable as an assertion like anything else. All of a terminating actor's assertions are automatically withdrawn in the dataspace model; those that describe a "liveness" property of interest can be monitored like any other. This dovetails with the notion of a conversational frame again, where presence of an assertion will frequently delimit a (sub)conversation. The assertions removed as a peer crashes act like exit signals in that they cause well-defined events to be delivered to conversational counterparties.

Beyond those three basic questions, some general issues with the observer pattern are worth highlighting. First, by encoding event dispatch among concurrent components as ordinary synchronous method call, common implementations make maintainability of and visibility into a design employing the observer pattern difficult. Programmers must determine for themselves the boundaries between ordinary code and event-driven code, and must reconstruct networks of interacting components by careful inspection of the details of their implementations. Syndicate separates protocol design into a separate programming phase, allowing maintenance of each protocol specification as an artifact of its own, and allowing development of tools specialized for visualization of dataspace traffic, thereby improving maintainability and visibility for concurrent programs. Second, the granularity of event selection in most implementations is coarse; in Smalltalk, for example, the granularity is usually at the level of an entire object. Observers must both filter and demultiplex their notifications to determine whether and, if so, how a particular change is relevant to them. Syndicate allows filtering of infor-mation to granularity limited only by the protocol design, and demultiplexes incoming events precisely to individual handler clauses in facet endpoints, thereby improving specificity and efficiency of communication in a concurrent program. Third, it is embarrassingly common when programming with the observer pattern in a synchronous, sequential object-oriented language to accidentally cause an infinite loop of mutual change-notifications, because such notifications do not include enough information to determine whether they are redundant. Syndicate only delivers notifications to observers when a true change is made to the contents of the dataspace; that is, updates in Syndicate are automatically idempotent, thereby improving robustness and reliability of concurrent programs. Finally, as we saw in the case of Java, “informal” implementations of the pattern lead to multiplication of effort with concomitant multiplication of bugs. By bringing state replication into the language once and for all, Syndicate rules out the possibility of competing, inconsistent implementations of the same idea, thereby improving understandability and maintainability of programs.

## 9.5 THE STATE PATTERN

The *state* pattern is a technique used in certain object-oriented languages to simulate the `become` operation from the original actor model (Hewitt, Bishop and Steiger 1973). Gamma et al. write that a use of the pattern allows an object “to alter its behavior when its internal state changes,” and that the object “will appear to change its class” (Gamma et al. 1994). Languages like Self that support dynamic inheritance do not need the pattern: an update to a so-called *parent* slot automatically adjusts the available state and behavior of an object (Ungar et al. 1991). This shows that it is possible for an “invisible” implementation of the pattern to exist. Languages like Java and C++, where an object’s interface and class are fixed for its lifetime, are where the pattern finds most application.

A state machine representing a video game character’s response to key press and release events exemplifies the pattern.[10] When the player is standing still, pressing the `JUMP` key causes the player to start a jump sequence. If, in mid-air, the `DOWN` key is pressed, the player should transition into a dive. However, when standing still, the `DOWN` key causes the player to move into a ducking stance. While ducking, release of the `DOWN` key reverts to the standing state. Each state should have associated with it a specific visual appearance (sprite) for the player character.

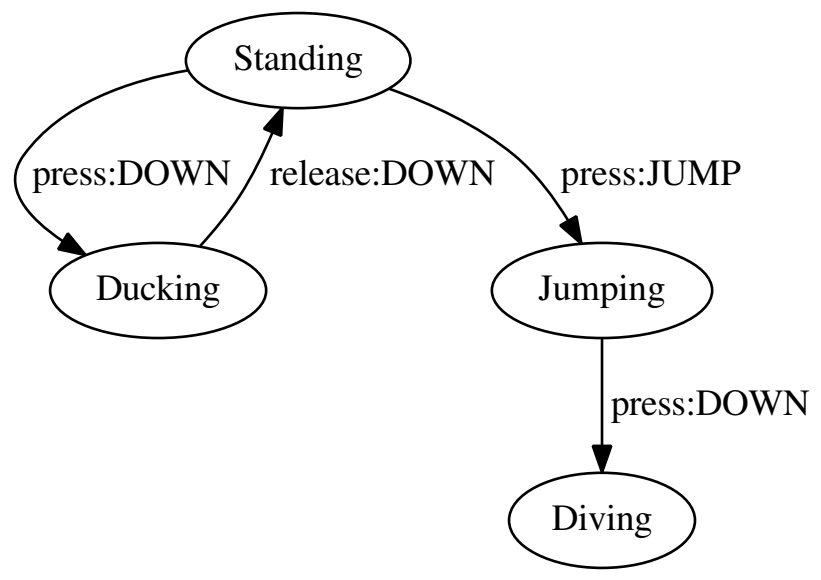

JAVA. Java implementations of the state pattern are “informal”. The Java program of figure 77 sketches a state pattern based implementation of the example state machine. Key state pattern characteristics are replication of a suite of methods, once in a “wrapper” class (`PlayerWrapper`), and again in an interface (`KeyHandler`) implemented by each “state” class. A separate “state” class is created for each state in the state machine. The interface (and “state” class) version of

10 The example is due to Nystrom (2014), also available at http://gameprogrammingpatterns.com/state.html.

```
interface KeyHandler {
    void handlePress(PlayerWrapper p, Key k);
    void handleRelease(PlayerWrapper p, Key k);
}

class PlayerWrapper {
    KeyHandler state = new StandingState(this);
    public void handlePress(Key k)   { state.handlePress(this, k); }
    public void handleRelease(Key k) { state.handleRelease(this, k); }
    public void setSprite(Sprite s) { /* ... */ }
}

class StandingState implements KeyHandler {
    public StandingState(PlayerWrapper p) { p.setSprite(Sprite.STANDING); }
    public void handlePress(PlayerWrapper p, Key k) {
        if (k == Key.JUMP) p.state = new JumpingState(p);
        if (k == Key.DOWN) p.state = new DuckingState(p);
    }
    public void handleRelease(PlayerWrapper p, Key k) {}
}

class JumpingState implements KeyHandler {
    public JumpingState(PlayerWrapper p) { p.setSprite(Sprite.JUMPING); }
    public void handlePress(PlayerWrapper p, Key k) {
        if (k == Key.DOWN) p.state = new DivingState(p);
    }
    public void handleRelease(PlayerWrapper p, Key k) {}
}

class DuckingState implements KeyHandler {
    public DuckingState(PlayerWrapper p) { p.setSprite(Sprite.DUCKING); }
    public void handlePress(PlayerWrapper p, Key k) {}
    public void handleRelease(PlayerWrapper p, Key k) {
        if (k == Key.DOWN) p.state = new StandingState(p);
    }
}

class DivingState implements KeyHandler {
    public DivingState(PlayerWrapper p) { p.setSprite(Sprite.DIVING); }
    public void handlePress(PlayerWrapper p, Key k) {}
    public void handleRelease(PlayerWrapper p, Key k) {}
}
```

Figure 77: State pattern example in Java

```
(assertion-struct key-down (key))
(assertion-struct player-sprite (variation))

(define (standing-state)
  (react (assert (player-sprite 'STANDING))
         (stop-when (asserted (key-down 'JUMP)) (jumping-state))
         (stop-when (asserted (key-down 'DOWN)) (ducking-state))))

(define (jumping-state)
  (react (assert (player-sprite 'JUMPING))
         (stop-when (asserted (key-down 'DOWN)) (diving-state))))

(define (ducking-state)
  (react (assert (player-sprite 'DUCKING))
         (stop-when (retracted (key-down 'DOWN)) (standing-state))))

(define (diving-state)
  (react (assert (player-sprite 'DIVING))))

(spawn #:name 'player
       (on-start (standing-state)))
```

Figure 78: State pattern example in Syndicate/rkt

each method takes an additional argument referencing the "wrapper". The "wrapper" class version of each method directly delegates to the current "state" object. In more complex situations, a "state" class may use instance variables of its own to keep track of information relevant during its tenure.

Syndicate. The Syndicate/rkt program shown in figure 78 implements the same state machine using facet *mixins*, abstractions of units of behavior and state that are named and reusable in multiple contexts. Recall from section 6.4 that Syndicate/rkt allows abstraction over facet creation using ordinary procedures. Here, each state becomes a separate *facet* rather than a separate *class*, abstracted into its own procedure so that it may be reused as state transitions take place. Events arriving from the dataspace trigger these transitions: use of `stop-when` ensures that the active state facet *terminates*, and the handler invokes a procedure that ensures that a new state facet replaces the old. The player actor dynamically includes an appropriate starting facet at initialization time. In more complex state machines, lexical variables and facet-local fields may be used freely for state-specific storage. As in Scheme, where states in a state machine are frequently implemented as mutually tail-calling procedures, the presentation of the state pattern here is "invisible".

analysis. The most noticeable difference between the two implementations is the ability of Syndicate/rkt to avoid the duplication that comes with mentioning each method in both the "wrapper", the interface, and each "state" class.[11] Where in Java the programmer must manually arrange for the "wrapper" class `PlayerWrapper` to delegate to matching methods on its current state object, the Syndicate/rkt program's facets directly extend the interaction surface of the containing actor. In essence, the language's built-in demultiplexing and dispatch mechanism is reused to perform the delegation implemented manually in the Java program.

A related difference is that in the Java program two objects must collaborate, with a reference to the "wrapper" passed to each "state" method, and a reference to the current state object held in the "wrapper". In the Syndicate/rkt program, only one object (actor) exists in the dataspace, and the instance variable required in the Java has disappeared, being replaced by the implicit state of the actor's facet tree. Sharing between the "wrapper" and a "state" in the Java program must be allowed for in terms of the visible interface of the "wrapper" object, while in Syndicate/rkt, sharing can be arranged by lexical closure, by passing references to shared fields, or by each state facet publishing shared assertions of its own, as in the example.

In this example, every state responds to the same kind of transition event, namely key presses and releases. If different states need to react to other situations in the simulated world, the situation in Java can quickly become complex. For example, the player character may respond to collisions with certain types of objects differently in different states, forcing addition of a `handleCollision()` method to the "wrapper" class, the interface, and all "state" classes—even those for which collisions are irrelevant. In Syndicate, only those facets reactive to an event need mention it, adding additional endpoints to attract and respond to the events concerned.

Finally, multiple facets may be active simultaneously in a single actor, allowing rich dynamic possibilities for mixing-in of state and behavior not available to the Java program. In object-oriented languages like Java, up-front planning is required to properly scope per-instance state and to install delegating "wrapper" method implementations.

## 9.6 THE CANCELLATION PATTERN

The *cancellation* pattern, known as *Cancel Task* in the business process modeling literature (Russell, van der Aalst and ter Hofstede 2016), appears whenever a long-running, asynchronous task may be interrupted. For example, one common reason for interrupting a task is that the party requesting its execution has lost interest in its outcome. Programming languages with support for asynchronous task execution often support the pattern; for example, the .NET library includes a class `CancellationToken` for use with its task execution machinery, and many implementations of "promises" for JavaScript include cancellability as a feature. A classic example may even be seen in the famous *parallel-or* operator (Plotkin 1977), provided we allow ourselves to imagine a realistic implementation which aborts the longer-running of the two branches of a use of parallel-or once the other yields a result.

Cancellation is similar to, but distinct from, an error or exception. As Denicola writes,

11 Syndicate cannot claim a unique ability to avoid the interface (`KeyHandler`) required by Java: languages like Smalltalk and Python get by without such interface declarations, while retaining many of the other features of the pattern seen in Java.

```
1  function makeCancellableRequest(url) {
2      return new Promise(function(resolve, reject, onCancel) {
3          var xhr = new XMLHttpRequest();
4          xhr.on("load", resolve);
5          xhr.on("error", reject);
6          xhr.open("GET", url, true);
7          xhr.send(null);
8          onCancel(function() { xhr.abort(); });
9      });
10 }
```

Figure 79: Cancellation pattern example in JavaScript+Bluebird. Adapted from http://bluebirdjs.com/docs/api/cancellation.html.

> A canceled operation is not "successful", but it did not really "fail" either. We want cancellation to propagate in the same way as an exception, but it is not an error. (Denicola 2016)

Syndicate propagates exceptions via automatic retraction of assertions on failure, and this is how it propagates cancellation as well.

To illustrate the pattern, we follow an example drawn from the API documentation of the Bluebird promise library for JavaScript.[12] In the example, an incremental search feature submits requests to an HTTP-based search service as the user types. Because the search service may not answer as quickly as the user can type, we wish to be able to abandon previously-started, not-yet-completed searches each time an update to the search term is given. An on-screen "spinning icon" display should appear whenever a search is in progress, disappearing again once results are available.

javascript. The JavaScript language has historically relied on *callbacks* for structuring its asynchronous tasks, but latterly has shifted to widespread use of *promises* instead. However, despite much discussion (Denicola 2016), the specification of the behavior of ES6 promises (ECMA 2015, section 25.4) does not include cancellation, leaving JavaScript itself with an "informal" implementation of the pattern each time it is required. Individual implementations of the specification, especially those developed prior to ratification of the standard such as the previously-mentioned Bluebird, include cancellation as an option, yielding "formal" implementations of the pattern when combining a JavaScript engine with a particular promise library.

Figure 79 shows a use of Bluebird promises to implement a cancellable HTTP `GET` request. The result of a call to `makeCancellableRequest` is a *promise* object with a `cancel` method in addition to the usual interface. Bluebird allows the configuration function given to the `Promise` constructor to accept an optional third `onCancel` argument (line 2), which itself is a callback that configures an action to be taken in case the client of the constructed promise decides to cancel it. Here, line 8 specifies that the ongoing `XMLHttpRequest` is to be `abort`ed if the promise returned by `makeCancellableRequest` is canceled.

12 The example can be seen at http://bluebirdjs.com/docs/api/cancellation.html.

```
1 var searchPromise = Promise.resolve();

2 function incrementalSearch(searchTerm) {
3     searchPromise.cancel();
4     showSpinner();
5     var thisSearch = makeCancellableRequest("/search?q=" + encodeURIComponent(searchTerm))
6         .then( function(results) { showResults(results); })
7         .catch(      function(e) { showSearchError(e); })
8         .finally(     function() { if (!thisSearch.isCancelled()) { hideSpinner(); } });
9     searchPromise = thisSearch;
10 }
```

Figure 80: Incremental search using JavaScript+Bluebird.
Adapted from http://bluebirdjs.com/docs/api/cancellation.html.

Figure 80 implements the core of our illustrative example. The function `incrementalSearch` is to be called with the current search term after every keystroke makes an alteration to the on-screen search field. A global variable, `searchPromise`, is used to remember the most recently-started interaction with the search service. Each time `incrementalSearch` is called, the previous search is canceled (line 3), and a new search is started (line 5) after making sure the "spinner" is displayed (line 4).

Each new search ends in one of three outcomes: success, in which case the callbacks of lines 6 and 8 execute; error, in which case lines 7 and 8 execute; or cancellation, in which case *only* line 8 runs. The test in the `if` statement on line 8 makes sure not to hide the "spinner" if the search has been canceled. After all, cancellation happens only when a new search replaces the currently-active one. If the user were able to also cancel the search without starting a new one, the test on line 8 would become more complex. In general, it can be difficult to decide on the correct placement and timing of such code.

Syndicate. The notion of an assertion-mediated *conversation frame* serves here as the heart of Syndicate's approach to task cancellation, making the pattern "invisible" in Norvig's terminology. A Syndicate/rkt equivalent to `makeCancellableRequest` is shown in figure 81. Clients trigger creation of a request-specific actor (line 2) by expressing interest in `http-request` tuples. As the actor starts up, it sends the request (line 3). When a response comes in, the actor terminates its root facet (line 5), replacing it with a facet that asserts the response body (line 6). The other situation that causes termination of the root facet is withdrawal of interest in the result. In either case, termination of the facet causes a message canceling the request to be issued (line 4).[13] The net effect of figure 81 is to adapt the imperative commands involved in web requests to the declarative, conversation-frame-based approach of assertion-mediated coordination.

Figure 82 implements the portion of the example that responds to changes in the text in the search term field. The UI component (not shown) maintains a (unique) `search-term-field`

13 Cancellation of a web-request is idempotent.

```
(assertion-struct http-request (id url body))

(spawn (during/spawn (observe (http-request $id $url _))
         (on-start (web-request-send! id url))
         (on-stop (web-request-cancel! id))
         (stop-when (message (web-response-complete id _ $body))
           (react (assert (http-request id url body))))))
```

Figure 81: Cancellation pattern example in SYNDICATE/RKT

```
1 (assertion-struct search-term-field (contents))
2 (assertion-struct search-results (results))

3 (spawn (during (search-term-field $term)
4          (define id (gensym 'search-id))
5          (during (http-request id (format "/search?q=~a" term) $results)
6            (assert (search-results results)))))
```

Figure 82: Incremental search using SYNDICATE/RKT

assertion, ensuring that it contains the up-to-date query text. In response, for each distinct `term` (line 3) a new request ID is generated (line 4) and an HTTP request is begun (line 5). Upon its completion, a `search-results` record is asserted. However, if the search term changes before the request completes, the entire facet constructed on lines 4–6 is terminated. In turn, this retracts interest in the results of the unwanted HTTP request, which then cancels itself via the code in figure 81.

No mention has been made of the "spinner" thus far. We may achieve the desired effect by introducing a `show-spinner` flag, asserting it only in the presence of a `search-term-field` assertion not paired with a `search-results` assertion. Figure 83 shows the technique. Because we wish to react to the *presence* of a `search-term-field` but the *absence* of a `search-results`, we must use the `assert!`/`retract!` commands to invert the sense of the inner `during`.[14]

The actual display of the outputs from our program—the "spinner" and any search results that might be available—can be done with separate actors responding to changes in the dataspace. Presence of a `show-spinner` flag assertion causes addition of the loading indicator to the UI; a change in asserted `search-results` causes an update to the relevant widget.

ANALYSIS. Central to the pattern is the idea of *chaining* cancellations: if task A has generated asynchronous sub-tasks B and C, then cancellation of A should automatically lead to the cancellation of B and C. In JavaScript, as in .NET, Go, E, Erlang and other languages where some analogue of the pattern appears, such chaining is arranged *manually* by the programmer. In SYNDICATE, however, the use of facets and assertions to frame (sub)conversations ensures

14 This issue is further discussed in section 11.3. In this specific case, a "`during/not`" macro could be introduced to abstract away from the details of implementing logical negation this way.

```
(assertion-struct show-spinner ())

(spawn (during (search-term-field _)
         (on-start (assert! (show-spinner)))
         (on-stop (retract! (show-spinner)))
         (during (search-results _)
           (on-start (retract! (show-spinner)))
           (on-stop (assert! (show-spinner))))))
```

Figure 83: Incremental search "loading" indicator in Syndicate/rkt

that when a facet terminates, its withdrawn assertions cause the termination of facets in peers engaged in conversation with it. The process is automatic, and follows without programmer effort as a side-effect of the Syndicate's protocol-centric design. That is, Syndicate's cancellation mechanism readily composes, often without explicit "wiring".

In languages with exceptions, interaction of cancellation with exception signaling is unclear. Our JavaScript example demonstrates error propagation along links between promises, but does not address JavaScript's exception mechanism. Solutions must be found on a case-by-case basis: the JavaScript promises API omits cancellation, so there is no standard way to connect cancellation to and from exception flow. In Syndicate, exceptions cause assertion withdrawal, which automatically triggers a cancellation cascade where necessary.

Finally, cancellation requires extra care in order to maintain consistency of global invariants. The conditions under which the "spinner" is hidden in the JavaScript example are subtle and could potentially become complicated under even quite simple additions to the scenario. By contrast, Syndicate encourages decomposition of the problem into two phases. First, actors contribute to an overall predicate deciding whether the "spinner" should be shown. Then, a single actor applies the decision from the first phase to the display. Aside from the awkwardness of `assert!`/`retract!`, the code in figure 83 can almost be read as the declarative statement, "show the spinner when a search is active but no results have yet appeared."

## 9.7 THE DEMAND-MATCHER PATTERN

A common pattern in concurrent systems is *demand-matching*: tracking of *demand* for some resource or service, and adding and removing corresponding *supply* in response to changes in demand. Despite its prevalence in many different types of software, this pattern has not, to my knowledge, previously been given a name. The structure of this section is therefore different to those that have preceded it. Instead of detailed examination of the pattern's appearance in several languages, this section describes the demand-matcher pattern in the style of Gamma et al. (1994) in order to bring the concept into focus. An understanding of the idea shows that Syndicate's design makes the pattern "invisible".

The demand-matcher pattern's purpose is *explicit lifetime management*. Even garbage-collected languages require patterned coding to deal with extra-linguistic resources such as sockets, file handles, or large memory buffers. These coding patterns track demand for a resource so that it may be created when required and destroyed in a timely manner when no longer relevant. On the construction side, *factory objects* and the *flyweight* pattern exemplify the concept; on the destruction side, close methods and reference-counting are two strategies commonly used.

Many instances can be found in a wide variety of programs. For example, every program offering TCP *server* functionality is an example of the demand-matcher pattern, in any programming language. Demand for the services of the server is indicated by the appearance of a new connection. On UNIX-like systems, this demand is signaled at a fundamental level by `accept(2)` returning a new connected TCP socket. The server provides supply by allocating resources to the new connection, maintaining connection-specific conversational state, and beginning the sub-conversation associated with the new connection. When the remote party disconnects, this *drop* in demand is signaled by the closing of the connected socket, and the server responds by releasing the connection's resources and cleaning up associated conversational state.

Correctly matching supply to demand—that is, correctly allocating and releasing a resource as the need for it waxes and wanes—is motivated by the same concerns that motivate automatic garbage collection. If a programmer does not correctly scope the lifetime of some resource to the lifetime of conversations involving that resource, the resulting program may suffer resource leaks (e.g. unclosed sockets) or dangling references (attempts to write to a closed socket).

In SYNDICATE, facet-based assertion-tracking obviates these patterns. To allocate a resource, a client asserts interest in it. The server responds by constructing a facet whose `on-start` action creates the underlying resource and whose `on-stop` action destroys it. The server's facet lifetime, and hence the resource's lifetime, is scoped by the consumer's continued assertion of interest. Once that interest is withdrawn, the server-side facet terminates, thereby releasing the resource. SYNDICATE's facets allow the programmer to bidirectionally associate resources with conversational frames—to bring external resources into the conversational state. Furthermore, dataspace programming allows a straightforward aggregation of interest in a resource because requests following the first one may simply discover the already existing instance of the resource. Similarly, dataspaces allow the easy realization of load-balancing schemes.

We have seen many instances of demand-matching in this dissertation already. Informative specimens may be found in example 8.12, figures 60 and 61, and particularly examples 8.27 and 8.31, in addition to many of the examples in this chapter.

INTENT Track *demand* for some resource or service; add and remove corresponding *supply* in response to changes in demand.

PARTICIPANTS

- Client: source of demand. Must signal changes in demand for service to the Demand Matcher.

- Demand Matcher: monitors demand and supply levels and brings them into balance when changes are detected; commonly spawns Service instances in response to increases in demand.

- Service: satisfies demand. Usually signals its existence and/or status to the Demand Matcher to help it perform its balancing task.

### KNOWN USES

- Every TCP service program treats incoming connections as demand, and allocates and deallocates internal resources as connections come and go.
- The UNIX `inetd` service can be configured to execute programs in response to incoming TCP connections. The recent `systemd` suite of programs can be configured similarly.
- The `dbus` service bus allows for applications and daemons to be instantiated "on demand when their services are needed."[15]
- Worker pools, e.g. Amazon's "Elastic Load Balancing" product, which "automatically scales its request handling capacity to meet the demands of application traffic."[16] "As Elastic Load Balancing sees changes in the traffic profile, it will scale up or down."[17]
- "Leasing" of resources and "renewal reminders" (Hohpe 2017) serve as a mechanism for tracking demand. For example, DHCP (Droms 1997) allocates IP addresses and issues a lease in response to a message from a client. Decrease in demand happens automatically when an issued lease expires. A second example can be seen in the WebSub protocol (W3C 2016), which associates a lease with each subscription and in addition sends a "renewal reminder" when a lease nears expiry.
- Work items in job queues are implicitly demands for allocation of some compute and/or IO resource, which is automatically freed once its work item completes. A frequently-seen embellishment is the notion of a *limited resource*, where only a certain number of resources may be in use at once, and if demand exceeds the possible supply, clients must wait for a resource to become free before they may proceed.

### RELATED CONCEPTS AND PATTERNS

- Flyweight (Gamma et al. 1994), as seen in e.g. symbol tables: interning a symbol is similar to demand for its existence; symbol tables that weakly hold their entries rely on the garbage-collector to lazily detect absence of demand, releasing the associated resource in response.
- The "Demand Matcher" participant is often an instance of Factory (Gamma et al. 1994).
- Garbage collection (distributed and non-distributed). Abstractly, garbage collection tracks demand for (references to) resources (objects), releasing each resource once its existence is no longer required. Garbage collection would be an example of the Demand Matcher pattern, except for the latter's ability to *manufacture* demanded resources when demand *increases*, a feature not offered by garbage collectors.

15 https://www.freedesktop.org/wiki/Software/dbus/
16 https://aws.amazon.com/elasticloadbalancing/
17 https://aws.amazon.com/articles/1636185810492479

```
1  do_call(Process, Request) ->
2      Mref = erlang:monitor(process, Process),
3      Process ! {'$gen_call', {self(), Mref}, Request},
4      receive
5          {Mref, Reply} ->
6              erlang:demonitor(Mref, [flush]),
7              {ok, Reply};
8          {'DOWN', Mref, _, _, Reason} ->
9              exit(Reason)
10     end.
```

Figure 84: Simple RPC client code in Erlang.
Adapted from gen:do_call from Erlang/OTP 19.2, ©1996–2017 Ericsson AB.

- Supervision in Erlang (Armstrong 2003). A supervisor monitors actors *supplying* some service, often created in response to some signal of demand. The supervisor takes action to ensure *stable supply*, restarting crashing actors as necessary.

## 9.8 ACTOR-LANGUAGE PATTERNS

In addition to the broadly-applicable programming patterns thus far discussed, the design of SYNDICATE eliminates the need for certain more narrowly-focused features seen in actor-based languages such as Erlang. In particular, the introduction of facets in combination with the `react/suspend` construct of section 6.5 eliminates some patterned uses of *selective receive*, in those languages with such a construct, and decoupling of presence information from actor identity eliminates complications in protocols making use of *request delegation*.

SELECTIVE RECEIVE. Erlang's *selective receive* facility allows a process to scan its mailbox for messages matching a certain pattern, yielding the first match or *blocking* until a matching message is later received. This is used to build RPC-style interaction as a library facility that appears to a client as an ordinary procedure call, hiding the details of communication. The chief drawback of the use of selective receive is a lack of *availability*. While the actor is in a state waiting for specific messages to arrive, it cannot attend to ordinary requests from peers, thereby increasing the risk of deadlock. SYNDICATE's facets eliminate RPC-like patterns involving selective receive entirely, allowing actors to remain available even when managing ongoing sub-conversations with peers.[18]

For example, figure 84 shows the essence of the library routine `gen:do_call` from Erlang/OTP 19.2.[19] Line 1 declares the routine as a function expecting the server's process ID and some

18 For those cases where selective-receive-like *unavailability* is required, it can be implemented as a library routine surrounding a few fields and an internal queue. This explicit representation of a queue of pending messages has some distant relationship to the "activators" of Frølund and Agha (1994) in that it marshals incoming events, suspending further activity until some palatable arrangement of them has been arrived at.

19 The code has been simplified, eliding timeout handling and support for cross-node calls in distributed Erlang.

```
(define (do-call self-pid process request)
  (define request-id (gensym 'request-id))
  (react/suspend (k)
    (on (asserted (present process))
        (send! (rpc-request process self-pid request-id request)))
    (stop-when (retracted (present process))
        (error 'do-call "Server exited before replying!"))
    (stop-when (message (rpc-reply self-pid request-id $reply))
        (k reply))))
```

Figure 85: Approximate Syndicate analogue of figure 84

value describing the RPC request to issue. Line 2 subscribes to lifecycle information about the server: if it crashes while the subscription is active, a `'DOWN'` message is delivered to subscribing processes. The call to `erlang:monitor` yields a globally unique reference, which here is cleverly used for two purposes: not only does it uniquely identify the subscription just established, it is also pressed into service as an identifier for the specific RPC request instance being executed. Line 3 delivers the request in an envelope. The envelope includes four things: the `Request` itself; an atom, `'$gen_call'`, identifying the RPC protocol that is expected; the sender's own process ID, `self()`; and the `Mref` unique identifier for the request instance.

Line 4 opens the selective receive expression. Here, the process expects one of two messages to arrive. If an envelope containing the reply, uniquely labeled with this request's `Mref`, arrives first (line 5), the process cancels its subscription to lifecycle information of the server process (line 6) and returns the result (line 7). If an indication that the server has crashed arrives first (line 8), with context again uniquely identified by `Mref`, then the routine causes *this* process to crash with the same "exit reason", thus propagating exceptions across process boundaries in a structured way.

If a `'DOWN'` message arrives first, the OTP library relies on the convention that no reply message will arrive later. This works well, though it does rule out patterns of delegation we return to below (section 9.8). However, if a *reply* arrives first, a `'DOWN'` message may still be issued in the window between control returning to line 5, and the unsubscription of line 6 taking effect. For this reason, after an unsubscription has taken effect, programmers must take care to use selective receive to discard any pending `'DOWN'` messages that might be queued.

Here, the programmer of the routine has done this by passing `flush` to `erlang:demonitor`. Quoting from the Erlang/OTP documentation (Ericsson AB 2017),

> Calling `demonitor(Mref, [flush])` is equivalent to the following, but more efficient:

```
demonitor(Mref),
receive
    {_, Mref, _, _, _} -> true
    after 0 -> true
end
```

Syndicate eliminates the need for these uses of selective receive. The rough equivalent of figure 84 is shown in figure 85. The use of `react/suspend` on line 3 reifies a partial continuation, making `do-call` a blocking procedure just like Erlang's `do_call`.[20] The protocol sketched here assumes that the server asserts (`present` *id*) as a placeholder for an appropriate domain-specific indication of presence; lines 6 and 7 in figure 85 react to retraction of presence, which corresponds to handling of the `'DOWN'` message on lines 8 and 9 of figure 84. Because the facet expressing interest in `present` assertions is terminated either when a reply is received or the server crashes, and the facet's interests are retracted at its termination, the situation of needing to flush a pending server termination message never arises.

Generally, the fact that Syndicate's facets combine subscriptions and event handlers, so that one never exists without the other, ensures that messages are delivered to relevant handlers, and conversely, that irrelevant messages are never handled at all.

TRANSPARENT REQUEST DELEGATION. An RPC server process in Erlang/OTP may, upon receiving an RPC request, forward the request to some other "worker" process, which replies on its behalf. Clients use Erlang's "monitor" facility to detect failures in server processes. However, if a server forwards a request, the client is left monitoring the original server process and not the worker process that has just been given responsibility for delivering the reply. If the worker process crashes, the client is left hanging; conversely, if the service process crashes, but the worker process is still running normally, the client will falsely assume its RPC request failed. Worse, because of the hard-coded assumption that a crash implies that no reply will subsequently arrive, the client will later be faced with a "martian packet"; that is, a reply from the worker will arrive after the necessary context information has already been discarded.

A few workarounds exist. The worker may forward the reply to the server process, which relays it in turn to the original requestor, thereby avoiding the "martian packet" scenario. The server may monitor the worker, perhaps crashing if it notices the worker crash, or perhaps remembering its obligations and synthesizing crash signals specifically for the clients affected by the crash. The worker may monitor the server, crashing if it notices the server crash, thereby at least avoiding wasted effort; however, it is important to note that a race condition exists here: the worker may finish its work and deliver the reply to the client before it receives notification that the server has crashed, bringing us back to a "martian packet" scenario. Finally, the transparency of delegation may be discarded, and the server may reply to the client with the identity of the worker, essentially telling it to expect a real reply later from a different source.[21]

Delegation in Syndicate does not suffer from any of these problems. Syndicate presence indications are not tied to actor identity: instead, presence may be as coarse- or fine-grained as the domain requires. For example, a Syndicate service may use a protocol that advertises the fact that a reply is on its way *for a certain request ID*:

---

20 Note, however, that `k` is not mentioned on line 7, which means that any exception handler in the context of `do-call` has no opportunity to catch the error signaled by line 7. The actor simply terminates. Local adjustment of `do-call` can rectify the problem; more sophisticated integration of Racket exceptions, partial continuations, and facets remains as future work.

21 This non-transparency is similar to the delegation involved in HTTP's redirect responses, with their `Location` headers. HTTP reverse-proxying, by contrast, is a form of transparent delegation.

```
(request-in-progress service-id request-id)
```

A service that does not delegate requests manages these request-specific assertions itself, while a delegating service passes responsibility for maintaining `request-in-progress` assertions along with the task at hand. In each case, the assertion of interest contains the same information: the client is unaware of the identity of the specific actor handling the request, but nonetheless is able to monitor the request's progress.

This benefit was achieved by generalizing away from using implementation-specific identifiers (process IDs) as proxies for domain-specific information (the possibility of receiving a reply to a request). Having broken this tight coupling, we are now free to explore additional possibilities not previously available. For example, we may extend `request-in-progress` assertions to include more detailed progress information for long-running tasks simply by adding a field; clients monitoring request progress are thus automatically informed as milestones go by.

# 10

## *Evaluation: Performance*

Established approaches to concurrency generally have well-understood performance models that programmers use when reasoning about the expected and actual behavior of their programs. The actor model, for example, naturally admits $\tilde{O}(1)$ message delivery and $\tilde{O}(1)$ process creation in most implementations; the abstract performance of the Relational database model can be understood in terms of table scans and index characteristics; and so on. The Syndicate model includes features such as multicast and state change notifications not present in other approaches to concurrency, and does not fit the established performance models. Therefore, we must develop a Syndicate-specific performance model that programmers can rely on in the design and evaluation of their Syndicate programs, and in the understanding of the Syndicate programs of others. We begin by considering the abstract costs of Syndicate actions (section 10.1), which we then confirm with measurements of asymptotic performance characteristics of representative protocols (section 10.2). Finally, we touch on the concrete performance of the Syndicate/rkt prototype implementation (section 10.3).

### 10.1 Reasoning about routing time and delivery time

The key to Syndicate's performance is the implementation of dataspace-model actions.[1] A model of performance must give the programmer a sense of the costs involved in the interpretation of the three possible types of action. Interpretation of a spawn action is like interpretation of a state change notification, because the initial assertion set conveyed with a leaf actor is transformed into just such an action. Interpretation of message and state change notification actions involves two steps, with associated costs: computation of the set of recipients ("routing") followed by delivery of an event to each identified recipient ("delivery").

Programmers might reasonably expect that the routing time of state change notifications should be bounded by the number of assertions in each notification, which is why the incremental semantics using patches instead of full sets is so important. A complication arises, however, when one considers that patches involving wildcards refer to *infinite* sets of assertions. The trie-based representation of assertion sets takes care to represent such infinite sets tractably, but the programmer cannot assume a routing time bounded by the size of the *representation* of the notification. To see this, consider that asserting $\star$ forces a traversal of the entirety of the ?-prefixed portion of the dataspace to discover *every* active interest.

1 Facets within actors are similar to actors within a dataspace, and the costs of routing *within* an actor can be understood by analogy to routing within a dataspace.

Fortunately, routing time of SCNs *can* be bounded by the size of the representation of the *intersection* of the patch with the dataspace itself. When processing a patch $\frac{\pi_i}{\pi_o}$ to a dataspace $R$, the function *combine* (figure 39) explores $R$ only along paths that are in $\pi_i$ or $\pi_o$. Thus, when reasoning about SCN routing time, programmers must set their performance expectations based on both the patches being issued and the assertions established in the environment to be modified by each patch. After routing has identified the actors to receive state change notifications, the associated delivery time should be linear in the number of recipients.

The costs of message actions are simpler to understand than those of SCN actions, allowing us to make more precise statements on expected upper bounds. The key variable is the fraction of each message that must be examined to route it to a set of destinations. For example, some Syndicate protocols treat messages as pairs of an address, used to select recipients, and a body that is not examined during routing. That is, messages are of the form $(\mathit{address}, \mathit{body})$, and assertions of interest are of the form $?(\mathit{address}, \star)$. For such protocols, the routing process should take time in $\tilde{O}(|\mathit{address}|)$. More general messaging protocols effectively use more of each message as address information. In such cases, routing time should be bounded by $\tilde{O}(|\mathit{message}|)$. In either case, noting that $|\mathit{address}| \leqslant |\mathit{message}|$, delivery to all $n$ interested recipients should take time in $\tilde{O}(n)$, for $\tilde{O}(|\mathit{message}| + n)$ overall processing time. Encoding actor-style unicast messaging is then a special case, where the address is a target process ID, $\tilde{O}(|\mathit{address}|) = \tilde{O}(1)$, the size of the message body is irrelevant, and $n = 1$, yielding $\tilde{O}(1)$ expected per-message cost.

## 10.2 MEASURING ABSTRACT SYNDICATE PERFORMANCE

Notwithstanding the remarks above, we cannot yet make precise statements about complexity bounds on routing and delivery costs in Syndicate in general. The difficulty is the complex interaction between the protocol chosen by the programmer and the data structures and algorithms used to represent and manipulate assertion sets in the Syndicate implementation.

We can, however, measure the performance of Syndicate/rkt on representative protocols. For example, we expect that:

- simple actor-style unicast messaging performs in $\tilde{O}(1)$;
- multicast messaging performs within $\tilde{O}(|\mathit{message}| + n)$;
- state change notification performance can be understood; and
- Syndicate programs can smoothly interoperate with the "real world."

UNICAST MESSAGING. We demonstrate a unicast, actor-like protocol using a simple "ping-pong" program. The program starts $k$ actors in a single Syndicate dataspace, with the $i$th peer asserting the subscription $?(\texttt{ping}, \star, i)$. When it receives a message $(\texttt{ping}, j, i)$, it replies by sending $(\texttt{ping}, i, j)$. Once all $k$ peers have started, a final process numbered $k + 1$ starts and exchanges messages with one of the others until ten seconds have elapsed. It then records the overall mean message delivery latency.

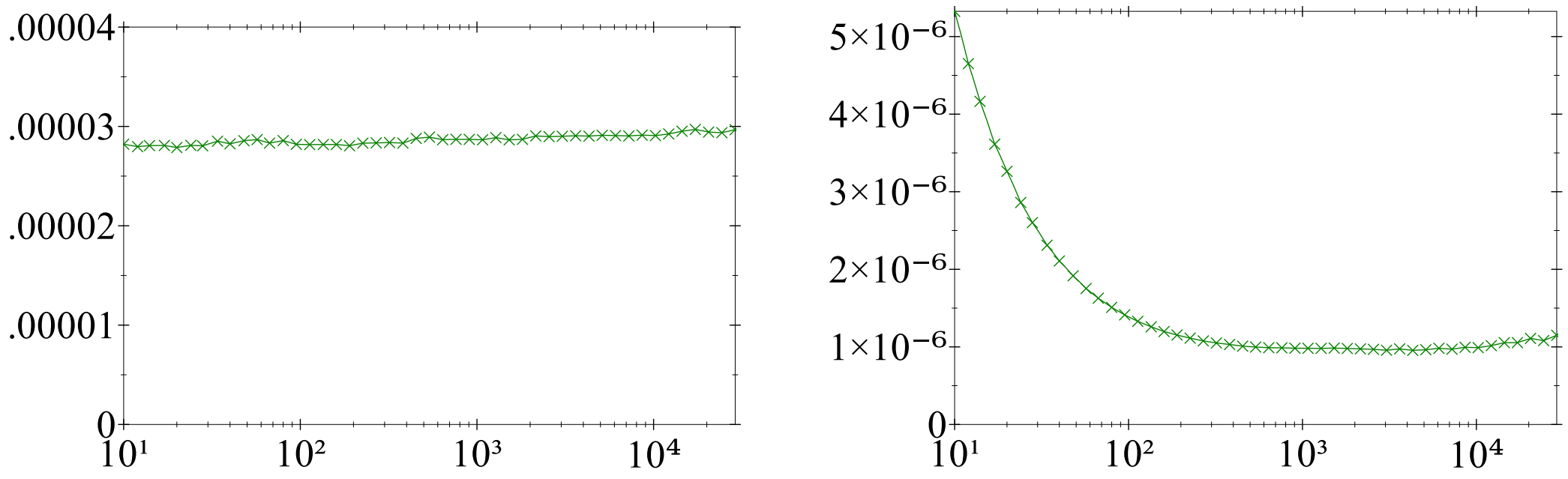

(a) Unicast (b) Broadcast

Figure 86: Message routing and delivery latencies, sec/msg vs. $k$

Figure 86a shows message latency as a function of the number of actors. Each point along the $x$-axis corresponds to a complete run with a specific value for $k$. It confirms that, as expected, total routing and delivery latency is roughly $\tilde{O}(1)$.

BROADCAST MESSAGING. To analyze the behavior of broadcasting, we measure a variation on the "ping-pong" program which broadcasts each ping to all $k$ participants. Each *sent* message results in $k$ *delivered* messages. Figure 86b shows mean latency of each delivery against $k$. This latency is comprised of a fixed per-delivery cost along with that delivery's share of a fixed *per-transmission* routing cost. In small groups, the fixed routing cost is divided among few actors, while in large groups it is divided among many, becoming an infinitesimal contributor to overall delivery latency. Latency of each delivery, then, is roughly $\tilde{O}(\frac{1}{k} + 1)$. Aggregating to yield latency for each transmission gives $\tilde{O}(1 + k)$, as expected.

STATE CHANGE NOTIFICATIONS. Protocols making use of state change notifications fall into one of two categories: either the number of assertions relevant to an actor's interests depends on the number of actors in the group, or it does not. Hence, we measure one of each kind of protocol.

The first program uses a protocol with assertion sets independent of group size. A single "publishing" actor asserts the set {`A`}, a single atom, and $k$ "subscribers" are started, each asserting {`?A`}. Exactly $k$ patch events $\frac{\{\mathtt{A}\}}{\emptyset}$ are delivered. Each event has constant, small size, no matter the value of $k$.

The second program demonstrates a protocol sensitive to group size, akin to a "chatroom" protocol. The program starts $k$ "peer" actors in total. The $i$th peer asserts a patch containing both $(\mathtt{presence}, i)$ and $?(\mathtt{presence}, \star)$. It thereby informs peers of its own existence while observing the presence of every other actor in the dataspace. Consequently, it initially receives a patch indicating its own presence along with that of the $i - 1$ previously-started peers, followed by $k - i - 1$ patches, one at a time as each subsequently-arriving peer starts up.

Measuring the time-to-inertness of differently-sized examples of each program and dividing by the number of state change notification events delivered shows that in both cases the pro-cessing required to compute and deliver each state change notification is roughly constant even as k varies (figure 87).

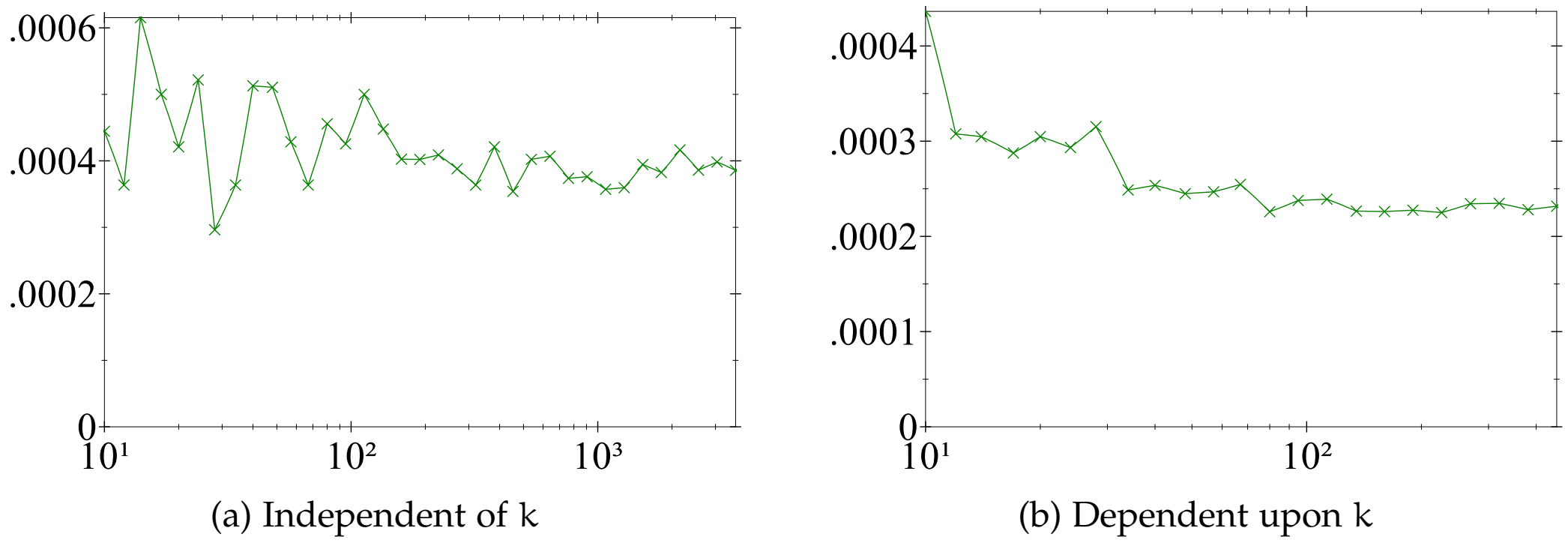

Figure 87: State Change Notification cost, sec/notification vs. k

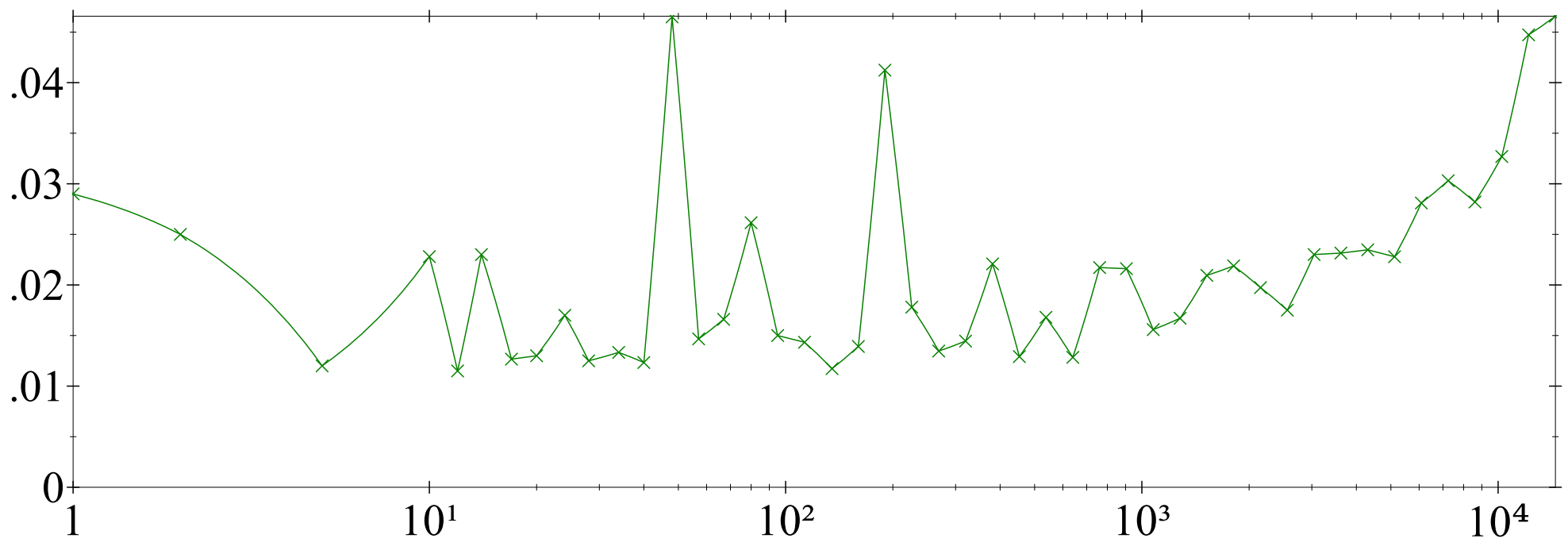

Figure 88: Marginal cost of additional connections, sec/conn. vs. k

COMMUNICATION WITH THE "OUTSIDE WORLD". An implementation of a TCP/IP "echo" service validates the claim that SYNDICATE can effectively structure a concurrent program that interacts with the wider world, because this service is a typical representative of many network server applications.

The implementation-provided TCP driver actor provides a pure SYNDICATE interface to socket functionality. A new connection is signaled by a new assertion. The program responds by spawning an actor for the connection. When the connection closes, the driver retracts the assertion, and the per-connection actor reacts by terminating.

The scalability of the server is demonstrated by gradually ramping up the number of active connections. The client program alternates between adding new connections and performing work spread evenly across all open connections. During each connection-opening phase, it computes the mean per-connection time taken for the server to become ready for work again after handling the batch of added connections. Figure 88 plots the value of k, the total number of connections at the end of a phase, on the (logarithmic) x-axis; on the y-axis, it records mean seconds taken for the server to handle each new arrival. The marginal cost of each additional connection remains essentially constant and small, though the results are noisy and subject to GC effects.

## 10.3 CONCRETE SYNDICATE PERFORMANCE

We have seen that the *abstract* (big-O) performance of Syndicate dataspace implementations using the trie structure satisfies our expectations. However, programmers rely not only on big-O evaluations, but also on absolute performance figures. In absolute terms, looking only at micro-benchmarks such as those explored above, we see that message-passing performance is quite respectable. The results of figures 86, 87 and 88 were all produced on my commodity 2015-era desktop Linux machine, an Intel Core i7-3770 running at 3.4 GHz; figure 86a shows that Syndicate/rkt can *route* approximately 30,000 messages per second, while figure 86b shows that it can *deliver* approximately 1,000,000 messages per second.[2] As a rough comparison, a crude "ping-pong" program written to Racket's built-in `thread`, `thread-send`, and `thread-receive` APIs yields approximately 780,000 point-to-point messages (i.e., each message involving both routing and delivery) per second. We may conclude that the Syndicate/rkt prototype's absolute routing performance is roughly a factor of twenty slower than the optimized point-to-point routing infrastructure available in a production language implementation. Assertion-set manipulation is new, so we have nothing to compare it against; that said, there are clear directions for improving the constant factors.[3]

More broadly, the speed of the prototype Syndicate implementations has not prevented effective use and evaluation of the programs written thus far. Of the larger Syndicate case studies, the 2D platformers are most challenging from a performance perspective: to achieve a 60 Hz frame rate, a program must never exceed a hard per-frame time budget of 16.67ms. Platformer "A", written primarily using protocols involving Syndicate *messages*, has no trouble maintaining a high frame rate even with tens of moving agents on-screen; platformer "B", written primarily using protocols manipulating *assertions*, only manages a high frame rate with one or two simultaneously moving agents on the screen. To see why, consider again figures 86 and 87: at 30,000 messages per second, we may send up to 500 messages per frame before exceeding our 60 Hz frame budget; but SCN processing is substantially slower. Even the very simple program whose measurements are shown in figure 87 does not exceed some 2,500 SCNs per second, which at 60 Hz gives us a budget of approximately 40 SCNs per frame. Clearly, future work on optimization of SCN processing will be of great benefit to applications like platformer "B" which have real-time constraints and make use of assertion-manipulation-heavy protocols. However, the platform games are an outlier; none of the other case studies places such extreme demands on the implementation. For example, the graphical window-layout

2 Both figures are for a single core. Syndicate/rkt does not yet take advantage of multiple cores because Racket requires special programming for multi-core operation.

3 The temptation to dismiss the Syndicate design on grounds of performance must be resisted. A relevant comparison is the development of Smalltalk, a dynamic object-oriented programming system. Early Smalltalk implementations (Kay 1993) required custom hardware for reasonable performance, and it was not until the Self research program bore fruit (Chambers 1992; Hölzle 1994; Hölzle and Ungar 1995) some twenty years later that dynamic object-oriented systems attained performance levels enabling their use in a wide array of production applications.

GUI system reliably responds to user input within one or two frame times, despite making use of assertion-based protocols; few things tend to change from frame to frame, unlike the animation-heavy platformer.

The next-most-challenging case study from a concrete performance perspective is the TCP/IP stack. On a wired 100GB Ethernet, an IPv4 ICMP "ping" round trip between two Linux machines adjacent on the network takes ~0.4ms; the highly-optimized C-language Linux kernel TCP/IP stack is used at both ends of the link. Approximately the same round trip times are achieved if we replace the responding party's kernel-based IP stack with a simple C program responding to pings using the Linux `packet(7)` packet-capture mechanism. Switching to the Syndicate/rkt implementation of TCP/IP, by contrast, yields round trips of ~3.5ms, suggesting that it adds ~3ms of round trip latency. In a more realistic setting, pinging the same machine from a computer on the other side of the city (around ten network hops away), we see ~22ms round trips via the Linux kernel's IP stack and ~25ms via the Syndicate/rkt stack: the extra 3ms from Syndicate starts to look less significant in context of ordinary network conditions. The Syndicate DNS resolver, which I have used for my day-to-day browsing since 2012, is certainly not as quick as the system's own resolver, written in C; but the ~11ms of latency it introduces is barely noticeable in the context of day-to-day web browsing.

# 11

# *Discussion*

Chapter 3 sketched a map of the design space of concurrency models, placing each at a point in a multi-dimensional landscape. The chapter concluded with a discussion of the properties of an interesting point in that landscape. In this chapter I show that Syndicate uniquely occupies this point in the design space and that it offers linguistic support for each of the categories C1–C12 defining the dimensions of the space. I also make connections between Syndicate and areas of related work not examined in chapter 3. Finally, no design can be perfect for all scenarios; therefore, I conclude the chapter with a discussion of limitations of the Syndicate design and the dataspace model.

## 11.1 Placing Syndicate on the map

Section 3.6 gave a collection of desiderata for a concurrency model, expressed in terms of characteristics C1–C12 described in section 3.1. Syndicate satisfies each of the particulars listed. No other model discussed in chapter 3 manages to satisfy all at once, though the fact space model (section 3.5) comes close. As discussed in section 2.5, Syndicate is like an integration of the fact space model into a programming language design (as opposed to middleware) and goes beyond the fact space model in its strong epistemic focus and support for explicit representation of conversations, conversational frames, and conversational state.

Each of the concurrency models of chapter 3 was illustrated with an implementation of a portion of a running example, a toy "chat server". Figure 89 gives a Syndicate equivalent at a similar level of abstraction. However, we do not have to be satisfied with pseudo-code: the real

```
(define (user-agent name socket-id)
  (assert (present name))
  (on (message (tcp-in-line socket-id $line))
      (send! (speak name line)))
  (during (present $who)
    (on-start (send! (tcp-out socket-id (format "~a arrived\n" who))))
    (on-stop  (send! (tcp-out socket-id (format "~a left\n" who))))
    (on (message (speak who $text))
        (send! (tcp-out socket-id (format "~a: ~a\n" who text))))))
```

Figure 89: Syndicate chat room user agent

```
1  #lang syndicate
2  (require/activate syndicate/drivers/tcp2)
3  (require racket/format)

4  (message-struct speak (who what))
5  (assertion-struct present (who))

6  (spawn #:name 'chat-server
7   (during/spawn (tcp-connection $id (tcp-listener 5999))
8     (assert (tcp-accepted id))

9     (define me (gensym 'user))
10    (assert (present me))
11    (on (message (tcp-in-line id $line))
12        (send! (speak me (bytes->string/utf-8 line))))

13    (during (present $user)
14      (on-start (send! (tcp-out id (string->bytes/utf-8 (~a user " arrived\n")))))
15      (on-stop  (send! (tcp-out id (string->bytes/utf-8 (~a user " left\n")))))
16      (on (message (speak user $text))
17          (send! (tcp-out id (string->bytes/utf-8 (~a user ": " text "\n")))))))))
```

Figure 90: A complete Syndicate TCP/IP chat server program

implementation is available to us. Figure 90 is a Racket source file that implements a complete TCP/IP chat server listening on port 5999. A fully-realized form of figure 89 appears within figure 90 as lines 10–17.

As the chat server program starts up, a single actor, `chat-server`, is created. It expresses interest in notifications of connections appearing on port 5999 (line 7). In response, the TCP driver activated on line 2 causes creation of a server socket listening on the specified port. If the `chat-server` actor were to terminate, the TCP driver would notice the drop in demand and stop accepting new connections. Each time a new connection arrives, an actor for the connection is spawned (line 7). The new actor signals the driver that the connection has been successfully accepted (line 8) and goes on to establish two related conversations. The first, lines 9–12, reacts to lines of input from the TCP socket, relaying them as speak messages to peers in the dataspace. The second, lines 13–17, reacts to each separate user present in the space. As the actor learns that a new user exists, it sends a notification over TCP (line 14); when that user departs, it sends a matching notification (line 15).[1] While connected, anything that user says is prefixed with the user's name and delivered via TCP (lines 16–17).

The Syndicate program shown is superficially similar to the sketch of a fact space program shown in figure 9 (page 38). An immediate difference is that the Syndicate program uses *messages* for utterances of connected users, while the fact space program uses tuples, roughly

1 Here we translate Syndicate assertions into messages sent via a non-Syndicate medium, the reverse of the situation discussed in example 8.17.

analogous to assertions. A second difference is that no matter the cause, each facet of the SYNDICATE program retracts its assertions from the shared space as it terminates, where the fact space program only does the same if the connection to the tuplespace is broken or closed. Fact space actors joining and leaving conversations dynamically must keep track of their published resources by hand. A more subtle difference between the two programs is that the SYNDICATE program attends only to utterances from users who have explicitly marked themselves as present. The fact space sketch responds to *all* `Message` tuples, no matter the existence of a corresponding `Presence` tuple. To adapt the fact space sketch to be equivalent in this way to the SYNDICATE program, we would have to add additional book-keeping to track which users the actor knows to be present. To adapt the SYNDICATE program to be equivalent to the fact space program, however, all we would have to do is hoist the endpoint of lines 16–17 above and outside the `during` clause of line 13, changing `user` to `$user` in the `speak` pattern. The alteration reflects a change in ontological relationship between `present` assertions and `speak` messages: after the change, the latter are no longer *framed* by the former.

(C1; C2; C7) Turning to the criteria described in section 3.1 (page 25), the dataspace model offers a single primitive mechanism that unifies one-to-one and many-to-many messaging and state exchange. SYNDICATE programmers design their assertions and messages to include correlation information that identifies relevant conversational context in domain-specific terms; because the language routes messages and assertions via pattern matching, the design directly supports arbitrary numbers of participants in a conversation. SYNDICATE's facets provide the associated control structure and directly express the correspondence between protocol-level conversational context and actor-level control context. The assertions of interest in an event handler endpoint serve as the interface between data and control, demultiplexing incoming events to the correct facets and event handlers. SYNDICATE programmers use nested sub-facets to capture sub-conversations within conversational contexts. Nesting of facets reflects nesting of contexts and captures relationships of ontological dependency between a containing and a contained conversational frame.

(C3; C4) Endpoints allow arbitrary reactions to changes in the dataspace, but an important special case is to maintain a local copy of shared information. SYNDICATE includes *streaming query forms* which take on the task of integrating changes from the dataspace with the contents of local fields. Conversely, SYNDICATE's assertion endpoints automatically transfer changes in local fields into the dataspace.

(C5; C7; C8) SYNDICATE automatically maintains dataspace integrity in case of partial failure because its dataspaces offer a *set* view of a *bag* of assertions. When an actor fails or terminates, the dataspace removes all assertions belonging to the actor from the bag. If this affects the set perspective on the bag, then the dataspace notifies the remaining actors. This *fate-sharing* (Clark 1988) of state and actor lifetime thus turns into a tool for maintaining application-level invariants. This idea extends beyond maintaining data invariants to maintaining *control* invariants. SYNDICATE's `during` and `during/spawn` forms create and destroy facets in response to assertions appearing and disappearing; in turn, those created facets assert derived knowledge back into the dataspace and establish related subscriptions and event handlers. Each such facet exists only as long as the assertion that led to its creation. This allows the programmer to rely on invariants connecting presence of assertions in the dataspace with existence of matching facets in an actor. For example, in our chat server, programmers may straightforwardly ensure that every connected user has an asserted `present` tuple and every `present` tuple describes a connected user. Exceptions fit the model smoothly, because they cause actor termination, which retracts all active assertions.

(C6) The core mechanism of the dataspace model, state replication over a lossless medium, offers an analog of *strong eventual consistency* (Viotti and Vukolić 2016) to the programmer. This allows reasoning about *common knowledge* (Fagin et al. 2004). An actor maintaining some assertion knows both that interested peers learn of the assertion and that each such peer knows that all others learn of the assertion. By providing this guarantee at the language level, Syndicate lets programmers rely on this additional form of epistemic reasoning in their protocol designs.

(C9) Syndicate programs may react to retraction of assertions as well as their establishment. This allows interpretation of particular assertions as demand for some resource. The TCP driver is a clear example: it allocates and releases sockets to match `tcp-connection` assertions in the dataspace. This approach to resource management is a form of garbage collection where domain-specific descriptions of resources take the place of pointers, and resources are released once the last expression of interest in them disappears. As such, this idiom is frequently used in Syndicate programs.[2] Even service startup ordering problems can be solved in this way, interpreting interest in *service presence* (Konieczny et al. 2009) as a request for service startup or shutdown.

In section 3.5, we discussed an enhancement to the running example where each user's presence record would also include a *status message*. Figure 10 (page 39) sketched the additional book-keeping required to track both presence and status of each user. The Syndicate equivalent makes use of the erasure of irrelevant information performed by the *inst* metafunction (definition 5.24):

```
(during (present $who _)
  (on-start (send! (tcp-out socket-id (format "~a arrived\n" who))))
  (on-stop  (send! (tcp-out socket-id (format "~a left\n" who)))))
(on (asserted (present $who $status))
    (send! (tcp-out socket-id (format "~a status: ~a\n" who status))))
```

(C10) As we saw in section 7.4, Syndicate implementations can capture program traces in terms of the actions and events of the underlying dataspace model. These traces can then be visualized in various ways, yielding insight into system state and activity. Similar trace information acts as the basis of an experimental unit-testing facility,[3] where executable specifications of expected behavior in terms of patterns over traces run alongside the program under test, signaling an error if an unwanted interaction is discovered.

(C11) Syndicate's facets and fields allow easy addition of new conversations and conversational state to an existing actor without affecting other conversations that actor might be engaged in. In Syndicate/rkt, Racket's own abstraction facilities (procedures and macros) allow programmers to extract facets into reusable chunks of functionality, allowing "mixin"

2 Unlike traditional GC, this resource management strategy allows *synthesis* of resources merely by naming them!

3 This unit-testing facility is a contribution of my colleague Sam Caldwell.

style augmentation of an actor's behavior. When it comes to altering a *conversation* to include more or fewer participants, programmers adjust their protocol definitions—if required. A protocol's schema may allow participants to freely express interest in certain assertions. Where such expressions of interest would interfere, however, the protocol must be revised, and the corresponding facets of actors must be updated to match. However, such changes are local to the facets concerned and do not affect neighboring facets.

(C12) Finally, tighter integration of Syndicate with existing experimental support for reloading of Racket modules is future work. Experience thus far is that the combination is promising: a Syndicate protocol for describing the availability of new code allows actors to serialize their own state and pass it to their post-upgrade replacements. Felleisen et al. (1988) raise the idea of abstract representations of continuations—that is, of ongoing tasks. It may be promising to explore this idea in the setting of serialization of actor state, since it may have benefits for code upgrade, orthogonal persistence, and program state visualization.

## 11.2 placing Syndicate in a wider context

The map of concurrency-model design space sketched in chapter 3 introduced many ideas, languages, designs and systems related to Syndicate. Here, we touch on other inspirational and related work. In particular, Syndicate and the underlying dataspace model invite comparison to general techniques for functional I/O as well as to process calculi, actor-based models of concurrency, and messaging middleware.

### 11.2.1 *Functional I/O*

Communication is intrinsically effectful. As a result, designers of functional languages (Peyton Jones 2001; Felleisen et al. 2009) have been faced with the challenge of reconciling effectful with pure programming when extending their ideas to functional *systems*.

worlds and universes. Felleisen et al. (2009) propose *Worlds*, one of the roots of this work. A World is a context within which a functional program responds to a fixed set of events chosen for teaching novice programmers. Concurrency is inherent in the model; the user's actions are interleaved with other events occurring in the system, making concurrency an integral part of the design process. The following sample World maintains a counter, incremented as time passes, that is reset on key-press and drawn to the screen each time it changes:

```
(big-bang 0 [on-tick (lambda (i) (+ i 1))]
            [on-key (lambda (i k) 0)]
            [to-draw (lambda (i) (text (~a i) 20 "black"))])
```

A World program is neither continuation-passing-style nor monadic. Rather than composing chains of sequential actions, the programmer focuses on formulating responses to asynchronous events. In this context, the programmer must keep in mind that the event following an action is not knowable.

Despite their concurrency, Worlds yield a functional model of I/O, since each transition function reacting to events is pure. Each transition function has roughly the type

$$\mathit{WorldState} \times \mathit{Event} \rightarrow \mathit{WorldState} \times \mathit{Actions}$$

sometimes omitting the *Event* input or *Actions* output. The insight that the specific details of each transition function's type signature can be generalized into instantiations of this general signature was one of the steps that led to Network Calculus and the dataspace model. Furthermore, the "operating system" underpinning a particular World keeps track of its state between invocations of transition functions; this, too, was early inspiration for Network Calculus. The behavior of a World program is all possible functional compositions of its event handlers with appropriate streams of events, again directly comparable to Network Calculus and the dataspace model.

World programs compose to form a *Universe*, communicating in a strict hub-and-spoke topology. Though each World runs in parallel with its neighbors in the Universe, Worlds are themselves single-threaded and cannot create or destroy Worlds dynamically. When Worlds fail, the hub is informed of their disappearance. The dataspace model can be seen as a generalization of this structure that also recursively demotes each Universe to a mere World in some larger Universe.

Worlds and Universes suffice for teaching novices, but they do not scale to "real" software. Following Hudak and Sundaresh in evaluating functional approaches to I/O (Hudak and Sundaresh 1988), we see that World programs enjoy good support for equational reasoning and interactive use but have only limited support for handling error situations. Furthermore, the fixed set of events offered to Worlds, the strictness of the communications topology, and the associated concurrency model in Universes impose serious restrictions that are lifted by the design of the dataspace model.

MONADIC I/O. Peyton Jones and Wadler (1993) propose *monadic I/O*, famously and successfully subsequently incorporated into Haskell (Peyton Jones 2001). The monadic approach to combination of functional programming and effectful programming is to reify side-effecting operations as values, and to use monadic type structure to enforce correct sequencing and other constraints on interpretation of these effect values. While a number of benefits flow from this design, there is a tendency for the resulting monads to mimic the familiar concepts, style, and techniques of imperative programming languages. For example, Haskell uses its monadic I/O facility to offer the programmer exceptions, mutable reference cells, threads, locks, file handles and so forth. These are accessed via *procedure calls* that can often be directly mapped to similar procedures in imperative languages.

The dataspace model shares the notion of reification of actions with the monadic approach. However, it differs strongly in two respects. First, it is event-based. The monadic model directly parallels traditional approaches to input, including blocking actions, callbacks and events. Second, the dataspace model uses a single language of general-purpose actions—state change notifications and messages—as a *lingua franca* through which many disparate protocols are expressed. The monadic approach uses many different monadic languages and interpreters. For example, Haskell's `IO` monad includes special-purpose representations for a fixed, large suite of actions, while the dataspace model offers only message- and assertion-based information exchange, and expects neighboring actors to interpret encoded descriptions of actions relayed via messages and assertions. On the one hand, the dataspace approach is modularly extensible, but on the other hand, it limits itself to a single form of interactivity, whereas monadic type structure can be used to encode a wide range of effects.

concurrent ml. CML (Reppy 1991, 1999) is a combinator language for coordinating I/O and concurrency, available in SML/NJ and Racket (Flatt and PLT 2010, version 6.2.1, S11.2.1). CML uses synchronous channels to coordinate preemptively-scheduled threads in a shared-memory environment. Like Syndicate, CML treats I/O, communication, and synchronization uniformly. In contrast to Syndicate, CML is at heart *transactional*. Where CML relies on garbage collection of threads and explicit "abort" handlers to release resources involved in rolled-back transactions, Syndicate monitors assertions of interest to detect situations when a counterparty is no longer interested in the outcome of a particular action. CML's threads inhabit a single, unstructured shared-memory space; it has no equivalent of Syndicate's process isolation and layered media.

### 11.2.2 *Functional operating systems*

The dataspace model harks back to early research on functional operating systems (Henderson 1982; Stoye 1986), as it is literally a message-based functional OS for coordinating concurrent components "in the large". Hudak and Sundaresh (1988) survey approaches to functional I/O; the dataspace model is distantly related to their "stream-based I/O" formulation. They suggest that a functional I/O system should provide support for (1) equational reasoning, (2) efficiency, (3) interactivity, (4) extensibility, and (5) handling of "anomalous situations," or errors.

equational reasoning. Like Worlds and Universes, the dataspace model allows for equational reasoning because event handlers are functional state transducers. When side-effects are absolutely required, they can be encapsulated in a process, limiting their scope. The state of the system as a whole can be partitioned into independent processes, allowing programmers to avoid global reasoning when designing and unit-testing their code (Eastlund and Felleisen 2009; Sullivan and Notkin 1990).

efficiency. A functional implementation of a dataspace manages both its own state and the state of its contained processes in a linear way. Hudak and Sundaresh, discussing their "stream" model of I/O, remark that the state of their kernel "is a single-threaded object, and so can be implemented efficiently". The dataspace model shares this advantage with streams.

There are no theoretical obstacles to providing more efficient and scalable implementations of the core dataspace model abstractions. Siena (Carzaniga, Rosenblum and Wolf 2000) and Hermes (Pietzuch and Bacon 2002) both use subscription and advertisement information to construct efficient routing *trees*. Using a similar technique for dataspace implementation would permit scale-out of the corresponding layer without changing any code in application processes.

INTERACTIVITY. The term "interactivity" in this context relates to the ability of the system to interleave communication and computation with other actors in the system, and in particular to permit user actions to affect the evolution of the system. The dataspace model naturally satisfies this requirement because all processes are concurrently-evolving, communicating entities.

EXTENSIBILITY. The dataspace model is extensible in that the ground dataspace multiplexes raw Racket events without abstracting away from them. Hence, driver processes can be written to adapt the system to any I/O facilities that Racket offers in the future. The collection of request and response types for the "stream" model given by Hudak and Sundaresh (Hudak and Sundaresh 1988, section 4.1) is static and non-extensible because their operating system is monolithic, with device drivers baked in to the kernel. On the one hand, monolithicity means that the possible communication failures are obvious from the set of device drivers available; on the other hand, its simplistic treatment of user-to-driver communication means that the system cannot express the kinds of failures that arise in microkernel or distributed systems. Put differently, a monolithic stream system is not suitable for a functional approach to systems programming.

The dataspace model action type (figure 12) appears to block future extensions because it consists of a finite set of variants. This appearance is deceiving. Actions are merely the interface between a program and its context. Extensibility is due to the information exchanged between a program and its peers. In other words, the action type is similar to the limited set of core forms in the lambda calculus, the limited set of methods in HTTP and the handful of core system calls in Unix: a finite kernel generating an infinite spectrum of possibilities.

ERRORS. In distributed systems, a request can fail in two distinct ways. Some "failures" are successful communications with a service, which just happens to fail at some requested task; but some failures are caused by the unreachability of the service requested. SYNDICATE represents the former kind of failure via protocols capable of expressing error responses to requests. For the latter kind of failure, it uses assertions as presence information to detect unavailability and crashes.

### 11.2.3 *Process calculi*

A major family of concurrency models is based on π-calculus (Milner, Parrow and Walker 1992) and its many derivatives.

THE CONVERSATION CALCULUS. Spiritually closest to the dataspace model is the Conversation Calculus (Caires and Vieira 2010; Vieira, Caires and Seco 2008), based on π-calculus. Its *conversational contexts* scope multi-party interactions. Named contexts nest hierarchically, forming a tree. Processes running within a context may communicate with others in the same context and processes running in their context's immediate container. Contexts on distinct tree branches may share a name and thus connect transparently through hyperlinks. The Conver-sation Calculus also provides a Lisp-style `throw` facility that aborts to the closest `catch` clause. This mechanism enables supervisor-like recovery strategies for exceptions.

Although the Conversation Calculus and the dataspace model serve different goals—the former is a calculus of services while the latter is the core of a language design—the two are similar. Like a dataspace, a conversational context has both a spatial meaning as a location for computation and a behavioral meaning as a delimiter for a session or protocol instance. Both models permit communication *within* their respective kinds of boundary as well as *across* them.

The two models starkly differ in two aspects. First, the dataspace model cannot *transparently* link subnets into logical overlay networks because its actors are nameless. Instead, inter-subnet routing has to be implemented in an explicit manner, based on state-change notifications. Proxy actors tunnel events and actions across links between subnets; once such a link is established, actors may ignore the actual route. Any implementation of Conversation Calculus must realize just such routing *within* the implementation; the dataspace model provides the same expressiveness as a library feature *external* to the implementation.

Second, Conversation Calculus lacks state-change notifications and does not automatically signal peers when conversations come to an end—normally or through failure. Normal termination in Conversation Calculus is a matter of convention, while exceptions signal failure to containing contexts but *not* to remote participants in the conversational context. In contrast, the state-change notification events of the dataspace model signal failure to all interested parties transparently.

MOBILE AMBIENTS. Cardelli and Gordon (2000) describe the *Mobile Ambient Calculus*. An *ambient* is a nestable grouping of processes, an "administrative domain" within which computation and communication occur.

At first glance, Mobile Ambients and the dataspace model are duals. While the dataspace model focuses on routing data between domains, from which code mobility can be derived via encodings, Mobile Ambients derives message routing by encoding it in terms of a primitive notion of process mobility. By restricting ourselves to transporting *data* rather than *code* from place to place, we avoid a large class of mobility-related complication and closely reflect real networks, which transport only first-order data. Moving higher-order data (functions, objects) happens via encodings. Furthermore, mobility of code is inherently point-to-point, and the $\pi$-calculus-like names attached to ambients reflect this fact. SYNDICATE's pattern-based routing is a natural fit for a more general class of conversational patterns in which duplication of messages is desired.

### 11.2.4 *Formal actor models*

Another major family of concurrency models has its roots in the actor model of Hewitt and others (Hewitt, Bishop and Steiger 1973; Agha et al. 1997; De Koster, Van Cutsem and De Meuter 2016). A particularly influential branch of the family, having some similarities to the dataspace model, is due to Agha and colleagues (Agha et al. 1997; Callsen and Agha 1994; Varela and Agha 1999).

Varela and Agha's variation on the actor model (Varela and Agha 1999) groups actors into hierarchical *casts* via *director* actors, which control some aspects of communication between their casts and other actors. If multicast is desired, it must be explicitly implemented by a director. While casts and directors have some semblance to the layered dataspace model, the two differ in many aspects. The availability of publish/subscribe to dataspace actors automatically provides multicast without forcing all members of a layer to use the same conversational pattern. Directors are computationally active, but dataspaces are not. In their place, the dataspace model employs *relay actors* that connect adjacent layers. Finally, Varela and Agha's system lacks state-change notification events and thus cannot deal with failures easily. They propose mobile *messenger* actors for localizing failure instead.

In Callsen and Agha's ActorSpace (Callsen and Agha 1994) actors join and leave *actorspaces*. Each actorspace provides a scoping mechanism for pattern-based multicast and anycast message delivery. Besides communication via actorspace, a separate mechanism exists to let actors address each other directly. In contrast, SYNDICATE performs all communication with interest-based routing and treats dataspaces as specialized actors, enforcing abstraction boundaries and making it impossible to distinguish between a single actor or an entire dataspace providing some service. Actors may join multiple actorspaces, whereas dataspace model actors may only inhabit a single dataspace, reflecting physical and logical layering and giving an account of locality. In the dataspace model, actors "join" multiple dataspaces by spawning proxy actors, which tunnel events and actions through intervening layered dataspaces. Finally, ActorSpace does not specify a failure model, whereas dataspaces signal failure with state-change notification events.

Peschanski et al. (2007) describe an actor-style system with multicast channels connecting actors, but do not consider layering of actor locations. They consider link failure, channel failure and location failure as distinct concepts, whereas the dataspace model unifies all of these in the state-change event mechanism. They describe their multicast routing technique as "additive", comparing to the "subtractive" discard-relation-based technique of Ene and Muntean (2001). The formal model of dataspaces is a hybrid of additive and subtractive: actors are grouped in a dataspace, providing a crisp scope for broadcast, within which a discard-like subtractive operation is applied.

Fiege et al. (2002) consider *scoping* of actors. Their notion of actor visibility differs from ours: dataspace model actors explicitly route messages and presence between layers using $\downharpoonleft \cdot$ and friends, whereas their actors are automatically visible to all other actors having a common super-scope. Their scopes form a directed acyclic graph rather than a tree, whereas dataspace layering is strictly tree-like. Their *event mappings* are similar in function to relay actors in dataspace model programs, translating between protocols at adjacent layers.

Finally, actor models to date lack an explicit interface to the outside world. I/O remains a brute-force side-effect instead of a messaging mechanism. The functional approach to messaging and recursive layers of the dataspace model empowers us to treat this question as an implementation decision.

### 11.2.5 *Messaging middleware*

A comparison of SYNDICATE with publish/subscribe brokers (Eugster et al. 2003) supplies an additional perspective. Essentially, a dataspace corresponds to a broker: the subset of assertions declaring interests is the subscription table of a broker; the event and action queues at each actor are broker "queues"; characteristic protocols are used for communication between parties connected to a broker; etc. In short, the dataspace model can be viewed as a formal semantics of brokers. The SYNDICATE/RKT "broker" actor (appendix C) exposes a WebSocket-based protocol connecting SYNDICATE/RKT programs with SYNDICATE/JS programs running in the browser, taking an important first step toward investigation of SYNDICATE in a distributed setting.

## 11.3 LIMITATIONS AND CHALLENGES

As we have seen, SYNDICATE is able to succinctly express solutions to a wide range of concurrency and coordination problems. However, it also suffers from weaknesses in addressing others. In this section, we explore some of the challenges to the dataspace model and the SYNDICATE language.

SECURITY PROPERTIES. Imagine a SYNDICATE encoding of the traditional actor model, where each actor knows its own identity and communications are conveyed as messages carrying tuples $(id, payload)$. An actor with id $x$ would assert $?(x, \star)$ in order to receive messages addressed to it; a peer $y$ would send messages $\langle x, \texttt{"hello"}\rangle$. However, nothing in the dataspace model prevents $y$, upon learning the name $x$, expressing interest $?(x, \star)$ itself, thereby receiving a carbon-copy of all messages addressed to $x$. To prevent this, some notion of *permissible assertions* for a given actor must be brought to bear.

Placing a trusted *firewall* between an untrusted actor and its dataspace can be used to enforce limits on the assertions made by an actor and its potential children. Key to the idea is that actor behaviors in the dataspace model are mere functions, opening the possibility of writing functions to *filter* the action and event streams traveling to and from an actor's behavior. In terms of the dataspace model formalism, a firewall can be as simple as shown in figure 91. The real SYNDICATE/RKT implementation is scarcely more complex. Use of a firewall to isolate untrustworthy actors such as $y$ can thus restore the security properties of the actor model. If our actor $y$, supposed to receive only messages addressed to $\texttt{id}_y$, is spawned with an action $P_y$, we can enforce its good behavior by interpreting

$$\textit{firewall}\ \ (((\star,\star) - ?\,(\star,\star)) \cup ?\,(\texttt{id}_y,\star))\ \ P_y$$

instead of $P_y$ itself. It remains future work to more closely integrate access controls with the dataspace model. Preliminary work toward a type system for the dataspace model also suggests that static enforcement of some of these properties is possible (Caldwell, Garnock-Jones and Felleisen 2017).

LOCKS AND MUTUAL EXCLUSION. Programs such as Dijkstra's "dining philosophers" require *locks* for controlling access to a shared, scarce resource. In the shared-memory concur-

$$f_{fw} : \forall\tau.(\mathbf{ASet} \times \mathcal{F}_\tau \to \mathcal{F}_\tau)$$

$$f_{fw}\ \pi\ f_{inner}\ e\ u = \begin{cases} \mathsf{exit}(\overrightarrow{\mathit{firewall}\ \pi\ a}) & \text{if } f_{inner}\ e\ u = \mathsf{exit}(\vec{a}) \\ \mathsf{continue}(\overrightarrow{\mathit{firewall}\ \pi\ a}, u') & \text{if } f_{inner}\ e\ u = \mathsf{continue}(\vec{a}, u') \end{cases}$$

$$\begin{aligned}
\mathit{firewall}\ \pi\ \langle c \rangle &= \begin{cases} \langle c \rangle & \text{if } c \in \pi \\ \cdot & \text{otherwise} \end{cases} \\
\mathit{firewall}\ \pi\ \frac{\pi_{in}}{\pi_{out}} &= \frac{\pi_{in} \cap \pi}{\pi_{out} \cap \pi} \\
\mathit{firewall}\ \pi\ \left(\mathsf{actor}\ f_{boot}\ \pi'\right) &= \mathsf{actor}\ f'\ \left(\pi' \cap \pi\right) \\
\text{where } f' = \lambda().&\begin{cases} \mathsf{init}(\overrightarrow{\mathit{firewall}\ \pi\ a}, \mathsf{pack}\ \langle \tau, (f_{fw}\ \pi\ f'', u) \rangle) & \text{if } f_{boot}() = \mathsf{init}(\vec{a}, \mathsf{pack}\ \langle \tau, (f'', u) \rangle) \\ \mathsf{exit}(\overrightarrow{\mathit{firewall}\ \pi\ a}) & \text{if } f_{boot}() = \mathsf{exit}(\vec{a}) \end{cases} \\
\mathit{firewall}\ \pi\ \left(\mathsf{dataspace}\ \vec{P}\right) &= \mathsf{dataspace}\ \left(\overrightarrow{\mathit{firewall}\ \pi\ P}\right)
\end{aligned}$$

Figure 91: "Firewalling" dataspace model actors

rency model, *semaphores* act as such locks. Performing a "down" operation on a semaphore is interpreted as claiming a resource; the corresponding "up" operation signifies resource release. The tuplespace model is able to implement the necessary mutual exclusion by assigning a specific tuple in the store as a quasi-physical representative of a lock. As tuples *move* from place to place, each tuple having an independent lifetime, the notion of a current *holder* of a tuple makes sense.[4] The locking protocol for a tuplespace, then, is to perform an `out(lock)` action to initialize the lock to its unlocked state, to perform `in(lock)` to claim the lock, and to release it by once more performing `out(lock)`. The actor model and other message-passing models must choose some other strategy, lacking shared state entirely; a common solution there is to *reify* a lock as an actor mediating access to the contested resource.

SYNDICATE's dataspaces are in some ways quite similar to tuplespaces. A key difference is that SYNDICATE assertions do not have the independent existence of tuples: multiple independent assertions of the same value cannot be distinguished from just one, and any observers expressing interest in a given assertion *all* receive updates regarding its presence in the dataspace. There is therefore no way for SYNDICATE's assertion-manipulation primitives to directly implement locking or mutual exclusion; the dataspace itself is not stateful in the right way. However, borrowing from the actor model, an *indirect* implementation of locking is perfectly possible, as we have seen already in figure 35.[5] The necessary state of each lock must be held

4 This property of tuples is known as "generativity" in the literature (Gelernter 1985, section 2, p. 82).

5 At heart, this problem is about *atomic transfer of ownership* of some resource. Interestingly, no realizable electronic computer network has any means of expressing such transfers. The actor model and SYNDICATE both share this within some actor, which is then able to make authoritative decisions about which peer is to be assigned the lock at any one time. This strategy is exactly the same as that required to implement locking in the actor model. A key improvement over both the equivalent actor model lock implementation and the tuplespace approach to locking is the ability of Syndicate protocols to rely on automatic assertion retraction on actor termination or failure: here, a lock-holder that crashes automatically releases the lock, freeing the lock-maintaining actor to assign the lock to some other waiting client.[6]

MESSAGE-BASED VS. ASSERTION-BASED SIGNALING. Syndicate not only allows but forces a kind of temporal decoupling of components: every time a request travels via the dataspace, the programmer may rely on eventually getting the answer needed, but does not know in general how soon. Other things may also happen in the meantime. Some protocol domains rely intrinsically on tight temporal bounds—sometimes on the "synchrony" hypothesis, on being able to access any part of the application's state in "no time at all"—and for these problems, Syndicate may be of limited application. Implementation of an IRC server makes for an interesting case study here: traditional implementations take advantage of being able to "stop the world" and query the global server state. However, even there we can adapt to the forced decoupling mandated by Syndicate and get some advantages. Appendix B presents the case study in more detail.

CAPTURING AND EMBEDDING OF SETS OF ASSERTIONS. Patterns in Syndicate match individual assertions from the dataspace, and pattern variables capture single host-language values. Similarly, the assertion *templates* in endpoints only allow embedding of fields holding single host-language values. From time to time, direct extraction or insertion of assertion *sets* would be valuable. For example, the "broker" program connecting Syndicate/rkt to Syndicate/js (appendix C) relays arbitrary assertion sets in bulk between the Racket and JavaScript sides of the network connection, whether those sets are finite or not. Given that the existing Syndicate pattern language only allows matching of single values, the broker relies on ad-hoc extensions to the language design in order to perform its task.

COMPLEX "JOINS" ON ASSERTIONS. In the IRC server example discussed in Appendix B, the program must communicate an initial set of channel members upon channel join. Setting aside interactions with the complications of `NICK/QUIT` tracking discussed in the appendix, one might imagine using Syndicate/rkt's `immediate-query` form to imperatively compute the initial set of channel members:

```
(define conns (immediate-query [query-set (ircd-channel-member Ch $c) c]))
(define nicks (immediate-query [query-set (ircd-connection-info conns $n) n]))
(send-initial-name-list! Ch nicks)
```

characteristic with such physical networks. This raises questions as to whether tuplespaces can be implemented "primitively" at all, or whether they must always be encoded in terms of some underlying message-passing system.

6 Erlang can use exit signals to achieve a similar outcome, as can other actor languages with an analogous construct. Various suggestions have been made to overcome the "lost tuple" problem in tuplespace languages (Bakken and Schlichting 1995).

Unfortunately, this doesn't work, because as we have just discussed, embedding sets of values like `conns` into an assertion set is not currently supported. An alternative is to iterate over `conns`, performing an `immediate-query` for each peer connection, making $n+1$ round trips to the dataspace. A future Syndicate design could perhaps include some way of specifying a *join*-like construct: a way of asserting "interest in all records `(ircd-connection-info c n)` where `c` is drawn from any record `(ircd-channel-member Ch c)`," retrieving the information of interest in a single round trip.

NEGATIVE KNOWLEDGE AND "SNAPSHOTS." It can be awkward to express programs that interpret the *absence* of a particular assertion as logically meaningful, a form of negation; recall the machinations that the code of figure 83 (page 227) was forced to engage in. There, the facet *assumed* absence of relevant knowledge at startup, acting as if no relevant assertions were present. It then updated its beliefs upon discovery of relevant knowledge, and altered its actions accordingly. This difficulty is related to the "open world" nature of Syndicate dataspaces. Relatedly, as discussed in section 4.8, in situations where one may validly make a closed-world assumption, it is awkward to gather a "complete" set of facts relevant to a given query. For example, consider again the task of the IRC server when a user joins an existing IRC channel. The server must collect and send the new user a list of all users already present in the channel before transitioning into an incremental membership-maintenance mode. This is the inverse of the IRC *client* example motivating the use of `assert!` and `retract!` seen in section 6.6. The IRC server solves the problem by establishing interest in assertions describing channel membership, then waiting for a rather arbitrary length of time—two dataspace-and-back round trip times—before calling the membership information it has gathered at that point "enough" and transmitting it. How long is long enough to wait? In this case, two round trips sufficed, but in general, *no limit can be placed*. At its root, the reason is that expression of interest in a record may result in lazy production of that record. A special, but important, case is that of a *relay* actor whose responsibility is to convey expressions of interest across some gap—be it a network link, or simply a bridge between two adjacent nested dataspaces—and to convey the resulting assertions back in the other direction. Each relay introduces latency between detection of interest in an assertion and production of the assertion itself. Actors interested in assertions cannot in general predict any upper bound on this latency.

# 12

## *Conclusion*

### 12.1 REVIEW

The thesis that this dissertation defends is

> SYNDICATE provides a new, effective, realizable linguistic mechanism for sharing state in a concurrent setting.

As in the introduction, we can examine this piece by piece.

MECHANISM FOR SHARING STATE. We have seen that, as promised, the dataspace model (chapter 4) directly focuses on the management, scoping, and sharing of conversational state among collaborating actors. Actors exchange state-change notification events with their surroundings, describing accumulated conversational knowledge. Dataspaces use epistemic knowledge of actors' interests to route information and record provenance to maintain integrity of the store after partial failure.

LINGUISTIC MECHANISM. The full SYNDICATE language design (chapter 5) equips a host language used to write leaf actors in a dataspace with new linguistic constructs: facets, endpoints and fields. Facets manifest conversations and conversational state within an actor. Each facet comprises a bundle of private state, shared state, subscriptions and event-handlers. Programmers tie facet lifetimes to the lifetimes of conversational frames. Facets nest, forming a structure that mirrors the logical structure of ongoing conversations.

REALIZABILITY. A new data structure, the assertion trie, provides efficient pattern matching and event routing at the heart of SYNDICATE/RKT and SYNDICATE/JS, the two prototype SYNDICATE implementations (chapters 6 and 7).

EFFECTIVENESS. The effectiveness of the design is shown through examination of programming idioms (chapter 8), discussion of programming patterns and design patterns eliminated from SYNDICATE programs (chapter 9), and through preliminary confirmation of the expected performance of the implementation approach taken (chapter 10).

NOVELTY. While SYNDICATE draws on prior work, it stands alone at an interesting point in the design space of concurrency models (chapter 11).

## 12.2 NEXT STEPS

The Syndicate design gives programmers a new tool and a new way of thinking about coordination of concurrent components in *non-distributed* programs. This dissertation has developed an intuition, a computational model, and the beginnings of a programming model for Syndicate. There are several possible paths forward from here.

ENHANCEMENTS TO THE FORMAL MODELS. First, development of a Syndicate type system could allow programmers to capture and check specifications not only for structural properties of the data to be placed in each dataspace, but behavioral properties of actors participating in Syndicate conversations, including their roles, responsibilities and obligations. Second, the core dataspace model does not include any kind of programmer-visible name or name-like entity, but many protocols depend on some notion of globally unique token; equipping the formal model with either *unique* or *unguessable* tokens would allow exploration of the formal properties of such protocols and the programs that implement them. Finally, as part of work toward a model of distributed Syndicate, separating the *grouping* aspect of dataspaces from their *layering* aspect would allow investigation of "subnets": fragmentary dataspaces that combine to form a logical whole.

SYSTEM MODEL. The few experiments exploring Syndicate tool support so far have been promising, suggesting that the design might offer a new perspective on broader systems questions. Development of protocols for process control, for generalized "streams" of assertions, and for console-based or graphical user interaction with programs would allow experimentation with operating systems design. The Syndicate/rkt prototype implementation already includes use of contracts (Dimoulas et al. 2016) to check field invariants; perhaps new kinds of contract could be employed to check actor, role, or conversation invariants within and between dataspaces. The "firewall" mechanism for securing access to the dataspace could be combined with ideas from certificate theory (Ellison et al. 1999) to explore multiuser Syndicate. Strategies for orthogonal persistence of Syndicate actors could allow investigation of database-like, long-lived dataspaces. The existing "broker" approach to integrating Syndicate/rkt with Syndicate/js could be generalized to support polyglot Syndicate programming more generally. Finally, the implementations of Syndicate to date have employed single-threaded, cooperative concurrency; introduction of true parallelism would be an important step toward a distributed Syndicate implementation.

DISTRIBUTED SYSTEMS. The centrality of state machine replication to distributed systems (Lamport 1984) is one of the reasons to hope Syndicate might work well in a distributed setting, given the centrality of state replication to the dataspace model. Communication via assertions, rather than messages, can lead to protocols that automatically withstand lost messages, even in the presence of certain kinds of "glitching". That is, replication by state-change notification is in some sense self-synchronizing. Syndicate programs must already cope with certain forms of partial failure familiar from distributed systems; for example, messages can be "lost" if they are routed through a relay actor that crashes at an inopportune moment. Even though the underlying dataspace itself guarantees reliable message delivery, this guarantee only applies on a "hop-by-hop" basis. It would be interesting to attempt to scale this nascent resilience up to a distributed setting, perhaps even transplanting some of the benefits of SYNDICATE back into the fact space model. Finally, since certain aspects of causal soundness (definition 4.31) are helpful but not essential, we are free to consider alternative "subnet"-based implementation strategies, such as making actors build copies of the whole routing table themselves, leaving the dataspace "empty" and stateless, and using Bloom filters or similar to narrowly overapproximate the interests of an actor or group of actors.

# A

# Syndicate/js *Syntax*

Figure 92 presents an Ohm (Warth, Dubroy and Garnock-Jones 2016) grammar that extends JavaScript with Syndicate's new language features. Support is provided for spawning new actors (lines 11–16), for creating (lines 17–18) and configuring (lines 19–30) facets, for managing fields (lines 35–37), sending messages (line 38) and matching incoming events (lines 39–48). The remainder of the compiler from the extended JavaScript dialect to the core language is placed alongside the grammar in a separate 460-line JavaScript file.

In order to keep the compiler simple, some of the tasks performed by the Syndicate/rkt macro-based compiler are deferred to runtime in the Syndicate/js implementation. In addition, the Ohm system is, at heart, a parsing toolkit, and does not offer an analogue of the intricately interwoven multi-phase expansion process available in Racket's syntactic extension system; therefore, features such as *event expanders*, which allow the Syndicate/rkt programmer to define custom event pattern forms, are precluded. This limits the Syndicate/js programmer to those event pattern forms built-in to the compiler.

Two entry points to the compiler are provided: a command-line tool, for ordinary batch compilation, and a browser-loadable package. The latter allows for rapid development of Syndicate/js-based web applications by on-the-fly translating HTML `script` tags with a `type` attribute of "`text/syndicate-js`" into plain JavaScript that the browser can understand.

## example

Figure 93 shows a complete example browser-based Syndicate/js program. Figure 93a specifies the HTML structure of the page loaded into the browser; figure 93b specifies Syndicate/js code giving the program its behavior. Lines 4 and 5 of the HTML load the latest versions of the Syndicate/js compiler and runtime, respectively, from the `syndicate-lang.org` domain. Line 6 connects the HTML to the Syndicate/js program, making sure to correctly label the `type` of the linked code as `text/syndicate-js` in order to arrange for it to be compiled to plain JavaScript on the fly. Lines 7 and 8 are the user-visible interface; in particular, two elements are given identifiers in order for them to be accessible from the script. The clickable button is named `counter`, and the span of text forming its label is named `button-label`.

Line 1 of the script in figure 93b opens a block declaring the boot script for the ground dataspace to be run. Line 2 activates the Syndicate/js "user interface" driver, responsible for mapping assertions describing HTML fragments into the page as well as responding to interest in DOM events by establishing subscriptions and relaying events from the page into the

```
Syndicate <: ES5 {
  Statement
    += ActorStatement
    | DataspaceStatement
    | ActorFacetStatement
    | ActorEndpointStatement
    | AssertionTypeDeclarationStatement
    | FieldDeclarationStatement
    | SendMessageStatement

  FunctionBodyBlock = "{" FunctionBody "}"

  ActorStatement
    = spawnStar (named Expression<withIn>)? FunctionBodyBlock -- noReact
    | spawn (named Expression<withIn>)? FunctionBodyBlock     -- withReact

  DataspaceStatement
    = ground dataspace identifier? FunctionBodyBlock -- ground
    | dataspace FunctionBodyBlock                    -- normal

  ActorFacetStatement
    = react FunctionBodyBlock

  ActorEndpointStatement
    = on start FunctionBodyBlock                            -- start
    | on stop FunctionBodyBlock                             -- stop
    | assert FacetPattern AssertWhenClause? #(sc)           -- assert
    | on FacetEventPattern FunctionBodyBlock                -- event
    | on event identifier FunctionBodyBlock                 -- onEvent
    | stop on FacetTransitionEventPattern FunctionBodyBlock -- stopOnWithK
    | stop on FacetTransitionEventPattern #(sc)             -- stopOnNoK
    | dataflow FunctionBodyBlock                            -- dataflow
    | during FacetPattern FunctionBodyBlock                 -- during
    | during FacetPattern spawn (named Expression<withIn>)?
        FunctionBodyBlock                                   -- duringSpawn

  AssertWhenClause = when "(" Expression<withIn> ")"

  AssertionTypeDeclarationStatement
    = (assertion | message) type identifier "(" FormalParameterList ")"
        ("=" stringLiteral)? #(sc)

  FieldDeclarationStatement = field MemberExpression ("=" AssignmentExpression<withIn>)? #(sc)
  MemberExpression += field MemberExpression -- fieldRefExp
  UnaryExpression += delete field MemberExpression -- fieldDelExp

  SendMessageStatement = "::" Expression<withIn> #(sc)

  FacetEventPattern
    = message FacetPattern   -- messageEvent
    | asserted FacetPattern  -- assertedEvent
    | retracted FacetPattern -- retractedEvent

  FacetTransitionEventPattern
    = FacetEventPattern          -- facetEvent
    | "(" Expression<withIn> ")" -- risingEdge

  FacetPattern
    = LeftHandSideExpression metalevel decimalIntegerLiteral -- withMetalevel
    | LeftHandSideExpression                                 -- noMetalevel

  // (Keyword definitions elided)
}
```

Figure 92: Ohm grammar for the SYNDICATE/JS extension to JavaScript

```
<!doctype html>
<html>
  <meta charset="utf-8">
  <script src="http://syndicate-lang.org/dist/syndicatecompiler.js"></script>
  <script src="http://syndicate-lang.org/dist/syndicate.js"></script>
  <script type="text/syndicate-js" src="index.js"></script>
  <h1>Button Example</h1>
  <button id="counter"><span id="button-label"></span></button>
</html>
```

(a) `index.html`, HTML page hosting the program

```
1  ground dataspace {
2    Syndicate.UI.spawnUIDriver();
3    spawn {
4      var ui = new Syndicate.UI.Anchor();
5      field this.counter = 0;
6      assert ui.html('#button-label', '' + this.counter);
7      on message Syndicate.UI.globalEvent('#counter', 'click', _) {
8        this.counter++;
9      }
10   }
11 }
```

(b) `index.js`, Syndicate/js source code, automatically translated to plain JavaScript

Figure 93: Example Syndicate/js in-browser program

dataspace. Lines 3–10 comprise the lone actor in this program. Line 4 constructs a JavaScript object offering convenience methods for constructing assertions and event patterns. On line 6, we see one of its uses. The actor asserts a record whose interpretation is, loosely, "please add the literal string representation of the value of `this.counter` to the collection of DOM nodes inside the element with ID `button-label`." The assertion make reference to the field `this.counter` declared on line 4. The dataflow mechanism ensures that as `this.counter` is updated, assertions and subscriptions depending on it are automatically updated to match. Lines 7–9 comprise the sole event handler endpoint in the program, soliciting notifications about mouse-clicks on the DOM element with ID `counter`. In response, the actor increments its `this.counter` field.

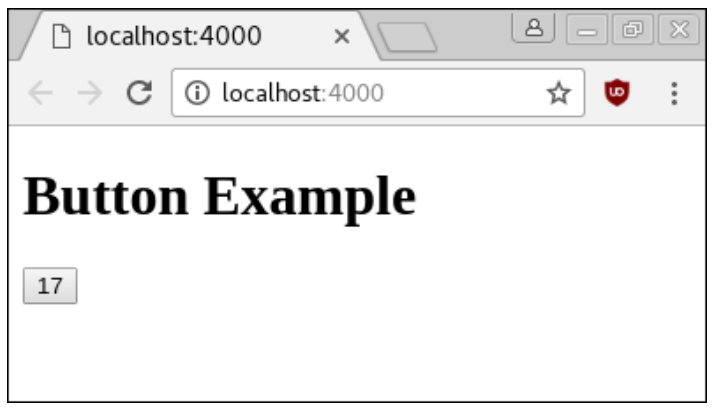

Figure 94: The running program

The net effect of all of this is shown in figure 94. Each time the user clicks the button, the number on the button's label is incremented.

# B

# *Case study: IRC server*

Syndicate encourages programmers to design protocols that use assertion signaling, rather than messages, to exchange information. In many cases, this results in a "logical" characterization of protocol progress that is robust in the face of unexpected processing latency and partial failure. Use of messages within a conversational frame established by assertions is also, in many cases, perfectly sensible. However, in some cases—predominantly integration with non-Syndicate protocols, where messages alone transfer changes in application state—the Syndicate programmer must still carefully reason about order of events and latency. The reasoning involved is in some ways similar to that used to design away races in languages with other approaches to concurrency, but is focused on epistemic questions rather than questions of state; programmers think about which components know certain facts, rather than which locks are in certain states. Solutions include tracking causality in exchanged messages, or explicitly serializing communications through a single actor performing the role of single-point-of-truth. Syndicate is no different to the actor model in this regard; programming in Erlang, for example, involves exactly the same kinds of considerations.[1]

A challenging example is found in the IRC protocol (Oikarinen and Reed 1993; Kalt 2000). Upon joining a channel, the server sends the client first an aggregate of all users previously present in the channel. Then, updates to that set are delivered via incremental `JOIN` and `PART` notifications; if a peer disconnects, `QUIT` replaces `PART`. However, if a channel member decides to change its nickname, this is to be reported by the server not as a `PART` of the old nickname followed by a `JOIN` of the new, but by a special `NICK` message. In the Syndicate IRC server case study,[2] the requirements thus far can be met with only modest contortions (figure 95). The challenge appears when we notice the requirement that if our connection is in two channels, and some peer `X` is in those same channels, and `X` renames itself to `Y`, the server should send only *one* `NICK` message; likewise, if `X` disconnects, only *one* `QUIT` message should be sent. That is, `NICK` and `QUIT` messages relate to connected users, not channels, but are only delivered to a client when they are relevant, namely when the client has one or more channel in common with the name-changing or disconnecting user. The code shown in figure 95 delivers redundant `NICK` and `QUIT` messages in these situations. A different approach is called for.

1 For example, the internal architecture of RabbitMQ had to be revised several times to avoid RPC-like interactions in favor of unidirectional streaming in order to avoid time-of-check-to-time-of-use problems stemming from the fact that messages between independent pairs of actors may legitimately be delivered in any order in Erlang (and the actor model). If multiple paths from a stateful component to some sink exist, then it is perfectly possible for updates involving the stateful component to arrive out-of-order.

2 Source code file `examples/ircd/session.rkt` in the Syndicate repository.

```
(during (ircd-channel-member $Ch this-conn)
  (field [initial-names-sent? #f]
         [initial-member-nicks (set)])

  (on-start (flush!)
            (flush!) ;; two round-trips to dataspace: gather current peers
            (define nicks (initial-member-nicks))
            (initial-names-sent? #t)
            (initial-member-nicks 'no-longer-valid)
            (send-initial-name-list! Ch nicks))

  (during (ircd-channel-member Ch $other-conn)
    (field [current-other-name #f])

    (define/query-value next-other-name #f
      (ircd-connection-info other-conn $N)
      N)

    (on (retracted (ircd-channel-member Ch other-conn))
      (when (current-other-name)
        (send-PART (current-other-name) Ch)))

    (begin/dataflow
     (when (not (equal? (current-other-name) (next-other-name)))
       (cond
        [(not (next-other-name))     ;; other-conn is disconnecting
         (send-QUIT (current-other-name))]
        [(not (initial-names-sent?)) ;; still gathering initial list
         (initial-member-nicks (set-add (initial-member-nicks)
                                        (next-other-name)))]
        [(not (current-other-name))  ;; other-conn is joining
         (send-JOIN (next-other-name) Ch)]
        [else                        ;; it's a nick change
         (send-NICK (current-other-name) (next-other-name))])
       (current-other-name (next-other-name))))))
```

Figure 95: Heart of the IRC server channel-membership-tracking code.

```
(field [peer-common-channels (hash)]
       [peer-names (hash)])

(define (add-peer-common-channel! other-conn Ch)
  (peer-common-channels
    (hashset-add (peer-common-channels) other-conn Ch)))

(define (remove-peer-common-channel! other-conn Ch)
  (peer-common-channels
    (hashset-remove (peer-common-channels) other-conn Ch)))

(define (no-common-channel-with-peer? other-conn)
  (not (hash-has-key? (peer-common-channels) other-conn)))

(define (forget-peer-name! other-conn)
  (peer-names (hash-remove (peer-names) other-conn)))

(define (most-recent-known-name other-conn)
  (hash-ref (peer-names) other-conn #f))

(define (remember-peer-name! other-conn name)
  (peer-names (hash-set (peer-names) other-conn name)))
```

Figure 96: Additional per-connection IRC server fields for `NICK`/`QUIT` deduplication.

Traditional IRC server implementations such as the original `ircd` (as of version `irc2.11.2p3`) and newer implementations such as `miniircd`[3] are able to avoid these concerns. Two differences in design interact to make this possible. First, they are single-threaded, event-driven programs. In effect, all state in the system is local to the active thread. Second, notification transmission is performed by the component responsible for the user being renamed or disconnecting, giving a convenient place to store a transient "checklist" of users to whom a particular `NICK` or `QUIT` notification has already been delivered. When preparing such notifications, these programs simply loop over all members of all the changing user's channels, making a note of peers to whom they have sent notifications as they go, in effect *deduplicating* the notifications.

Nothing prevents us from writing a Syndicate IRC server in this style: a single "server" actor could hold all relevant state, with a facet for each connected user; in its event handlers, it would be able to interrogate the instantaneous state of the server as a whole without having to make allowance for the temporal decoupling that arises every time a Syndicate actor accesses its dataspace. However, taking this approach forfeits the advantages offered by idiomatic Syndicate design. In the Syndicate IRC implementation, authoritative aggregate system state lives in the dataspace, not in individual actors, and notification transmission is the responsibility of the component representing the party to be notified; deduplication must happen there.

3 https://github.com/jrosdahl/miniircd

```
(during (ircd-channel-member $Ch this-conn)
  (field [initial-names-sent? #f]
         [initial-member-nicks (set)])

  (on-start (flush!)
            (flush!) ;; two round-trips to dataspace: gather current peers
            (define nicks (initial-member-nicks))
            (initial-names-sent? #t)
            (initial-member-nicks 'no-longer-valid)
            (send-initial-name-list! Ch nicks))

  (during (ircd-channel-member Ch $other-conn)
*   (on-start (add-peer-common-channel! other-conn Ch))
*   (on-stop (remove-peer-common-channel! other-conn Ch)
*            (when (no-common-channel-with-peer? other-conn)
*              (forget-peer-name! other-conn)))

    (field [current-other-name #f])

    (define/query-value next-other-name #f
      (ircd-connection-info other-conn $N)
      N)

    (on (retracted (ircd-channel-member Ch other-conn))
      (when (current-other-name)
        (send-PART (current-other-name) Ch)))

    (begin/dataflow
     (when (not (equal? (current-other-name) (next-other-name)))
       (cond
        [(not (next-other-name))     ;; other-conn is disconnecting
*        (when (most-recent-known-name other-conn)
*          (send-QUIT (current-other-name))
*          (forget-peer-name! other-conn))]
        [(not (initial-names-sent?)) ;; still gathering initial list
         (initial-member-nicks (set-add (initial-member-nicks)
                                        (next-other-name)))
*        (remember-peer-name! other-conn (next-other-name))]
        [(not (current-other-name))  ;; other-conn is joining
         (send-JOIN (next-other-name) Ch)
*        (remember-peer-name! other-conn (next-other-name))]
        [else                        ;; it's a nick change
*        (when (not (equal? (next-other-name)
*                           (most-recent-known-name other-conn)))
*          (send-NICK (current-other-name) (next-other-name))
*          (remember-peer-name! other-conn (next-other-name)))])
       (current-other-name (next-other-name))))))
```

Figure 97: IRC server channel-membership-tracking with NICK/QUIT deduplication.

To perform this deduplication, the actor must track exactly the names of peers with whom we share a channel. The simplest approach I could come up with uses two new connection-scoped fields to do this. Figure 96 shows the new fields and their use. The changes to the code of figure 95 are the lines marked with * in figure 97. Lines 10–11 manage the connection's view of which other connections have a channel in common with this connection. The actual deduplication, the purpose of the exercise, occurs on lines 25–27 and 36–39.

One noteworthy feature of the code in figure 96 is its similarity to a special-purpose representation of a local dataspace containing "virtual assertions" of the form

```
(ircd-common-channel this-conn other-conn)
(ircd-connection-info other-conn name)
```

The fact that the program already relies on `ircd-connection-info` assertions in the dataspace raises the question of why we do not simply assert

```
(ircd-common-channel this-conn other-conn)
```

within the `during` clause starting on line 9 of figure 97, and add a new facet to the connection actor reacting to `ircd-common-channel`, tracking `ircd-connection-info` and issuing `NICK` and `QUIT` messages when required. The answer is that building an initial summary of names is a stateful procedure that is part of joining an individual channel, while tracking `NICK` changes and `QUIT` events is done on a per-connection basis. It would be possible for the summary-construction process to add a nickname `X` to its set, for `X` to rename itself `Y`, and for the corresponding "`:X NICK Y`" message to be transmitted *before* the summary list, containing the already-obsolete `X`. Absent the requirement to summarize channel members in a manner syntactically distinct from subsequent changes to channel membership, this assertion-based approach of "following the logic" would work well.

# C

# *Polyglot* SYNDICATE

Many of the programs developed in SYNDICATE have involved multiple separate processes, some running SYNDICATE/RKT and others SYNDICATE/JS code, communicating via a simple JSON-based encoding of SYNDICATE events carried over WebSockets. Informally, imagine a function $enc(\cdot)$ which maps SYNDICATE objects to JSON terms. We might encode tries like this:

$$\begin{aligned} enc(\mathsf{mt}) &= [] \\ enc(\mathsf{ok}(\alpha)) &= [enc(\alpha)] \\ enc(\mathsf{br}(T', \{s \mapsto T, \ldots\})) &= [enc(T'), [[enc(s), enc(T)], \ldots]] \end{aligned}$$

and might encode SYNDICATE events like this:

$$\begin{aligned} enc(\langle c \rangle) &= [\texttt{"message"}, enc(c)] \\ enc(\frac{\pi_i}{\pi_o}) &= [\texttt{"patch"}, [enc(\pi_i), enc(\pi_o)]] \end{aligned}$$

Interoperation between Racket and JavaScript requires some agreement on the atoms and structure-types exchanged. I have chosen a conservative approach of identifying corresponding strings, numbers and booleans in each of Racket, JavaScript and JSON. Racket lists map to JSON and JavaScript arrays. Racket "prefab" structs map to JSON objects with special `@type` and `fields` members, which in turn map to the "structs" used extensively in the JavaScript dataspace implementation. JSON's objects—key/value dictionaries—are not otherwise supported, consonant with the restrictions on SYNDICATE/JS assertions discussed in section 7.2.1.

At each end of a connected WebSocket, a SYNDICATE actor maps between events arriving from its dataspace and JSON-encoded packets arriving from the socket. Depending on the details of the transformation between events and packets, a number of different effects can be obtained.

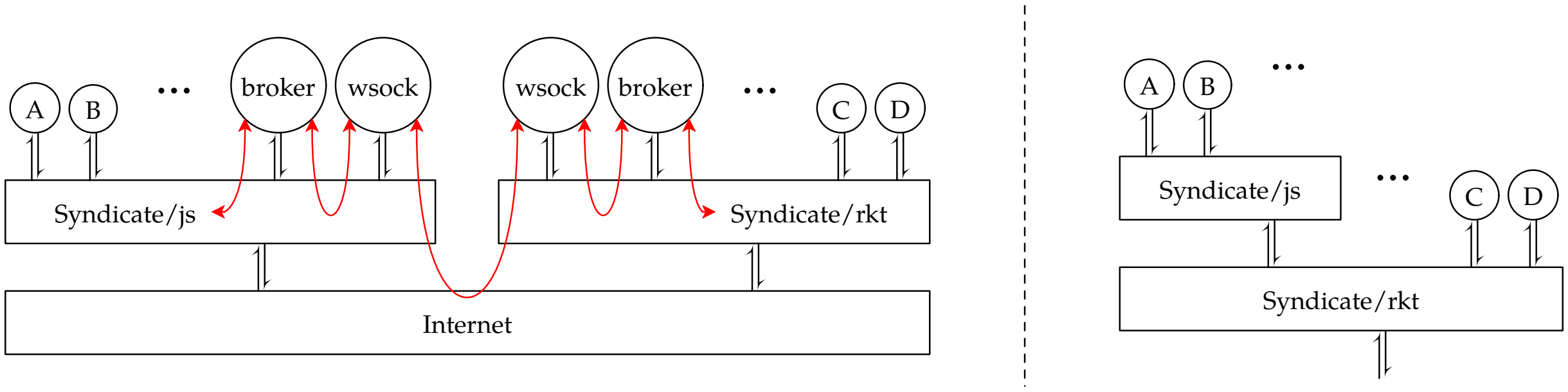

Figure 98: Physical (left) and logical (right) arrangement of connected SYNDICATE processes.

Figure 98 shows two separate Syndicate processes communicating via WebSockets. The left-hand portion of the figure illustrates the "physical" arrangement: two processes, connected via the Internet, with Syndicate actors contained in each; in particular, with one actor ("wsock") on each side dedicated to managing a WebSocket connection, and one actor ("broker") dedicated to relaying between local dataspace events and transmitted WebSocket JSON messages.[1] The right-hand portion of the figure shows one possible logical arrangement that can be achieved.

The illustrated configuration is asymmetric, despite the seeming symmetry of the "physical" arrangement; the key is in the transformations applied in the "broker" actors at each end of the link. If the Racket-side "broker" wraps received assertions in a `shared()` constructor, and the JavaScript-side "broker" relays *out* assertions labeled with a `toServer()` constructor, and labels assertions with a `fromServer()` constructor when relaying them *in*, the resulting logical arrangement has the shape depicted. Imagine that D in the diagram has expressed interest in some assertion $\texttt{shared}(x)$, and that A wishes to assert $x$ such that D can see it. D simply asserts $?\texttt{shared}(x)$ as usual, and A asserts $\texttt{toServer}(x)$. The "broker" on A's side has previously asserted $\{?\texttt{toServer}(\star), ??\texttt{fromServer}(\star)\}$, thereby expressing interest in *outbound* assertions as well as interest in *interest in inbound* assertions.[2] After A's action, the broker thus learns that $\texttt{toServer}(x)$ has been asserted, and accordingly sends $enc(\frac{\{x\}}{\emptyset})$ along the WebSocket. The broker on the Racket side receives and decodes this event, and then transforms the assertions carried within it by wrapping them with the `shared()` constructor. It then sends the resulting event, $\frac{\{\texttt{shared}(x)\}}{\emptyset}$, to its dataspace as if the event were endogenous. D then learns of the assertion as usual. Assertions may also flow in the reverse direction: if B asserts $?\texttt{fromServer}(y)$, then the JavaScript-side broker sends $enc(\frac{\{?y\}}{\emptyset})$ through the WebSocket, and the Racket-side broker asserts $\{\texttt{shared}(?y), ?\texttt{shared}(y)\}$. Note that the Racket-side broker has now expressed interest in $\texttt{shared}(y)$ assertions as if it were interested in such assertions itself. If C then asserts $\texttt{shared}(y)$, the Racket-side broker receives an event $\frac{\{\texttt{shared}(y)\}}{\emptyset}$, transforms it to $\frac{\{y\}}{\emptyset}$, and relays it to the JavaScript-side broker, which transforms it to $\frac{\{\texttt{fromServer}(y)\}}{\emptyset}$ before delivering it to the JavaScript-side dataspace, again as if it were endogenous. B then learns of the assertion as usual.

Using transformations similar to these allows us to effectively embed labeled portions of a dataspace within other dataspaces in a virtual hierarchy. If more than one JavaScript client is connected at the same time, it appears alongside the other connected (and local) actors in the Racket-side dataspace. Naturally, when the client disconnects, be it cleanly or as the result of a crash or networking problem, this manifests to the Racket-side broker as a WebSocket disconnection; the broker terminates itself in response, thereby automatically retracting the assertions from the remote dataspace.

The specific transformation scheme sketched above wraps assertions received from *all* clients with the *same* constructor; in practice, we often wish to be able to securely distinguish between

1 Source code file `racket/syndicate/broker/server.rkt` in the Syndicate repository.

2 Note the strong similarity to the out metafunction (definition 4.14), used to translate between assertions in adjacent dataspaces within a Syndicate program.

assertions made by individual connected clients: the implemented broker therefore allows customization of the wrappers on a per-connection basis.

By labeling assertions received from connected clients, the broker enforces a kind of *spatial separation* between the remote party and local actors. This can be used for *sandboxing*, among other things. The "web chat" case study takes advantage of this sandboxing, carefully checking labeled, untrusted assertions from each connected client before relaying them to peers in the server-side dataspace. This is a core element in the enforcement of the application's security policy, closely related to the "firewalls" described in section 11.3.

Labeling of received assertions has a second benefit: it eliminates any ambiguity between assertions pertaining to the operation of the broker itself and its websocket connection (which, recall, is just another actor, communicating with the broker via assertions and messages) and assertions pertaining to the dataspace on the other end of the websocket link. In particular, events bearing assertions describing local websocket activity are clearly separated from events describing remote assertions. The per-connection constructor used to label received assertions acts as a form of quotation.

# D

## *Racket Dataflow Library*

This appendix presents a listing of the Racket dataflow library discussed in section 7.3.3.

The `dataflow.rkt` source file implements the dataflow mechanism proper.

```
#lang racket/base

(provide dataflow-graph?
         make-dataflow-graph
         dataflow-graph-edges-forward

         current-dataflow-subject-id

         dataflow-record-observation!
         dataflow-record-damage!
         dataflow-forget-subject!
         dataflow-repair-damage!)

(require racket/set)
(require "support/hash.rkt")

(struct dataflow-graph (edges-forward  ;; object-id -> (Setof subject-id)
                        edges-reverse  ;; subject-id -> (Setof object-id)
                        damaged-nodes) ;; Setof object-id
  #:mutable)

(define current-dataflow-subject-id (make-parameter #f))

(define (make-dataflow-graph)
  (dataflow-graph (hash)
                  (hash)
                  (set)))

(define (dataflow-record-observation! g object-id)
  (define subject-id (current-dataflow-subject-id))
  (when subject-id
    (define fwd (dataflow-graph-edges-forward g))
    (set-dataflow-graph-edges-forward! g (hashset-add fwd object-id subject-id))
    (define rev (dataflow-graph-edges-reverse g))
    (set-dataflow-graph-edges-reverse! g (hashset-add rev subject-id object-id))))
```

```
(define (dataflow-record-damage! g object-id)
  (set-dataflow-graph-damaged-nodes! g
    (set-add (dataflow-graph-damaged-nodes g) object-id)))

(define (dataflow-forget-subject! g subject-id)
  (define rev (dataflow-graph-edges-reverse g))
  (define subject-objects (hash-ref rev subject-id set))
  (set-dataflow-graph-edges-reverse! g (hash-remove rev subject-id))
  (for [(object-id (in-set subject-objects))]
    (define fwd (dataflow-graph-edges-forward g))
    (set-dataflow-graph-edges-forward! g (hashset-remove fwd object-id subject-id))))

(define (dataflow-repair-damage! g repair-node!)
  (define repaired-this-round (set))
  (let loop ()
    (define workset (dataflow-graph-damaged-nodes g))
    (set-dataflow-graph-damaged-nodes! g (set))

    (let ((already-damaged (set-intersect workset repaired-this-round)))
      (when (not (set-empty? already-damaged))
        (log-warning "Cyclic dependencies involving ids ~v\n" already-damaged)))

    (set! workset (set-subtract workset repaired-this-round))
    (set! repaired-this-round (set-union repaired-this-round workset))

    (when (not (set-empty? workset))
      (for [(object-id (in-set workset))]
        (define subjects (hash-ref (dataflow-graph-edges-forward g) object-id set))
        (for [(subject-id (in-set subjects))]
          (dataflow-forget-subject! g subject-id)
          (parameterize ((current-dataflow-subject-id subject-id))
            (repair-node! subject-id))))
      (loop))))
```

The `support/hash.rkt` source file implements support routines for maintaining hash-tables mapping keys to sets of values.

```
#lang racket/base

(provide hash-set/remove
         hashset-member?
         hashset-add
         hashset-remove)

(require racket/set)

(define (hash-set/remove ht key val [default-val #f] #:compare [compare equal?])
  (if (compare val default-val)
```

```
      (hash-remove ht key)
      (hash-set ht key val)))

(define (hashset-member? ht key val)
  (define s (hash-ref ht key #f))
  (and s (set-member? s val)))

(define (hashset-add ht key val #:set [set set])
  (hash-set ht key (set-add (hash-ref ht key set) val)))

(define (hashset-remove ht k v)
  (define old (hash-ref ht k #f))
  (if old
      (let ((new (set-remove old v)))
        (if (set-empty? new)
            (hash-remove ht k)
            (hash-set ht k new)))
      ht))
```